\documentclass[12pt,letterpaper]{article}

\usepackage[T1]{fontenc}
\usepackage[utf8]{inputenc}
\usepackage{epsfig}
\usepackage{amsmath,amssymb,amsthm,amsfonts}
\usepackage{a4}
\usepackage{tikz}
\usepackage{color}
\usepackage{natbib}
\usepackage{graphicx}       
\usepackage{caption}
\usepackage[labelformat=simple]{subcaption}

\usepackage{listings}
\usepackage{url}
\usepackage{hyperref}
\usepackage{listings}
\usepackage{enumerate}
\usepackage{wrapfig}
\usepackage{pgfplots}

\pgfmathdeclarefunction{densTheta}{1}{%
  \pgfmathparse{(1 / 0.9840912) * exp(\x)*exp(-((\x)^2/(2*(#1^2)))) / ((1+exp(\x))^2)}%
}

\pgfmathdeclarefunction{densP}{1}{%
  \pgfmathparse{(1 / 0.9840912) * exp(-(1/(2*(#1)^2))*(ln((1-\x)/(\x)))^2)}%
}

\begin{document}

\begin{center}
{\noindent{\LARGE \textbf{Prior specification for binary\\ Markov mesh models} \vspace{1cm} \\} 
{\large \textsc{Xin Luo}} and {\large \textsc{H\aa kon Tjelmeland}}\\{\it Department of
    Mathematical Sciences, Norwegian University of Science and
    Technology}}
\vspace{0.4cm}
\end{center}

\begin{abstract}
We propose prior distributions for all parts of the specification of a Markov mesh model. In the formulation we
define priors for the sequential neighborhood, for the parametric form of the conditional distributions and for 
the parameter values. By simulating from the resulting posterior distribution when conditioning on an observed 
scene, we thereby obtain an automatic model selection procedure for Markov mesh models. To sample from such a
posterior distribution, we construct a reversible jump Markov chain Monte Carlo algorithm (RJMCMC). We demonstrate the 
usefulness of our prior formulation and the limitations of our RJMCMC algorithm in two examples.
\end{abstract}

\vspace{0.5cm}
\vspace{-0.1cm}

\renewcommand{\baselinestretch}{1.3}   \small\normalsize

\section{Introduction}
Discrete Markov random fields (MRFs) and Markov mesh models (MMMs) defined on rectangular lattices
are popular model classes in spatial statistics, see for example
\citet{book28} and \citet{col6} for MRFs, and \citet{art133} and \citet{art119} for MMMs. 
Discrete MRFs are frequently used to 
model available prior information about an unobserved scene $x$ of a discrete variable. This 
prior is then combined with a likelihood function describing the relation between $x$
and some observed data $y$ into a posterior distribution, and this posterior is the basis for
making inference about $x$. When specifying the MRF prior, the most frequent approach is to 
fix the neighborhood and parametric model structures and also to specify the values
of the model parameters a priori. However, some authors have also explored a more fully Bayesian approach 
\citep{art109,art144,art117,art156}. In particular, \citet{art157} formulate a prior
for all parts of the MRF prior and demonstrate how MCMC
sampling from the corresponding posterior distribution when conditioning on an observed 
scene produces MRFs that give realizations with a similar spatial structure as present
in the scene used to define the posterior. 

The class of Markov mesh models, and the partially ordered Markov model (POMM) generalization 
\citep{art119} of this class, is much less used in the literature. We think the main reason for this
is that it is much harder to manually choose an MMM than an MRF that reflects given prior information.
It is neither an easy task to specify an MRF that is 
consistent with given prior information, but except for boundary effects it is for an MRF
easy to ensure that the field is stationary. This is an important practical 
argument when completely specifying the prior a priori, 
but it is not so important when a fully Bayesian model is adopted as in \citet{art157}. 
It should also be noted that MRFs contain a computationally intractable normalizing
constant which severely limits the practicability of MRFs in a fully Bayesian context,
see for example the discussion in \citet{art117}. In contrast, the normalizing constant
for an MMM is explicitly given in an easy to compute form. Also for this reason an MMM
is much better suited as a prior than an MRF when adopting the fully Bayesian approach.

Our goal in the present article is to formulate a fully Bayesian MMM. In particular, we
would like the hyper-prior to include distributions for the neighborhood structure, for the
interaction structure of the conditional distributions defining the MMM, and for the parameter values.
We should thereby obtain a flexible prior that is able to adapt to a wide variety 
of scenes. To specify the MMM hyper-prior, we adapt the general strategy used in 
\citet{art157} for the MRF to our MMM situation. 
Given such a Bayesian model, we also want to formulate an MCMC algorithm 
to simulate from the resulting posterior distribution conditioned on an observed 
scene. It should thereby be possible to learn both the form of the parametric model
and the values of the model parameters from an observed scene. For simplicity we here limit our attention 
to binary MMMs, but our approach can quite easily be generalized to a 
situation where each node has more than two possible values. 

The remainder of this article is organized as follows. In Section \ref{sec:preliminaries} we
introduce most of the notations we use for defining our Bayesian Markov mesh model, and in particular
discuss pseudo-Boolean functions. In Section \ref{sec:mmm} we use this to formulate 
the Markov mesh model class. In Section \ref{sec:prior} we construct our prior
distribution, and in Section \ref{sec:simulation algorithm} we formulate proposal 
distributions that we use in a reversible jump Markov chain Monte Carlo algorithm to 
simulate from the corresponding posterior when conditioning on an observed scene.
In Section \ref{sec:examples} we present two simulation examples and lastly we give 
some closing remarks in Section \ref{sec:closing remarks}.

\section{\label{sec:preliminaries}Preliminaries}
In this section we first introduce the notation we use to represent a rectangular lattice, the 
variables associated to this lattice and some quantities we use to formulate our 
Markov mesh model defined on this lattice. Thereafter, we define the class of pseudo-Boolean 
functions and explain how a pseudo-Boolean function can be used to represent a conditional 
distribution for binary variables.

\subsection{\label{sec:notation}Notation}
Consider a rectangular $m\times n$ lattice. Let $v=(i,j)$ denote a node in this lattice, where 
$i$ and $j$ specify the vertical and horizontal positions of the node in the lattice,
respectively. We let $i=1$ be at the top of the lattice and $i=m$ at the bottom, and $j=1$ 
and $j=n$ are at the left and right ends of the lattice, respectively. We use lowercase Greek 
letters to denote sets of nodes, and in particular we let $\chi=\{(i,j):i=1,\ldots,m,j=1,\ldots,n\}$
be the set of all nodes in the lattice. Occasionally we also consider 
an infinite lattice $\mathbb{Z}^2$, where $\mathbb{Z}=\{ 0,\pm 1,\pm 2,\ldots\}$ is the set of 
all integers, and we use $v=(i,j)\in \mathbb{Z}^2$ also to denote a node in such an infinite
lattice. We use $\lambda,\lambda^\star\subseteq \mathbb{Z}^2$ to denote
arbitrary sets of nodes. To translate a node $(i,j)\in \mathbb{Z}^2$ by an amount $(k,l)\in\chi$,
we adopt the notation
\begin{equation}
(i,j) \oplus (k,l) = (i+k,j+l).
\end{equation}
One should note that even if $(i,j)\in\chi$, $(i,j)\oplus (k,l)$ may fall outside the finite 
$m\times n$ lattice. 
To translate all nodes in a set $\lambda\subseteq \mathbb{Z}^2$ by the same amount $(k,l)\in\chi$,
we write
\begin{equation}
\lambda \oplus (k,l) = \{ (i,j) \oplus (k,l): (i,j) \in \lambda\}.
\end{equation}
To denote sets of subsets of nodes, we use uppercase Greek letters, and in particular, we let 
$\Omega(\chi)=\{\lambda:\lambda\subseteq\chi\}$ denote the set of all subsets of $\chi$, often 
called the power set of $\chi$. One should note that $\Omega(\chi)$ in particular includes
the empty set and $\chi$ itself. We use $\Lambda,\Lambda^\star\subseteq \Omega(\chi)$ to denote 
arbitrary sets of subsets of nodes.

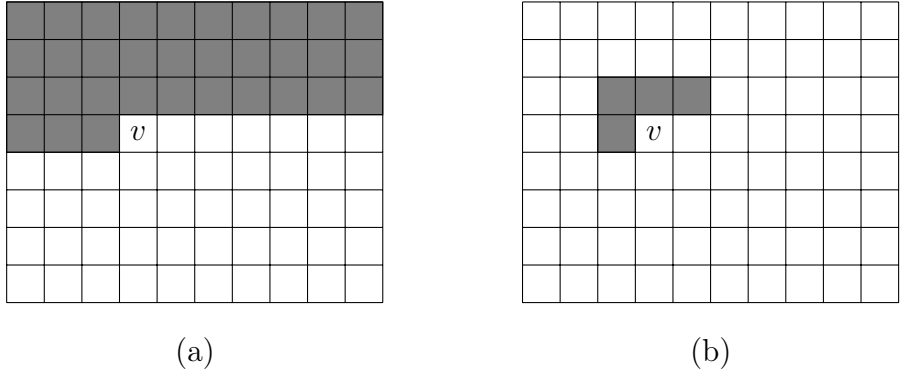
\begin{figure}
  \begin{center}
    \begin{tabular}{ccc}
      \begin{tikzpicture}
        \path[draw,fill=gray] (0,8/2) -- (10/2,8/2) -- (10/2,5/2) -- (3/2,5/2) -- (3/2,4/2) -- (0,4/2) -- cycle;
        \node[] at (3.5/2, 4.5/2) (n1) {$v$};
        \draw[step=0.5cm,color=black, scale=1, thin] (0,0) grid (5,4);
      \end{tikzpicture}
      & 
      \hspace*{1.0cm}
      & 
      \begin{tikzpicture}
        \node[] at (3.5/2, 4.5/2) (n1) {$v$};
        \path[draw,fill=gray] (3/2,4/2) -- (2/2,4/2) -- (2/2,6/2) -- (5/2,6/2) -- (5/2,5/2) -- (3/2,5/2) -- cycle;
        \draw[step=0.5cm,color=black, scale=1, thin] (0,0) grid (5,4);
      \end{tikzpicture}
      \\[0.1cm]
      (a) & & (b)
    \end{tabular}
  \end{center}
  \caption{\label{fig:seqn}Illustration of the predecessor set $\rho_v$ and a possible sequential neighborhood
    $\nu_v$ for node $v=(4,4)$ in a $8\times 10$ lattice. (a) The nodes in $\rho_v$ are shown in gray. (b)
    The nodes in a possible sequential neighborhood $\nu_v=\{(4,3),(3,3),(3,4),(3,5)\}$ are shown in gray.}
\end{figure}

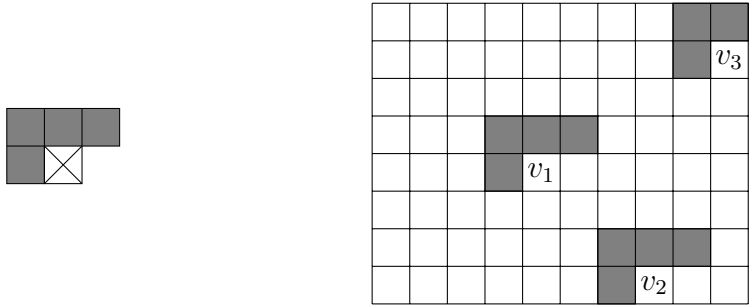
\begin{figure}
  \begin{center}
    \begin{tabular}{ccc}
      \raisebox{1.6cm}{\begin{tikzpicture}
        \path[draw, fill=gray] (3,0) -- (3,1) -- (4.5,1) -- (4.5,0.5) -- (3.5,0.5) -- (3.5,0) -- cycle;
        \draw (3.5,0) -- (3.5,1);
        \draw (4,0.5) -- (4,1);
        \draw (3.0,0.5) -- (3.5,0.5);
        \draw (3.5,0) -- (4.0,0) -- (4.0,0.5);
        \draw (3.5,0) -- (4.0,0.5);
        \draw (4.0,0) -- (3.5,0.5);
      \end{tikzpicture}}
      &
      \hspace*{2.5cm}
      &
      \begin{tikzpicture}
        
        \node[] at (2.25, 1.75) (n1) {$v_1$};
        \node[] at (3.75, 0.25) (n2) {$v_2$};
        \node[] at (4.75, 3.25) (n2) {$v_3$};
        \path[draw, fill=gray] (3,0) -- (3,1) -- (4.5,1) -- (4.5,0.5) -- (3.5,0.5) -- (3.5,0) -- cycle;
        \path[draw, fill=gray] (1.5,1.5) -- (1.5,2.5) -- (3,2.5) -- (3,2) -- (2,2) -- (2,1.5) -- cycle;
        \path[draw, fill=gray] (4.5,3) -- (4,3) -- (4,4) -- (5,4) -- (5,3.5) -- (4.5,3.5) -- cycle;
        \draw[step=0.5cm,color=black, scale=1, thin] (0,0) grid (5,4);
      \end{tikzpicture}
    \end{tabular}
  \end{center}	
  \caption{\label{fig:neighbourhoods}Illustration of the construction of sequential neighborhoods
    from a template $\tau$. The left figure shows a possible template $\tau=\{(0,-1),(-1,-1),(-1,0),(-1,1)\}$, 
    where the node $(0,0)$ is represented with $\boxtimes$ and the elements of $\tau$ are shown in gray.
    The right figure shows the resulting sequential neighborhoods (again in gray) for nodes $v_1=(5,5)$, $v_2=(8,8)$ and 
    $v_3=(2,10)$ in a $8\times 10$ lattice.}
\end{figure}

To define a Markov mesh model one must, for each node $v=(i,j)$, define a so-called predecessor set 
and a sequential neighborhood. After numbering the nodes in the lattice from one to $mn$ in the lexicographical 
order, we let the predecessor set of a node $(i,j)$ consist of all nodes with a lower number than 
the number of $(i,j)$.
We let $\rho_v = \rho_{(i,j)}\subset \chi$ denote the predecessor set of a node $v=(i,j)\in \chi$, i.e.
\begin{equation}
\rho_{(i,j)} = \{(k,l)\in \chi: nk+l < ni+j\},
\end{equation}
see the illustration in Figure \ref{fig:seqn}(a).
We let $\nu_v=\nu_{(i,j)}\subseteq \rho_{(i,j)}$ denote the sequential neighborhood for node $v=(i,j)\in\chi$
as illustrated in Figure \ref{fig:seqn}(b).
In Section \ref{sec:mmm} we consider a Markov mesh model where all the sequential neighborhoods are defined
by a translation of a template sequential neighborhood $\tau$. The $\tau$ can be thought of as the 
sequential neighborhood of node $(0,0)$ in the infinite lattice. More precisely,
$\tau$ is required to include a finite number of elements and 
\begin{equation}\label{eq:tau}
\tau \subset \psi = \{ (i,j): i\in\mathbb{Z},j\in\mathbb{Z}^-\} \cup 
\{ (0,j): j\in \mathbb{Z}^-\},
\end{equation}
where $\mathbb{Z}^- = \{ -1,-2,\ldots\}$ is the set of all negative integers. 
The sequential neighborhood of a node $v\in\chi$ is then 
defined as 
\begin{equation}\label{eq:nu}
\nu_v = \left( \tau \oplus v\right) \cap \chi.
\end{equation}
As illustrated in Figure \ref{fig:neighbourhoods},
sequential neighborhoods for all nodes sufficiently 
far away from the lattice borders then have the same form, whereas nodes close to the borders have
fewer sequential neighbors. One can note that with this construction one always has
$\nu_{(1,1)}=\emptyset$.

To each node $v=(i,j)\in\chi$, we also associate a corresponding binary variable which we denote by 
$x_v=x_{(i,j)}\in \{0,1\}$. The collection of all these binary variables we denote by $x=(x_v; x\in\chi)$ 
and we let $x_\lambda=(x_v;v\in\lambda)$ represent the collection of variables associated to the nodes
in a set $\lambda\subseteq \chi$. In particular, $x_{\nu_v}$ is the collection of variables associated to 
the sequential neighborhood of node $v$. If $x_v=1$ we say node $v$ is on, and if $x_v=0$ we say the 
node is off. We let $\xi(x)\subseteq \chi$ denote the set of all nodes that are on, i.e. 
\begin{equation}
\xi(x) = \{ v\in\chi: x_v=1\}.
\end{equation}
In particular, the set of nodes in the sequential neighborhood of node $v$ that is on is then,
using (\ref{eq:nu}) and that $\xi(x)\subseteq \chi$,
\begin{equation}\label{eq:neighboursOn}
\xi(x)\cap \nu_v = \xi(x)\cap (\tau\oplus v).
\end{equation}
In the next section, we define the class of pseudo-Boolean functions which we in Section \ref{sec:mmm}
use to define the class of Markov mesh models.

\subsection{Pseudo-Boolean functions\label{sec:pseudoboolean functions}}
When defining pseudo-Boolean functions, we reuse some of the symbols introduced when discussing concepts
related to the rectangular $m\times n$ lattice above. In particular, we define a pseudo-Boolean function 
with respect to some finite set, denoted by $\tau$. In the definition, this $\tau$ has no relation to the 
template sequential neighborhood $\tau$ introduced above. However, when applying a pseudo-Boolean function
to represent the conditional distribution of $x_v$ for a node $v\in \chi$ given the values of the nodes
in the sequential neighborhood $\nu_v$, the set $\tau$ used to define a pseudo-Boolean function is
equal to the template sequential neighborhood $\tau$. In particular, the elements of $\tau$ is then the nodes
in the lattice $\chi$, and therefore we use $\lambda$ and $\lambda^\star$ to represent subsets of $\tau$ also 
when discussing pseudo-Boolean functions in general.

A pseudo-Boolean function $\theta(\cdot )$ defined on a finite set $\tau$ is a function that associates
a real value to each subset of $\tau$, i.e. 
\begin{equation}\label{eq:theta}
\theta: \Omega(\tau) \rightarrow \mathbb{R},
\end{equation}
where $\Omega(\tau)$ is the power set of $\tau$. Thereby, for any $\lambda\subseteq \tau$ the value of 
the pseudo-Boolean function is $\theta(\lambda)$. Equivalently, one may think of a pseudo-Boolean 
function as a function that associates a real value to each vector $z\in \{0,1\}^{|\tau|}$, where
$|\tau|$ is the number of elements in the set $\tau$. To see the correspondence, one should set an 
element in $z$ equal to one if and only if the corresponding element in $\tau$ is in the set $\lambda$.
This last formulation of pseudo-Boolean functions is the more popular one, see for example \citet{art137} and 
\citet{art138}, but in the present article we adopt the formulation in (\ref{eq:theta}) as this gives 
simpler expressions when formulating our Markov mesh model in Section \ref{sec:mmm} and the 
corresponding prior distribution in Section \ref{sec:prior}.

\citet{book35} show that any pseudo-Boolean function 
can be uniquely represented by a collection of 
interaction parameters $(\beta(\lambda),\lambda\in\Omega(\tau))$ by the relation
\begin{equation}\label{eq:m1}
\theta(\lambda) = \beta(\lambda) + \sum_{\lambda^\star\subset \lambda} \beta(\lambda^\star)\mbox{~~for $\lambda\subseteq\tau$.}
\end{equation}
The corresponding inverse relation is given by
\begin{equation}\label{eq:m2}
\beta(\lambda) = \theta(\lambda) + \sum_{\lambda^\star\subset \lambda} (-1)^{|\lambda\setminus\lambda^\star|} \theta(\lambda^\star)
\mbox{~for $\lambda\subseteq \tau$.}
\end{equation}
The one-to-one relation in (\ref{eq:m1}) and (\ref{eq:m2}) is known as Moebious inversion, see for example
\citet{book39}.

If one or more of the interaction parameters $\beta(\lambda)$ are restricted to be zero, a reduced representation 
of the pseudo-Boolean function can be defined. For some $\Lambda\subseteq\Omega(\tau)$ assume now that one restricts
$\beta(\lambda)=0$ for all $\lambda\not\in\Lambda$. One can then represent the pseudo-Boolean function $\theta(\cdot )$
by the interaction parameters $\{\beta(\lambda),\lambda\in\Lambda\}$, and the relation in (\ref{eq:m1})
becomes
\begin{equation}\label{eq:pseudoboolean_reduced}
\theta(\lambda) = \sum_{\lambda^\star\in \Lambda\cap \Omega(\lambda)}\beta(\lambda^\star) \mbox{~for $\lambda\in \Omega(\tau)$,}
\end{equation}
where $\Omega(\lambda)$ is the power set of $\lambda$. We then say that $\theta(\cdot)$ is represented on 
$\Lambda$. Moreover, we say that the set $\Lambda$ is dense if for all $\lambda\in\Lambda$, all subsets of 
$\lambda$ is also included in $\Lambda$, and that the template sequential neighborhood $\tau$
is minimal for $\Lambda$ if all nodes $v\in\tau$ are included in at least one of the elements of $\Lambda$.
One should note that if $\Lambda$ is dense and $\tau$ is minimal for $\Lambda$ then there is a one-to-one 
relation between the elements in $\tau$ and the sets $\lambda\in\Lambda$ which contains only one node, 
\begin{equation}\label{eq:Lambda1}
\{ \{v\}: v\in \tau\} = \{\lambda\in \Lambda: |\lambda| = 1\}.
\end{equation}
Throughout this paper, we restrict attention to pseudo-Boolean functions that are represented on a
$\Lambda$ that is dense and the template sequential neighborhood $\tau$ that is minimal
for this $\Lambda$. A $\lambda\in\Omega(\tau)$ we term an interaction, we say the interaction is active if $\lambda\in\Lambda$
and otherwise we say it is inactive. The $\Lambda$ is thereby the set of active interactions.

\begin{figure}
\begin{center}
\begin{tikzpicture}[scale=2.5]
  \def\radius{38.9};
  \coordinate (Pempty) at (0,0);
  \coordinate (P1) at (-1.5,1);
  \coordinate (P2) at (-0.5,1);
  \coordinate (P3) at (0.5,1);
  \coordinate (P4) at (1.5,1);
  \coordinate (P12) at (-1,2);
  \coordinate (P14) at (0,2);

  \coordinate (c1) at (-0.05,-0.05);
  \coordinate (c2) at (0.05,-0.05);
  \coordinate (c3) at (0.05,0.05);
  \coordinate (c4) at (-0.05,0.05);

  \coordinate (r1) at (-0.1,0);
  \coordinate (r2) at (0,0.1);
  \coordinate (r3) at (-0.1,0.1);
  \coordinate (r4) at (0.1,0.1);

  \node[draw,circle,inner sep=0pt,minimum size=\radius,name=Nempty] at (Pempty) {};
  \draw[thick] (Pempty) +(c1) -- +(c2) -- +(c3) -- +(c4) -- +(c1) +(c1) -- +(c3) +(c2) -- +(c4);

  \node[draw,circle,inner sep=0pt,minimum size=\radius,name=N1] at (P1) {};
  \draw[thick] (P1) +(c1) -- +(c2) -- +(c3) -- +(c4) -- +(c1) +(c1) -- +(c3) +(c2) -- +(c4);
  \draw[thick] (P1) ++(r1) +(c1) -- +(c2) -- +(c3) -- +(c4) -- cycle;
  \draw[thick,->] (Nempty) -- (N1);
   
  \node[draw,circle,inner sep=0pt,minimum size=\radius,name=N2] at (P2) {};
  \draw[thick] (P2) +(c1) -- +(c2) -- +(c3) -- +(c4) -- +(c1) +(c1) -- +(c3) +(c2) -- +(c4);
  \draw[thick] (P2) ++(r2) +(c1) -- +(c2) -- +(c3) -- +(c4) -- cycle;
  \draw[thick,->] (Nempty) -- (N2);

  \node[draw,circle,inner sep=0pt,minimum size=\radius,name=N3] at (P3) {};
  \draw[thick] (P3) +(c1) -- +(c2) -- +(c3) -- +(c4) -- +(c1) +(c1) -- +(c3) +(c2) -- +(c4);
  \draw[thick] (P3) ++(r3) +(c1) -- +(c2) -- +(c3) -- +(c4) -- cycle;
  \draw[thick,->] (Nempty) -- (N3);
 
  \node[draw,circle,inner sep=0pt,minimum size=\radius,name=N4] at (P4) {};
  \draw[thick] (P4) +(c1) -- +(c2) -- +(c3) -- +(c4) -- +(c1) +(c1) -- +(c3) +(c2) -- +(c4);
  \draw[thick] (P4) ++(r4) +(c1) -- +(c2) -- +(c3) -- +(c4) -- cycle;
  \draw[thick,->] (Nempty) -- (N4);
  
  \node[draw,circle,inner sep=0pt,minimum size=\radius,name=N12] at (P12) {};
  \draw[thick] (P12) +(c1) -- +(c2) -- +(c3) -- +(c4) -- +(c1) +(c1) -- +(c3) +(c2) -- +(c4);
  \draw[thick] (P12) ++(r1) +(c1) -- +(c2) -- +(c3) -- +(c4) -- cycle;
  \draw[thick] (P12) ++(r2) +(c1) -- +(c2) -- +(c3) -- +(c4) -- cycle;
  \draw[thick,->] (N1) -- (N12);
  \draw[thick,->] (N2) -- (N12);
  
  \node[draw,circle,inner sep=0pt,minimum size=\radius,name=N14] at (P14) {};
  \draw[thick] (P14) +(c1) -- +(c2) -- +(c3) -- +(c4) -- +(c1) +(c1) -- +(c3) +(c2) -- +(c4);
  \draw[thick] (P14) ++(r1) +(c1) -- +(c2) -- +(c3) -- +(c4) -- cycle;
  \draw[thick] (P14) ++(r4) +(c1) -- +(c2) -- +(c3) -- +(c4) -- cycle;
  \draw[thick,->] (N1) -- (N14);
  \draw[thick,->] (N4) -- (N14);


\end{tikzpicture}
\end{center}
\caption{\label{fig:dag1}DAG visualization of the set 
$\Lambda=\{\emptyset,\{(0,-1)\},\{(-1,0)\},\{(-1,-1)\},\{(-1,1)\}$, $\{(0,-1),(-1,0)\},\{(0,-1),(-1,1)\}\}$ based on 
$\tau=\{(0,-1),(-1,-1),(-1,0),(-1,1)\}$. Thinking of the elements of $\tau$ as a finite set of nodes in a lattice,
$\boxtimes$ is used in the vertices of the DAG to represent the node $(0,0)$, whereas each node $(i,j)\in\lambda$ for
each $\lambda\in\Lambda$ is represented by a $\square$ placed in position $(i,j)$ relative to $\boxtimes$.}
\end{figure}
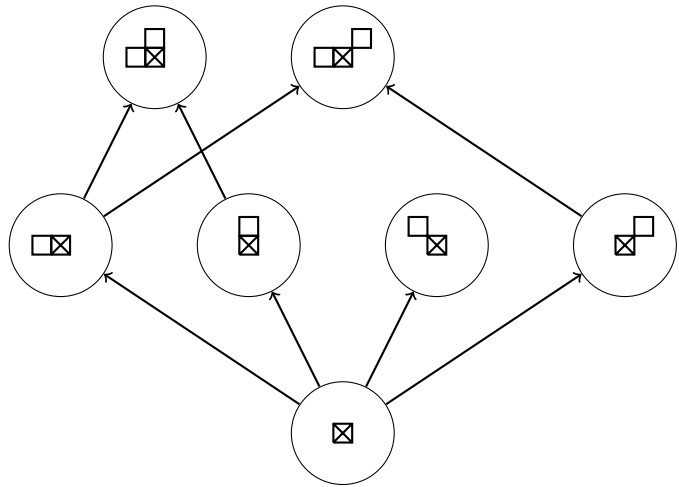

As also discussed in \citet{art158}, the set of active interactions $\Lambda$ can be visualized by a 
directed acyclic graph (DAG), where we have one vertex for each active interaction $\lambda\in\Lambda$ and a 
vertex $\lambda\in\Lambda$ is a 
child of another vertex $\lambda^\star\in\Lambda$ if and only if $\lambda=\lambda^\star\cup \{ v\}$ for some 
$v\in\tau\setminus\lambda^\star$. Figure \ref{fig:dag1}
shows such a DAG for $\Lambda=\{\emptyset$, $\{(0,-1)\}$, $\{(-1,0)\}$, $\{(-1,-1)\}$, $\{(-1,1)\}$, $\{(0,-1),(-1,0)\}$,
$\{(0,-1),(-1,1)\}\}$, which is based on $\tau=\{(0,-1),(-1,-1),(-1,0),(-1,1)\}$. This $\tau$
can be used to define the sequential neighborhoods for nodes in a rectangular lattice as 
discussed in Section \ref{sec:notation}. In the vertices of the DAG shown in the figure, node $(0,0)$ is represented by the symbol 
$\boxtimes$, whereas each of the nodes in $\lambda\in\Lambda$ is represented by the symbol $\square$. Thinking of $\tau$
as a finite set of nodes in a lattice, the position of the $\square$ representing node $(i,j)\in\lambda$ is placed 
at position $(i,j)$ relative to $\boxtimes$.

As also discussed in \citet{art157}, one should note that a pseudo-Boolean function $\theta(\cdot)$ that 
is represented on a dense set $\Lambda\subseteq\Omega(\tau)$ can be uniquely specified by 
the values of $\{\theta(\lambda):\lambda\in\Lambda\}$. The remaining values of the pseudo-Boolean function,
$\theta(\lambda),\lambda\in \Omega(\tau)\setminus\Lambda$, are then given by \eqref{eq:m1} and \eqref{eq:m2}
and the restriction $\beta(\lambda)=0$ for $\lambda\not\in\Lambda$. Moreover, as the relations in 
\eqref{eq:m1} and \eqref{eq:m2} are linear, each $\theta(\lambda), \lambda\in \Omega(\tau)\setminus\Lambda$ is
a linear function of  $\{\theta(\lambda):\lambda\in\Lambda\}$.

\section{Markov mesh model\label{sec:mmm}}
In this section we formulate a homogeneous binary Markov mesh model \citep{art133} for a
rectangular $m\times n$ lattice. We adopt the notation introduced in Section \ref{sec:preliminaries}, 
so in particular $\chi$ denotes the set of 
all nodes in the $m\times n$ lattice and $x = (x_v,v\in \chi)$ is the collection of 
the binary variables associated to $\chi$. In a Markov mesh model the 
distribution of $x$ is expressed as
\begin{equation}
f(x) = \prod_{v\in\chi} f(x_v|x_{\rho_v}),
\end{equation}
where $f(x_v|x_{\rho_v})$ is the conditional distribution for $x_v$ given the values of the 
variables in the predecessor nodes. Moreover, one assumes the Markov property 
\begin{equation}\label{eq:markov property}
f(x_v|x_{\rho_v}) = f(x_v|x_{\nu_v}),
\end{equation}
i.e. the conditional distribution of $x_v$ given the values in all predecessors of $v$ 
only depends on the values in the nodes in the sequential neighborhood of $v$. As discussed in 
Section \ref{sec:notation}, we assume the sequential neighborhoods $\nu_v,v\in\chi$ to be defined 
as translations of a template sequential neighborhood $\tau$ 
as specified in \eqref{eq:tau} and \eqref{eq:nu}.
Using the result in \eqref{eq:neighboursOn}, the conditional distribution 
$f(x_v|x_{\nu_v})$ can then be uniquely represented by a pseudo-Boolean function
$\theta_v(\lambda),\lambda\subseteq\tau$ by the relation 
\begin{equation}\label{eq:cond_dist}
f(x_v|x_{\rho_v})=\frac{\exp\left\{ x_v\cdot  \theta_v\left( \xi(x) \cap (\tau\oplus v)\right)\right\}}
{1 + \exp\left\{\theta_v\left( \xi(x) \cap (\tau\oplus v)\right)\right\}}.
\end{equation}
In general, one may have one pseudo-Boolean function $\theta_v(\lambda)$ for 
each $v\in\chi$, but in the following we limit the attention to homogeneous models, so 
we require all $\theta_v(\cdot),v\in\chi$ to be equal. We let $\theta(\cdot)$ denote this
common pseudo-Boolean function, i.e. $\theta_v(\lambda)=\theta(\lambda)$ for all 
$\lambda\subseteq\tau$ and $v\in\chi$ and, without loss of generality, we assume 
$\theta(\cdot )$ to have a dense representation on a set $\Lambda\subseteq \Omega(\tau)$
and $\tau$ to be minimal for $\Lambda$.
Thus, the distribution of our homogeneous binary Markov mesh model is
\begin{equation}\label{eq:mmm}
f(x) = \prod_{v\in \chi} \frac{\exp\left\{ x_v\cdot  \theta\left( \xi(x) \cap (\tau\oplus v)\right)\right\}}
{1 + \exp\left\{\theta\left( \xi(x) \cap (\tau\oplus v)\right)\right\}}.
\end{equation}
Assuming, as we do, the Markov mesh model to be homogeneous is convenient in
that we do not need to specify a separate pseudo-Boolean function for each node $v\in\chi$,
and it is also statistically favorable as it limits the number of parameters in the model.
However, one should note that this choice implies that for a node $v\in\chi$ close to the 
boundary of the lattice so that the set $(\tau\oplus v)\setminus\chi$ is non-empty, the 
conditional distribution $f(x_v|x_{\nu_v})$ is as if the nodes (for the infinite lattice) in 
the translation of $\tau$ that fall outside the lattice $\chi$ are all zero. Thus, 
even if the model is homogeneous it is not stationary, and in particular one should 
expect strong edge effects since we in some sense are conditioning on everything outside 
the lattice $\chi$ to be zero. When estimating or fitting the model to an observed scene, it is crucial to take this edge effect into account.

Having defined our class of homogeneous Markov mesh models as above,
a model is specified by the template sequential neighborhood $\tau$,
the set of active interactions $\Lambda\subseteq\Omega(\tau)$ on which 
the pseudo-Boolean function $\theta(\cdot)$
is represented, and the parameter values $\{\theta(\lambda):\lambda\in\Lambda\}$. Thus, to 
adopt a fully Bayesian approach, we need to formulate prior distributions for 
$\tau$, $\Lambda$ and $\{\theta(\lambda):\lambda\in\Lambda\}$, and this is 
the focus of the next section.

\section{Prior distribution\label{sec:prior}}
When constructing our prior distribution for the template sequential neighborhood $\tau$, the set of active 
interactions $\Lambda$ 
and the parameter values $\left\lbrace\theta(\lambda):\lambda\in\Lambda\right\rbrace$, we have two properties in mind. Firstly,
the prior should be vague so that the Markov mesh model manages to adapt to a large variety of scenes. To obtain this,
the number of elements in $\tau$ should be allowed to be reasonably large and higher-order interactions should be allowed in 
the model. Secondly, to avoid overfitting, the prior should favor parsimonious Markov mesh models, and in particular this
implies that the highest prior probabilities should be assigned to models with just a few higher-order interactions.

We define the prior as a product of three factors
\begin{equation}\label{eq:prior}
f(\tau,\Lambda,\{\theta(\lambda): \lambda\in\Lambda\}) = f(\tau) f(\Lambda|\tau) f(\{\theta(\lambda):\lambda \in \Lambda\}|\tau,\Lambda),
\end{equation}
where $f(\tau)$ is a prior for the template sequential neighborhood $\tau$, $f(\Lambda|\tau)$ is a prior for the set of active 
interactions 
$\Lambda$ when $\tau$ is given, and $f(\{\theta(\lambda):\lambda\in\Lambda\}|\Lambda)$ is a prior for the parameter values given 
$\tau$ and $\Lambda$. In the following we discuss each of these factors in turn.

\subsection{Prior for the template sequential neighborhood $\tau$\label{sec:prior1}}
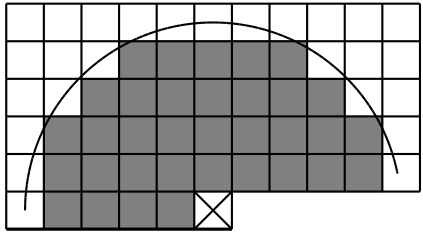
\begin{figure}
\begin{center}
\begin{tikzpicture}[scale=0.5]

\draw[black,fill=gray] (-0.5,-0.5) -- (-4.5,-0.5) -- (-4.5,2.5) -- (-3.5,2.5) -- (-3.5,3.5) -- (-2.5,3.5) -- (-2.5,4.5) -- (2.5,4.5) -- (2.5,3.5) -- (3.5,3.5) -- (3.5,2.5) -- (4.5,2.5) -- (4.5,0.5) -- (-0.5,0.5) -- cycle;

\foreach \x in {-5.5,-4.5,-3.5,-2.5,-1.5,-0.5,0.5} {
  \draw[black,thick] (\x,-0.5) -- (\x,5.5);
}
\foreach \x in {1.5,2.5,3.5,4.5,5.5} {
  \draw[black,thick] (\x,0.5) -- (\x,5.5);
}

\draw[black,very thick] (-5.5,-0.5) -- (0.5,-0.5);
\foreach \y in {0.5,1.5,2.5,3.5,4.5,5.5} {
  \draw[black,thick] (-5.5,\y) -- (5.5,\y);
}

\draw[black,thick] ({5*cos(11.31)},{5*sin(11.31)}) -- ({5*cos(12)},{5*sin(12)});
\foreach \angle in {12,14,16,18,20,22,24,26,28,30,32,34,36,38,40,42,44,46,48,50,52,54,56,58,60,62,64,66,68,70,72,74,76,78,80,82,84,86,88,90,92,94,96,98,100,102,104,106,108,110,112,114,116,118,120,122,124,126,128,130,132,134,136,138,140,142,144,146,148,150,152,154,156,158,160,162,164,166,168,170,172,174,176,178} {
  \draw[black,thick] ({5*cos(\angle)},{5*sin(\angle)}) -- ({5*cos(\angle+2)},{5*sin(\angle+2)});
}

\draw[black,thick] (-0.5,-0.5) -- (0.5,0.5);
\draw[black,thick] (-0.5,0.5) -- (0.5,-0.5);

\end{tikzpicture}
\end{center}
\caption{Illustration of the $\tau_0$ used in the simulation examples in Section \ref{sec:examples}. $\boxtimes$ is node $(0,0)$, and 
gray nodes are elements of $\tau_0$. The black curve is a part of the circle centered at $(0,0)$ and with radius $r=5$.}
\label{fig:potentialneighbors}
\end{figure}

We restrict the template sequential neighborhood to be a subset of a given finite set $\tau_0\subset \psi$, where $\psi$
is defined in (\ref{eq:tau}). The $\tau_0$ can be though 
of as a set of possible sequential neighbors for node $(0,0)$. To get a flexible prior it is important that the number of 
elements in $\tau_0$ is not too small, and it is natural to let $\tau_0$ include nodes close to $(0,0)$. For example, one may let 
$\psi$ include all nodes that are inside the circle centered at $(0,0)$ with some specified radius $r$. In the examples
discussed in Section \ref{sec:examples} we use this with $r=5$, see the illustration in Figure \ref{fig:potentialneighbors}.

Given the set $\tau_0$ we specify the prior for $\tau \subseteq \tau_0$ by first choosing a prior distribution for the number of elements in $\tau$, and 
thereafter a prior for $\tau$ given the number of elements in $\tau$. Letting $n_\tau=|\tau|$ denote the number of elements in $\tau$
we thereby have
\begin{equation}
f(\tau) = f(n_\tau) f(\tau|n_\tau).
\end{equation}
For simplicity we choose both $f(n_\tau)$ and $f(\tau|n_\tau)$ to be uniform distributions. The possible values for $n_\tau$ are all integers from 
$0$ to $|\tau_0|$, so we get
\begin{equation}\label{eq:priorTau1}
f(n_\tau) = \frac{1}{n_\tau + 1} \mbox{~~for $n_\tau=0,1,\ldots,|\tau_0|$}.
\end{equation}
Moreover, having chosen $\tau$ to be uniform given $n_\tau=|\tau|$, we get
\begin{equation}\label{eq:priorTau2}
f(\tau|n_\tau) = \dfrac{1}{\binom{|\tau_0|}{n_\tau}},
\end{equation}
where the binomial coefficient in the numerator is the number of possible sets $\tau$'s with $n_\tau$ elements.

One should note that our choice of the two uniforms above is very different from adopting a uniform prior for $\tau$ directly. 
A uniform prior on $\tau$ would have resulted in very high a priori probabilities for $n_\tau$ being close to $|\tau_0|/2$ and very 
small a priori probabilities for values of $n_\tau$ close to zero, which is clearly not desirable.

One can easily construct other reasonable priors for $\tau$ than the one defined above. For example, one could want to build into the 
prior $f(\tau|n_\tau)$ that nodes close to $(0,0)$ are more likely to be in $\tau$ than nodes further away. Recalling
that we want to simulate from a corresponding posterior distribution by a reversible jump Markov chain Monte Carlo
algorithm (RJMCMC) \citep{art22}, the 
challenge is to formulate a prior with this property so that we are able to compute the (normalized)
probability $f(\tau|n_\tau)$, as this is needed to evaluate the Metropolis--Hastings acceptance probability. For the 
data sets discussed in Section \ref{sec:examples}, we have also tried a prior $f(\tau|n_\tau)$ in which we split the nodes in 
$\tau_0$ into two or three zones dependent on their distances from $(0,0)$ and have a different prior probability for 
a node to be in $\tau$ dependent on which zone it is in. As long as the number of zones is reasonably small, it is then 
possible to compute the normalizing constant of $f(\tau|n_\tau)$ efficiently. However, in our simulation examples
this gave essentially the same posterior results as the very simple double uniform prior specified above.

\subsection{Prior for the set of active interactions $\Lambda$\label{sec:prior2}}
To specify a prior for the set of active interactions $\Lambda$, we first split $\Lambda$ into several subsets dependent on how many nodes an element
$\lambda\in\Lambda$ contains. More precisely, for $k=0,1,\ldots,|\tau|$ we define 
\begin{equation}
\Omega_k(\tau) = \{\lambda\in \Omega(\tau): |\lambda|=k\}
\mbox{~~~and~~~}
\Lambda_k = \{ \lambda\in \Lambda: |\lambda|=k\}.
\end{equation}
Thus, $\Omega_k(\tau)$ contains all $k$'th order interactions, and $\Lambda_k\subseteq \Omega_k(\tau)$ is the set of all $k$'th order active interactions. 
As we have assumed $\tau$ to be minimal for $\Lambda$, $\tau$ is uniquely 
specifying $\Lambda_1=\{ \lambda\in\Lambda: |\lambda|=1\}$, see the discussion in 
Section \ref{sec:pseudoboolean functions} and in particular (\ref{eq:Lambda1}). 
Moreover, we restrict  $\emptyset$ always to be active, i.e. $\emptyset\in\Lambda$ with probability one, which implies that we force the pseudo-Boolean 
function $\theta(\cdot)$ always to include a constant term. As we have already assumed $\Lambda$ to be 
dense and $\tau$ to be minimal for $\Lambda$ this is only an extra restriction when $\tau=\emptyset$. Thus, for given $\tau$ the sets $\Lambda_0$ and $\Lambda_1$ are 
known, so to formulate a prior for $\Lambda$ we only need to define a prior for $\Lambda_k,k=2,\ldots,|\tau|$.
We assume a Markov property for these sets in that 
\begin{equation}\label{eq:interaction0}
f(\Lambda|\tau) = \prod_{k=2}^{|\tau|} f(\Lambda_k |\Lambda_{k-1}).
\end{equation}
Thus, to choose a prior $f(\Lambda|\tau)$ we only need to formulate $f(\Lambda_k|\Lambda_{k-1})$, and to do so we adopt the 
same strategy for all values of $k$. In the specification process of $f(\Lambda_k|\lambda_{k-1})$ we should remember that we have already 
restricted $\Lambda$ to be dense, so the chosen prior needs to be consistent with this. For a given $\Lambda_{k-1}$, an interaction
$\lambda\in\Omega_k(\tau)$ can then be active only if all 
$k-1$'th order interactions $\lambda^\star\in\Omega_{k-1}(\lambda)$ are active. We let $\Pi_k$ denote this set of 
possible active $k$'th order interactions, i.e. we must have
\begin{equation}
\Lambda_k \subseteq \Pi_k = \{\lambda\in\Omega_k(\tau): \lambda^\star\in\Lambda_{k-1} \mbox{ for all } \lambda^\star\subset\lambda\}.
\end{equation}
We assume each interaction $\lambda\in\Pi_k$ to be active with some probability $p_k$, independently of each other, and get
\begin{equation}\label{eq:interaction1}
f\left(\Lambda_k|\Lambda_{k-1}\right)=p_k^{|\Lambda_k|}(1-p_k)^{|\Pi_k|-|\Lambda_k|}
\mbox{~~~for~~~} \Lambda_k\subseteq\Pi_k.
\end{equation}
One should note that if $\Lambda_{k-1}=\emptyset$ one gets $\Pi_k=\emptyset$ and thereby also $f(\Lambda_k=\emptyset|\Lambda_{k-1})=1$.

The probabilities $p_k,k=2,\ldots,|\tau|$ should be chosen to get a reasonable number of higher-order active interactions. To obtain
a parsimonious model, one need to adopt a small value for $p_k$ if the number of elements in $\Pi_k$ is large, but to favor a
model to include some higher-order interactions, the value of $p_k$ can be large when the number of elements in $\Pi_k$ is small.
We choose
\begin{equation}\label{eq:interaction2}
p_k = \begin{cases}
p^\star & \mbox{if } |\Pi_k| \leq |\Lambda_{k-1}|,\\
p^\star\cdot \dfrac{|\Lambda_{k-1}|}{|\Pi_k|} & \mbox{otherwise,} \\
\end{cases}
\end{equation}
where $p^\star\in (0,1)$ is a hyper-parameter to be specified. One should note that this choice in particular ensures the 
expected number of active $k$'th order interactions to be smaller than $|\Lambda_{k-1}|$.

\subsection{Prior for the parameter values $\{\theta(\lambda):\lambda\in\Lambda\}$\label{sec:prior3}}
Given $\tau$ and the set of active interactions $\Lambda$, the set of model parameters for which we need to formulate a prior
is $\{\theta(\lambda):\lambda\in\Lambda\}$. From the model assumptions in (\ref{eq:cond_dist}) and (\ref{eq:mmm}),
we have that each 
$\theta(\lambda),\lambda\in\Lambda$ have a one-to-one correspondence with the conditional probability
\begin{equation}\label{eq:theta_p}
p(\lambda) = f(x_v=1|x_{\rho_v}) = \frac{\exp\{\theta(\lambda)\}}{1+\exp\{\theta(\lambda)\}}
\mbox{~~for $\lambda = \xi(x)\cap (\tau\oplus v)$}.
\end{equation}
Since the $\theta(\lambda)$'s define probabilities conditioning on different values for $x_{\rho_v}$, we find 
it reasonable, unless particular prior information is available and suggests otherwise, to assume the 
$\theta(\lambda),\lambda\in\Lambda$ to be independent. In the following we adopt this independence assumption.
Moreover, as we do not have a particular class of scenes in mind but want the prior to be reasonable for a wide 
variety of scenes, we adopt the same prior density for all parameters $\theta(\lambda),\lambda\in\Lambda$.

\begin{figure}
  \begin{center}
    \begin{tabular}{ccc}
          \begin{tikzpicture}[xscale=0.75,yscale=0.65]
  
\begin{axis}[very thick,
  axis x line=center,
  axis y line=center,
  samples=10000,
  xtick={0.2,0.4,0.6,0.8,1.0},
  ytick={0.2,0.4,0.6,0.8,1.0},
  xlabel={$p(\lambda)$},
  ylabel={$f(p(\lambda)|\tau,\Lambda)$},
  xlabel style={at={(1.1,-0.01)},anchor=north east},
  ylabel style={at={(0.18,1.13)},anchor=north east},
  xmin=-0.035,
  xmax=1.1,
  ymin=-0.035,
  ymax=1.1]
  \addplot [very thick,black,opacity=0.5] {densP(10.0)};
\end{axis}

\end{tikzpicture} 
      & 
      \hspace*{0.5cm}
      & 
      \begin{tikzpicture}[xscale=0.75,yscale=0.65]
  
\begin{axis}[very thick,
  axis x line=center,
  axis y line=center,
  samples=500,
  xtick={-4,-2,2,4},
  ytick={0.10,0.15,0.2,0.25},
  xlabel={$\theta(\lambda)$},
  ylabel={$f(\theta(\lambda)|\tau,\Lambda)$},
  xlabel style={at={(1.1,-0.01)},anchor=north east},
  ylabel style={at={(0.64,1.13)},anchor=north east},
  xmin=-6,
  xmax=6,
  ymin=-0.01,
  ymax=0.3]
  \addplot [very thick,black,opacity=0.5] {densTheta(10.0)};
\end{axis}

\end{tikzpicture}
      \\[-0.1cm]
      (a) & & (b)
    \end{tabular}
  \end{center}
  \caption{\label{fig:prior_theta_p}The prior distributions for $p(\lambda)$ and $\theta(\lambda)$. 
(a) The density curve of $f(p(\lambda)|\tau,\Lambda)$ when $\sigma=10$, and (b) the corresponding density curve $f(\theta|\tau,\Lambda)$ 
given in \eqref{eq:prior_theta}.}
\end{figure}
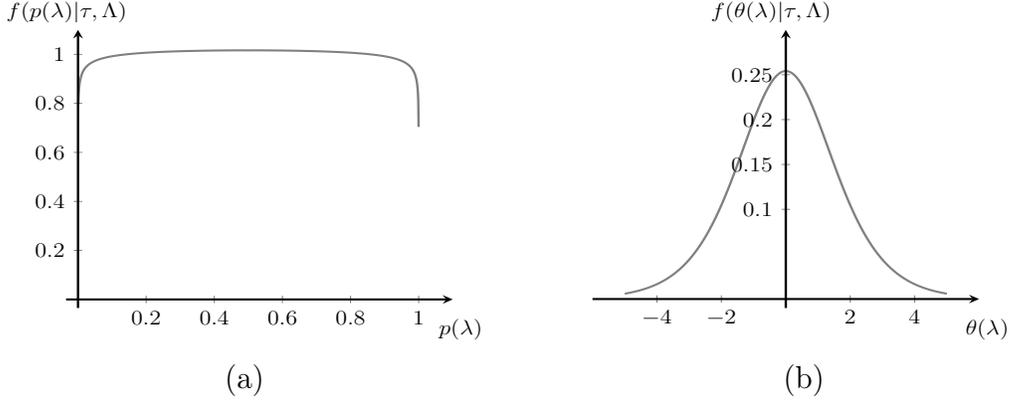

To formulate a reasonable and vague prior for $\theta(\lambda)$, we use the one-to-one correspondence between $\theta(\lambda)$ and 
the probability $p(\lambda)$. The interpretation for $p(\lambda)$ is much simpler than that of $\theta(\lambda)$, so our
strategy is first to choose a prior for $p(\lambda)$ and from this derive the corresponding prior for $\theta(\lambda)$.
As we do not have a particular class of scenes in mind but want our prior to be reasonable for a wide variety of scenes, we
find it most natural to adopt a uniform prior on $[0,1]$ for $p(\lambda)$. However, as previously mentioned we want to explore a 
corresponding posterior distribution by running a reversible jump Metropolis--Hastings algorithm, and in particular we 
want to use adaptive rejection sampling \citep{pro21} to update $\theta(\lambda)$. For this to work,
the full conditional for $\theta(\lambda)$ needs to be log-concave.
Adopting the uniform on $[0,1]$ prior for $p(\lambda)$ the resulting posterior full conditional becomes
log-concave, but the second derivative of the log full conditional converges to zero when $\theta(\lambda)$ goes to 
plus or minus infinity. As this may generate numerical problems when running the adaptive rejection sampling algorithm,
we adopt a prior for $p(\lambda)$ slightly modified relative to the uniform and obtain 
a posterior distribution where the second derivative of the log full conditional for $\theta(\lambda)$ converges to 
a value strictly less than zero. More precisely, we adopt the following prior for $\theta(\lambda)$,
\begin{equation}\label{eq:prior_theta}
f(\theta(\lambda)|\tau,\Lambda) \propto \dfrac{e^{\theta(\lambda)}}{(1+e^{\theta(\lambda)})^2}\cdot e^{-\frac{\theta(\lambda)^2}{2\sigma^2}},
\end{equation}
where the first factor is the prior resulting from assuming $p(\lambda)$ to be uniform, the second factor is
the modification we adopt to avoid numerical problems when running the adaptive rejection sampling algorithm, and 
$\sigma > 0$ is a hyper-parameter to be specified. The resulting priors for $p(\lambda)$ and $\theta(\lambda)$ 
when $\sigma = 10$ are shown in Figure \ref{fig:prior_theta_p}.
We see that the prior for $p(\lambda)$ is close to the uniform. One can also note that $f(\theta(\lambda)|\Lambda)$ have heavier tails 
than a normal distribution with the same variance. One should note that the normalizing constant in (\ref{eq:prior_theta}) is 
required when updating $\Lambda$ in a reversible jump Metropolis--Hastings algorithm targeting a corresponding posterior distribution, but since 
(\ref{eq:prior_theta}) is a univariate distribution this normalizing constant can easily be found by numerical integration.
Letting $c(\sigma)$ denote the normalizing constant of $f(\theta(\lambda)|\tau,\Lambda)$ the complete expression for the prior for 
$\{\theta(\lambda):\lambda\in\Lambda\}$ is
\begin{equation}\label{eq:prior_Theta}
f\left(\left\lbrace \theta(\lambda):\lambda\in\Lambda\right\rbrace|\tau,\Lambda\right)=
\prod_{\lambda\in\Lambda}\left[ c(\sigma) \cdot \dfrac{e^{\theta(\lambda)}}{(1+e^{\theta(\lambda)})^2}\cdot e^{-\frac{\theta(\lambda)^2}{2\sigma^2}}\right].
\end{equation}

Having specified priors for $\tau$, $\Lambda$ and $\left\lbrace \theta(\lambda):\lambda\in\Lambda\right\rbrace$ we formulate in the next section
a reversible jump Metropolis--Hastings algorithm for simulating from the corresponding posterior when a scene $x$ is observed.

\section{Simulation algorithm\label{sec:simulation algorithm}}
In the section we assume we have observed a complete scene $x=(x_v;v\in\chi)$ and assume this to be a realization from 
the Markov mesh model defined in Section \ref{sec:mmm}. We adopt the prior defined in Section \ref{sec:prior} and 
want to explore the resulting posterior distribution 
\begin{equation}\label{eq:posterior}
f\left(\tau,\Lambda,\left\lbrace \theta(\lambda):\lambda\in\Lambda\right\rbrace|x\right) \propto 
f(\tau,\Lambda,\{ \theta(\lambda):\lambda\in\Lambda\}) f(x|\tau,\Lambda,\{ \theta(\lambda):\lambda\in\Lambda\}),
\end{equation}
by a reversible jump Markov chain Monte Carlo algorithm (RJMCMC), see \citet{art22}. We combine two types of updates. 
In the first update class, we keep $\tau$ and $\Lambda$ unchanged and update the parameter 
vector $\{\theta(\lambda):\lambda\in\Lambda\}$ by a Gibbs step along a direction sampled uniformly at random. 
In the second update class, 
we propose a trans-dimensional move by adding an inactive interaction to $\Lambda$ or removing an active interaction from 
$\Lambda$, and proposing corresponding changes for the parameter vector $\{\theta(\lambda):\lambda\in\Lambda\}$.

It is clearly of interest to consider also the resulting posterior distribution when parts of the scene $x$ is 
unobserved or when $x$ is an unobserved latent field. The former is of interest if one wants to reduce the 
boundary effects of the Markov mesh model by letting $x$ include an unobserved boundary  around the observed area,
and the latter is a common situation in image analysis applications. 
However, to simplify the discussion of the simulation algorithm in this section, 
we assume the complete scene $x$ to be observed. In Section \ref{sec:examples}, where we present a number of 
simulation examples, we describe how to adapt the simulation algorithm to situation in which a part of $x$ is unobserved.

In the following we describe each of the two update types in turn, starting with the Gibbs update for the parameter values.
We only discuss the proposal distribution, as the acceptance probabilities is then given by standard formulas. 

\subsection{Gibbs update for the parameter values 
$\left\lbrace \theta(\lambda):\lambda\in\Lambda\right\rbrace$ \label{sec:propose par}}
Let $\tau$, $\Lambda$ and $\{\theta(\lambda):\lambda\in\Lambda\}$ be the current state. In this update, we keep $\tau$ and 
$\Lambda$ unchanged and generate new parameter values $\{\theta^\star(\lambda):\lambda\in\Lambda\}$. 
To generate the new parameter values we first draw a random direction $\{\Delta(\lambda):\lambda\in\Lambda\}$
by sampling $\Delta(\lambda)$ from a standard normal distribution, independently for each $\lambda\in\Lambda$. We then set
\begin{equation}\label{eq:theta_old_new}
\theta^\ast(\lambda) = \theta(\lambda) + \alpha\Delta(\lambda),
\end{equation}
where $\alpha\in\mathbb{R}$ is sampled from the full conditional 
\begin{equation}
\begin{split}
f(\alpha|\tau,\Lambda,\{\theta(\lambda) + \alpha \Delta(\lambda):\lambda\in\Lambda\},x) &\propto
f(\{\theta(\lambda) + \alpha\Delta(\lambda):\lambda\in\Lambda\}|\tau,\Lambda) \\
&\cdot f(x|\tau,\Lambda,\{ \theta(\lambda)+\alpha\Delta(\lambda):\lambda\in\Lambda\}).\label{eq:fullAlpha}
\end{split}
\end{equation}
As $\alpha$ is sampled from its full conditional, this is a Gibbs update and the Metropolis--Hastings acceptance 
probability is one. The full conditional \eqref{eq:fullAlpha} for $\alpha$ is not of a standard form, but 
in Appendix \ref{app:logConcave} we show that it is log-concave, so to generate samples from it we adopt
the adaptive rejection sampling algorithm of \citet{pro21}.

\subsection{Updating the set of active interactions\label{sec:propose interactions}}
Again let again $\tau$, $\Lambda$ and $\{\theta(\lambda):\lambda\in\Lambda\}$ be the current state. In this update we
modify $\Lambda$, and possibly also $\tau$, by adding an inactive interaction to $\Lambda$ or by removing 
an active interaction from $\Lambda$. We let $\tau^\star$ and $\Lambda^\star$ denote the potential new values for 
$\tau$ and $\Lambda$, respectively. With a change in $\Lambda$, the number of parameter values 
$\{\theta(\lambda):\lambda\in\Lambda\}$ is also changed, and to try to obtain a high acceptance rate, we in fact 
propose a change also in some of the parameter values that are in both the current and potential new states. 
We let $\{\theta^\star(\lambda):\lambda\in\Lambda^\star\}$
denote the set of potential parameter values.

To generate $\tau^\star$, $\Lambda^\star$ and $\{\theta^\star(\lambda):\lambda\in\Lambda^\star\}$, we first draw at random 
whether to add an inactive interaction to $\Lambda$ or to remove an active interaction from $\Lambda$.
In the following we specify in turn our procedures for proposing to remove and add an interaction. 

\subsubsection{Proposing to remove an active interaction from $\Lambda$\label{sec:proposeRemove}}
Having decided that an interaction should be removed, the next step is to decide what 
interaction $\lambda^\star\in\Lambda$ to remove. As the potential new $\Lambda^\star=\Lambda\setminus\{\lambda^\star\}$
should be dense, we first find the set of active interactions $\lambda^\star$ that fulfill this requirement, 
\begin{equation}
\Lambda_r=\{\lambda \in\Lambda\setminus\{\emptyset\} : \Lambda\setminus\{\lambda\} \mbox{ is dense}\}.
\end{equation}
Thereafter we draw what interaction $\lambda^\star\in\Lambda_r$ to be removed, with probabilities
\begin{equation}\label{eq:lambdastar}
q(\lambda^\star) = \frac{\exp\left\{-\nu d(\lambda^\star,\tau,\Lambda,\{\theta(\lambda):\lambda\in\Lambda\})\right\}}
{\sum_{\widetilde{\lambda}\in\Lambda_r} \exp\left\{ -\nu d(\widetilde{\lambda},\tau,\Lambda,\{\theta(\lambda):\lambda\in\Lambda\})\right\}}
\mbox{~~for~~}
\lambda^\star\in\Lambda_r,
\end{equation}
where $\nu \geq 0$ is an algorithmic tuning parameter to be specified, and 
$d(\lambda^\star,\tau,\Lambda,\{\theta(\lambda):\lambda\in\Lambda\})$
is a function that should measure the difference between the current pseudo-Boolean function defined by 
$\tau$, $\Lambda$ and $\{\theta(\lambda):\lambda\in\Lambda\}$ and the potential new pseudo-Boolean function 
defined by $\tau^\star$, $\Lambda^\star$ and $\{\theta^\star(\lambda):\lambda\in\Lambda^\star\}$. The precise
formula we use for $d(\lambda^\star,\tau,\Lambda,\{\theta(\lambda):\lambda\in\Lambda\})$ we specify below,
after having specified how to set the potential new parameter values $\{\theta^\star(\lambda):\lambda\in\Lambda^\star\}$.
By setting the algorithmic tuning parameter $\nu=0$, we draw $\Lambda^\star$ uniformly at random from the elements in 
$\Lambda_r$. With a larger value for $\nu$, we get higher probability for proposing to remove an 
interaction $\lambda^\star$ that gives a small change in the pseudo-Boolean function.
If it should happen that $\Lambda_r=\emptyset$, 
we simply propose an unchanged state. Assuming we have sampled a $\lambda^\star$ to remove,
we have two possibilities.  If $\lambda^\star$
is a higher-order interaction the sequential neighborhood is unchanged, i.e. $\tau^\star=\tau$, whereas if $\lambda^\star$ is 
a first-order interaction the sequential neighborhood is reduced to $\tau^\star = \tau \setminus \lambda^\star$. 

Having decided $\tau^\star$ and $\Lambda^\star$, the next step is to specify the potential new parameter values 
$\{\theta^\star(\lambda):\lambda\in\Lambda^\star\}$. To understand our procedure for doing this,
one should remember that there is a one-to-one 
relation between the current parameter values $\{\theta(\lambda):\lambda\in\Lambda\}$ and a set of current
interaction parameters $\{\beta(\lambda):\lambda\in\Lambda\}$, where the relation is given by \eqref{eq:m1} and \eqref{eq:m2}.
Moreover, together with the restriction $\beta(\lambda)=0$ for $\lambda\not\in\Lambda$, this defines a
pseudo-Boolean function $\{\theta(\lambda):\lambda\in\Omega(\tau_0)\}$. Correspondingly, there is a one-to-one relation 
between the potential new parameter values $\{\theta^\star(\lambda):\lambda\in\Lambda\}$ and a set of potential 
new interaction parameters $\{\beta^\star(\lambda):\lambda\in\Lambda^\star\}$, and together with the restrictions
$\beta^\star(\lambda)=0$ for $\lambda\not\in\Lambda^\star$ this defines a potential new pseudo-Boolean function 
$\{\theta^\star(\lambda):\lambda\in\Omega(\tau_0)\}$. To get a high acceptance probability for the proposed change,
it is reasonable to choose the potential new parameter values $\{\theta^\star(\lambda):\lambda\in\Lambda^\star\}$ so that 
the difference between the two pseudo-Boolean functions $\{\theta(\lambda):\lambda\in\Omega(\tau_0)\}$ and 
$\{\theta^\star(\lambda):\lambda\in\Omega(\tau_0)\}$ is small. One may consider the potential new pseudo-Boolean function 
$\{\theta^\star(\lambda):\lambda\in\Omega(\tau_0)\}$ as an approximation to the current 
$\{\theta(\lambda):\lambda\in\Omega(\tau_0)\}$ and, adopting a minimum sum of squares criterion, minimize
\begin{equation}\label{eq:SSE}
\mbox{SSE}(\{\theta^\star(\lambda):\lambda\in\Lambda^\star\}) = 
\sum_{\lambda\in\Omega(\tau_0)} \left( \theta^\star(\lambda) - \theta(\lambda)\right)^2
\end{equation}
with respect to $\{\theta^\star(\lambda):\lambda\in\Omega(\tau_0)\}$. \citet{art138} solved this  
minimization problem. Expressed in terms of the corresponding interaction parameters 
$\{\beta(\lambda): \lambda\in\Lambda\}$, the optimal potential new parameter values are 
\begin{equation}\label{eq:proposeRemove}
\beta^\star(\lambda) = \left\{ \begin{array}{ll} \beta(\lambda) - 
\left(-\frac{1}{2}\right)^{|\lambda^\star|-|\lambda|}\beta(\lambda^\star) & 
\mbox{~~if $\lambda\subset\lambda^\star$,}\\[0.1cm]
\beta(\lambda) & \mbox{~~otherwise,}\end{array}\right.
\end{equation}
and the obtained minimum sum of squares is
\begin{equation}\label{eq:minSSE}
\min\left\{ \mbox{SSE}(\{\theta^\star(\lambda):\lambda\in\Lambda\})\right\} = \frac{\beta(\lambda^\star)}{2^{|\lambda^\star|}}.
\end{equation}
We use the latter to define the function $d(\lambda^\star,\tau,\Lambda,\{\theta(\lambda):\lambda\in\Lambda\})$,
used in \eqref{eq:lambdastar} to define the distribution for what interaction $\lambda^\star$ to remove. We simply set
\begin{equation}
d(\lambda^\star,\tau,\Lambda,\{\theta(\lambda):\lambda\in\Lambda\}) = \frac{\beta(\lambda^\star)}{2^{|\lambda^\star|}}.
\end{equation}
Combining the  expression in \eqref{eq:proposeRemove} with the one-to-one relations in \eqref{eq:m1} and \eqref{eq:m2}, 
one can find the potential new parameters $\{\theta^\star(\lambda):\lambda\in\Lambda^\star\}$ in terms of the current
parameters $\{\theta(\lambda):\lambda\in\Lambda\}$. In particular, we see that this relation is 
linear and we have a $|\Lambda|\times |\Lambda|$ matrix $A$ so that 
\begin{equation}\label{eq:1-1}
\left[\begin{array}{c}\theta^\star \\ \beta(\lambda^\star) \end{array}\right] = A \theta \mbox{~~~~~}
\Leftrightarrow\mbox{~~~~~} \theta = A^{-1}\left[ \begin{array}{c}\theta^\star\\\beta(\lambda^\star)\end{array}\right],
\end{equation}
where $\theta^\star=(\theta^\star(\lambda):\lambda\in\Lambda)^T$ and 
$\theta = (\theta(\lambda):\lambda\in\Lambda)^T$ are column vectors of the potential new and current parameter values, respectively.
As the number of elements in $\theta^\star$ is one less than the number of elements in $\theta$, we 
use $\beta(\lambda^\star)$ to obtain the one-to-one relation we need for a reversible 
jump proposal. The Jacobian determinant in the expression for 
the corresponding acceptance probability is clearly $\det(A)$, and in Appendix \ref{app:determinant} we show that the absolute 
value of this determinant is always equal to one, i.e. $|\det(A)| = 1$.

\subsubsection{Proposing to add an inactive interaction to $\Lambda$}

If it is decided that an inactive interaction should be added to $\Lambda$, the next step is to decide what interaction 
$\lambda^\star\in \Omega(\tau_0)\setminus\Lambda$ to add. We do this in two steps, first we draw at random whether a 
first-order or a higher-order interaction should be added to $\Lambda$. If a first-order interaction should be added, we
draw uniformly at random a node $v^\star$ from $\tau_0\setminus\tau$ and set $\lambda^\star=\{v^\star\}$. Then 
$\tau^\star = \tau\cup \lambda^\star$ and $\Lambda^\star = \Lambda \cup \{\lambda^\star\}$. If $\tau=\tau_0$, so that no such 
$v^\star$ exist, we simply propose an unchanged state. If a higher-order interaction should be added we need to ensure that 
$\Lambda\cup \{\lambda^\star\}$ is dense. We therefore first find
\begin{equation}
\Lambda_a = \{ \lambda\in \Omega(\tau_0)\setminus\Lambda: |\lambda| > 1 \mbox{~and~} \Lambda\cup\{\lambda\} \mbox{~is dense}\}
\end{equation}
and thereafter draw $\lambda^\star$ uniformly at random from $\Lambda_a$. Then $\tau^\star=\tau$ and 
$\Lambda^\star=\Lambda\cup\{\lambda^\star\}$. If it should happen that $\Lambda_a=\emptyset$,
we again simply propose an unchanged state. 

Having decided $\tau^\star$ and $\Lambda^\star$, the next step is to generate the potential new parameter values
$\{\theta^\star(\lambda): \lambda\in\Lambda^\star\}$. When doing this, one should remember that this adding 
a potential new interaction proposal must be one-to-one with the reverse removing an interaction proposal 
discussed in Section \ref{sec:proposeRemove}. Therefore, the proposal distribution for the potential new
parameter values $\{\theta^\star(\lambda): \lambda\in\Lambda^\star\}$ must conform with \eqref{eq:proposeRemove},
and thereby also with \eqref{eq:1-1}. A natural way to achieve this is to draw a value $\beta^\star(\lambda^\star)$
from some distribution and define the potential new interaction parameters by the inverse transformation of 
\eqref{eq:proposeRemove}, i.e. 
\begin{equation}\label{eq:proposeAdd}
\beta^\star(\lambda) = \left\{\begin{array}{ll} 
\beta(\lambda) + \left(-\frac{1}{2}\right)^{|\lambda^\star|-|\lambda|} \beta^\star(\lambda^\star) 
& \mbox{~~if $\lambda\subset\lambda^\star$},\\[0.1cm]
\beta(\lambda) & \mbox{~~otherwise.}\end{array}\right.
\end{equation}
It now just remains to specify from what distribution to sample $\beta^\star(\lambda^\star)$.
The potential new parameter values $\{\theta^\star(\lambda):\lambda\in\Lambda^\star\}$ are linear functions
of $\beta^\star(\lambda^\star)$, and by setting $\beta^\star(\lambda^\star)=\alpha$ it can be expressed as in 
\eqref{eq:theta_old_new} for the Gibbs update. The difference between what we now have to do and what is done
in the Gibbs update is that in the Gibbs update the values $\Delta(\lambda)$ are sampled independently 
from a Gaussian distribution,
whereas here these are implicitly defined by \eqref{eq:proposeAdd} together with the one-to-one
relations \eqref{eq:m1} and \eqref{eq:m2}. It is tempting to sample $\alpha=\beta^\star(\lambda^\star)$ from the resulting full
conditional, as this would give a high density for values of $\beta^\star(\lambda^\star)$ that corresponds to models with 
a high posterior probability. As discussed in Section \ref{sec:propose par} for the Gibbs update, it is 
computationally feasible to sample from this full conditional by
adaptive rejection sampling. However, the 
normalizing constant of this full conditional is not computationally available, and for computing the associated 
acceptance probability the normalizing constant of the distribution of $\beta^\star(\lambda^\star)$ must be available. 
To construct a proposal distribution for $\beta^\star(\lambda^\star)=\alpha$, we therefore 
instead first generate $r$ (say) independent samples
$\alpha_1,\ldots,\alpha_r$ from the full conditional for $\alpha$, by adaptive rejection sampling, and thereafter draw 
$\alpha=\beta^\star(\lambda^\star)$ from a Gaussian distribution with mean value $\bar{\alpha}=\frac{1}{n}\sum_{i=1}^r\alpha_i$ and 
variance $s_{\alpha}^2 = \frac{1}{r-1}\sum_{i=1}^r (\alpha_i-\bar{\alpha})^2$. Our proposal distribution 
for $\beta^\star(\lambda^\star)$ is thereby an approximation to its full conditional.

As this is a reversible jump proposal, the associated acceptance probability includes a Jacobian determinant. 
By construction the Jacobian determinant for this proposal is the inverse of the Jacobian determinant
for the removing an interaction proposal discussed in Section \ref{sec:proposeRemove}. As we have 
$|\det(A)|=1$, we also get $|\det(A^{-1})|=1$.

\section{Simulation examples\label{sec:examples}}
In this section we investigate our prior and proposal distributions on two binary example scenes. 
Firstly, we consider a mortality map for 
liver and gallbladder cancers for white males from 1950 to 1959 in the eastern United States, compiled by \citet{book36}. 
Using Markov random field models, this data set has previously been analyzed by \citet{art148}, \citet{art149} and \citet{art158}, see 
also \citet{book37}. Secondly, we consider a data set 
previously considered by \cite{art132}. They also fitted a Markov mesh model to this data set, but with manually
chosen neighborhood and interaction structures. In the following we first discuss some general aspects relevant
for both the two examples and thereafter present details of each of the two examples in turn.

As also briefly discussed in Section \ref{sec:simulation algorithm}, we reduce the boundary effects of the Markov mesh model by
letting $x$ include an unobserved boundary around the observed area. We choose the unobserved boundary large enough so 
that each of the observed nodes are at least $20$ nodes away from the extended lattice boundary.
We let $\chi$ denote the set of nodes in the extended
lattice and let $x = (x_v,v\in \chi)$ be the corresponding collection of binary variables. We assume $x$ to be distributed 
according to the Markov mesh model defined in Section \ref{sec:mmm}, and for $\tau$, $\Lambda$ and 
$\{\theta(\lambda) : \lambda\in \Lambda\}$ we adopt the prior specified in Section \ref{sec:prior}. We let 
$\chi_o\subset \chi$ denote the set of nodes for which we have observed values. Thereby $\chi_u=\chi\setminus\chi_o$
is the set of unobserved nodes. Correspondingly, we let $x_o=(x_v, v\in\chi_o)$ be the observed values and 
$x_u=(x_v,v\in\chi_u)$ the unobserved values. The posterior distribution of interest is thereby 
$f(\tau,\Lambda,\{\theta(\lambda),\lambda\in\Lambda\}|x_o)$. To simplify the posterior simulation, we include $x_u$ as 
auxiliary variables and adopt the reversible jump Metropolis--Hastings algorithm to simulate from 
\begin{equation}\label{eq:postAll}
f(\tau,\Lambda,\{\theta(\lambda),\lambda\in\Lambda\},x_u|x_o) \propto
f(\tau,\Lambda,\{\theta(\lambda),\lambda\in\Lambda\}) f(x_o,x_u|\tau,\Lambda,\{\theta(\lambda),\lambda\in\Lambda\}).
\end{equation}
To simulate from this distribution, we adopt the updates discussed in Section \ref{sec:simulation algorithm} to update
$\tau$, $\Lambda$ and $\{\theta(\lambda),\lambda\in\Lambda\}$ conditioned on $x=(x_o,x_u)$, and we use single-site 
Gibbs updates for each unobserved node $v\in\chi_u$ given $\tau$, $\Lambda$, $\{\theta(\lambda),\lambda\in\Lambda\}$ 
and $x_{\chi\setminus\{ v\}}$. We define one iteration of the algorithm to include $|\chi_u|$ single-site Gibbs updates
for randomly chosen nodes in $\chi_u$ followed by either one Gibbs update of the parameter values 
$\{\theta(\lambda),\lambda\in\Lambda\}$ as discussed in Section \ref{sec:propose par} or one update of the active 
interactions as discussed in Section \ref{sec:propose interactions}. In each iteration we independently update 
the parameter values or the active interactions with probabilities $0.55$ and $0.45$ respectively.

The prior defined in Section \ref{sec:prior} contains three hyper-parameters, the radius $r$ which defines the set of 
possible neighbors, the probability $p^\star$ in \eqref{eq:interaction2}, and the parameter $\sigma$ in \eqref{eq:prior_theta}.
In both examples, we use $r=5$ which gives the $34$ possible neighbors shown in Figure \ref{fig:potentialneighbors}.
To get a prior where the probability for a Markov mesh model with higher-order interactions is reasonably high,
we set the value of $p^\star$ as high as $0.9$, and to get an essentially uniform prior distribution for $p(\lambda)$,
we set $\sigma=100$. The proposal distribution discussed in Section \ref{sec:propose interactions} has one 
algorithmic tuning parameter, $\nu$, and based on simulation results in preliminary runs we set $\nu=0.5$.

In the following we present the example scene and discuss corresponding simulation results for each of our two 
examples. We start with the cancer mortality map compiled by \citet{book36}. 

\subsection{Cancer mortality map\label{sec:cancer}}
\begin{figure}
        \begin{subfigure}[b]{0.5\textwidth}
                	\vspace*{-0.6cm}
                \includegraphics[width=\linewidth]{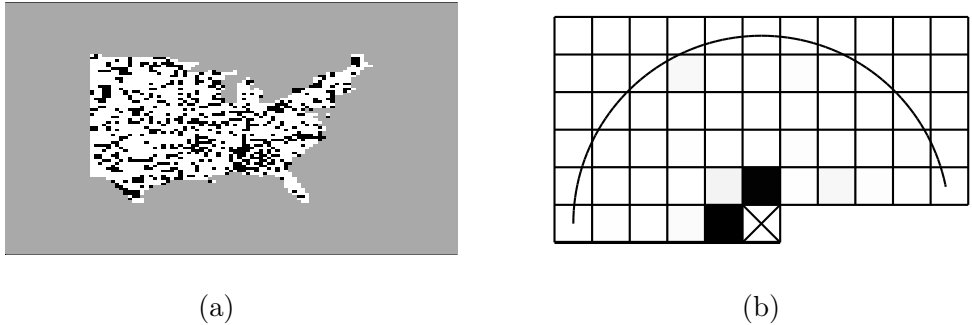}
                	\vspace*{-1.5cm}
                \caption{ }
                \label{fig:cancer}
        \end{subfigure}%
        \hfill
        \begin{subfigure}[b]{0.5\textwidth}
             
                \begin{center}
\begin{tikzpicture}[scale=0.5]

\draw[black,fill=gray!5.55016](-1.5,3.5)--(-2.5,3.5)--(-2.5,4.5)--(-1.5,4.5)--cycle; 
\draw[black,fill=gray!0.1814624](-0.5,3.5)--(-1.5,3.5)--(-1.5,4.5)--(-0.5,4.5)--cycle; 
\draw[black,fill=gray!0.22294](0.5,3.5)--(-0.5,3.5)--(-0.5,4.5)--(0.5,4.5)--cycle; 
\draw[black,fill=gray!2.0246](1.5,3.5)--(0.5,3.5)--(0.5,4.5)--(1.5,4.5)--cycle; 
\draw[black,fill=gray!0.1555392](2.5,3.5)--(1.5,3.5)--(1.5,4.5)--(2.5,4.5)--cycle; 
\draw[black,fill=gray!0.1685008](-2.5,2.5)--(-3.5,2.5)--(-3.5,3.5)--(-2.5,3.5)--cycle; 
\draw[black,fill=gray!0.497726](-1.5,2.5)--(-2.5,2.5)--(-2.5,3.5)--(-1.5,3.5)--cycle; 
\draw[black,fill=gray!0.1996086](-0.5,2.5)--(-1.5,2.5)--(-1.5,3.5)--(-0.5,3.5)--cycle; 
\draw[black,fill=gray!0.225532](0.5,2.5)--(-0.5,2.5)--(-0.5,3.5)--(0.5,3.5)--cycle; 
\draw[black,fill=gray!0.432918](1.5,2.5)--(0.5,2.5)--(0.5,3.5)--(1.5,3.5)--cycle; 
\draw[black,fill=gray!0.17887](2.5,2.5)--(1.5,2.5)--(1.5,3.5)--(2.5,3.5)--cycle; 
\draw[black,fill=gray!0.1607238](3.5,2.5)--(2.5,2.5)--(2.5,3.5)--(3.5,3.5)--cycle; 
\draw[black,fill=gray!0.1348006](-3.5,1.5)--(-4.5,1.5)--(-4.5,2.5)--(-3.5,2.5)--cycle; 
\draw[black,fill=gray!0.274786](-2.5,1.5)--(-3.5,1.5)--(-3.5,2.5)--(-2.5,2.5)--cycle; 
\draw[black,fill=gray!0.985082](-1.5,1.5)--(-2.5,1.5)--(-2.5,2.5)--(-1.5,2.5)--cycle; 
\draw[black,fill=gray!0.264416](-0.5,1.5)--(-1.5,1.5)--(-1.5,2.5)--(-0.5,2.5)--cycle; 
\draw[black,fill=gray!0.207386](0.5,1.5)--(-0.5,1.5)--(-0.5,2.5)--(0.5,2.5)--cycle; 
\draw[black,fill=gray!1.423184](1.5,1.5)--(0.5,1.5)--(0.5,2.5)--(1.5,2.5)--cycle; 
\draw[black,fill=gray!0.215162](2.5,1.5)--(1.5,1.5)--(1.5,2.5)--(2.5,2.5)--cycle; 
\draw[black,fill=gray!0.316262](3.5,1.5)--(2.5,1.5)--(2.5,2.5)--(3.5,2.5)--cycle; 
\draw[black,fill=gray!0.891758](4.5,1.5)--(3.5,1.5)--(3.5,2.5)--(4.5,2.5)--cycle; 
\draw[black,fill=gray!0.225532](-3.5,0.5)--(-4.5,0.5)--(-4.5,1.5)--(-3.5,1.5)--cycle; 
\draw[black,fill=gray!0.209978](-2.5,0.5)--(-3.5,0.5)--(-3.5,1.5)--(-2.5,1.5)--cycle; 
\draw[black,fill=gray!0.298116](-1.5,0.5)--(-2.5,0.5)--(-2.5,1.5)--(-1.5,1.5)--cycle; 
\draw[black,fill=gray!9.64342](-0.5,0.5)--(-1.5,0.5)--(-1.5,1.5)--(-0.5,1.5)--cycle; 
\draw[black,fill=gray!198.1154](0.5,0.5)--(-0.5,0.5)--(-0.5,1.5)--(0.5,1.5)--cycle; 
\draw[black,fill=gray!0.741404](1.5,0.5)--(0.5,0.5)--(0.5,1.5)--(1.5,1.5)--cycle; 
\draw[black,fill=gray!6.23194](2.5,0.5)--(1.5,0.5)--(1.5,1.5)--(2.5,1.5)--cycle; 
\draw[black,fill=gray!1.412814](3.5,0.5)--(2.5,0.5)--(2.5,1.5)--(3.5,1.5)--cycle; 
\draw[black,fill=gray!0.36811](4.5,0.5)--(3.5,0.5)--(3.5,1.5)--(4.5,1.5)--cycle; 
\draw[black,fill=gray!0.588456](-3.5,-0.5)--(-4.5,-0.5)--(-4.5,0.5)--(-3.5,0.5)--cycle; 
\draw[black,fill=gray!0.305894](-2.5,-0.5)--(-3.5,-0.5)--(-3.5,0.5)--(-2.5,0.5)--cycle; 
\draw[black,fill=gray!3.4063](-1.5,-0.5)--(-2.5,-0.5)--(-2.5,0.5)--(-1.5,0.5)--cycle; 
\draw[black,fill=gray!199.9638](-0.5,-0.5)--(-1.5,-0.5)--(-1.5,0.5)--(-0.5,0.5)--cycle;

\foreach \x in {-5.5,-4.5,-3.5,-2.5,-1.5,-0.5,0.5} {
  \draw[black,thick] (\x,-0.5) -- (\x,5.5);
}
\foreach \x in {1.5,2.5,3.5,4.5,5.5} {
  \draw[black,thick] (\x,0.5) -- (\x,5.5);
}

\draw[black,very thick] (-5.5,-0.5) -- (0.5,-0.5);
\foreach \y in {0.5,1.5,2.5,3.5,4.5,5.5} {
  \draw[black,thick] (-5.5,\y) -- (5.5,\y);
}

\draw[black,thick] ({5*cos(11.31)},{5*sin(11.31)}) -- ({5*cos(12)},{5*sin(12)});
\foreach \angle in {12,14,16,18,20,22,24,26,28,30,32,34,36,38,40,42,44,46,48,50,52,54,56,58,60,62,64,66,68,70,72,74,76,78,80,82,84,86,88,90,92,94,96,98,100,102,104,106,108,110,112,114,116,118,120,122,124,126,128,130,132,134,136,138,140,142,144,146,148,150,152,154,156,158,160,162,164,166,168,170,172,174,176,178} {
  \draw[black,thick] ({5*cos(\angle)},{5*sin(\angle)}) -- ({5*cos(\angle+2)},{5*sin(\angle+2)});
}

\draw[black,thick] (-0.5,-0.5) -- (0.5,0.5);
\draw[black,thick] (-0.5,0.5) -- (0.5,-0.5);

\end{tikzpicture}
\end{center}
                	\vspace*{-0.15cm}
                \caption{ }
                \label{fig:probneigh_cancer}
        \end{subfigure}
        \caption{Cancer mortality map example: (a) Observed cancer mortality map. Black and white nodes represent
counties with high and low cancer mortality rates, respectively. The nodes added to the lattice to reduce 
the boundary effects of the Markov mesh model is shown in gray. (b) Map of estimated a posteriori marginal
probabilities for each node $v\in \tau_0$ to be a neighbor. A grayscale is used to visualize the probabilities, where black and white
represents one and zero, respectively.}\label{fig:TI_probneigh_cancer}
\end{figure}

\begin{figure}   
        \begin{subfigure}[b]{0.5\textwidth}
        \vspace*{-0.5cm}
                \includegraphics[width=\linewidth,height=5.0cm]{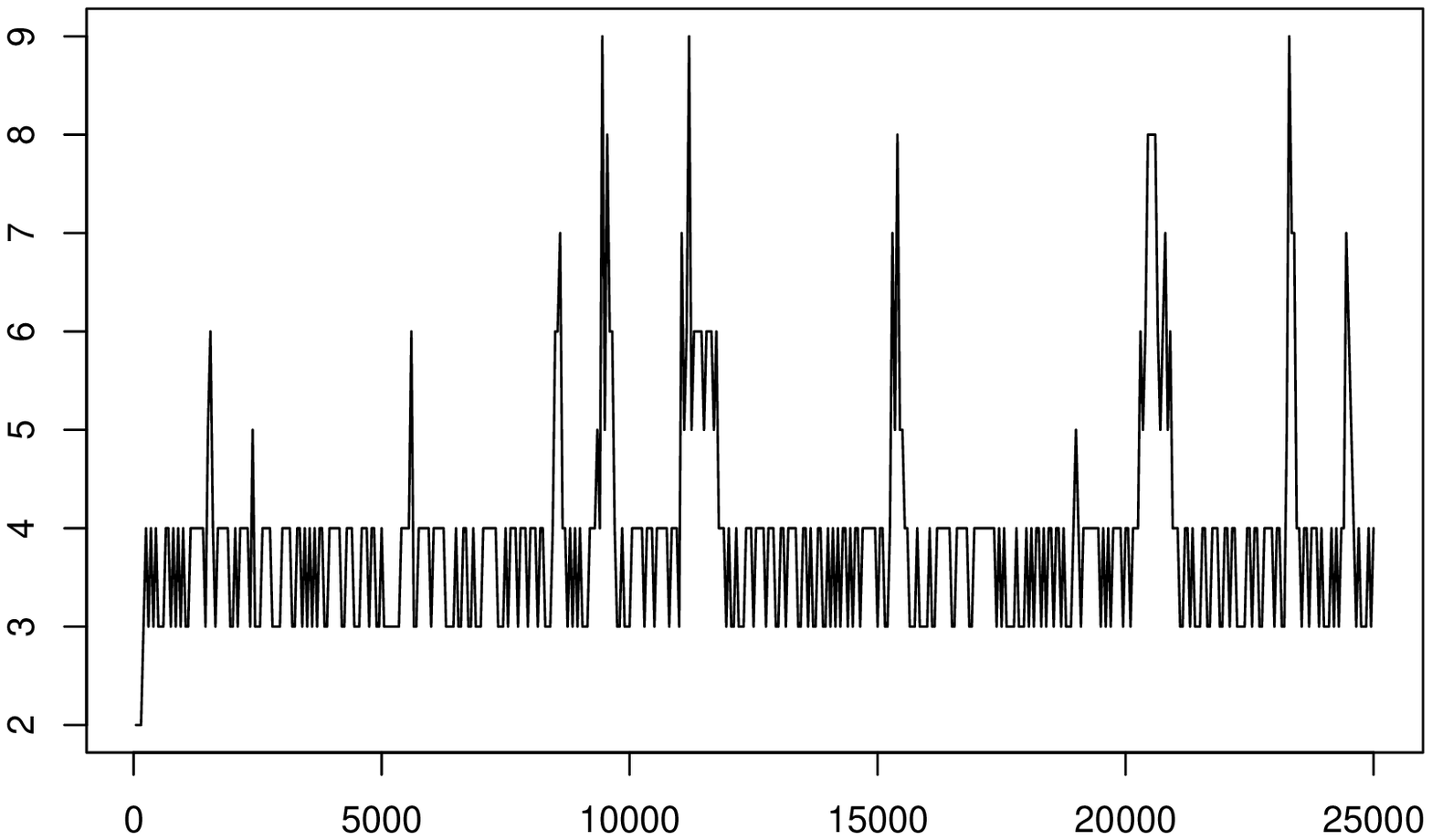}
        \vspace*{-1.0cm}
                \caption{ }
                \label{fig:traceplot_num_short_cancer}
        \end{subfigure}%
        \hfill
        \begin{subfigure}[b]{0.5\textwidth}
         \vspace*{-0.5cm}
                \includegraphics[width=\linewidth,height=5.0cm]{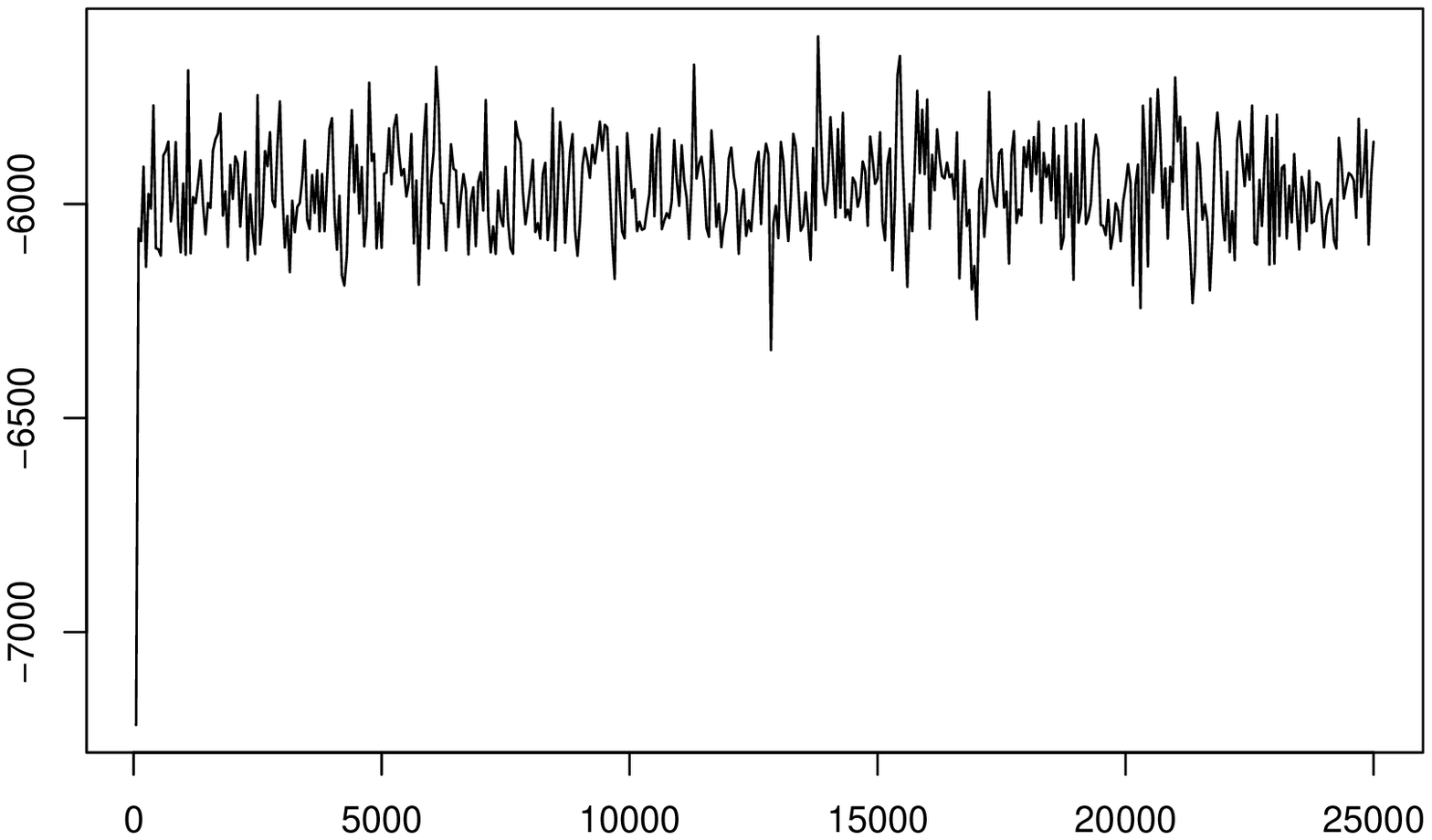}
        \vspace*{-1.0cm}
                \caption{ }
                \label{fig:traceplot_log_short_cancer}
        \end{subfigure}
        
        \caption{Cancer mortality map example: Trace plots of the first $25\,000$ iterations of the RJMCMCM run. 
(a) Number of interactions $|\Lambda|$, (b) logarithm of the posterior density 
$\log\left[f\left(\tau,\Lambda,\left\lbrace \theta(\lambda):\lambda\in\Lambda\right\rbrace,x_u|x_o\right)\right]$}\label{fig:traceplots_cancer}
\end{figure}

The cancer mortality map data are shown in Figure \ref{fig:cancer}, where black ($x_v=1$) and white ($x_v=0$)
pixels represent 
counties with high and low cancer mortality rates, respectively. The gray area around 
the observed map represents unobserved nodes which we included in the model to reduce the boundary effects of the Markov mesh model.
Adopting the Markov mesh and prior models discussed in Sections \ref{sec:mmm} and \ref{sec:prior}, respectively,
with the hyper-parameters defined above, we use the RJMCMC setup discussed above to explore the resulting 
posterior distribution. We run the Markov chain for $2\,500\,000$ iterations, and study trace plots of different 
scalar quantities to evaluate the convergence and mixing properties of the simulated Markov chain.
Figure \ref{fig:traceplots_cancer} 
shows trace plots of the first $25\,000$ iterations for the
number of interactions and for the logarithm of the posterior density.
From these two and the other trace plots we have studied, we conclude that the simulated chain has converged
at least within the first $10\,000-15\,000$ iterations. As an extra precaution we discard the first 
$25\,000$ iterations when estimating posterior properties.

To study the posterior distribution we first estimate, for each of the $34$ apriori potential neighbors in $\tau_0$,
the posterior probability for $v\in \tau_0$ to be a neighbor. To estimate this
we simply use the fraction of simulated models where $v$ is in the template sequential neighborhood $\tau$.
The result is shown in Figure \ref{fig:probneigh_cancer}, where we use a gray scale to visualize the probabilities.
Nodes $(0,-1)$ and $(-1,0)$ have high estimated posterior probabilities, equal to $0.999819$ and $0.990577$, respectively. 
The third and fourth most probable neighbor nodes are $(-1,-1)$ and $(-1,2)$, where the estimated 
probabilities are 0.049388 and 0.030353, respectively. From the data set shown in Figure \ref{fig:cancer}, we see that the dependence 
between neighbor nodes seems to be quite weak, so the low number of simulated neighbors should come as no 
surprise.

Next we correspondingly estimate the posterior probabilities for each possible interaction to be included in the model. 
Table \ref{tab:probInt} shows the top $10$ a posteriori most likely interactions and the corresponding estimated probabilities. 
\begin{table}
\centering
\caption{Cancer mortality map example: Table with the top $10$ a posteriori most likely interactions and their 
estimated posterior probabilities.}
\label{tab:probInt}
\begin{tabular}{|c|c|c|c|c|c|}
\hline
& & & & & \\[-0.4cm]
Interaction &   
 \begin{tikzpicture}[scale=2]
  \def\radius{38.9};
  \coordinate (Pempty) at (0,0);

  \coordinate (c1) at (-0.05,-0.05);
  \coordinate (c2) at (0.05,-0.05);
  \coordinate (c3) at (0.05,0.05);
  \coordinate (c4) at (-0.05,0.05);

  \draw[thick] (Pempty) +(c1) -- +(c2) -- +(c3) -- +(c4) -- +(c1) +(c1) -- +(c3) +(c2) -- +(c4);

\end{tikzpicture}  
&   
   \begin{tikzpicture}[scale=2]
  \def\radius{38.9};
  \coordinate (Pempty) at (0,0);
  \coordinate (P1) at (-1.5,1);
  \coordinate (P2) at (-0.5,1);
  \coordinate (P3) at (0.5,1);
  \coordinate (P4) at (1.5,1);
  \coordinate (P12) at (-1,2);
  \coordinate (P14) at (0,2);

  \coordinate (c1) at (-0.05,-0.05);
  \coordinate (c2) at (0.05,-0.05);
  \coordinate (c3) at (0.05,0.05);
  \coordinate (c4) at (-0.05,0.05);
  
   \coordinate (r1) at (-0.1,0);
  \coordinate (r2) at (0,0.1);
  \coordinate (r3) at (-0.1,0.1);
  \coordinate (r4) at (0.1,0.1);

  \draw[thick] (P1) +(c1) -- +(c2) -- +(c3) -- +(c4) -- +(c1) +(c1) -- +(c3) +(c2) -- +(c4);
  \draw[thick] (P1) ++(r1) +(c1) -- +(c2) -- +(c3) -- +(c4) -- cycle;

\end{tikzpicture}  
&     
\begin{tikzpicture}[scale=2]
  \def\radius{38.9};
  \coordinate (Pempty) at (0,0);
  \coordinate (P1) at (-1.5,1);
  \coordinate (P2) at (-0.5,1);
  \coordinate (P3) at (0.5,1);
  \coordinate (P4) at (1.5,1);
  \coordinate (P12) at (-1,2);
  \coordinate (P14) at (0,2);

  \coordinate (c1) at (-0.05,-0.05);
  \coordinate (c2) at (0.05,-0.05);
  \coordinate (c3) at (0.05,0.05);
  \coordinate (c4) at (-0.05,0.05);

  \coordinate (r1) at (-0.1,0);
  \coordinate (r2) at (0,0.1);
  \coordinate (r3) at (-0.1,0.1);
  \coordinate (r4) at (0.1,0.1);  

  \draw[thick] (P2) +(c1) -- +(c2) -- +(c3) -- +(c4) -- +(c1) +(c1) -- +(c3) +(c2) -- +(c4);
  \draw[thick] (P2) ++(r2) +(c1) -- +(c2) -- +(c3) -- +(c4) -- cycle;

\end{tikzpicture}
&    
\begin{tikzpicture}[scale=2]
  \def\radius{38.9};
  \coordinate (Pempty) at (0,0);
  \coordinate (P1) at (-1.5,1);
  \coordinate (P2) at (-0.5,1);
  \coordinate (P3) at (0.5,1);
  \coordinate (P4) at (1.5,1);
  \coordinate (P12) at (-1,2);
  \coordinate (P14) at (0,2);

  \coordinate (c1) at (-0.05,-0.05);
  \coordinate (c2) at (0.05,-0.05);
  \coordinate (c3) at (0.05,0.05);
  \coordinate (c4) at (-0.05,0.05);

  \coordinate (r1) at (-0.1,0);
  \coordinate (r2) at (0,0.1);
  \coordinate (r3) at (-0.1,0.1);
  \coordinate (r4) at (0.1,0.1);

  \draw[thick] (P12) +(c1) -- +(c2) -- +(c3) -- +(c4) -- +(c1) +(c1) -- +(c3) +(c2) -- +(c4);
  \draw[thick] (P12) ++(r1) +(c1) -- +(c2) -- +(c3) -- +(c4) -- cycle;
  \draw[thick] (P12) ++(r2) +(c1) -- +(c2) -- +(c3) -- +(c4) -- cycle;

\end{tikzpicture}   
&    
\begin{tikzpicture}[scale=2]
  \def\radius{38.9};
  \coordinate (Pempty) at (0,0);
  \coordinate (P1) at (-1.5,1);
  \coordinate (P2) at (-0.5,1);
  \coordinate (P3) at (0.5,1);
  \coordinate (P4) at (1.5,1);
  \coordinate (P12) at (-1,2);
  \coordinate (P14) at (0,2);

  \coordinate (c1) at (-0.05,-0.05);
  \coordinate (c2) at (0.05,-0.05);
  \coordinate (c3) at (0.05,0.05);
  \coordinate (c4) at (-0.05,0.05);

  \coordinate (r1) at (-0.1,0);
  \coordinate (r2) at (0,0.1);
  \coordinate (r3) at (-0.1,0.1);
  \coordinate (r4) at (0.1,0.1);

  \draw[thick] (P3) +(c1) -- +(c2) -- +(c3) -- +(c4) -- +(c1) +(c1) -- +(c3) +(c2) -- +(c4);
  \draw[thick] (P3) ++(r4) +(c1) -- +(c2) -- +(c3) -- +(c4) -- cycle;

\end{tikzpicture}       
           \\
Probability & 1.0000 & 0.9998 & 0.9902 & 0.5452 & 0.0489 \\ \hline
& & & & & \\[-0.4cm]
Interaction &      
\begin{tikzpicture}[scale=2]
  \def\radius{38.9};
  \coordinate (Pempty) at (0,0);
  \coordinate (P1) at (-1.5,1);
  \coordinate (P2) at (-0.5,1);
  \coordinate (P3) at (0.5,1);
  \coordinate (P4) at (1.5,1);
  \coordinate (P12) at (-1,2);
  \coordinate (P14) at (0,2);

  \coordinate (c1) at (-0.05,-0.05);
  \coordinate (c2) at (0.05,-0.05);
  \coordinate (c3) at (0.05,0.05);
  \coordinate (c4) at (-0.05,0.05);

  \coordinate (r1) at (-0.1,0);
  \coordinate (r2) at (0,0.1);
  \coordinate (r3) at (-0.1,0.1);
  \coordinate (r4) at (0.1,0.1);
  \coordinate (r5) at (0.2,0.1);

  \draw[thick] (P3) +(c1) -- +(c2) -- +(c3) -- +(c4) -- +(c1) +(c1) -- +(c3) +(c2) -- +(c4);
  \draw[thick] (P3) ++(r5) +(c1) -- +(c2) -- +(c3) -- +(c4) -- cycle;
\end{tikzpicture}
&     
 \begin{tikzpicture}[scale=2]
  \def\radius{38.9};
  \coordinate (Pempty) at (0,0);
  \coordinate (P1) at (-1.5,1);
  \coordinate (P2) at (-0.5,1);
  \coordinate (P3) at (0.5,1);
  \coordinate (P4) at (1.5,1);
  \coordinate (P12) at (-1,2);
  \coordinate (P14) at (0,2);

  \coordinate (c1) at (-0.05,-0.05);
  \coordinate (c2) at (0.05,-0.05);
  \coordinate (c3) at (0.05,0.05);
  \coordinate (c4) at (-0.05,0.05);

  \coordinate (r1) at (-0.1,0);
  \coordinate (r2) at (0,0.1);
  \coordinate (r3) at (-0.1,0.1);
  \coordinate (r4) at (0.1,0.1);
  \coordinate (r5) at (0.2,0.1);

  \draw[thick] (P3) +(c1) -- +(c2) -- +(c3) -- +(c4) -- +(c1) +(c1) -- +(c3) +(c2) -- +(c4);
  \draw[thick] (P3) ++(r1) +(c1) -- +(c2) -- +(c3) -- +(c4) -- cycle;
  \draw[thick] (P3) ++(r3) +(c1) -- +(c2) -- +(c3) -- +(c4) -- cycle;
\end{tikzpicture} 
&    
\begin{tikzpicture}[scale=2]
  \def\radius{38.9};
  \coordinate (Pempty) at (0,0);
  \coordinate (P1) at (-1.5,1);
  \coordinate (P2) at (-0.5,1);
  \coordinate (P3) at (0.5,1);
  \coordinate (P4) at (1.5,1);
  \coordinate (P12) at (-1,2);
  \coordinate (P14) at (0,2);

  \coordinate (c1) at (-0.05,-0.05);
  \coordinate (c2) at (0.05,-0.05);
  \coordinate (c3) at (0.05,0.05);
  \coordinate (c4) at (-0.05,0.05);

  \coordinate (r1) at (-0.1,0);
  \coordinate (r2) at (0,0.1);
  \coordinate (r3) at (-0.1,0.1);
  \coordinate (r4) at (0.1,0.1);
  \coordinate (r5) at (0.2,0.1);
  \coordinate (r6) at (-0.4,0.2);

  \draw[thick] (P3) +(c1) -- +(c2) -- +(c3) -- +(c4) -- +(c1) +(c1) -- +(c3) +(c2) -- +(c4);
  \draw[thick] (P3) ++(r6) +(c1) -- +(c2) -- +(c3) -- +(c4) -- cycle;
   \end{tikzpicture} 
&    
 \begin{tikzpicture}[scale=2]
  \def\radius{38.9};
  \coordinate (Pempty) at (0,0);
  \coordinate (P1) at (-1.5,1);
  \coordinate (P2) at (-0.5,1);
  \coordinate (P3) at (0.5,1);
  \coordinate (P4) at (1.5,1);
  \coordinate (P12) at (-1,2);
  \coordinate (P14) at (0,2);

  \coordinate (c1) at (-0.05,-0.05);
  \coordinate (c2) at (0.05,-0.05);
  \coordinate (c3) at (0.05,0.05);
  \coordinate (c4) at (-0.05,0.05);

  \coordinate (r1) at (-0.1,0);
  \coordinate (r2) at (0,0.1);
  \coordinate (r3) at (-0.1,0.1);
  \coordinate (r4) at (0.1,0.1);
  \coordinate (r5) at (0.2,0.1);

  \draw[thick] (P3) +(c1) -- +(c2) -- +(c3) -- +(c4) -- +(c1) +(c1) -- +(c3) +(c2) -- +(c4);
  \draw[thick] (P3) ++(r2) +(c1) -- +(c2) -- +(c3) -- +(c4) -- cycle;
  \draw[thick] (P3) ++(r3) +(c1) -- +(c2) -- +(c3) -- +(c4) -- cycle;
\end{tikzpicture}      
&    
 \begin{tikzpicture}[scale=2]
  \def\radius{38.9};
  \coordinate (Pempty) at (0,0);
  \coordinate (P1) at (-1.5,1);
  \coordinate (P2) at (-0.5,1);
  \coordinate (P3) at (0.5,1);
  \coordinate (P4) at (1.5,1);
  \coordinate (P12) at (-1,2);
  \coordinate (P14) at (0,2);

  \coordinate (c1) at (-0.05,-0.05);
  \coordinate (c2) at (0.05,-0.05);
  \coordinate (c3) at (0.05,0.05);
  \coordinate (c4) at (-0.05,0.05);

  \coordinate (r1) at (-0.1,0);
  \coordinate (r2) at (0,0.1);
  \coordinate (r3) at (-0.1,0.1);
  \coordinate (r4) at (0.1,0.1);
  \coordinate (r5) at (0.2,0.1);

  \draw[thick] (P3) +(c1) -- +(c2) -- +(c3) -- +(c4) -- +(c1) +(c1) -- +(c3) +(c2) -- +(c4);
  \draw[thick] (P3) ++(r1) +(c1) -- +(c2) -- +(c3) -- +(c4) -- cycle;
  \draw[thick] (P3) ++(r5) +(c1) -- +(c2) -- +(c3) -- +(c4) -- cycle;
\end{tikzpicture} 
    \\
Probability & 0.0303 & 0.0280 & 0.0274 & 0.0270 & 0.0251 \\ \hline
\end{tabular}
\end{table}
We see that the first four interactions have high posterior probabilities while the others have low probabilities. 
In addition, the four most likely interactions only include the high probability neighbor nodes $(0,-1)$ and $(-1,0)$.

\begin{figure}
        \begin{subfigure}[b]{0.5\textwidth}
         \vspace*{-0.1cm}
                \includegraphics[width=\linewidth]{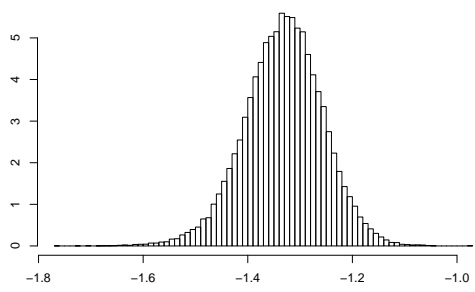}
         \vspace*{-1.0cm}
                \caption[]{$\theta\hspace*{-0.08cm}\left(\hspace*{0.02cm}
                \begin{tikzpicture}[scale=2]
	  \def\radius{38.9};	
  		\coordinate (Pempty) at (0,0);

  \coordinate (c1) at (-0.05,-0.05);
  \coordinate (c2) at (0.05,-0.05);
  \coordinate (c3) at (0.05,0.05);
  \coordinate (c4) at (-0.05,0.05);

  \draw[thick] (Pempty) +(c1) -- +(c2) -- +(c3) -- +(c4) -- +(c1) +(c1) -- +(c3) +(c2) -- +(c4);

\end{tikzpicture}
\hspace*{0.02cm}\right)$}
                \label{fig:histpar1_cancer}
        \end{subfigure}%
        \hfill
        \begin{subfigure}[b]{0.5\textwidth}
        \vspace*{-0.1cm}
                \includegraphics[width=\linewidth]{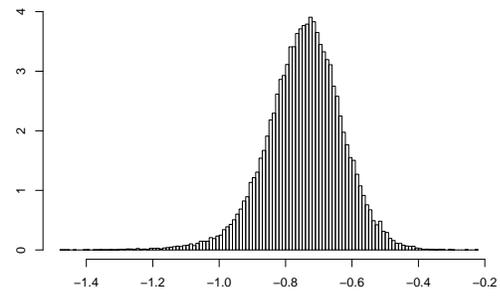}
         \vspace*{-1.0cm}
                \caption[]{$\theta\hspace*{-0.05cm}\left(\hspace*{0.02cm}
                \begin{tikzpicture}[scale=2]
  \def\radius{38.9};
  \coordinate (Pempty) at (0,0);
  \coordinate (P1) at (-1.5,1);
  \coordinate (P2) at (-0.5,1);
  \coordinate (P3) at (0.5,1);
  \coordinate (P4) at (1.5,1);
  \coordinate (P12) at (-1,2);
  \coordinate (P14) at (0,2);

  \coordinate (c1) at (-0.05,-0.05);
  \coordinate (c2) at (0.05,-0.05);
  \coordinate (c3) at (0.05,0.05);
  \coordinate (c4) at (-0.05,0.05);
  
  \coordinate (r1) at (-0.1,0);
  \coordinate (r2) at (0,0.1);
  \coordinate (r3) at (-0.1,0.1);
  \coordinate (r4) at (0.1,0.1);

  \draw[thick] (P12) +(c1) -- +(c2) -- +(c3) -- +(c4) -- +(c1) +(c1) -- +(c3) +(c2) -- +(c4);
  \draw[thick] (P12) ++(r1) +(c1) -- +(c2) -- +(c3) -- +(c4) -- cycle;

\end{tikzpicture}
\hspace*{0.02cm}\right)$}
                \label{fig:histpar2_cancer}
        \end{subfigure}
        
        \begin{subfigure}[b]{0.5\textwidth}
        \vspace*{-0.1cm}
                \includegraphics[width=\linewidth]{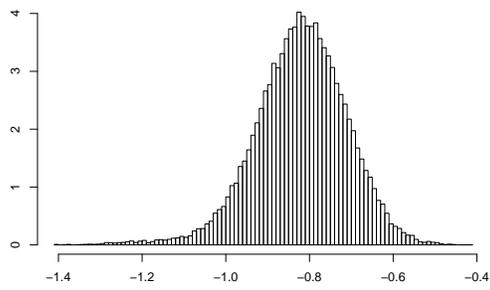}
         \vspace*{-1.0cm}
            \caption[]{$\theta\hspace*{-0.06cm}\left(\hspace*{-0.15cm}\raisebox{-0.1cm}{
               \begin{tikzpicture}[scale=2]
  \def\radius{38.9};
  \coordinate (Pempty) at (0,0);
  \coordinate (P1) at (-1.5,1);
  \coordinate (P2) at (-0.5,1);
  \coordinate (P3) at (0.5,1);
  \coordinate (P4) at (1.5,1);
  \coordinate (P12) at (-1,2);
  \coordinate (P14) at (0,2);

  \coordinate (c1) at (-0.05,-0.05);
  \coordinate (c2) at (0.05,-0.05);
  \coordinate (c3) at (0.05,0.05);
  \coordinate (c4) at (-0.05,0.05);

  \coordinate (r1) at (-0.1,0);
  \coordinate (r2) at (0,0.1);
  \coordinate (r3) at (-0.1,0.1);
  \coordinate (r4) at (0.1,0.1);  

  \draw[thick] (P2) +(c1) -- +(c2) -- +(c3) -- +(c4) -- +(c1) +(c1) -- +(c3) +(c2) -- +(c4);
  \draw[thick] (P2) ++(r2) +(c1) -- +(c2) -- +(c3) -- +(c4) -- cycle;

\end{tikzpicture}}
\right)$}
                \label{fig:histpar3_cancer}
        \end{subfigure}%
        \hfill
        \begin{subfigure}[b]{0.5\textwidth}
        \vspace*{-0.1cm}
                \includegraphics[width=\linewidth]{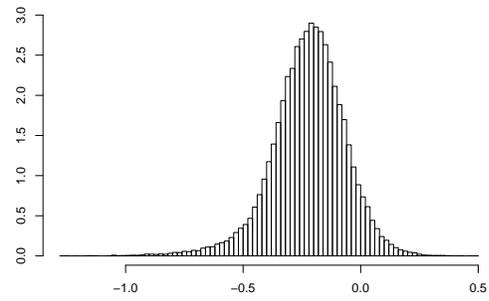}
        \vspace*{-1.0cm}
                \caption[]{$\theta\hspace*{-0.08cm}\left(\hspace*{-0.15cm}\raisebox{-0.1cm}{
              \begin{tikzpicture}[scale=2]
  \def\radius{38.9};
  \coordinate (Pempty) at (0,0);
  \coordinate (P1) at (-1.5,1);
  \coordinate (P2) at (-0.5,1);
  \coordinate (P3) at (0.5,1);
  \coordinate (P4) at (1.5,1);
  \coordinate (P12) at (-1,2);
  \coordinate (P14) at (0,2);

  \coordinate (c1) at (-0.05,-0.05);
  \coordinate (c2) at (0.05,-0.05);
  \coordinate (c3) at (0.05,0.05);
  \coordinate (c4) at (-0.05,0.05);

  \coordinate (r1) at (-0.1,0);
  \coordinate (r2) at (0,0.1);
  \coordinate (r3) at (-0.1,0.1);
  \coordinate (r4) at (0.1,0.1);

  \draw[thick] (P12) +(c1) -- +(c2) -- +(c3) -- +(c4) -- +(c1) +(c1) -- +(c3) +(c2) -- +(c4);
  \draw[thick] (P12) ++(r1) +(c1) -- +(c2) -- +(c3) -- +(c4) -- cycle;
  \draw[thick] (P12) ++(r2) +(c1) -- +(c2) -- +(c3) -- +(c4) -- cycle;

\end{tikzpicture}}   
\right)$}
                \label{fig:histpar4_cancer}
        \end{subfigure}
        \caption{Cancer mortality map example: Histograms of the simulated parameter values $\theta(\cdot)$ for the top four 
a posteriori most likely interactions.}\label{fig:histpar_cancer}
\end{figure}

\begin{figure}
        \begin{subfigure}[b]{0.5\textwidth}
        \begin{center}
                \begin{tikzpicture}[scale=2.0]
  \def\radius{38.9};
  \coordinate (Pempty) at (-1,0);
  \coordinate (P1) at (-1.5,1);
  \coordinate (P2) at (-0.5,1);
  \coordinate (P3) at (0.5,1);
  \coordinate (P4) at (1.5,1);
  \coordinate (P12) at (-1,2);
  \coordinate (P14) at (0,2);

  \coordinate (c1) at (-0.05,-0.05);
  \coordinate (c2) at (0.05,-0.05);
  \coordinate (c3) at (0.05,0.05);
  \coordinate (c4) at (-0.05,0.05);

  \coordinate (r1) at (-0.1,0);
  \coordinate (r2) at (0,0.1);
  \coordinate (r3) at (-0.1,0.1);
  \coordinate (r4) at (0.1,0.1);

  \node[draw,circle,inner sep=0pt,minimum size=\radius,name=Nempty] at (Pempty) {};
  \draw[thick] (Pempty) +(c1) -- +(c2) -- +(c3) -- +(c4) -- +(c1) +(c1) -- +(c3) +(c2) -- +(c4);
  
  \node[draw,circle,inner sep=0pt,minimum size=\radius,name=N1] at (P1) {};
  \draw[thick] (P1) +(c1) -- +(c2) -- +(c3) -- +(c4) -- +(c1) +(c1) -- +(c3) +(c2) -- +(c4);
  \draw[thick] (P1) ++(r1) +(c1) -- +(c2) -- +(c3) -- +(c4) -- cycle;
  \draw[thick,->] (Nempty) -- (N1);
   
  \node[draw,circle,inner sep=0pt,minimum size=\radius,name=N2] at (P2) {};
  \draw[thick] (P2) +(c1) -- +(c2) -- +(c3) -- +(c4) -- +(c1) +(c1) -- +(c3) +(c2) -- +(c4);
  \draw[thick] (P2) ++(r2) +(c1) -- +(c2) -- +(c3) -- +(c4) -- cycle;
  \draw[thick,->] (Nempty) -- (N2);
  
  \node[draw,circle,inner sep=0pt,minimum size=\radius,name=N12] at (P12) {};
  \draw[thick] (P12) +(c1) -- +(c2) -- +(c3) -- +(c4) -- +(c1) +(c1) -- +(c3) +(c2) -- +(c4);
  \draw[thick] (P12) ++(r1) +(c1) -- +(c2) -- +(c3) -- +(c4) -- cycle;
  \draw[thick] (P12) ++(r2) +(c1) -- +(c2) -- +(c3) -- +(c4) -- cycle;
  \draw[thick,->] (N1) -- (N12);
  \draw[thick,->] (N2) -- (N12);
  
\end{tikzpicture}
\end{center}
                \caption{Posterior probability: $0.475$}
                \label{fig:dag1_cancer}
        \end{subfigure}%
        \hfill
        \begin{subfigure}[b]{0.5\textwidth}        
                \begin{center}
\begin{tikzpicture}[scale=2.0]
  \def\radius{38.9};
  \coordinate (Pempty) at (-1,0);
  \coordinate (P1) at (-1.5,1);
  \coordinate (P2) at (-0.5,1);
  \coordinate (P3) at (0.5,1);
  \coordinate (P4) at (1.5,1);
  \coordinate (P12) at (-1,2);
  \coordinate (P14) at (0,2);

  \coordinate (c1) at (-0.05,-0.05);
  \coordinate (c2) at (0.05,-0.05);
  \coordinate (c3) at (0.05,0.05);
  \coordinate (c4) at (-0.05,0.05);

  \coordinate (r1) at (-0.1,0);
  \coordinate (r2) at (0,0.1);
  \coordinate (r3) at (-0.1,0.1);
  \coordinate (r4) at (0.1,0.1);

  \node[draw,circle,inner sep=0pt,minimum size=\radius,name=Nempty] at (Pempty) {};
  \draw[thick] (Pempty) +(c1) -- +(c2) -- +(c3) -- +(c4) -- +(c1) +(c1) -- +(c3) +(c2) -- +(c4);

  \node[draw,circle,inner sep=0pt,minimum size=\radius,name=N1] at (P1) {};
  \draw[thick] (P1) +(c1) -- +(c2) -- +(c3) -- +(c4) -- +(c1) +(c1) -- +(c3) +(c2) -- +(c4);
  \draw[thick] (P1) ++(r1) +(c1) -- +(c2) -- +(c3) -- +(c4) -- cycle;
  \draw[thick,->] (Nempty) -- (N1);
   
  \node[draw,circle,inner sep=0pt,minimum size=\radius,name=N2] at (P2) {};
  \draw[thick] (P2) +(c1) -- +(c2) -- +(c3) -- +(c4) -- +(c1) +(c1) -- +(c3) +(c2) -- +(c4);
  \draw[thick] (P2) ++(r2) +(c1) -- +(c2) -- +(c3) -- +(c4) -- cycle;
  \draw[thick,->] (Nempty) -- (N2);
  
\end{tikzpicture}
\end{center}
                \caption{Posterior probability: $0.381$}
                \label{fig:dag2_cancer}
        \end{subfigure}
        \caption{Cancer mortality map example: The two a posteriori most likely models and the corresponding 
estimated posterior probabilities.}\label{fig:dag_cancer}
\end{figure}
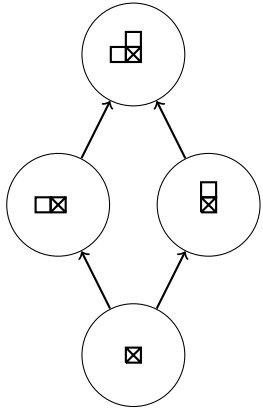
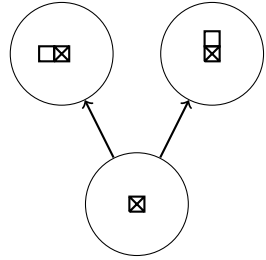

We also estimate the a posteriori marginal distributions for the parameter values $\theta(\cdot)$ corresponding to 
the four high probable interactions. Note that some of the interactions do not exist in some of the simulated models, 
but the $\theta(\cdot)$ value is still well defined and can be computed as discussed in Section \ref{sec:pseudoboolean functions}. 
Figure \ref{fig:histpar_cancer} depicts the histograms of the simulated parameter values $\theta(\cdot)$.
From the simulation we also estimate the posterior probability for each of the possible models. The two most probable models
are shown in Figure \ref{fig:dag_cancer}. These two models have posterior probabilities as high as $0.475$ and 
$0.381$ while the remaining probability mass is spread out on a very large number of models.

\begin{figure}
        \begin{subfigure}[b]{0.5\textwidth}
        	\vspace*{-0.5cm}
                \includegraphics[width=\linewidth]{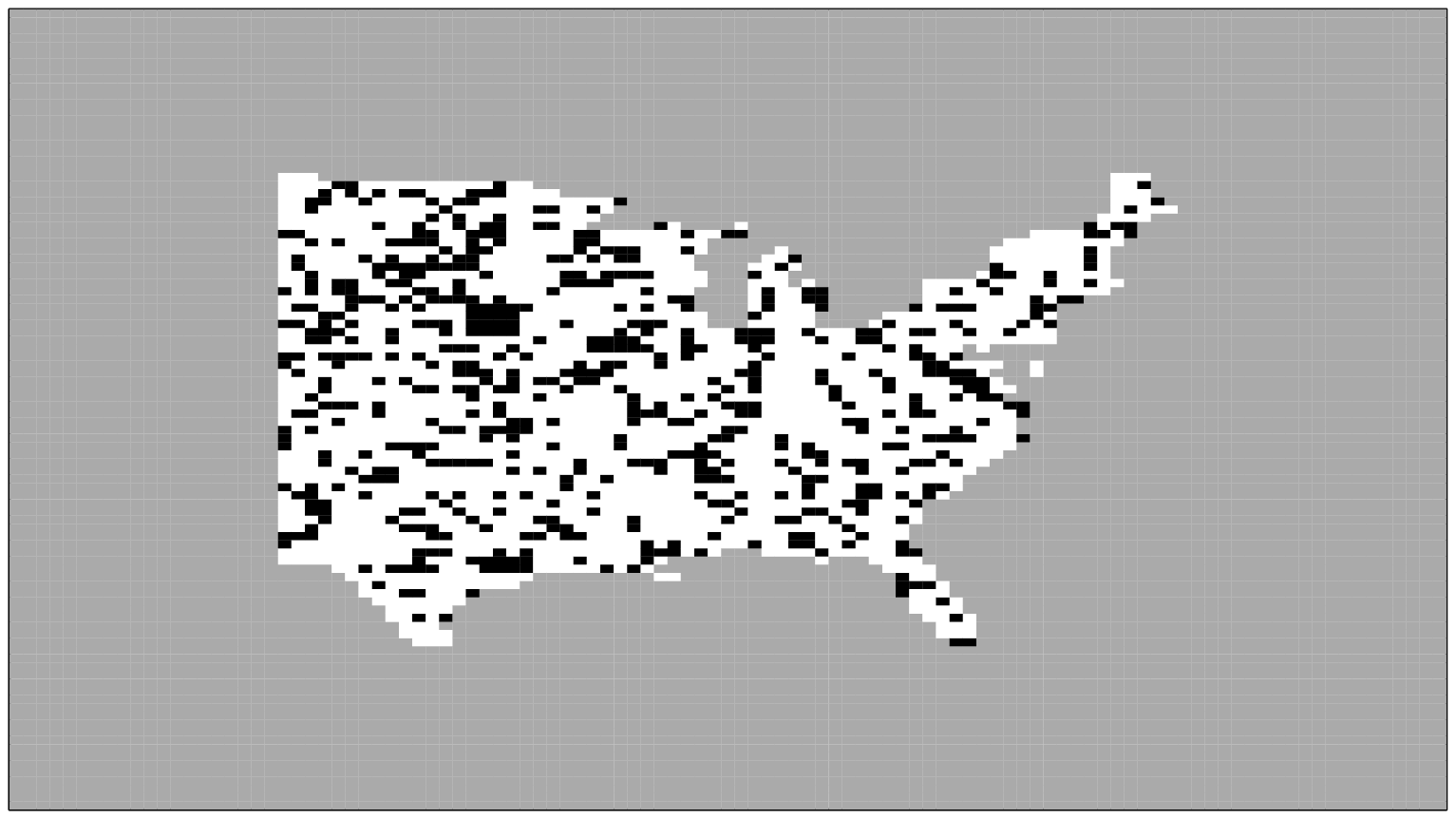}
         \vspace*{-1.5cm}
        \end{subfigure}%
        \hfill
        \begin{subfigure}[b]{0.5\textwidth}
        \vspace*{-0.5cm}
                \includegraphics[width=\linewidth]{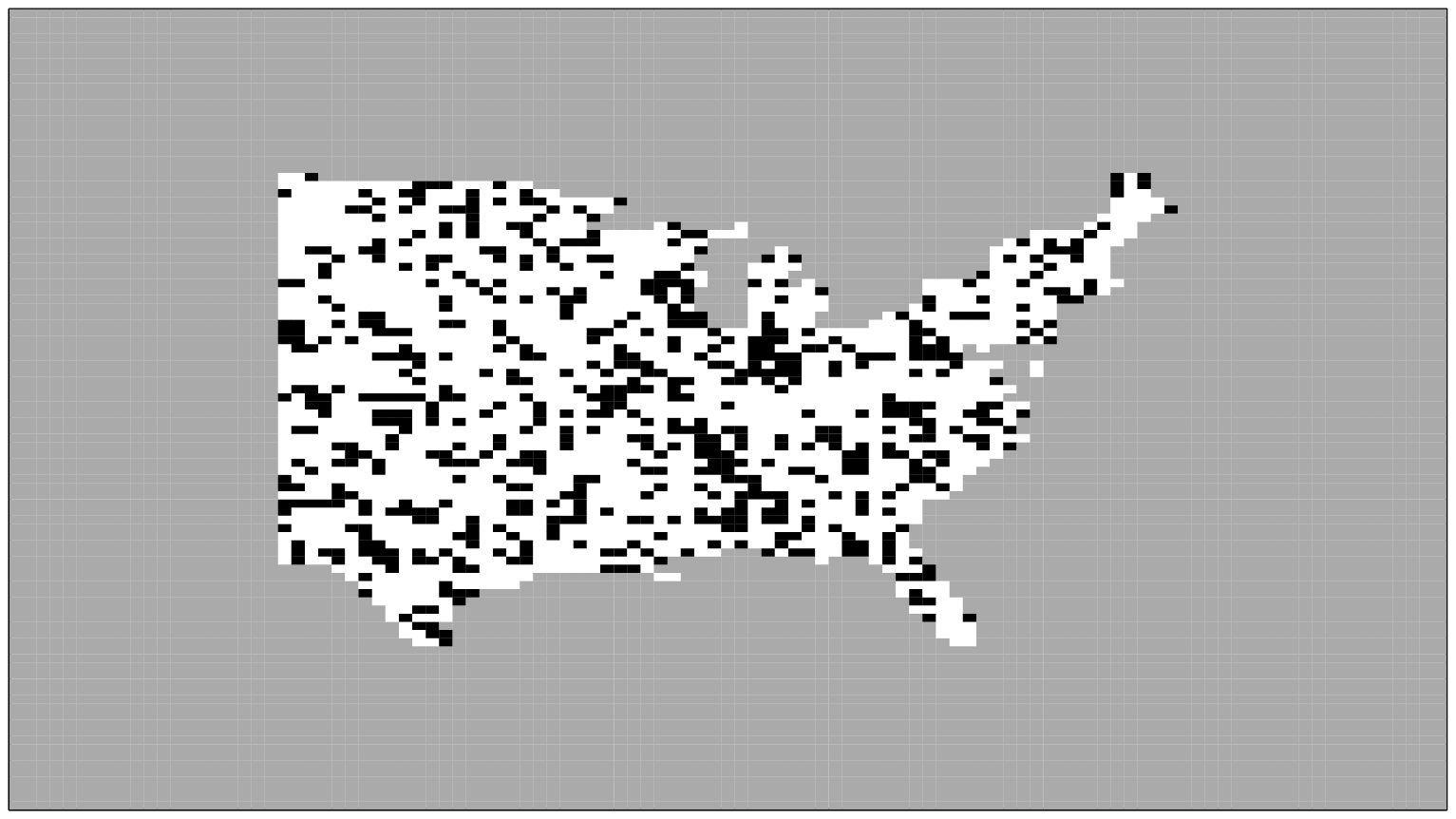}
        \vspace*{-1.5cm}
        \end{subfigure}
     
        \begin{subfigure}[b]{0.5\textwidth}
        \vspace*{-0.1cm}
                \includegraphics[width=\linewidth]{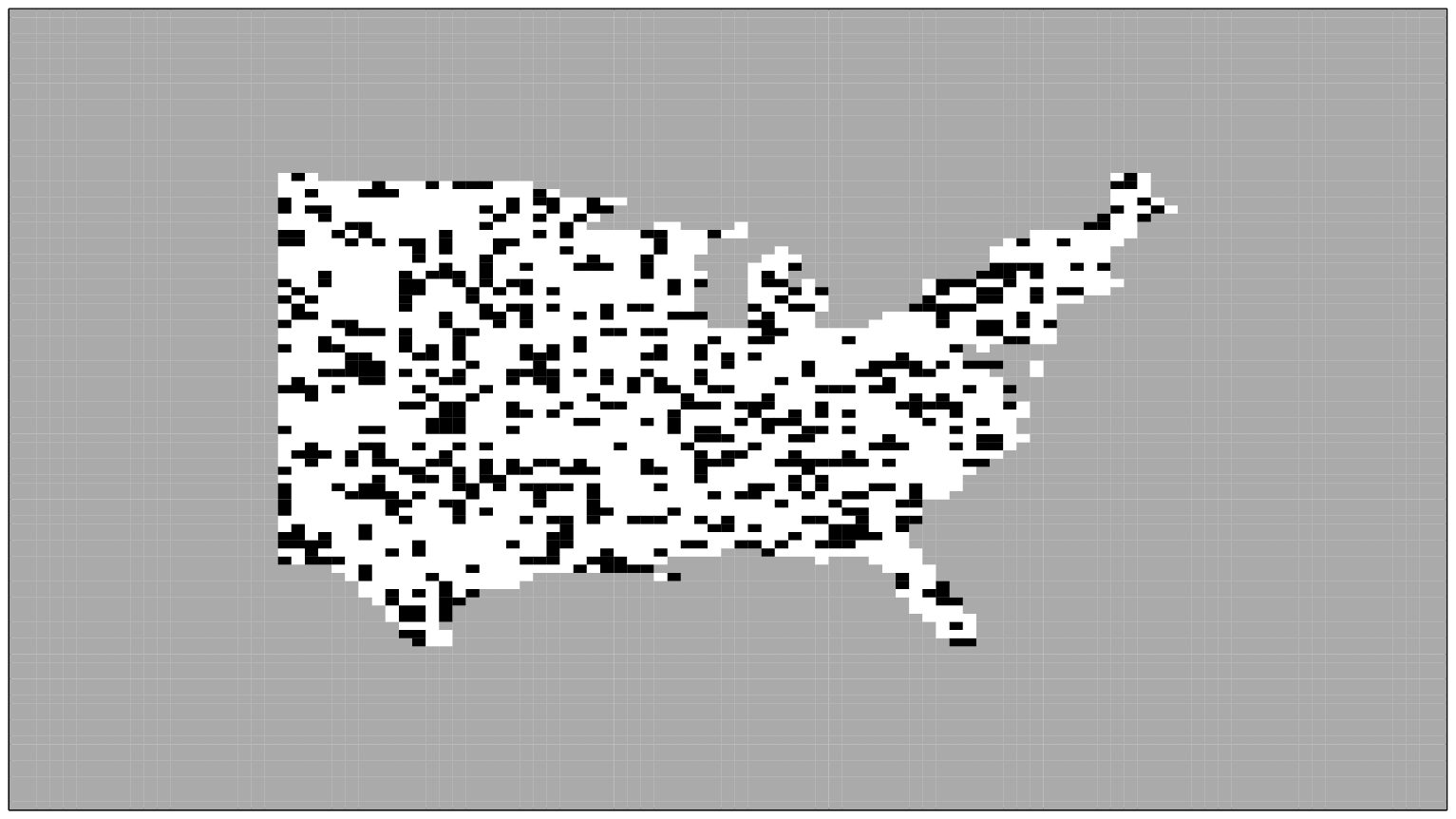}
        \vspace*{-1.5cm}
        \end{subfigure}%
        \hfill
        \begin{subfigure}[b]{0.5\textwidth}
         \vspace*{-0.1cm}
                \includegraphics[width=\linewidth]{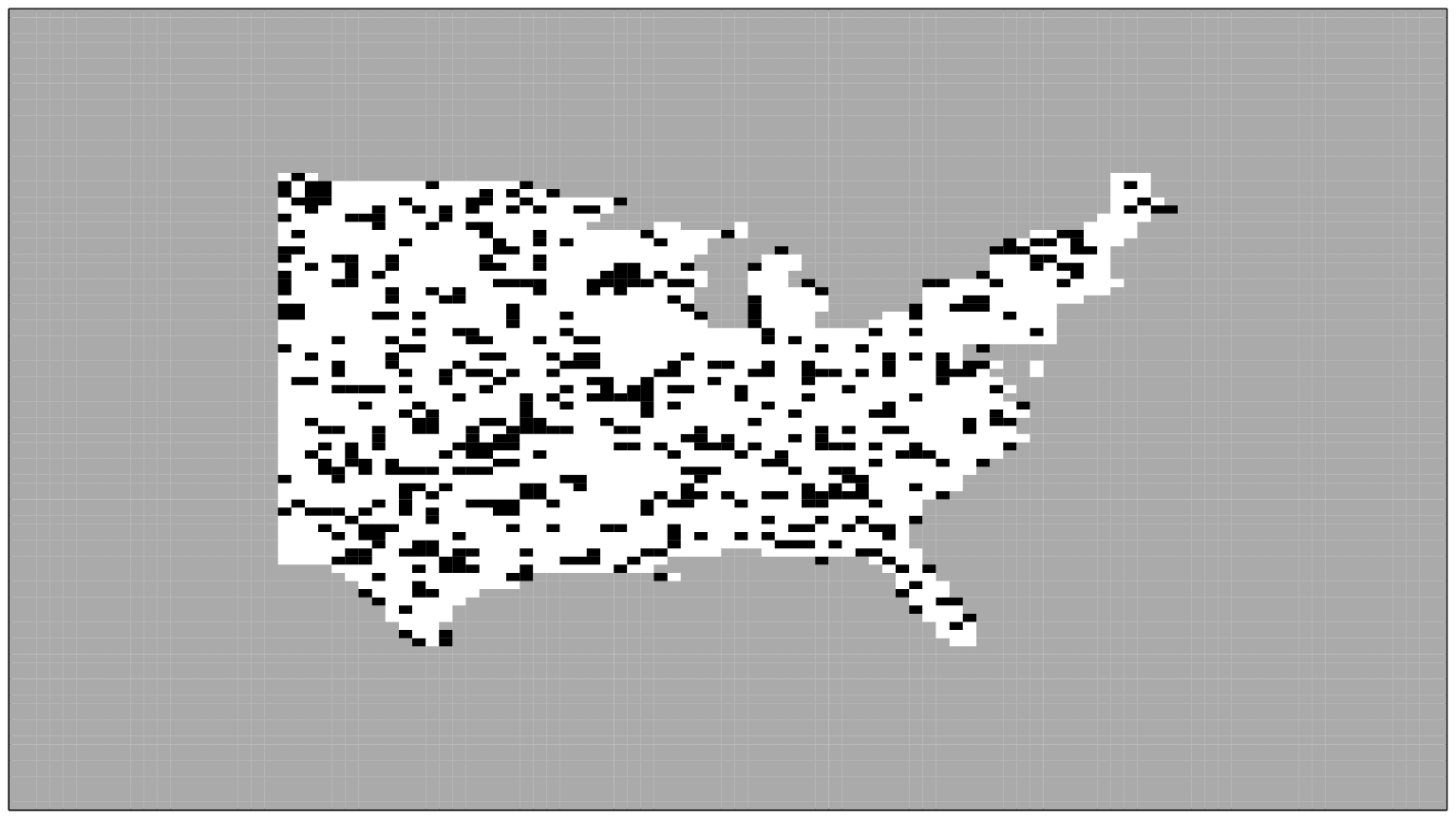}
        \vspace*{-1.5cm}
        \end{subfigure}      
        \caption{Cancer mortality map example: Four Markov mesh model realizations where the models used are randomly sampled 
from all models simulated in the RJMCMC (after the specified burn-in). The color coding is the same as in Figure 
\ref{fig:cancer}.}\label{fig:real_cancer}
\end{figure}

\begin{figure}
\begin{tabular}{@{}c@{}c@{}c@{}}
\\[-0.9cm]
  \includegraphics[width=0.33\linewidth,height=0.20\linewidth]{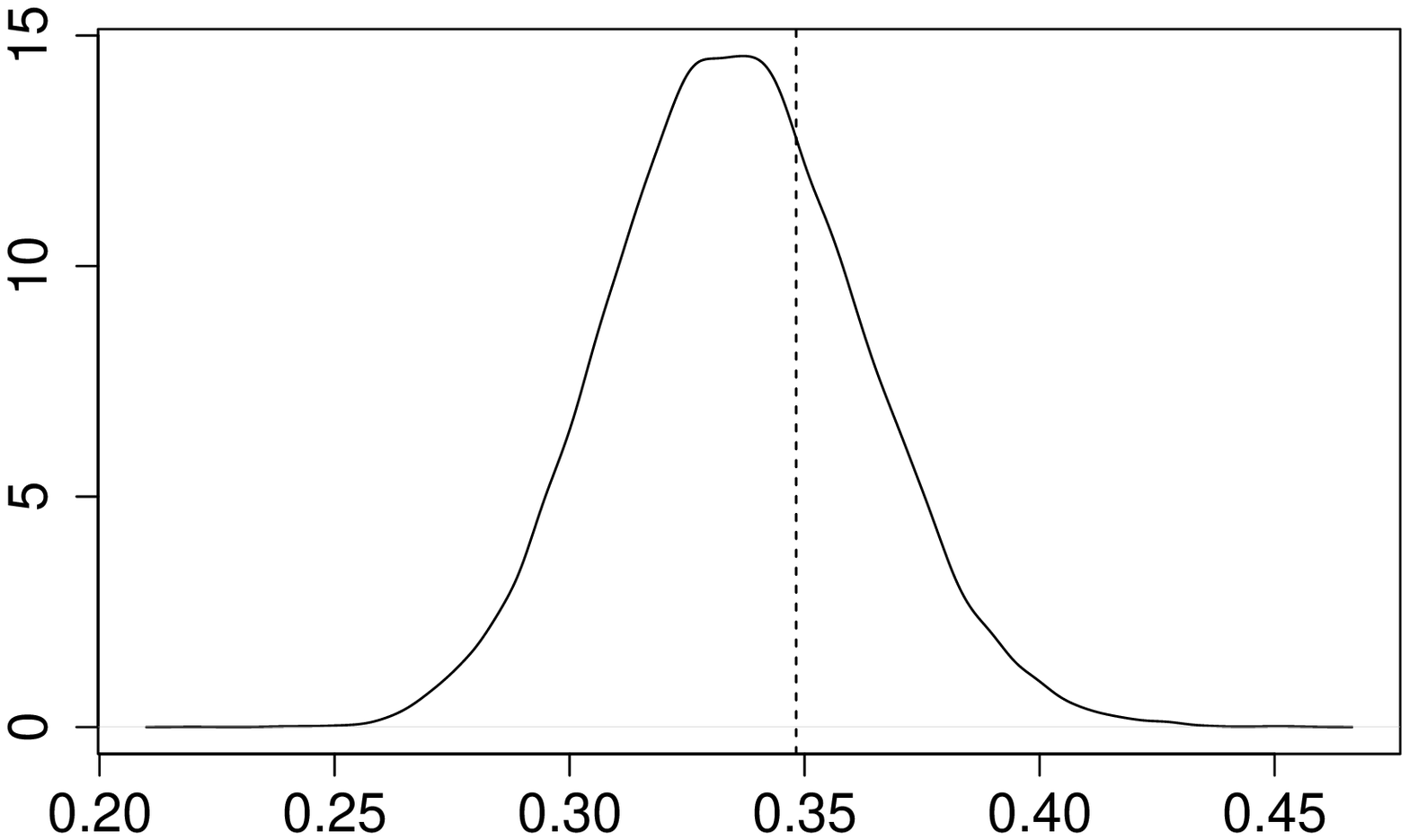} &
  \includegraphics[width=0.33\linewidth,height=0.20\linewidth]{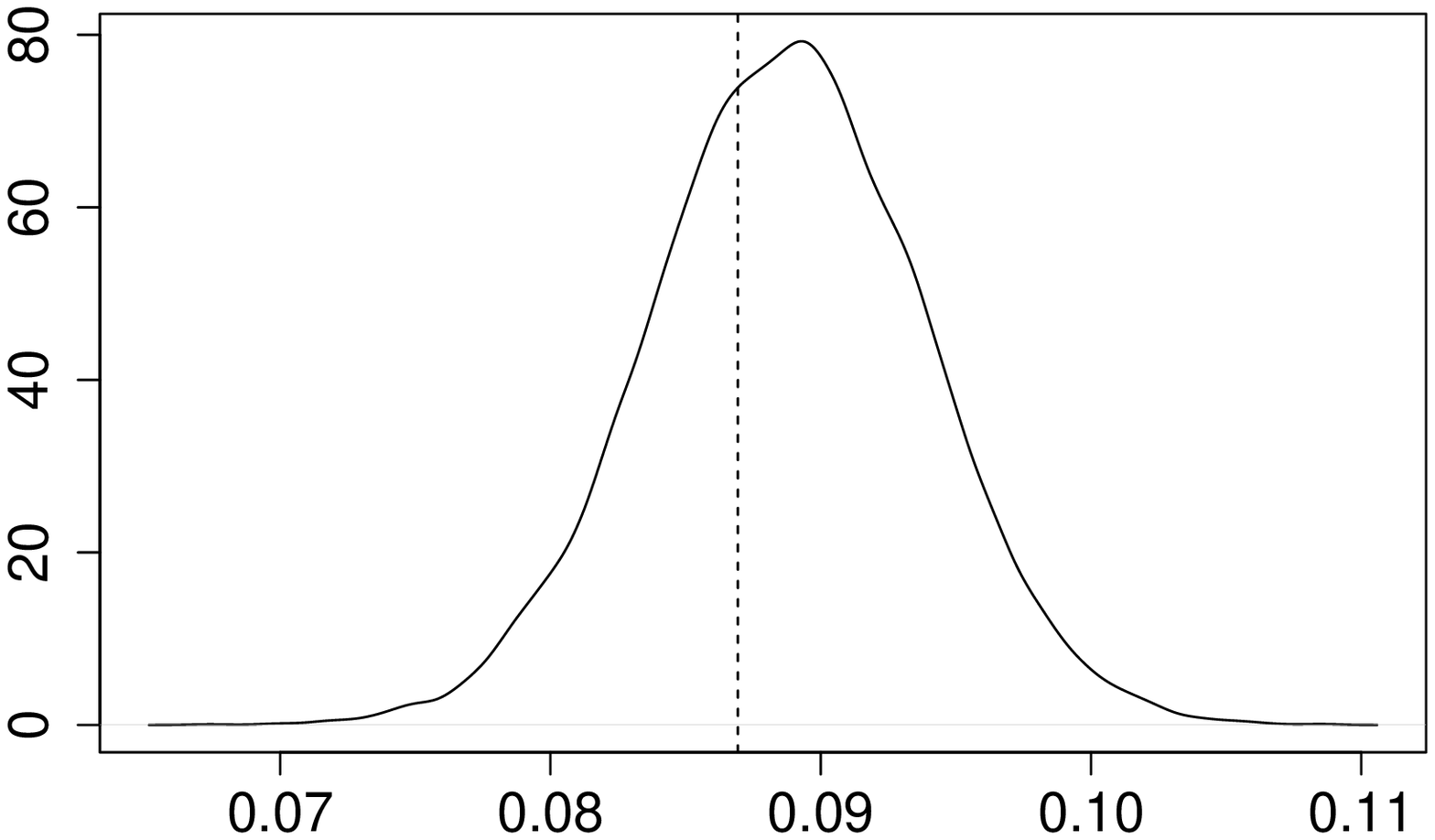} &
  \includegraphics[width=0.33\linewidth,height=0.20\linewidth]{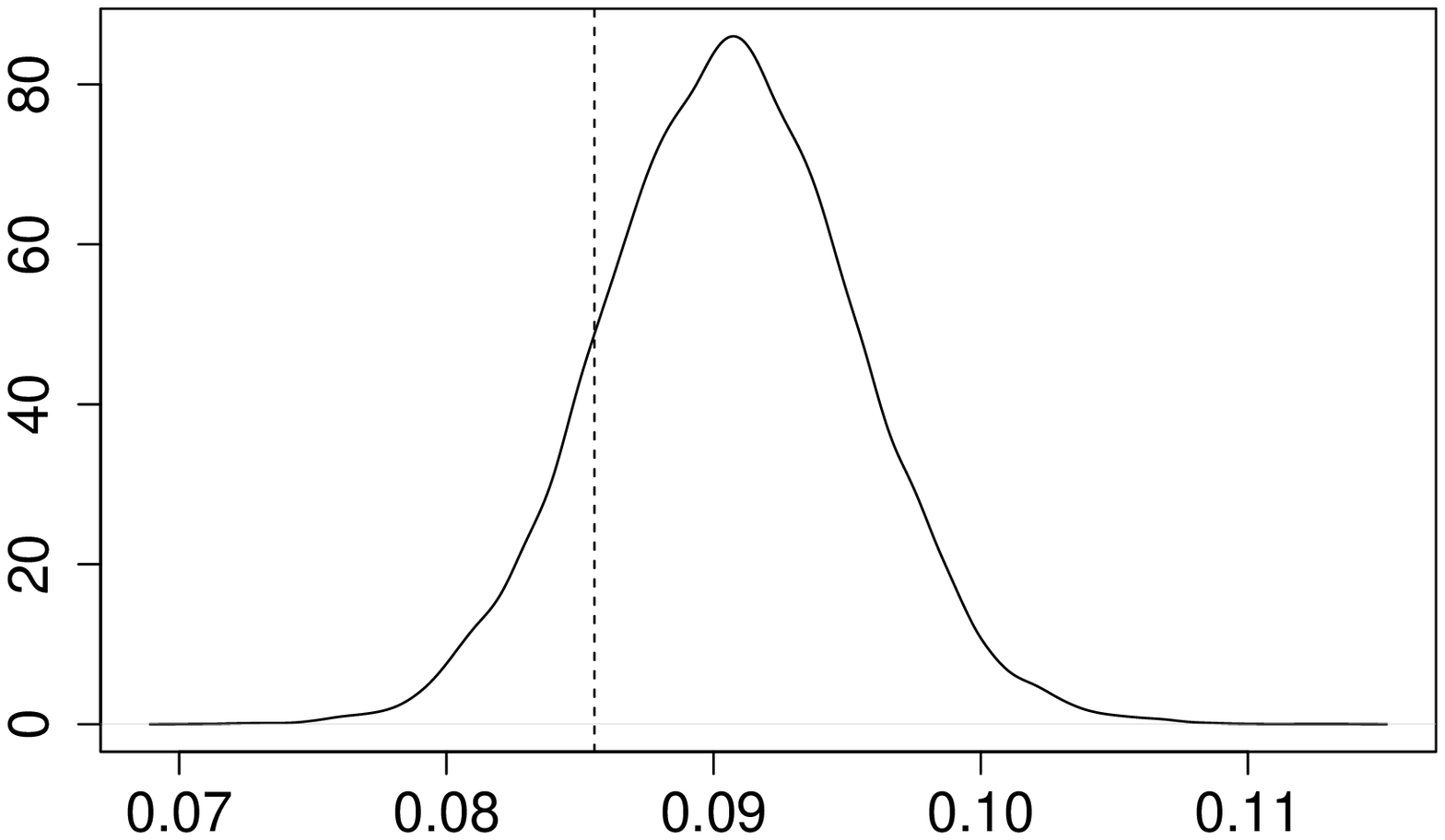} \\[-0.4cm]
(a)  \begin{tikzpicture}[scale=1.5]
  \def\radius{38.9};
  \coordinate (Pempty) at (0,0);
  \coordinate (P1) at (-1.5,1);
  \coordinate (P2) at (-0.5,1);
  \coordinate (P3) at (0.5,1);
  \coordinate (P4) at (1.5,1);
  \coordinate (P12) at (-1,2);
  \coordinate (P14) at (0,2);

  \coordinate (c1) at (-0.05,-0.05);
  \coordinate (c2) at (0.05,-0.05);
  \coordinate (c3) at (0.05,0.05);
  \coordinate (c4) at (-0.05,0.05);

  \coordinate (r1) at (-0.1,0);
  \coordinate (r2) at (0,0.1);
  \coordinate (r3) at (-0.1,0.1);
  \coordinate (r4) at (0.1,0.1);

   \draw[thick] (P12) ++(r3) +(c1) -- +(c2) -- +(c3) -- +(c4) -- cycle;
   \draw[thick] (P12) ++(r2) +(c1) -- +(c2) -- +(c3) -- +(c4) -- cycle;
   \draw[thick] (P12) ++(r1) +(c1) -- +(c2) -- +(c3) -- +(c4) -- cycle;
   \draw[thick] (P12) ++(Pempty) +(c1) -- +(c2) -- +(c3) -- +(c4) -- cycle;
   
\end{tikzpicture}   &          
(b) \begin{tikzpicture}[scale=1.5]
  \def\radius{38.9};
  \coordinate (Pempty) at (0,0);
  \coordinate (P1) at (-1.5,1);
  \coordinate (P2) at (-0.5,1);
  \coordinate (P3) at (0.5,1);
  \coordinate (P4) at (1.5,1);
  \coordinate (P12) at (-1,2);
  \coordinate (P14) at (0,2);

  \coordinate (c1) at (-0.05,-0.05);
  \coordinate (c2) at (0.05,-0.05);
  \coordinate (c3) at (0.05,0.05);
  \coordinate (c4) at (-0.05,0.05);

  \coordinate (r1) at (-0.1,0);
  \coordinate (r2) at (0,0.1);
  \coordinate (r3) at (-0.1,0.1);
  \coordinate (r4) at (0.1,0.1);

   \draw[thick] (P12) ++(r3) +(c1) -- +(c2) -- +(c3) -- +(c4) -- cycle;
   \draw[thick] (P12) ++(r2) +(c1) -- +(c2) -- +(c3) -- +(c4) -- cycle;
   \draw[thick] (P12) ++(r1) +(c1) -- +(c2) -- +(c3) -- +(c4) -- cycle;
   \draw[thick, fill=black] (P12) ++(Pempty) +(c1) -- +(c2) -- +(c3) -- +(c4) -- cycle;
   
\end{tikzpicture}    &         
(c) \begin{tikzpicture}[scale=1.5]
  \def\radius{38.9};
  \coordinate (Pempty) at (0,0);
  \coordinate (P1) at (-1.5,1);
  \coordinate (P2) at (-0.5,1);
  \coordinate (P3) at (0.5,1);
  \coordinate (P4) at (1.5,1);
  \coordinate (P12) at (-1,2);
  \coordinate (P14) at (0,2);

  \coordinate (c1) at (-0.05,-0.05);
  \coordinate (c2) at (0.05,-0.05);
  \coordinate (c3) at (0.05,0.05);
  \coordinate (c4) at (-0.05,0.05);

  \coordinate (r1) at (-0.1,0);
  \coordinate (r2) at (0,0.1);
  \coordinate (r3) at (-0.1,0.1);
  \coordinate (r4) at (0.1,0.1);

   \draw[thick] (P12) ++(r3) +(c1) -- +(c2) -- +(c3) -- +(c4) -- cycle;
   \draw[thick] (P12) ++(r2) +(c1) -- +(c2) -- +(c3) -- +(c4) -- cycle;
   \draw[thick, fill=black] (P12) ++(r1) +(c1) -- +(c2) -- +(c3) -- +(c4) -- cycle;
   \draw[thick] (P12) ++(Pempty) +(c1) -- +(c2) -- +(c3) -- +(c4) -- cycle;
   
\end{tikzpicture}     \\[-0.05cm]        
  \includegraphics[width=0.33\linewidth,height=0.20\linewidth]{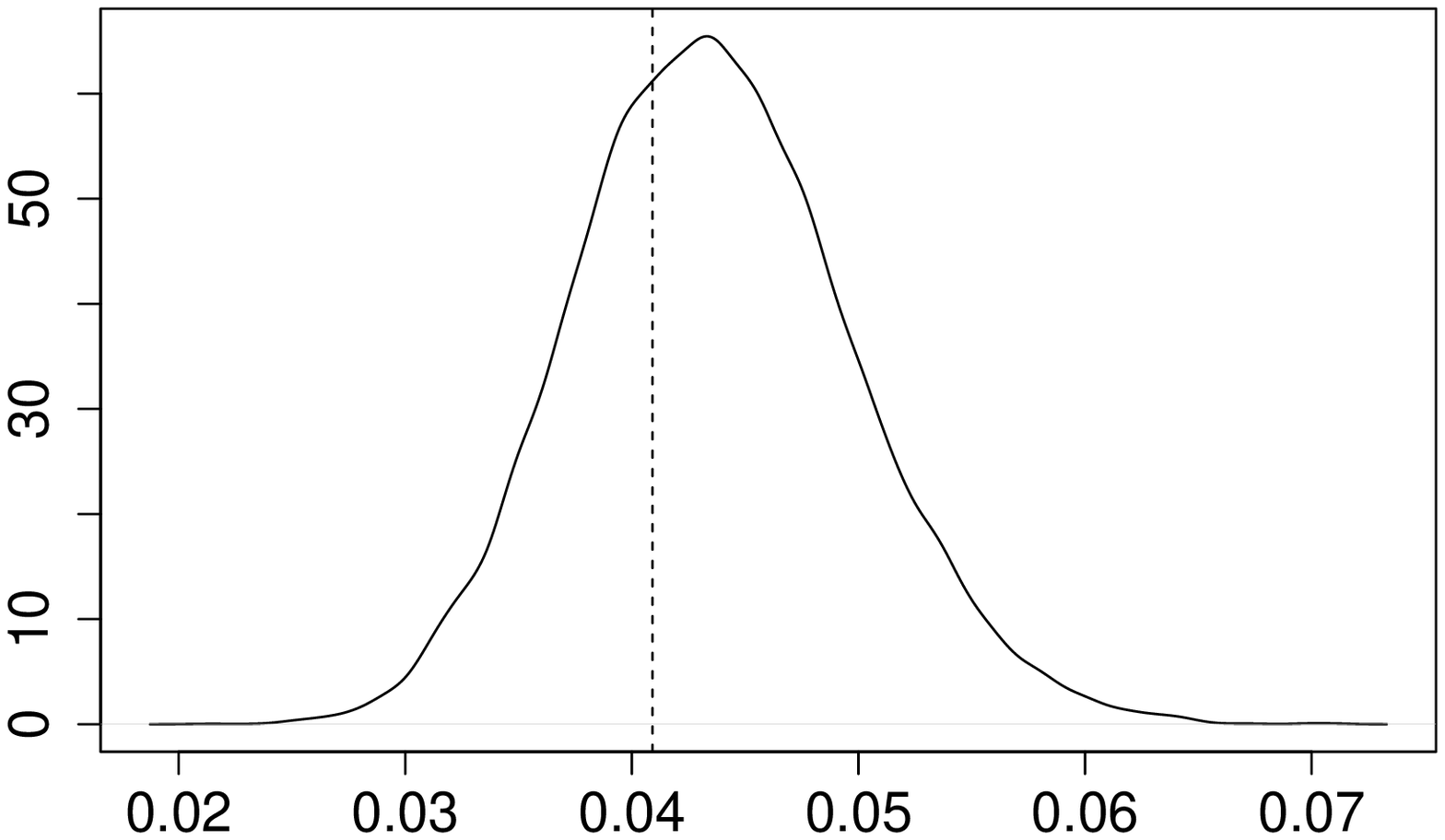} &
  \includegraphics[width=0.33\linewidth,height=0.20\linewidth]{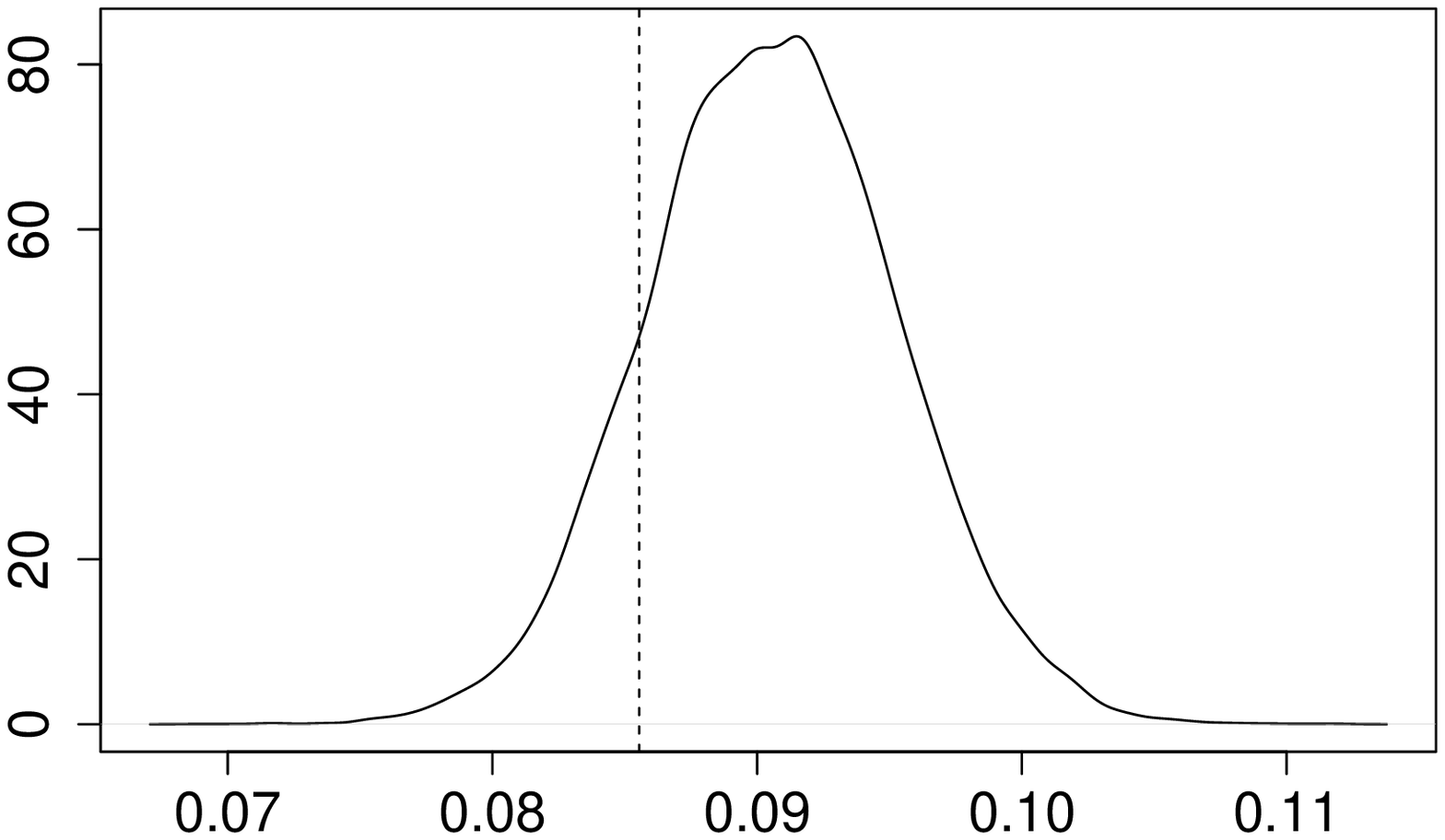} &
  \includegraphics[width=0.33\linewidth,height=0.20\linewidth]{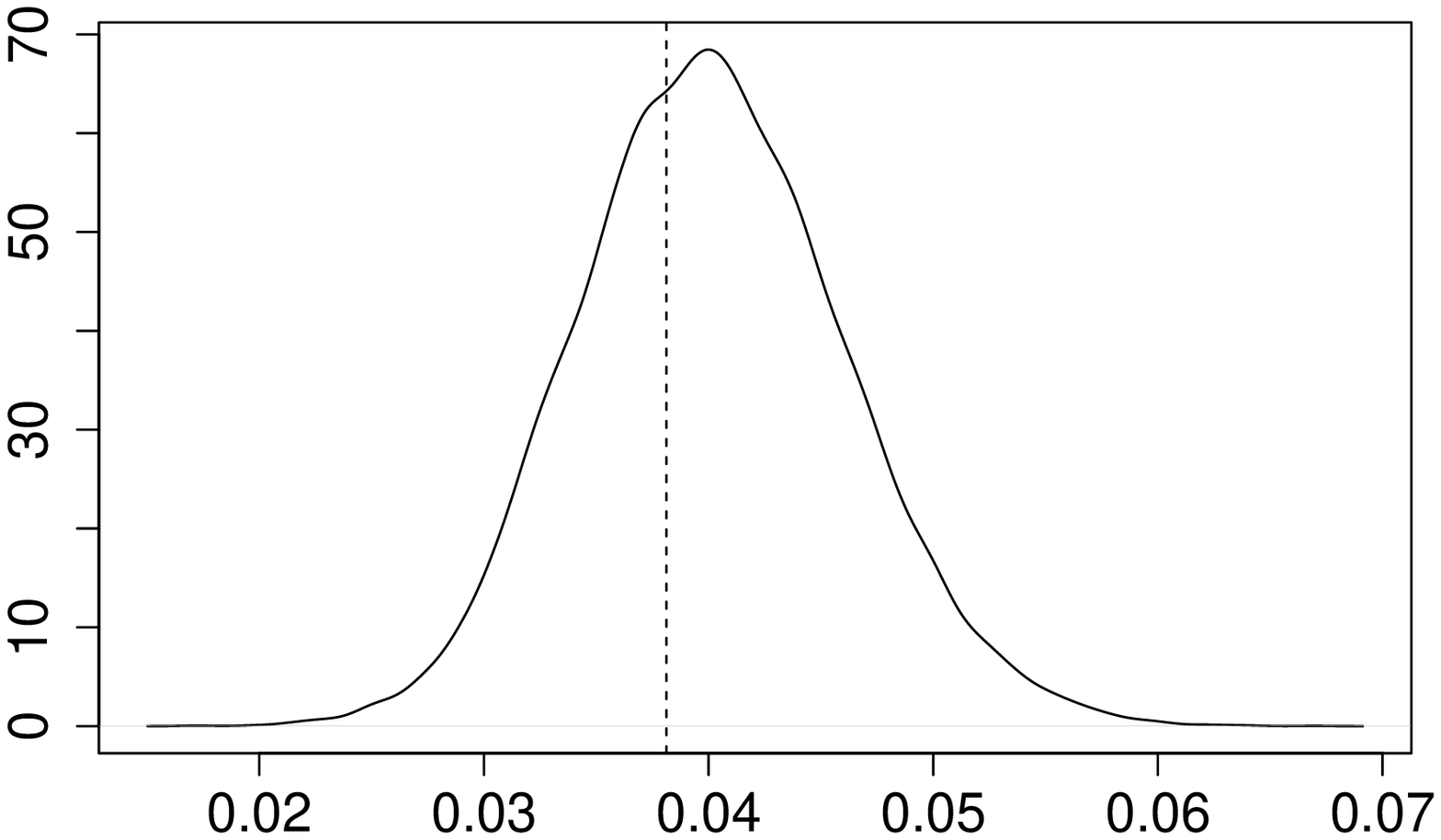} \\[-0.4cm]
(d)  \begin{tikzpicture}[scale=1.5]
  \def\radius{38.9};
  \coordinate (Pempty) at (0,0);
  \coordinate (P1) at (-1.5,1);
  \coordinate (P2) at (-0.5,1);
  \coordinate (P3) at (0.5,1);
  \coordinate (P4) at (1.5,1);
  \coordinate (P12) at (-1,2);
  \coordinate (P14) at (0,2);

  \coordinate (c1) at (-0.05,-0.05);
  \coordinate (c2) at (0.05,-0.05);
  \coordinate (c3) at (0.05,0.05);
  \coordinate (c4) at (-0.05,0.05);

  \coordinate (r1) at (-0.1,0);
  \coordinate (r2) at (0,0.1);
  \coordinate (r3) at (-0.1,0.1);
  \coordinate (r4) at (0.1,0.1);

   \draw[thick] (P12) ++(r3) +(c1) -- +(c2) -- +(c3) -- +(c4) -- cycle;
   \draw[thick] (P12) ++(r2) +(c1) -- +(c2) -- +(c3) -- +(c4) -- cycle;
   \draw[thick, fill=black] (P12) ++(r1) +(c1) -- +(c2) -- +(c3) -- +(c4) -- cycle;
   \draw[thick, fill=black] (P12) ++(Pempty) +(c1) -- +(c2) -- +(c3) -- +(c4) -- cycle;
   
\end{tikzpicture}             &
(e)  \begin{tikzpicture}[scale=1.5]
  \def\radius{38.9};
  \coordinate (Pempty) at (0,0);
  \coordinate (P1) at (-1.5,1);
  \coordinate (P2) at (-0.5,1);
  \coordinate (P3) at (0.5,1);
  \coordinate (P4) at (1.5,1);
  \coordinate (P12) at (-1,2);
  \coordinate (P14) at (0,2);

  \coordinate (c1) at (-0.05,-0.05);
  \coordinate (c2) at (0.05,-0.05);
  \coordinate (c3) at (0.05,0.05);
  \coordinate (c4) at (-0.05,0.05);

  \coordinate (r1) at (-0.1,0);
  \coordinate (r2) at (0,0.1);
  \coordinate (r3) at (-0.1,0.1);
  \coordinate (r4) at (0.1,0.1);

   \draw[thick] (P12) ++(r3) +(c1) -- +(c2) -- +(c3) -- +(c4) -- cycle;
   \draw[thick, fill=black] (P12) ++(r2) +(c1) -- +(c2) -- +(c3) -- +(c4) -- cycle;
   \draw[thick] (P12) ++(r1) +(c1) -- +(c2) -- +(c3) -- +(c4) -- cycle;
   \draw[thick] (P12) ++(Pempty) +(c1) -- +(c2) -- +(c3) -- +(c4) -- cycle;
   
\end{tikzpicture}             &
(f)  \begin{tikzpicture}[scale=1.5]
  \def\radius{38.9};
  \coordinate (Pempty) at (0,0);
  \coordinate (P1) at (-1.5,1);
  \coordinate (P2) at (-0.5,1);
  \coordinate (P3) at (0.5,1);
  \coordinate (P4) at (1.5,1);
  \coordinate (P12) at (-1,2);
  \coordinate (P14) at (0,2);

  \coordinate (c1) at (-0.05,-0.05);
  \coordinate (c2) at (0.05,-0.05);
  \coordinate (c3) at (0.05,0.05);
  \coordinate (c4) at (-0.05,0.05);

  \coordinate (r1) at (-0.1,0);
  \coordinate (r2) at (0,0.1);
  \coordinate (r3) at (-0.1,0.1);
  \coordinate (r4) at (0.1,0.1);

   \draw[thick] (P12) ++(r3) +(c1) -- +(c2) -- +(c3) -- +(c4) -- cycle;
   \draw[thick, fill=black] (P12) ++(r2) +(c1) -- +(c2) -- +(c3) -- +(c4) -- cycle;
   \draw[thick] (P12) ++(r1) +(c1) -- +(c2) -- +(c3) -- +(c4) -- cycle;
   \draw[thick, fill=black] (P12) ++(Pempty) +(c1) -- +(c2) -- +(c3) -- +(c4) -- cycle;
   
\end{tikzpicture}          \\[-0.05cm]
  \includegraphics[width=0.33\linewidth,height=0.20\linewidth]{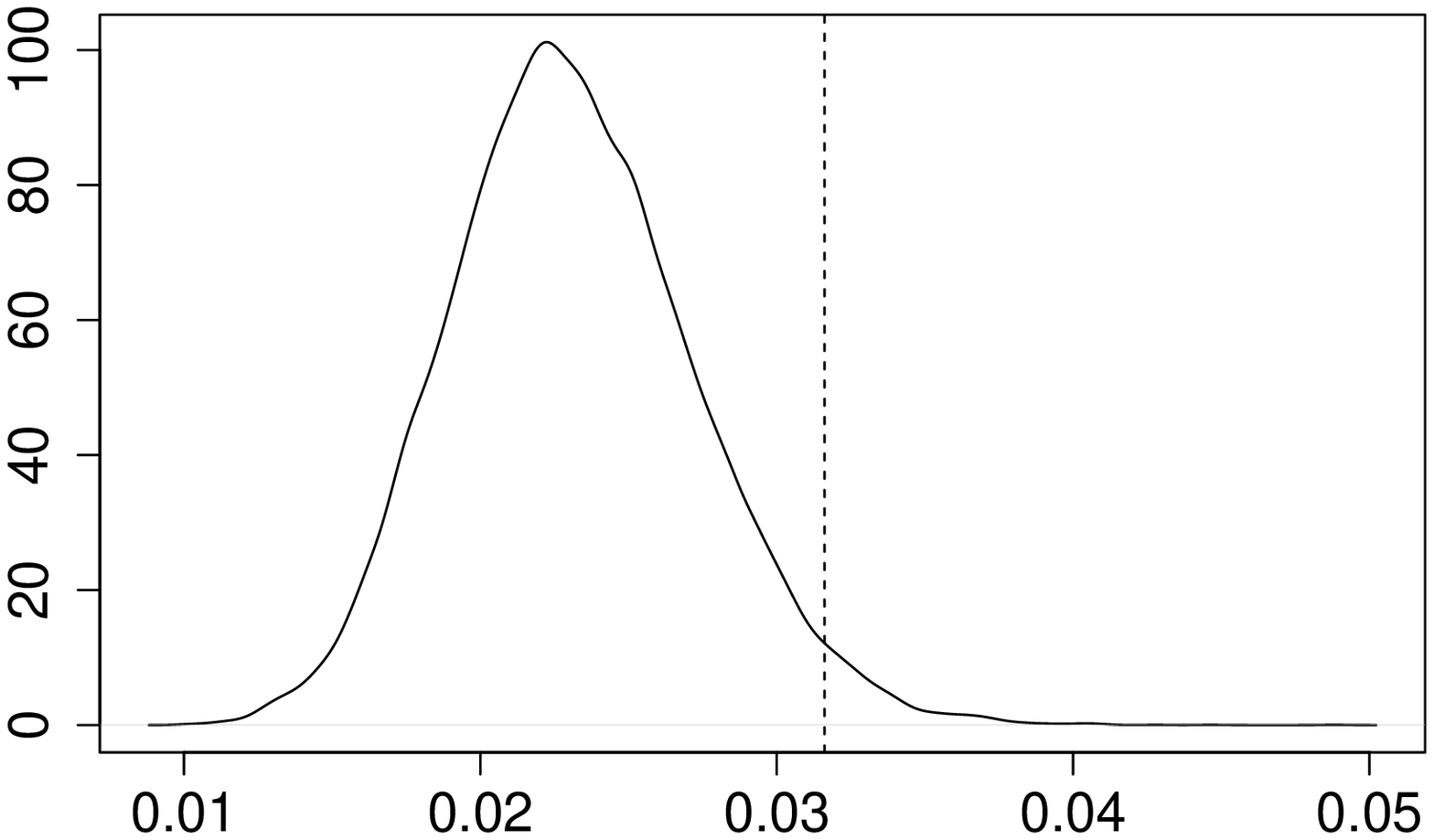} &
  \includegraphics[width=0.33\linewidth,height=0.20\linewidth]{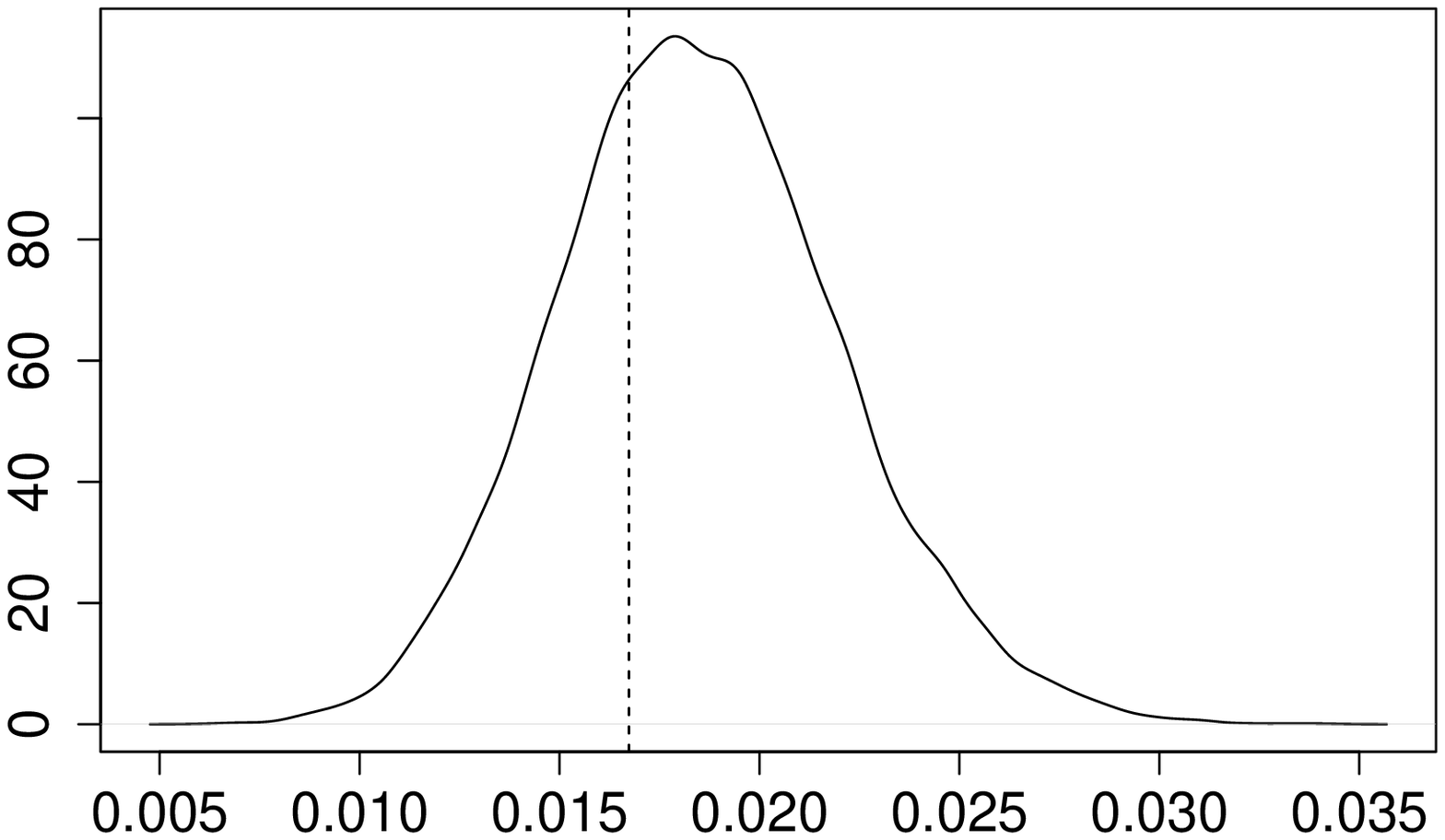} &
  \includegraphics[width=0.33\linewidth,height=0.20\linewidth]{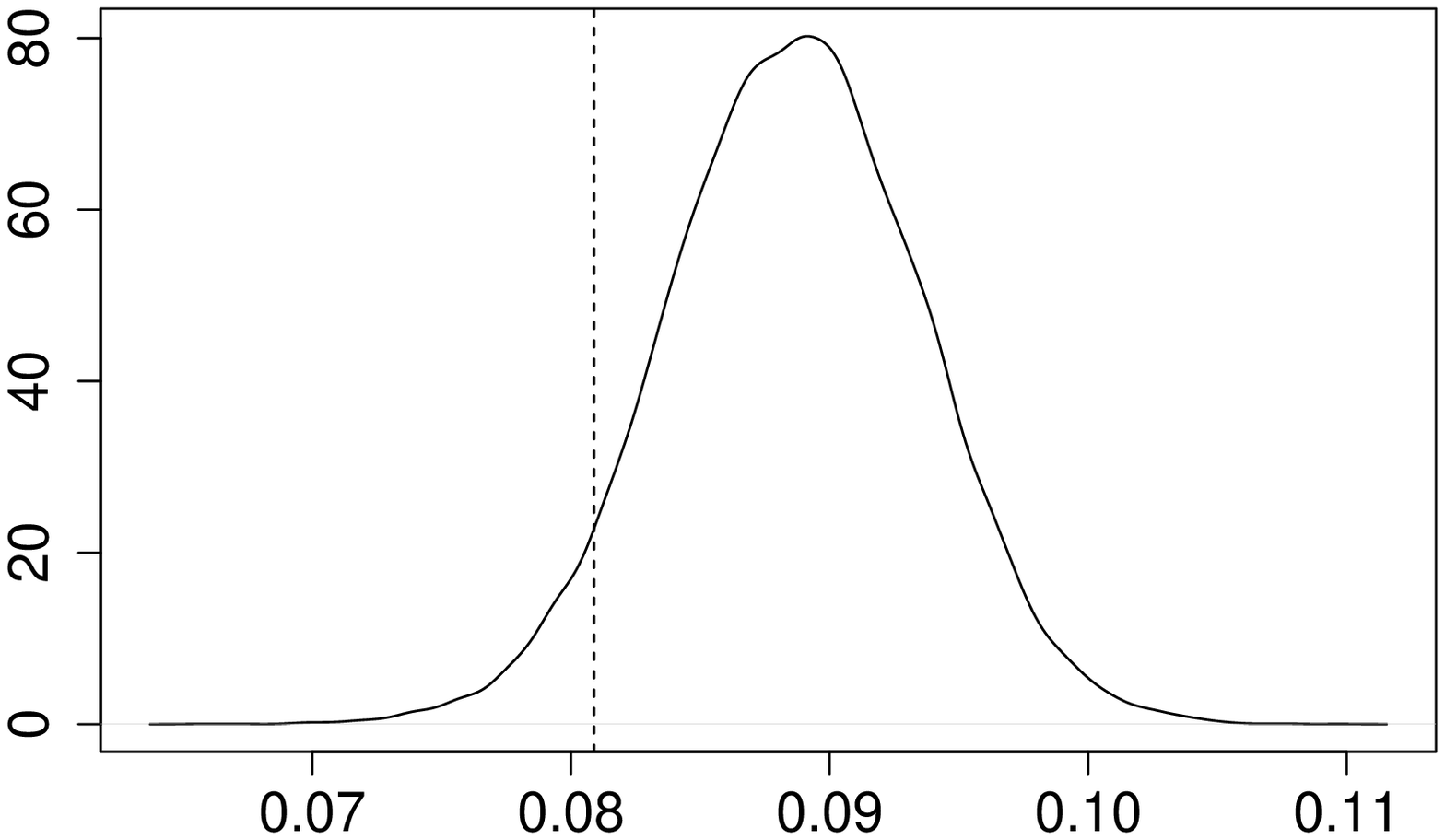} \\[-0.4cm]
(g)  \begin{tikzpicture}[scale=1.5]
  \def\radius{38.9};
  \coordinate (Pempty) at (0,0);
  \coordinate (P1) at (-1.5,1);
  \coordinate (P2) at (-0.5,1);
  \coordinate (P3) at (0.5,1);
  \coordinate (P4) at (1.5,1);
  \coordinate (P12) at (-1,2);
  \coordinate (P14) at (0,2);

  \coordinate (c1) at (-0.05,-0.05);
  \coordinate (c2) at (0.05,-0.05);
  \coordinate (c3) at (0.05,0.05);
  \coordinate (c4) at (-0.05,0.05);

  \coordinate (r1) at (-0.1,0);
  \coordinate (r2) at (0,0.1);
  \coordinate (r3) at (-0.1,0.1);
  \coordinate (r4) at (0.1,0.1);

   \draw[thick] (P12) ++(r3) +(c1) -- +(c2) -- +(c3) -- +(c4) -- cycle;
   \draw[thick, fill=black] (P12) ++(r2) +(c1) -- +(c2) -- +(c3) -- +(c4) -- cycle;
   \draw[thick, fill=black] (P12) ++(r1) +(c1) -- +(c2) -- +(c3) -- +(c4) -- cycle;
   \draw[thick] (P12) ++(Pempty) +(c1) -- +(c2) -- +(c3) -- +(c4) -- cycle;
   
\end{tikzpicture}             &
(h)  \begin{tikzpicture}[scale=1.5]
  \def\radius{38.9};
  \coordinate (Pempty) at (0,0);
  \coordinate (P1) at (-1.5,1);
  \coordinate (P2) at (-0.5,1);
  \coordinate (P3) at (0.5,1);
  \coordinate (P4) at (1.5,1);
  \coordinate (P12) at (-1,2);
  \coordinate (P14) at (0,2);

  \coordinate (c1) at (-0.05,-0.05);
  \coordinate (c2) at (0.05,-0.05);
  \coordinate (c3) at (0.05,0.05);
  \coordinate (c4) at (-0.05,0.05);

  \coordinate (r1) at (-0.1,0);
  \coordinate (r2) at (0,0.1);
  \coordinate (r3) at (-0.1,0.1);
  \coordinate (r4) at (0.1,0.1);

   \draw[thick] (P12) ++(r3) +(c1) -- +(c2) -- +(c3) -- +(c4) -- cycle;
   \draw[thick, fill=black] (P12) ++(r2) +(c1) -- +(c2) -- +(c3) -- +(c4) -- cycle;
   \draw[thick, fill=black] (P12) ++(r1) +(c1) -- +(c2) -- +(c3) -- +(c4) -- cycle;
   \draw[thick, fill=black] (P12) ++(Pempty) +(c1) -- +(c2) -- +(c3) -- +(c4) -- cycle;
   
\end{tikzpicture}             &
(i)  \begin{tikzpicture}[scale=1.5]
  \def\radius{38.9};
  \coordinate (Pempty) at (0,0);
  \coordinate (P1) at (-1.5,1);
  \coordinate (P2) at (-0.5,1);
  \coordinate (P3) at (0.5,1);
  \coordinate (P4) at (1.5,1);
  \coordinate (P12) at (-1,2);
  \coordinate (P14) at (0,2);

  \coordinate (c1) at (-0.05,-0.05);
  \coordinate (c2) at (0.05,-0.05);
  \coordinate (c3) at (0.05,0.05);
  \coordinate (c4) at (-0.05,0.05);

  \coordinate (r1) at (-0.1,0);
  \coordinate (r2) at (0,0.1);
  \coordinate (r3) at (-0.1,0.1);
  \coordinate (r4) at (0.1,0.1);

   \draw[thick, fill=black] (P12) ++(r3) +(c1) -- +(c2) -- +(c3) -- +(c4) -- cycle;
   \draw[thick] (P12) ++(r2) +(c1) -- +(c2) -- +(c3) -- +(c4) -- cycle;
   \draw[thick] (P12) ++(r1) +(c1) -- +(c2) -- +(c3) -- +(c4) -- cycle;
   \draw[thick] (P12) ++(Pempty) +(c1) -- +(c2) -- +(c3) -- +(c4) -- cycle;
   
\end{tikzpicture}             \\[-0.05cm]
  \includegraphics[width=0.33\linewidth,height=0.20\linewidth]{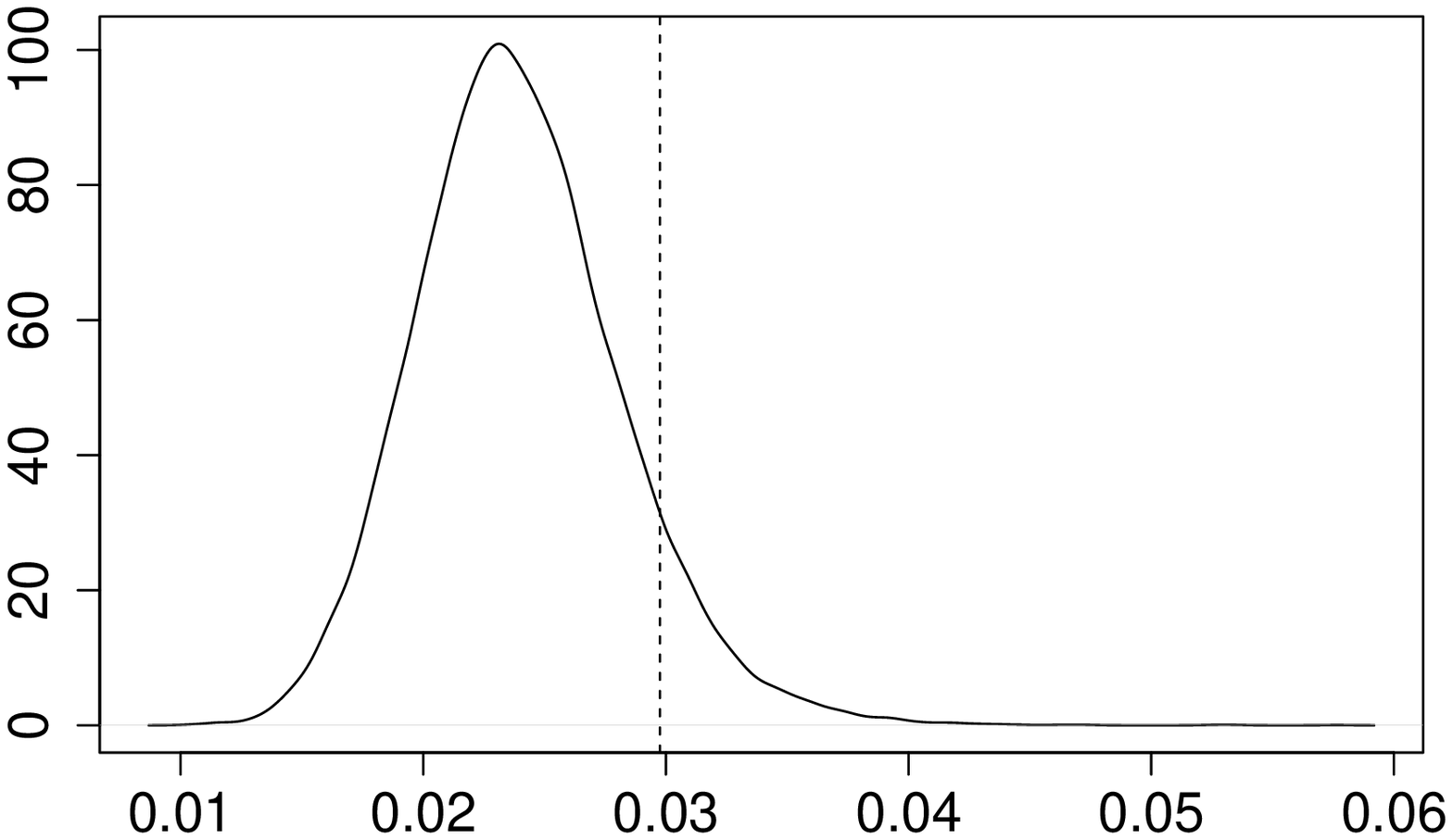} &
  \includegraphics[width=0.33\linewidth,height=0.20\linewidth]{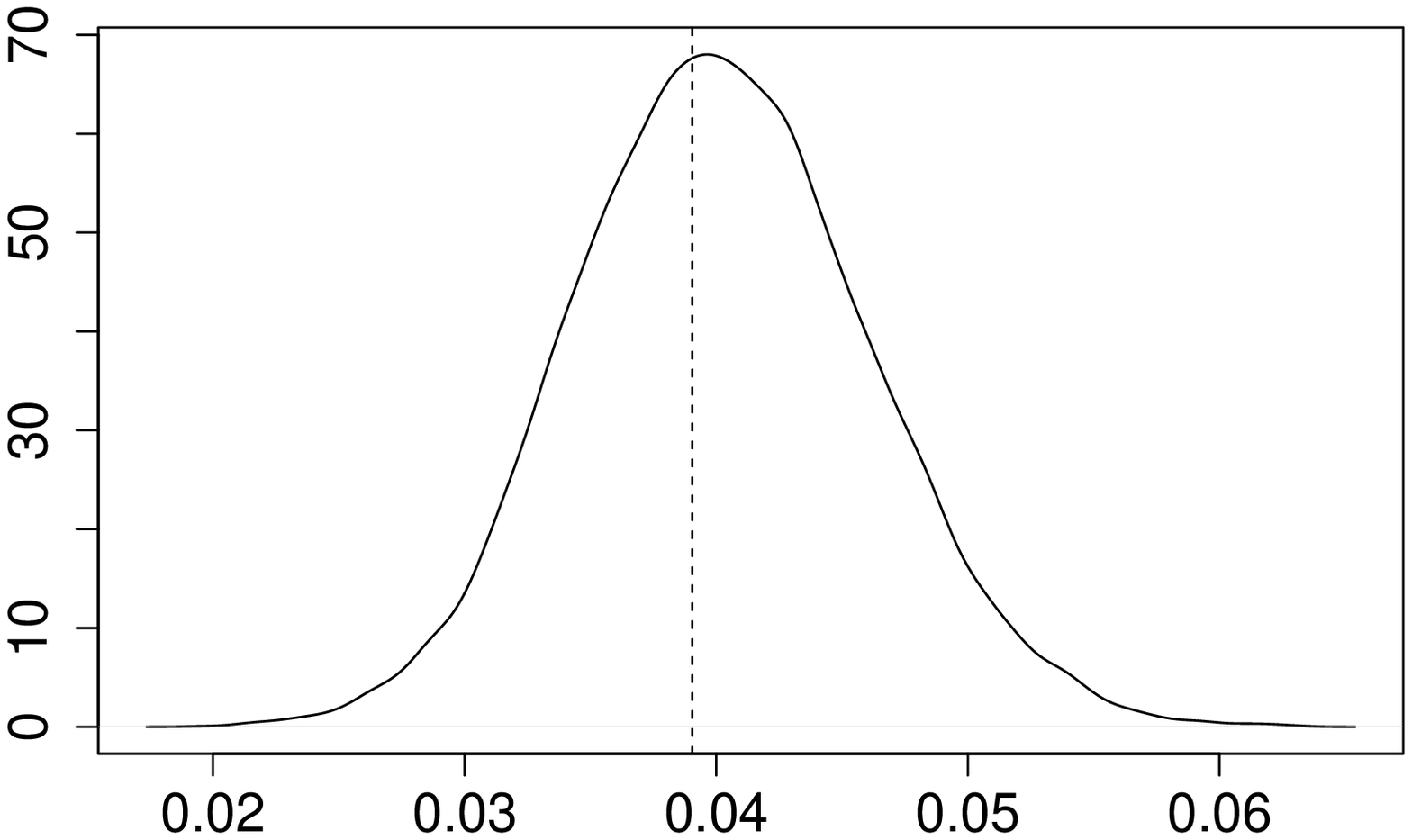} &
  \includegraphics[width=0.33\linewidth,height=0.20\linewidth]{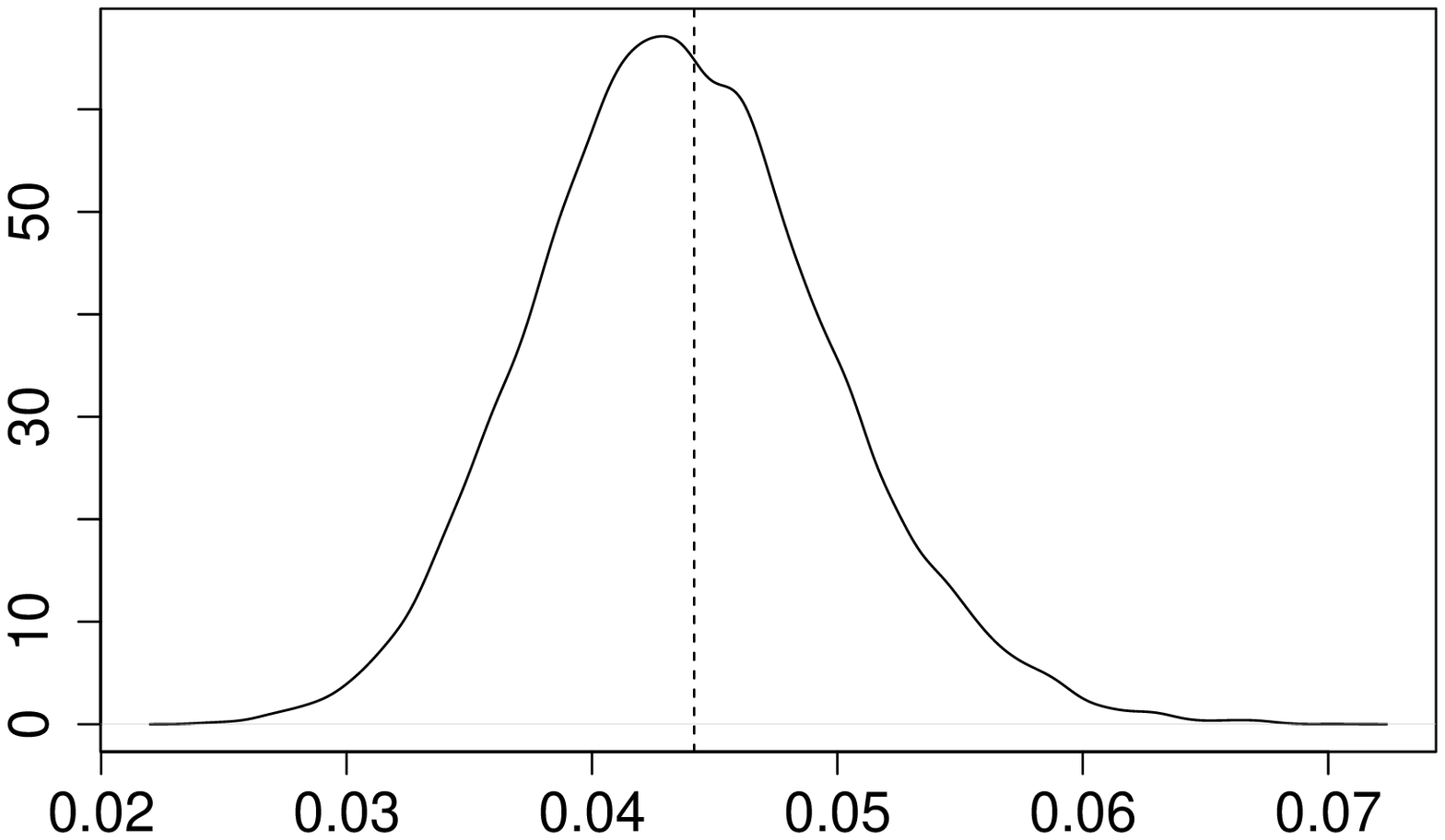} \\[-0.4cm]
(j)  \begin{tikzpicture}[scale=1.5]
  \def\radius{38.9};
  \coordinate (Pempty) at (0,0);
  \coordinate (P1) at (-1.5,1);
  \coordinate (P2) at (-0.5,1);
  \coordinate (P3) at (0.5,1);
  \coordinate (P4) at (1.5,1);
  \coordinate (P12) at (-1,2);
  \coordinate (P14) at (0,2);

  \coordinate (c1) at (-0.05,-0.05);
  \coordinate (c2) at (0.05,-0.05);
  \coordinate (c3) at (0.05,0.05);
  \coordinate (c4) at (-0.05,0.05);

  \coordinate (r1) at (-0.1,0);
  \coordinate (r2) at (0,0.1);
  \coordinate (r3) at (-0.1,0.1);
  \coordinate (r4) at (0.1,0.1);

   \draw[thick, fill=black] (P12) ++(r3) +(c1) -- +(c2) -- +(c3) -- +(c4) -- cycle;
   \draw[thick] (P12) ++(r2) +(c1) -- +(c2) -- +(c3) -- +(c4) -- cycle;
   \draw[thick] (P12) ++(r1) +(c1) -- +(c2) -- +(c3) -- +(c4) -- cycle;
   \draw[thick, fill=black] (P12) ++(Pempty) +(c1) -- +(c2) -- +(c3) -- +(c4) -- cycle;
   
\end{tikzpicture}             &
(k)  \begin{tikzpicture}[scale=1.5]
  \def\radius{38.9};
  \coordinate (Pempty) at (0,0);
  \coordinate (P1) at (-1.5,1);
  \coordinate (P2) at (-0.5,1);
  \coordinate (P3) at (0.5,1);
  \coordinate (P4) at (1.5,1);
  \coordinate (P12) at (-1,2);
  \coordinate (P14) at (0,2);

  \coordinate (c1) at (-0.05,-0.05);
  \coordinate (c2) at (0.05,-0.05);
  \coordinate (c3) at (0.05,0.05);
  \coordinate (c4) at (-0.05,0.05);

  \coordinate (r1) at (-0.1,0);
  \coordinate (r2) at (0,0.1);
  \coordinate (r3) at (-0.1,0.1);
  \coordinate (r4) at (0.1,0.1);

   \draw[thick, fill=black] (P12) ++(r3) +(c1) -- +(c2) -- +(c3) -- +(c4) -- cycle;
   \draw[thick] (P12) ++(r2) +(c1) -- +(c2) -- +(c3) -- +(c4) -- cycle;
   \draw[thick, fill=black] (P12) ++(r1) +(c1) -- +(c2) -- +(c3) -- +(c4) -- cycle;
   \draw[thick] (P12) ++(Pempty) +(c1) -- +(c2) -- +(c3) -- +(c4) -- cycle;
   
\end{tikzpicture}             &
(l)  \begin{tikzpicture}[scale=1.5]
  \def\radius{38.9};
  \coordinate (Pempty) at (0,0);
  \coordinate (P1) at (-1.5,1);
  \coordinate (P2) at (-0.5,1);
  \coordinate (P3) at (0.5,1);
  \coordinate (P4) at (1.5,1);
  \coordinate (P12) at (-1,2);
  \coordinate (P14) at (0,2);

  \coordinate (c1) at (-0.05,-0.05);
  \coordinate (c2) at (0.05,-0.05);
  \coordinate (c3) at (0.05,0.05);
  \coordinate (c4) at (-0.05,0.05);

  \coordinate (r1) at (-0.1,0);
  \coordinate (r2) at (0,0.1);
  \coordinate (r3) at (-0.1,0.1);
  \coordinate (r4) at (0.1,0.1);

   \draw[thick, fill=black] (P12) ++(r3) +(c1) -- +(c2) -- +(c3) -- +(c4) -- cycle;
   \draw[thick, fill=black] (P12) ++(r2) +(c1) -- +(c2) -- +(c3) -- +(c4) -- cycle;
   \draw[thick] (P12) ++(r1) +(c1) -- +(c2) -- +(c3) -- +(c4) -- cycle;
   \draw[thick] (P12) ++(Pempty) +(c1) -- +(c2) -- +(c3) -- +(c4) -- cycle;
   
\end{tikzpicture}             \\[-0.05cm]
  \includegraphics[width=0.33\linewidth,height=0.20\linewidth]{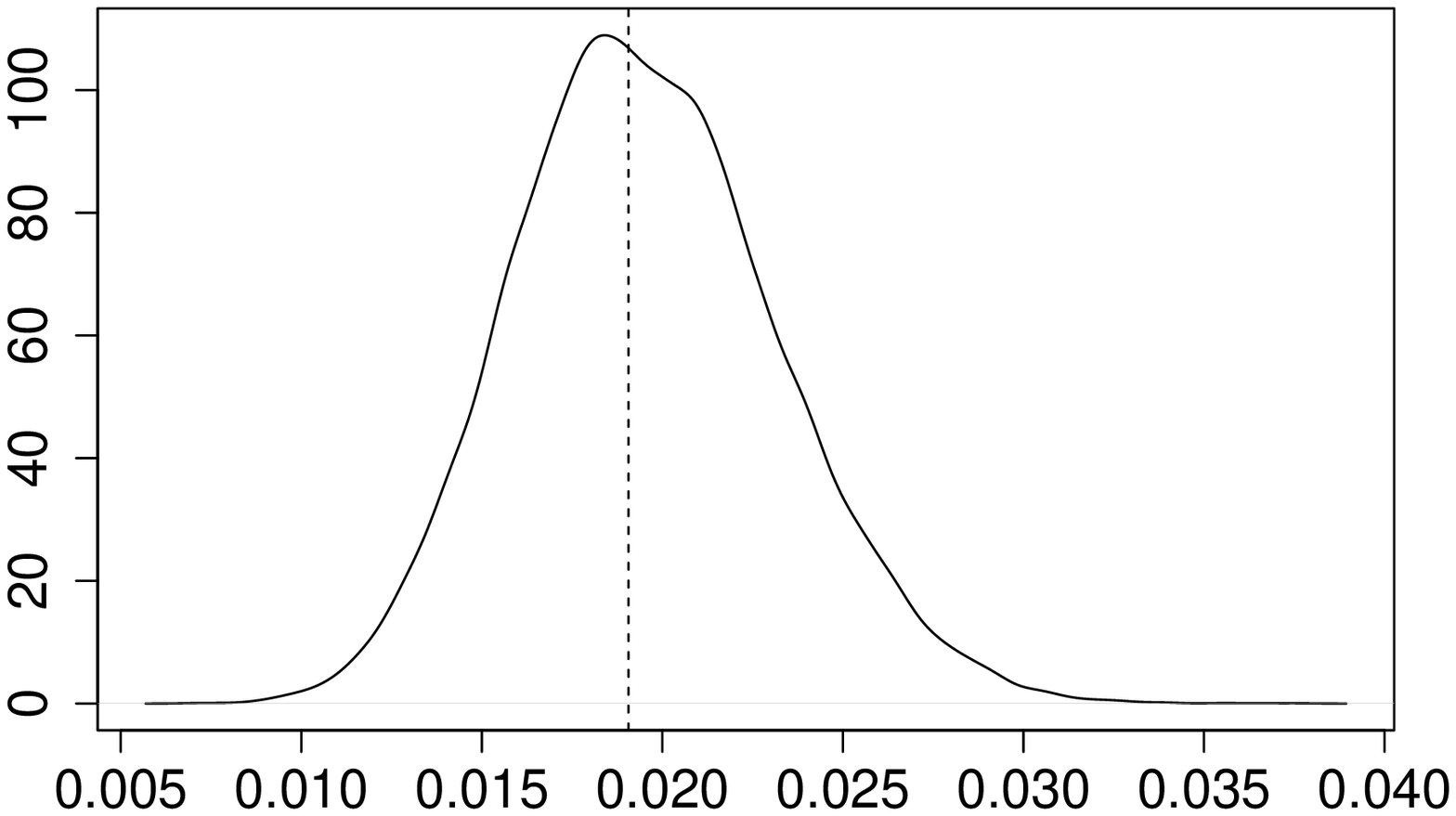} &
  \includegraphics[width=0.33\linewidth,height=0.20\linewidth]{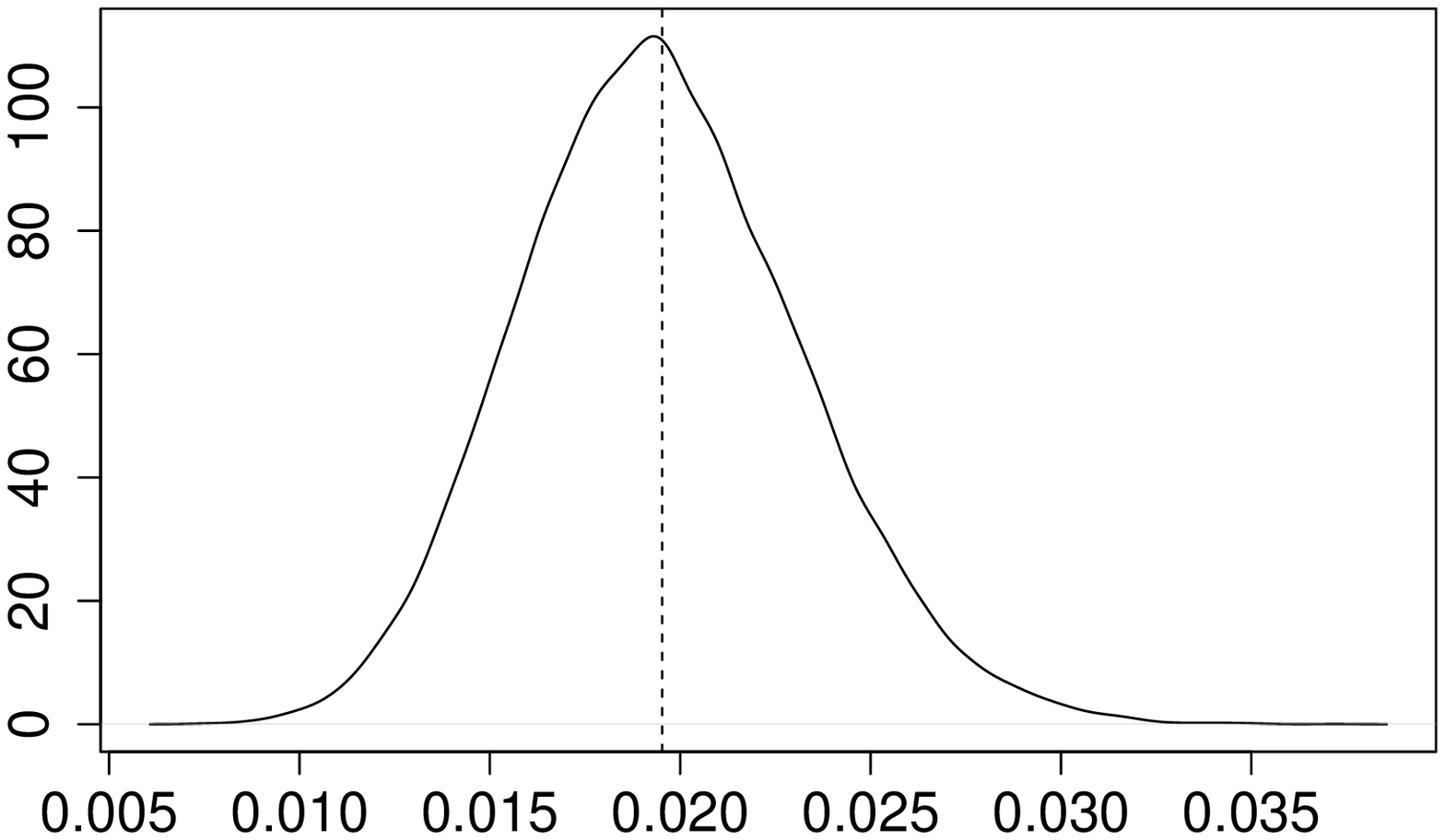} &
  \includegraphics[width=0.33\linewidth,height=0.20\linewidth]{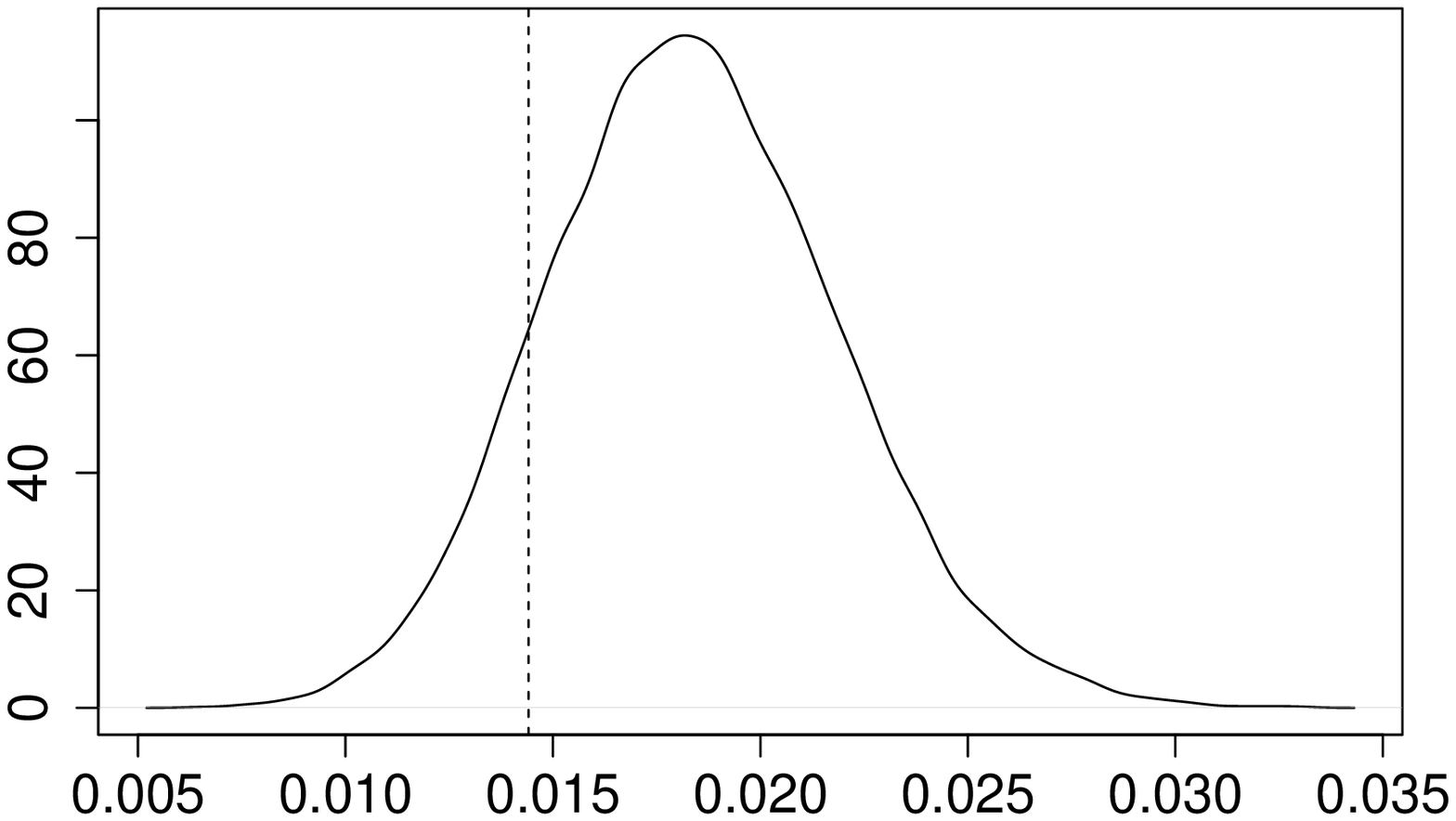} \\[-0.4cm]
(m)  \begin{tikzpicture}[scale=1.5]
  \def\radius{38.9};
  \coordinate (Pempty) at (0,0);
  \coordinate (P1) at (-1.5,1);
  \coordinate (P2) at (-0.5,1);
  \coordinate (P3) at (0.5,1);
  \coordinate (P4) at (1.5,1);
  \coordinate (P12) at (-1,2);
  \coordinate (P14) at (0,2);

  \coordinate (c1) at (-0.05,-0.05);
  \coordinate (c2) at (0.05,-0.05);
  \coordinate (c3) at (0.05,0.05);
  \coordinate (c4) at (-0.05,0.05);

  \coordinate (r1) at (-0.1,0);
  \coordinate (r2) at (0,0.1);
  \coordinate (r3) at (-0.1,0.1);
  \coordinate (r4) at (0.1,0.1);

   \draw[thick, fill=black] (P12) ++(r3) +(c1) -- +(c2) -- +(c3) -- +(c4) -- cycle;
   \draw[thick] (P12) ++(r2) +(c1) -- +(c2) -- +(c3) -- +(c4) -- cycle;
   \draw[thick, fill=black] (P12) ++(r1) +(c1) -- +(c2) -- +(c3) -- +(c4) -- cycle;
   \draw[thick, fill=black] (P12) ++(Pempty) +(c1) -- +(c2) -- +(c3) -- +(c4) -- cycle;
   
\end{tikzpicture}             &
(n)  \begin{tikzpicture}[scale=1.5]
  \def\radius{38.9};
  \coordinate (Pempty) at (0,0);
  \coordinate (P1) at (-1.5,1);
  \coordinate (P2) at (-0.5,1);
  \coordinate (P3) at (0.5,1);
  \coordinate (P4) at (1.5,1);
  \coordinate (P12) at (-1,2);
  \coordinate (P14) at (0,2);

  \coordinate (c1) at (-0.05,-0.05);
  \coordinate (c2) at (0.05,-0.05);
  \coordinate (c3) at (0.05,0.05);
  \coordinate (c4) at (-0.05,0.05);

  \coordinate (r1) at (-0.1,0);
  \coordinate (r2) at (0,0.1);
  \coordinate (r3) at (-0.1,0.1);
  \coordinate (r4) at (0.1,0.1);

   \draw[thick, fill=black] (P12) ++(r3) +(c1) -- +(c2) -- +(c3) -- +(c4) -- cycle;
   \draw[thick, fill=black] (P12) ++(r2) +(c1) -- +(c2) -- +(c3) -- +(c4) -- cycle;
   \draw[thick] (P12) ++(r1) +(c1) -- +(c2) -- +(c3) -- +(c4) -- cycle;
   \draw[thick, fill=black] (P12) ++(Pempty) +(c1) -- +(c2) -- +(c3) -- +(c4) -- cycle;
   
\end{tikzpicture}             &
(o)  \begin{tikzpicture}[scale=1.5]
  \def\radius{38.9};
  \coordinate (Pempty) at (0,0);
  \coordinate (P1) at (-1.5,1);
  \coordinate (P2) at (-0.5,1);
  \coordinate (P3) at (0.5,1);
  \coordinate (P4) at (1.5,1);
  \coordinate (P12) at (-1,2);
  \coordinate (P14) at (0,2);

  \coordinate (c1) at (-0.05,-0.05);
  \coordinate (c2) at (0.05,-0.05);
  \coordinate (c3) at (0.05,0.05);
  \coordinate (c4) at (-0.05,0.05);

  \coordinate (r1) at (-0.1,0);
  \coordinate (r2) at (0,0.1);
  \coordinate (r3) at (-0.1,0.1);
  \coordinate (r4) at (0.1,0.1);

   \draw[thick, fill=black] (P12) ++(r3) +(c1) -- +(c2) -- +(c3) -- +(c4) -- cycle;
   \draw[thick, fill=black] (P12) ++(r2) +(c1) -- +(c2) -- +(c3) -- +(c4) -- cycle;
   \draw[thick, fill=black] (P12) ++(r1) +(c1) -- +(c2) -- +(c3) -- +(c4) -- cycle;
   \draw[thick] (P12) ++(Pempty) +(c1) -- +(c2) -- +(c3) -- +(c4) -- cycle;
   
\end{tikzpicture}             \\[-0.05cm]
 &
  \includegraphics[width=0.33\linewidth,height=0.20\linewidth]{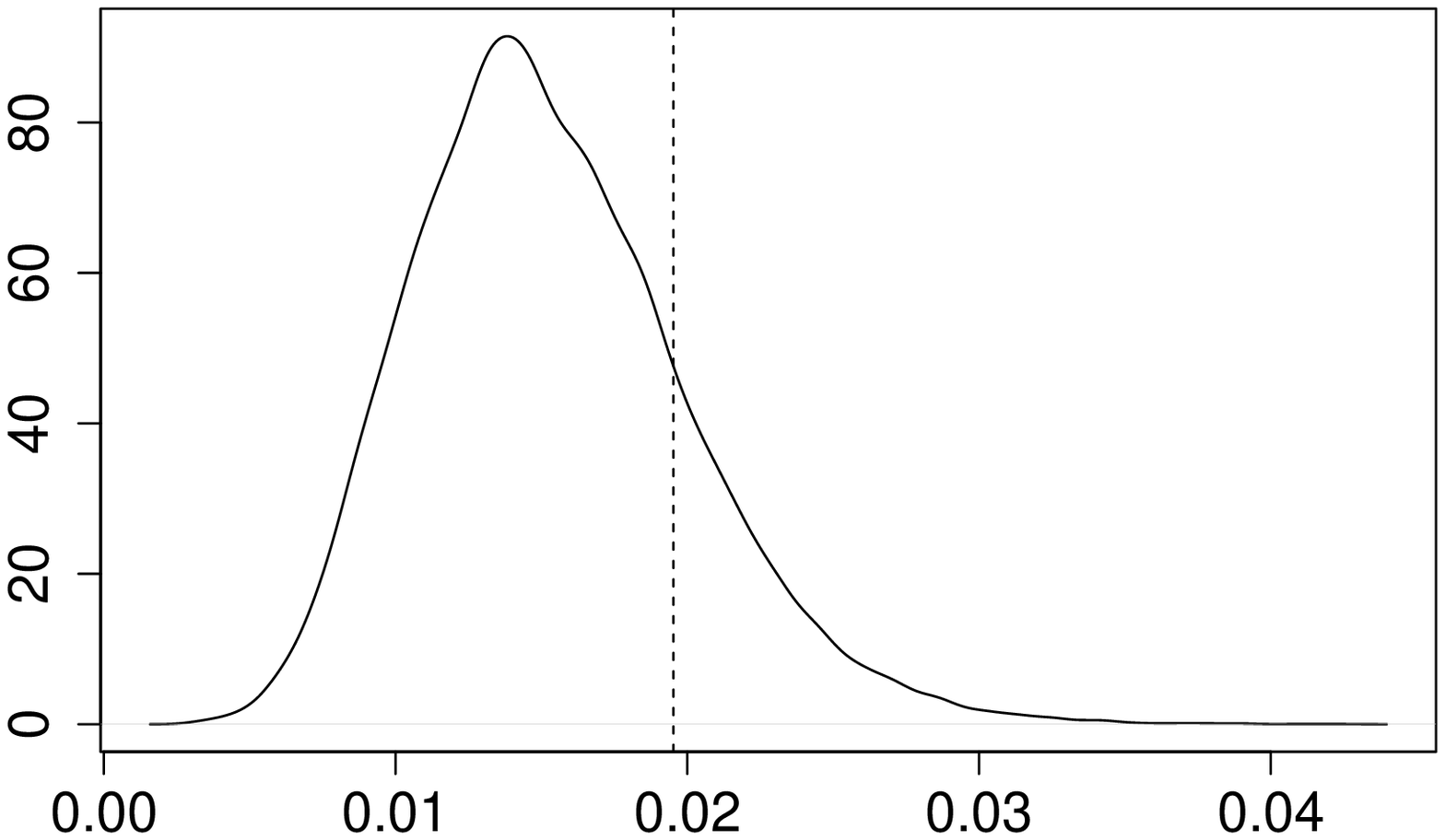} &
 \\[-0.4cm]
 & (p)  \begin{tikzpicture}[scale=1.5]
  \def\radius{38.9};
  \coordinate (Pempty) at (0,0);
  \coordinate (P1) at (-1.5,1);
  \coordinate (P2) at (-0.5,1);
  \coordinate (P3) at (0.5,1);
  \coordinate (P4) at (1.5,1);
  \coordinate (P12) at (-1,2);
  \coordinate (P14) at (0,2);

  \coordinate (c1) at (-0.05,-0.05);
  \coordinate (c2) at (0.05,-0.05);
  \coordinate (c3) at (0.05,0.05);
  \coordinate (c4) at (-0.05,0.05);

  \coordinate (r1) at (-0.1,0);
  \coordinate (r2) at (0,0.1);
  \coordinate (r3) at (-0.1,0.1);
  \coordinate (r4) at (0.1,0.1);

   \draw[thick, fill=black] (P12) ++(r3) +(c1) -- +(c2) -- +(c3) -- +(c4) -- cycle;
   \draw[thick, fill=black] (P12) ++(r2) +(c1) -- +(c2) -- +(c3) -- +(c4) -- cycle;
   \draw[thick, fill=black] (P12) ++(r1) +(c1) -- +(c2) -- +(c3) -- +(c4) -- cycle;
   \draw[thick, fill=black] (P12) ++(Pempty) +(c1) -- +(c2) -- +(c3) -- +(c4) -- cycle;
   
\end{tikzpicture}             
& 
\end{tabular}
        \caption{Cancer mortality map example: Estimated a posteriori marginal densities for each of the possible $16$ configurations 
in a $2\times 2$ block of nodes. Corresponding values computed from the cancer map data set is shown as a vertical dotted line. 
The configuration corresponding to an estimated density is shown below each figure, where black and white nodes
represent one and zero, respectively.}\label{fig:frac_cancer}
\end{figure}

Finally, we generate realizations from simulated Markov mesh models. Figure \ref{fig:real_cancer} contains realizations 
simulated from four randomly chosen models simulated in the Markov chain (after the specified burn-in).
As in Figure \ref{fig:cancer}, showing the observed data set, black and white nodes $v$ represent $x_v=1$ and $0$, respectively.
Comparing the realizations with the data set in Figure \ref{fig:cancer}, we can get a visual impression of to what degree
the simulated models have captured the dependence structure in the data set. To study this also more
quantitatively, we consider the $16$ possible configurations in a $2\times2$ block of nodes. For each of these 
configurations, we find in a realization the fraction of such blocks that has the specified configuration.
By repeating this for a large number of realizations we estimate the posterior distribution for the 
fraction of $2\times 2$ blocks with a specified configuration in a realization. This distribution should be compared
with the corresponding fraction in the observed data set.
Figure \ref{fig:frac_cancer} 
shows the estimated density for each of the $16$ configurations. The corresponding fractions 
for the observed data set are marked by vertical dotted lines. Note that for most of these distributions the corresponding 
fractions for the observed data set are centrally located in the distribution. The exceptions are (g) and partly (i) and (j), 
where the observed quantity is more in the tail of the distribution.

\subsection{Sisim data set\label{sec:sisim}}
\begin{figure}
        \begin{subfigure}[b]{0.5\textwidth}
                	\vspace*{-1.0cm}
                \includegraphics[width=\linewidth]{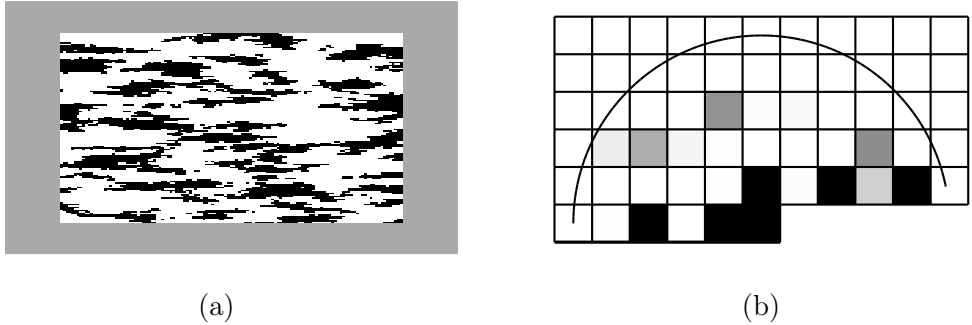}
                	\vspace*{-1.5cm}
                \caption{ }
                \label{fig:sisim}
        \end{subfigure}%
        \hfill
        \begin{subfigure}[b]{0.5\textwidth}
             
                \begin{center}
\begin{tikzpicture}[scale=0.5]

\draw[black,fill=gray!0.0626146](-1.5,3.5)--(-2.5,3.5)--(-2.5,4.5)--(-1.5,4.5)--cycle; 
\draw[black,fill=gray!0.1252292](-0.5,3.5)--(-1.5,3.5)--(-1.5,4.5)--(-0.5,4.5)--cycle; 
\draw[black,fill=gray!0.161009](0.5,3.5)--(-0.5,3.5)--(-0.5,4.5)--(0.5,4.5)--cycle; 
\draw[black,fill=gray!0.152064](1.5,3.5)--(0.5,3.5)--(0.5,4.5)--(1.5,4.5)--cycle; 
\draw[black,fill=gray!0.304128](2.5,3.5)--(1.5,3.5)--(1.5,4.5)--(2.5,4.5)--cycle; 
\draw[black,fill=gray!0.268348](-2.5,2.5)--(-3.5,2.5)--(-3.5,3.5)--(-2.5,3.5)--cycle; 
\draw[black,fill=gray!0.509862](-1.5,2.5)--(-2.5,2.5)--(-2.5,3.5)--(-1.5,3.5)--cycle; 
\draw[black,fill=gray!85.1558](-0.5,2.5)--(-1.5,2.5)--(-1.5,3.5)--(-0.5,3.5)--cycle; 
\draw[black,fill=gray!2.19152](0.5,2.5)--(-0.5,2.5)--(-0.5,3.5)--(0.5,3.5)--cycle; 
\draw[black,fill=gray!4.22202](1.5,2.5)--(0.5,2.5)--(0.5,3.5)--(1.5,3.5)--cycle; 
\draw[black,fill=gray!0.1431192](2.5,2.5)--(1.5,2.5)--(1.5,3.5)--(2.5,3.5)--cycle; 
\draw[black,fill=gray!0.0805044](3.5,2.5)--(2.5,2.5)--(2.5,3.5)--(3.5,3.5)--cycle; 
\draw[black,fill=gray!14.13302](-3.5,1.5)--(-4.5,1.5)--(-4.5,2.5)--(-3.5,2.5)--cycle; 
\draw[black,fill=gray!64.6362](-2.5,1.5)--(-3.5,1.5)--(-3.5,2.5)--(-2.5,2.5)--cycle; 
\draw[black,fill=gray!7.75526](-1.5,1.5)--(-2.5,1.5)--(-2.5,2.5)--(-1.5,2.5)--cycle; 
\draw[black,fill=gray!1.162842](-0.5,1.5)--(-1.5,1.5)--(-1.5,2.5)--(-0.5,2.5)--cycle; 
\draw[black,fill=gray!0.0805044](0.5,1.5)--(-0.5,1.5)--(-0.5,2.5)--(0.5,2.5)--cycle; 
\draw[black,fill=gray!0.1162842](1.5,1.5)--(0.5,1.5)--(0.5,2.5)--(1.5,2.5)--cycle; 
\draw[black,fill=gray!0.214678](2.5,1.5)--(1.5,1.5)--(1.5,2.5)--(2.5,2.5)--cycle; 
\draw[black,fill=gray!88.9216](3.5,1.5)--(2.5,1.5)--(2.5,2.5)--(3.5,2.5)--cycle; 
\draw[black,fill=gray!0.304128](4.5,1.5)--(3.5,1.5)--(3.5,2.5)--(4.5,2.5)--cycle; 
\draw[black,fill=gray!0.0536696](-3.5,0.5)--(-4.5,0.5)--(-4.5,1.5)--(-3.5,1.5)--cycle; 
\draw[black,fill=gray!0.268348](-2.5,0.5)--(-3.5,0.5)--(-3.5,1.5)--(-2.5,1.5)--cycle; 
\draw[black,fill=gray!0.205734](-1.5,0.5)--(-2.5,0.5)--(-2.5,1.5)--(-1.5,1.5)--cycle; 
\draw[black,fill=gray!0.67087](-0.5,0.5)--(-1.5,0.5)--(-1.5,1.5)--(-0.5,1.5)--cycle; 
\draw[black,fill=gray!200](0.5,0.5)--(-0.5,0.5)--(-0.5,1.5)--(0.5,1.5)--cycle; 
\draw[black,fill=gray!2.26308](1.5,0.5)--(0.5,0.5)--(0.5,1.5)--(1.5,1.5)--cycle; 
\draw[black,fill=gray!200](2.5,0.5)--(1.5,0.5)--(1.5,1.5)--(2.5,1.5)--cycle; 
\draw[black,fill=gray!36.5758](3.5,0.5)--(2.5,0.5)--(2.5,1.5)--(3.5,1.5)--cycle; 
\draw[black,fill=gray!200](4.5,0.5)--(3.5,0.5)--(3.5,1.5)--(4.5,1.5)--cycle; 
\draw[black,fill=gray!0.590366](-3.5,-0.5)--(-4.5,-0.5)--(-4.5,0.5)--(-3.5,0.5)--cycle; 
\draw[black,fill=gray!200](-2.5,-0.5)--(-3.5,-0.5)--(-3.5,0.5)--(-2.5,0.5)--cycle; 
\draw[black,fill=gray!0.0805044](-1.5,-0.5)--(-2.5,-0.5)--(-2.5,0.5)--(-1.5,0.5)--cycle; 
\draw[black,fill=gray!200](-0.5,-0.5)--(-1.5,-0.5)--(-1.5,0.5)--(-0.5,0.5)--cycle; 
\draw[black,fill=gray!200](0.5,-0.5)--(-0.5,-0.5)--(-0.5,0.5)--(0.5,0.5)--cycle;

\foreach \x in {-5.5,-4.5,-3.5,-2.5,-1.5,-0.5,0.5} {
  \draw[black,thick] (\x,-0.5) -- (\x,5.5);
}
\foreach \x in {1.5,2.5,3.5,4.5,5.5} {
  \draw[black,thick] (\x,0.5) -- (\x,5.5);
}

\draw[black,very thick] (-5.5,-0.5) -- (0.5,-0.5);
\foreach \y in {0.5,1.5,2.5,3.5,4.5,5.5} {
  \draw[black,thick] (-5.5,\y) -- (5.5,\y);
}

\draw[black,thick] ({5*cos(11.31)},{5*sin(11.31)}) -- ({5*cos(12)},{5*sin(12)});
\foreach \angle in {12,14,16,18,20,22,24,26,28,30,32,34,36,38,40,42,44,46,48,50,52,54,56,58,60,62,64,66,68,70,72,74,76,78,80,82,84,86,88,90,92,94,96,98,100,102,104,106,108,110,112,114,116,118,120,122,124,126,128,130,132,134,136,138,140,142,144,146,148,150,152,154,156,158,160,162,164,166,168,170,172,174,176,178} {
  \draw[black,thick] ({5*cos(\angle)},{5*sin(\angle)}) -- ({5*cos(\angle+2)},{5*sin(\angle+2)});
}

\draw[black,thick] (-0.5,-0.5) -- (0.5,0.5);
\draw[black,thick] (-0.5,0.5) -- (0.5,-0.5);

\end{tikzpicture}
\end{center}
                	\vspace*{-0.15cm}
                \caption{ }
                \label{fig:probneigh_sisim}
        \end{subfigure}
        \caption{Sisim data set example: (a) Given scene. Nodes added to the lattice to reduce the boundary effects of the 
Markov mesh model is shown in gray. (b) Map of estimated a posteriori
probabilities for each node $v\in \tau_0$ to be a neighbor. A grayscale is used to visualize the probabilities, where black and white
represents one and zero, respectively.}\label{fig:TI_probneigh_sisim}
\end{figure}

In this example we reconsider a data set previously studied in 
\cite{art132}. The scene, shown in Figure \ref{fig:sisim}, 
is simulated by the sequential indicator
simulation procedure \citep{pro22,book40} and it is a much used example scene in the geostatistical community.
We name the data set "sisim". The sisim scene is represented on a $121\times 121$ lattice. To reduce the boundary effects
of the Markov mesh model we again include unobserved nodes around the observed area, shown as gray in Figure 
\ref{fig:sisim}.

\begin{figure}   
        \begin{subfigure}[b]{0.5\textwidth}
        \vspace*{-0.5cm}
                \includegraphics[width=\linewidth,height=5.0cm]{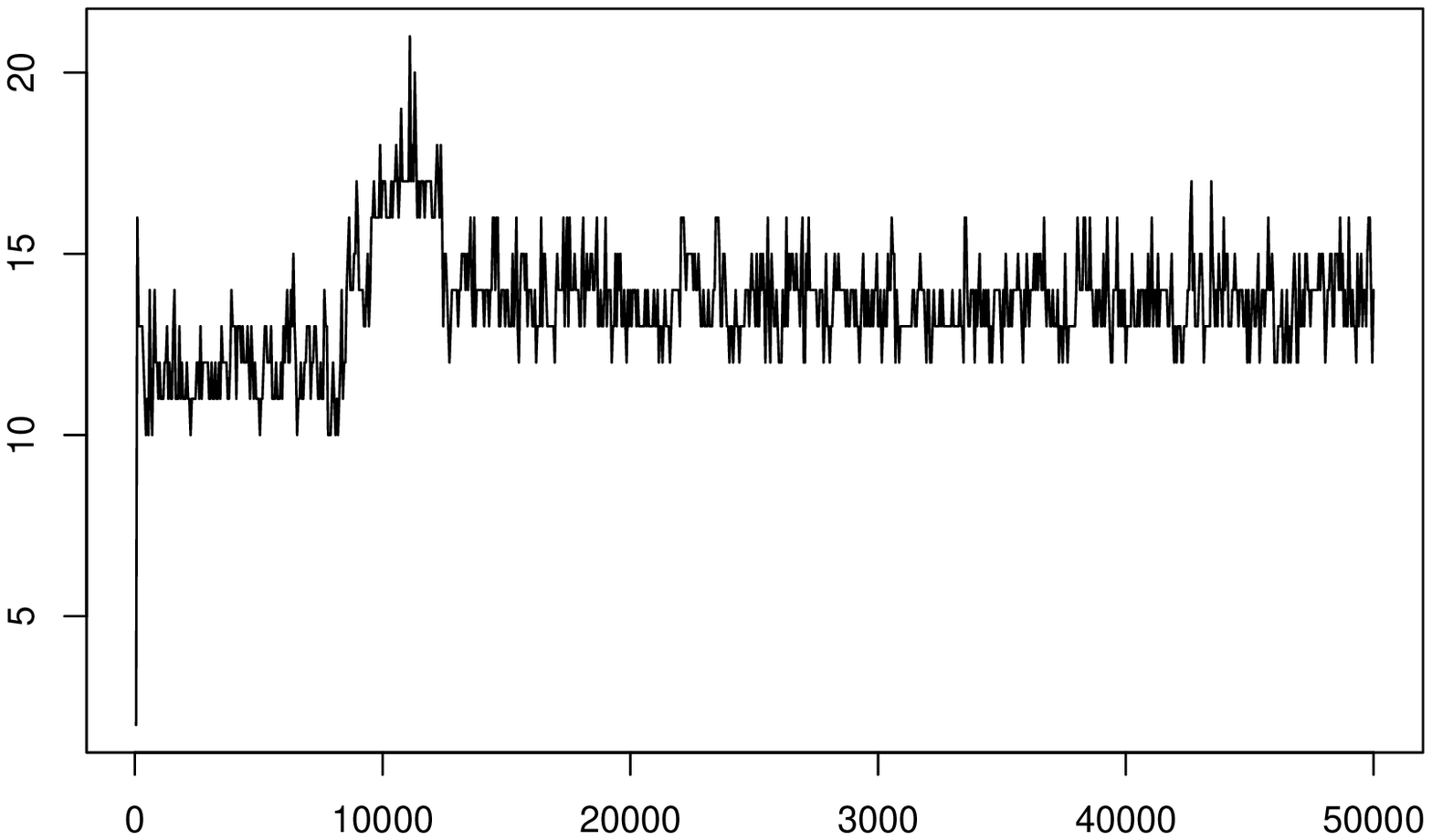}
        \vspace*{-1.0cm}
                \caption{ }
                \label{fig:traceplot_num_short_sisim}
        \end{subfigure}%
        \hfill
        \begin{subfigure}[b]{0.5\textwidth}
         \vspace*{-0.5cm}
                \includegraphics[width=\linewidth,height=5.0cm]{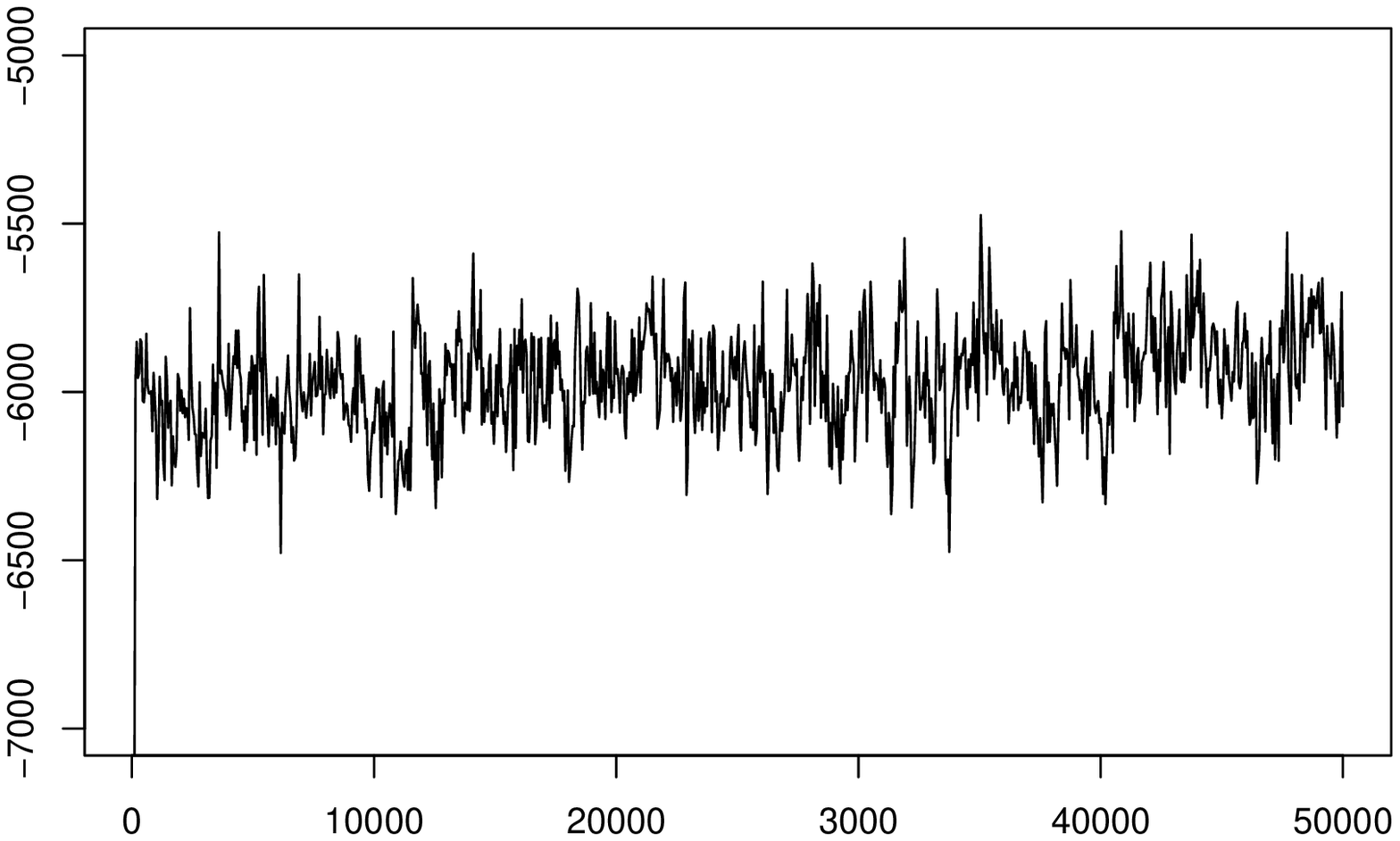}
        \vspace*{-1.0cm}
                \caption{ }
                \label{fig:traceplot_log_short_sisim}
        \end{subfigure}
        
        \caption{Sisim data set example: Trace plots of the first $50\,000$ iterations of the RJMCMCM run. 
(a) Number of interactions $|\Lambda|$, (b) logarithm of the posterior density 
$\log\left[f\left(\tau,\Lambda,\left\lbrace \theta(\lambda):\lambda\in\Lambda\right\rbrace,x_u|x_o\right)\right]$.} %
\label{fig:traceplots_sisim}
\end{figure}

Again adopting the Markov mesh and prior models defined in Sections \ref{sec:mmm} and \ref{sec:prior} and 
the hyper-parameters defined above, we use the RJMCMC setup discussed above to explore the resulting 
posterior distribution. For this data set each iteration of the algorithm requires more computation time than 
in the cancer mortality map data, so we run the Markov chain for only $1\,250\,000$ iterations. 
To evaluate the convergence properties of the simulated Markov chain, we study trace plots of different scalar quantities 
in the same way as in Section \ref{sec:cancer}.
Figure \ref{fig:traceplots_sisim} 
shows trace plots of the first $50\,000$ iterations for the
number of interactions and for the logarithm of the posterior density.
At first glance at these two and the other trace plots we have studied, we asserted 
that the simulated chain had converged at least within the first $30\,000-40\,000$ iterations. As an extra precaution we discarded 
the first $250\,000$ iterations when estimating posterior properties.

As in Section \ref{sec:cancer}, we estimate the posterior probability for $v\in \tau_0$ to be in the template
sequential neighborhood $\tau$. The results are shown in Figure \ref{fig:probneigh_sisim}, 
where we use a gray scale to visualize the probabilities. There are five nodes whose estimated posterior 
probabilities are essentially equal to $1$, and these are $(0,-1)$, $(-1,0)$, $(-1,2)$, $(0,-3)$ and $(-1,4)$. Four more nodes have 
estimated posterior probabilities higher than $0.1$. These are $(-2,3)$, $(-3,-1)$, $(-2,-3)$ and $(-1,3)$ with estimated probabilities $0.444608$, $0.425779$, $0.323181$ and $0.182879$, respectively. It is interesting to note the spatial locations of the high probability nodes. At least for a part of the area
every second node is chosen as a neighbor with high probability. To understand this effect, we must remember that the values of
two nodes that are lying next to each other are highly correlated, so one would not gain much extra information by including 
both of them in the template sequential neighborhood. Moreover, the prior prefers parsimonious models, which we
obtain by not including too many nodes in the template sequential neighborhood. 

Next, as for the cancer mortality map data set, we correspondingly estimate the posterior probabilities for each possible 
interaction to be included in the model.
Table \ref{tab:probInt_sisim} shows the top $20$ a posteriori most likely interactions and corresponding estimated probabilities. 
\begin{table}
\centering
\caption{Sisim data set example: The top $20$ a posteriori most likely interactions and their estimated posterior probabilities.}
\label{tab:probInt_sisim}
\begin{tabular}{|c|c|c|c|c|c|}
\hline
& & & & & \\
Interaction &   
 \begin{tikzpicture}[scale=2]
  \def\radius{38.9};
  \coordinate (Pempty) at (0,0);

  \coordinate (c1) at (-0.05,-0.05);
  \coordinate (c2) at (0.05,-0.05);
  \coordinate (c3) at (0.05,0.05);
  \coordinate (c4) at (-0.05,0.05);

  \draw[thick] (Pempty) +(c1) -- +(c2) -- +(c3) -- +(c4) -- +(c1) +(c1) -- +(c3) +(c2) -- +(c4);

\end{tikzpicture}  
&   
   \begin{tikzpicture}[scale=2]
  \def\radius{38.9};
  \coordinate (Pempty) at (0,0);
  \coordinate (P1) at (-1.5,1);
  \coordinate (P2) at (-0.5,1);
  \coordinate (P3) at (0.5,1);
  \coordinate (P4) at (1.5,1);
  \coordinate (P12) at (-1,2);
  \coordinate (P14) at (0,2);

  \coordinate (c1) at (-0.05,-0.05);
  \coordinate (c2) at (0.05,-0.05);
  \coordinate (c3) at (0.05,0.05);
  \coordinate (c4) at (-0.05,0.05);
  
   \coordinate (r1) at (-0.1,0);

  \draw[thick] (P1) +(c1) -- +(c2) -- +(c3) -- +(c4) -- +(c1) +(c1) -- +(c3) +(c2) -- +(c4);
  \draw[thick] (P1) ++(r1) +(c1) -- +(c2) -- +(c3) -- +(c4) -- cycle;

\end{tikzpicture}  
&     
\begin{tikzpicture}[scale=2]
  \def\radius{38.9};
  \coordinate (Pempty) at (0,0);
  \coordinate (P1) at (-1.5,1);
  \coordinate (P2) at (-0.5,1);
  \coordinate (P3) at (0.5,1);
  \coordinate (P4) at (1.5,1);
  \coordinate (P12) at (-1,2);
  \coordinate (P14) at (0,2);

  \coordinate (c1) at (-0.05,-0.05);
  \coordinate (c2) at (0.05,-0.05);
  \coordinate (c3) at (0.05,0.05);
  \coordinate (c4) at (-0.05,0.05);

  \coordinate (r1) at (-0.1,0);
  \coordinate (r2) at (0,0.1);

  \draw[thick] (P2) +(c1) -- +(c2) -- +(c3) -- +(c4) -- +(c1) +(c1) -- +(c3) +(c2) -- +(c4);
  \draw[thick] (P2) ++(r2) +(c1) -- +(c2) -- +(c3) -- +(c4) -- cycle;

\end{tikzpicture}
&    
\begin{tikzpicture}[scale=2]
  \def\radius{38.9};
  \coordinate (Pempty) at (0,0);
  \coordinate (P1) at (-1.5,1);
  \coordinate (P2) at (-0.5,1);
  \coordinate (P3) at (0.5,1);
  \coordinate (P4) at (1.5,1);
  \coordinate (P12) at (-1,2);
  \coordinate (P14) at (0,2);

  \coordinate (c1) at (-0.05,-0.05);
  \coordinate (c2) at (0.05,-0.05);
  \coordinate (c3) at (0.05,0.05);
  \coordinate (c4) at (-0.05,0.05);

  \coordinate (r1) at (-0.1,0);
  \coordinate (r2) at (0,0.1);
  \coordinate (r3) at (0.2,0.1);
 
  \draw[thick] (P12) +(c1) -- +(c2) -- +(c3) -- +(c4) -- +(c1) +(c1) -- +(c3) +(c2) -- +(c4);
  \draw[thick] (P12) ++(r3) +(c1) -- +(c2) -- +(c3) -- +(c4) -- cycle;

\end{tikzpicture}   
&    
\begin{tikzpicture}[scale=2]
  \def\radius{38.9};
  \coordinate (Pempty) at (0,0);
  \coordinate (P1) at (-1.5,1);
  \coordinate (P2) at (-0.5,1);
  \coordinate (P3) at (0.5,1);
  \coordinate (P4) at (1.5,1);
  \coordinate (P12) at (-1,2);
  \coordinate (P14) at (0,2);

  \coordinate (c1) at (-0.05,-0.05);
  \coordinate (c2) at (0.05,-0.05);
  \coordinate (c3) at (0.05,0.05);
  \coordinate (c4) at (-0.05,0.05);

  \coordinate (r1) at (-0.1,0);
  \coordinate (r2) at (0,0.1);
  \coordinate (r3) at (0.2,0.1);
  \coordinate (r4) at (-0.3,0);

  \draw[thick] (P3) +(c1) -- +(c2) -- +(c3) -- +(c4) -- +(c1) +(c1) -- +(c3) +(c2) -- +(c4);
  \draw[thick] (P3) ++(r4) +(c1) -- +(c2) -- +(c3) -- +(c4) -- cycle;

\end{tikzpicture}       
           \\
Probability & 1.0000 & 1.0000 & 1.0000 & 1.0000 & 1.0000 \\ \hline
& & & & & \\
Interaction &      
\begin{tikzpicture}[scale=2]
  \def\radius{38.9};
  \coordinate (Pempty) at (0,0);
  \coordinate (P1) at (-1.5,1);
  \coordinate (P2) at (-0.5,1);
  \coordinate (P3) at (0.5,1);
  \coordinate (P4) at (1.5,1);
  \coordinate (P12) at (-1,2);
  \coordinate (P14) at (0,2);

  \coordinate (c1) at (-0.05,-0.05);
  \coordinate (c2) at (0.05,-0.05);
  \coordinate (c3) at (0.05,0.05);
  \coordinate (c4) at (-0.05,0.05);

  \coordinate (r1) at (-0.1,0);
  \coordinate (r2) at (0,0.1);
  \coordinate (r3) at (0.2,0.1);
  \coordinate (r4) at (-0.3,0);

  \draw[thick] (P3) +(c1) -- +(c2) -- +(c3) -- +(c4) -- +(c1) +(c1) -- +(c3) +(c2) -- +(c4);
  \draw[thick] (P3) ++(r1) +(c1) -- +(c2) -- +(c3) -- +(c4) -- cycle;
  \draw[thick] (P3) ++(r4) +(c1) -- +(c2) -- +(c3) -- +(c4) -- cycle;
\end{tikzpicture}
&     
 \begin{tikzpicture}[scale=2]
  \def\radius{38.9};
  \coordinate (Pempty) at (0,0);
  \coordinate (P1) at (-1.5,1);
  \coordinate (P2) at (-0.5,1);
  \coordinate (P3) at (0.5,1);
  \coordinate (P4) at (1.5,1);
  \coordinate (P12) at (-1,2);
  \coordinate (P14) at (0,2);

  \coordinate (c1) at (-0.05,-0.05);
  \coordinate (c2) at (0.05,-0.05);
  \coordinate (c3) at (0.05,0.05);
  \coordinate (c4) at (-0.05,0.05);

  \coordinate (r1) at (-0.1,0);
  \coordinate (r2) at (0,0.1);
  \coordinate (r3) at (0.2,0.1);
  \coordinate (r4) at (-0.3,0);
  \coordinate (r5) at (0.4,0.1);

  \draw[thick] (P3) +(c1) -- +(c2) -- +(c3) -- +(c4) -- +(c1) +(c1) -- +(c3) +(c2) -- +(c4);
  \draw[thick] (P3) ++(r5) +(c1) -- +(c2) -- +(c3) -- +(c4) -- cycle;

\end{tikzpicture} 
&    
\begin{tikzpicture}[scale=2]
  \def\radius{38.9};
  \coordinate (Pempty) at (0,0);
  \coordinate (P1) at (-1.5,1);
  \coordinate (P2) at (-0.5,1);
  \coordinate (P3) at (0.5,1);
  \coordinate (P4) at (1.5,1);
  \coordinate (P12) at (-1,2);
  \coordinate (P14) at (0,2);

  \coordinate (c1) at (-0.05,-0.05);
  \coordinate (c2) at (0.05,-0.05);
  \coordinate (c3) at (0.05,0.05);
  \coordinate (c4) at (-0.05,0.05);

  \coordinate (r1) at (-0.1,0);
  \coordinate (r2) at (0,0.1);
  \coordinate (r3) at (0.2,0.1);
  \coordinate (r4) at (-0.3,0);

  \draw[thick] (P3) +(c1) -- +(c2) -- +(c3) -- +(c4) -- +(c1) +(c1) -- +(c3) +(c2) -- +(c4);
  \draw[thick] (P3) ++(r3) +(c1) -- +(c2) -- +(c3) -- +(c4) -- cycle;
  \draw[thick] (P3) ++(r4) +(c1) -- +(c2) -- +(c3) -- +(c4) -- cycle;
   \end{tikzpicture} 
&    
 \begin{tikzpicture}[scale=2]
 \def\radius{38.9};
  \coordinate (Pempty) at (0,0);
  \coordinate (P1) at (-1.5,1);
  \coordinate (P2) at (-0.5,1);
  \coordinate (P3) at (0.5,1);
  \coordinate (P4) at (1.5,1);
  \coordinate (P12) at (-1,2);
  \coordinate (P14) at (0,2);

  \coordinate (c1) at (-0.05,-0.05);
  \coordinate (c2) at (0.05,-0.05);
  \coordinate (c3) at (0.05,0.05);
  \coordinate (c4) at (-0.05,0.05);

  \coordinate (r1) at (-0.1,0);
  \coordinate (r2) at (0,0.1);
  \coordinate (r3) at (0.2,0.1);
  \coordinate (r4) at (-0.3,0);
  \coordinate (r5) at (0.4,0.1);

  \draw[thick] (P3) +(c1) -- +(c2) -- +(c3) -- +(c4) -- +(c1) +(c1) -- +(c3) +(c2) -- +(c4);
  \draw[thick] (P3) ++(r1) +(c1) -- +(c2) -- +(c3) -- +(c4) -- cycle;
  \draw[thick] (P3) ++(r3) +(c1) -- +(c2) -- +(c3) -- +(c4) -- cycle;
\end{tikzpicture}      
&    
 \begin{tikzpicture}[scale=2]
 \def\radius{38.9};
  \coordinate (Pempty) at (0,0);
  \coordinate (P1) at (-1.5,1);
  \coordinate (P2) at (-0.5,1);
  \coordinate (P3) at (0.5,1);
  \coordinate (P4) at (1.5,1);
  \coordinate (P12) at (-1,2);
  \coordinate (P14) at (0,2);

  \coordinate (c1) at (-0.05,-0.05);
  \coordinate (c2) at (0.05,-0.05);
  \coordinate (c3) at (0.05,0.05);
  \coordinate (c4) at (-0.05,0.05);

  \coordinate (r1) at (-0.1,0);
  \coordinate (r2) at (0,0.1);
  \coordinate (r3) at (0.2,0.1);
  \coordinate (r4) at (-0.3,0);
  \coordinate (r5) at (0.4,0.1);

  \draw[thick] (P3) +(c1) -- +(c2) -- +(c3) -- +(c4) -- +(c1) +(c1) -- +(c3) +(c2) -- +(c4);
  \draw[thick] (P3) ++(r1) +(c1) -- +(c2) -- +(c3) -- +(c4) -- cycle;
  \draw[thick] (P3) ++(r3) +(c1) -- +(c2) -- +(c3) -- +(c4) -- cycle;
  \draw[thick] (P3) ++(r4) +(c1) -- +(c2) -- +(c3) -- +(c4) -- cycle;
\end{tikzpicture} 
    \\
Probability & 1.0000 & 1.0000 & 0.8572 & 0.8525 & 0.8484 \\ \hline
& & & & & \\[-0.4cm]
Interaction &      
\begin{tikzpicture}[scale=2]
  \def\radius{38.9};
  \coordinate (Pempty) at (0,0);
  \coordinate (P1) at (-1.5,1);
  \coordinate (P2) at (-0.5,1);
  \coordinate (P3) at (0.5,1);
  \coordinate (P4) at (1.5,1);
  \coordinate (P12) at (-1,2);
  \coordinate (P14) at (0,2);

  \coordinate (c1) at (-0.05,-0.05);
  \coordinate (c2) at (0.05,-0.05);
  \coordinate (c3) at (0.05,0.05);
  \coordinate (c4) at (-0.05,0.05);

  \coordinate (r1) at (-0.1,0);
  \coordinate (r2) at (0,0.1);
  \coordinate (r3) at (0.2,0.1);
  \coordinate (r4) at (-0.3,0);
  \coordinate (r5) at (0.4,0.1);

  \draw[thick] (P3) +(c1) -- +(c2) -- +(c3) -- +(c4) -- +(c1) +(c1) -- +(c3) +(c2) -- +(c4);
  \draw[thick] (P3) ++(r2) +(c1) -- +(c2) -- +(c3) -- +(c4) -- cycle;
  \draw[thick] (P3) ++(r5) +(c1) -- +(c2) -- +(c3) -- +(c4) -- cycle;
\end{tikzpicture}
&     
 \begin{tikzpicture}[scale=2]
  \def\radius{38.9};
  \coordinate (Pempty) at (0,0);
  \coordinate (P1) at (-1.5,1);
  \coordinate (P2) at (-0.5,1);
  \coordinate (P3) at (0.5,1);
  \coordinate (P4) at (1.5,1);
  \coordinate (P12) at (-1,2);
  \coordinate (P14) at (0,2);

  \coordinate (c1) at (-0.05,-0.05);
  \coordinate (c2) at (0.05,-0.05);
  \coordinate (c3) at (0.05,0.05);
  \coordinate (c4) at (-0.05,0.05);

  \coordinate (r1) at (-0.1,0);
  \coordinate (r2) at (0,0.1);
  \coordinate (r3) at (0.2,0.1);
  \coordinate (r4) at (-0.3,0);
  \coordinate (r5) at (0.4,0.1);
  \coordinate (r6) at (0.3,0.2);

  \draw[thick] (P3) +(c1) -- +(c2) -- +(c3) -- +(c4) -- +(c1) +(c1) -- +(c3) +(c2) -- +(c4);
  \draw[thick] (P3) ++(r6) +(c1) -- +(c2) -- +(c3) -- +(c4) -- cycle;

\end{tikzpicture} 
&    
\begin{tikzpicture}[scale=2]
  \def\radius{38.9};
  \coordinate (Pempty) at (0,0);
  \coordinate (P1) at (-1.5,1);
  \coordinate (P2) at (-0.5,1);
  \coordinate (P3) at (0.5,1);
  \coordinate (P4) at (1.5,1);
  \coordinate (P12) at (-1,2);
  \coordinate (P14) at (0,2);

  \coordinate (c1) at (-0.05,-0.05);
  \coordinate (c2) at (0.05,-0.05);
  \coordinate (c3) at (0.05,0.05);
  \coordinate (c4) at (-0.05,0.05);

  \coordinate (r1) at (-0.1,0);
  \coordinate (r2) at (0,0.1);
  \coordinate (r3) at (0.2,0.1);
  \coordinate (r4) at (-0.3,0);
  \coordinate (r7) at (-0.1,0.3);

  \draw[thick] (P3) +(c1) -- +(c2) -- +(c3) -- +(c4) -- +(c1) +(c1) -- +(c3) +(c2) -- +(c4);
  \draw[thick] (P3) ++(r7) +(c1) -- +(c2) -- +(c3) -- +(c4) -- cycle;

   \end{tikzpicture} 
&    
 \begin{tikzpicture}[scale=2]
 \def\radius{38.9};
  \coordinate (Pempty) at (0,0);
  \coordinate (P1) at (-1.5,1);
  \coordinate (P2) at (-0.5,1);
  \coordinate (P3) at (0.5,1);
  \coordinate (P4) at (1.5,1);
  \coordinate (P12) at (-1,2);
  \coordinate (P14) at (0,2);

  \coordinate (c1) at (-0.05,-0.05);
  \coordinate (c2) at (0.05,-0.05);
  \coordinate (c3) at (0.05,0.05);
  \coordinate (c4) at (-0.05,0.05);

  \coordinate (r1) at (-0.1,0);
  \coordinate (r2) at (0,0.1);
  \coordinate (r3) at (0.2,0.1);
  \coordinate (r4) at (-0.3,0);
  \coordinate (r8) at (-0.3,0.2);

  \draw[thick] (P3) +(c1) -- +(c2) -- +(c3) -- +(c4) -- +(c1) +(c1) -- +(c3) +(c2) -- +(c4);
  \draw[thick] (P3) ++(r8) +(c1) -- +(c2) -- +(c3) -- +(c4) -- cycle;

\end{tikzpicture}      
&    
 \begin{tikzpicture}[scale=2]
 \def\radius{38.9};
  \coordinate (Pempty) at (0,0);
  \coordinate (P1) at (-1.5,1);
  \coordinate (P2) at (-0.5,1);
  \coordinate (P3) at (0.5,1);
  \coordinate (P4) at (1.5,1);
  \coordinate (P12) at (-1,2);
  \coordinate (P14) at (0,2);

  \coordinate (c1) at (-0.05,-0.05);
  \coordinate (c2) at (0.05,-0.05);
  \coordinate (c3) at (0.05,0.05);
  \coordinate (c4) at (-0.05,0.05);

  \coordinate (r1) at (-0.1,0);
  \coordinate (r2) at (0,0.1);
  \coordinate (r3) at (0.2,0.1);
  \coordinate (r4) at (-0.3,0);
  \coordinate (r9) at (-0.1,0.1);

  \draw[thick] (P3) +(c1) -- +(c2) -- +(c3) -- +(c4) -- +(c1) +(c1) -- +(c3) +(c2) -- +(c4);
  \draw[thick] (P3) ++(r2) +(c1) -- +(c2) -- +(c3) -- +(c4) -- cycle;
  \draw[thick] (P3) ++(r9) +(c1) -- +(c2) -- +(c3) -- +(c4) -- cycle;

\end{tikzpicture} 
    \\
Probability & 0.7351 & 0.4446 & 0.4258 & 0.3232 & 0.1949 \\ \hline
& & & & & \\
Interaction &      
\begin{tikzpicture}[scale=2]
  \def\radius{38.9};
  \coordinate (Pempty) at (0,0);
  \coordinate (P1) at (-1.5,1);
  \coordinate (P2) at (-0.5,1);
  \coordinate (P3) at (0.5,1);
  \coordinate (P4) at (1.5,1);
  \coordinate (P12) at (-1,2);
  \coordinate (P14) at (0,2);

  \coordinate (c1) at (-0.05,-0.05);
  \coordinate (c2) at (0.05,-0.05);
  \coordinate (c3) at (0.05,0.05);
  \coordinate (c4) at (-0.05,0.05);

  \coordinate (r1) at (-0.1,0);
  \coordinate (r2) at (0,0.1);
  \coordinate (r3) at (0.2,0.1);
  \coordinate (r4) at (-0.3,0);

  \draw[thick] (P3) +(c1) -- +(c2) -- +(c3) -- +(c4) -- +(c1) +(c1) -- +(c3) +(c2) -- +(c4);
  \draw[thick] (P3) ++(r2) +(c1) -- +(c2) -- +(c3) -- +(c4) -- cycle;
  \draw[thick] (P3) ++(r4) +(c1) -- +(c2) -- +(c3) -- +(c4) -- cycle;
\end{tikzpicture}
&     
 \begin{tikzpicture}[scale=2]
  \def\radius{38.9};
  \coordinate (Pempty) at (0,0);
  \coordinate (P1) at (-1.5,1);
  \coordinate (P2) at (-0.5,1);
  \coordinate (P3) at (0.5,1);
  \coordinate (P4) at (1.5,1);
  \coordinate (P12) at (-1,2);
  \coordinate (P14) at (0,2);

  \coordinate (c1) at (-0.05,-0.05);
  \coordinate (c2) at (0.05,-0.05);
  \coordinate (c3) at (0.05,0.05);
  \coordinate (c4) at (-0.05,0.05);

  \coordinate (r1) at (-0.1,0);
  \coordinate (r2) at (0,0.1);
  \coordinate (r3) at (0.2,0.1);
  \coordinate (r4) at (-0.3,0);
  \coordinate (r10) at (0.3,0.1);

  \draw[thick] (P3) +(c1) -- +(c2) -- +(c3) -- +(c4) -- +(c1) +(c1) -- +(c3) +(c2) -- +(c4);
  \draw[thick] (P3) ++(r10) +(c1) -- +(c2) -- +(c3) -- +(c4) -- cycle;

\end{tikzpicture} 
&    
\begin{tikzpicture}[scale=2]
  \def\radius{38.9};
  \coordinate (Pempty) at (0,0);
  \coordinate (P1) at (-1.5,1);
  \coordinate (P2) at (-0.5,1);
  \coordinate (P3) at (0.5,1);
  \coordinate (P4) at (1.5,1);
  \coordinate (P12) at (-1,2);
  \coordinate (P14) at (0,2);

  \coordinate (c1) at (-0.05,-0.05);
  \coordinate (c2) at (0.05,-0.05);
  \coordinate (c3) at (0.05,0.05);
  \coordinate (c4) at (-0.05,0.05);

  \coordinate (r1) at (-0.1,0);
  \coordinate (r2) at (0,0.1);
  \coordinate (r3) at (0.2,0.1);
  \coordinate (r4) at (-0.3,0);
  \coordinate (r5) at (0.4,0.1);

  \draw[thick] (P3) +(c1) -- +(c2) -- +(c3) -- +(c4) -- +(c1) +(c1) -- +(c3) +(c2) -- +(c4);
  \draw[thick] (P3) ++(r4) +(c1) -- +(c2) -- +(c3) -- +(c4) -- cycle;
  \draw[thick] (P3) ++(r5) +(c1) -- +(c2) -- +(c3) -- +(c4) -- cycle;
   \end{tikzpicture} 
&    
 \begin{tikzpicture}[scale=2]
 \def\radius{38.9};
  \coordinate (Pempty) at (0,0);
  \coordinate (P1) at (-1.5,1);
  \coordinate (P2) at (-0.5,1);
  \coordinate (P3) at (0.5,1);
  \coordinate (P4) at (1.5,1);
  \coordinate (P12) at (-1,2);
  \coordinate (P14) at (0,2);

  \coordinate (c1) at (-0.05,-0.05);
  \coordinate (c2) at (0.05,-0.05);
  \coordinate (c3) at (0.05,0.05);
  \coordinate (c4) at (-0.05,0.05);

  \coordinate (r1) at (-0.1,0);
  \coordinate (r2) at (0,0.1);
  \coordinate (r3) at (0.2,0.1);
  \coordinate (r4) at (-0.3,0);
  \coordinate (r5) at (0.4,0.1);

  \draw[thick] (P3) +(c1) -- +(c2) -- +(c3) -- +(c4) -- +(c1) +(c1) -- +(c3) +(c2) -- +(c4);
  \draw[thick] (P3) ++(r1) +(c1) -- +(c2) -- +(c3) -- +(c4) -- cycle;
  \draw[thick] (P3) ++(r2) +(c1) -- +(c2) -- +(c3) -- +(c4) -- cycle;
  \draw[thick] (P3) ++(r4) +(c1) -- +(c2) -- +(c3) -- +(c4) -- cycle;
\end{tikzpicture}      
&    
 \begin{tikzpicture}[scale=2]
 \def\radius{38.9};
  \coordinate (Pempty) at (0,0);
  \coordinate (P1) at (-1.5,1);
  \coordinate (P2) at (-0.5,1);
  \coordinate (P3) at (0.5,1);
  \coordinate (P4) at (1.5,1);
  \coordinate (P12) at (-1,2);
  \coordinate (P14) at (0,2);

  \coordinate (c1) at (-0.05,-0.05);
  \coordinate (c2) at (0.05,-0.05);
  \coordinate (c3) at (0.05,0.05);
  \coordinate (c4) at (-0.05,0.05);

  \coordinate (r1) at (-0.1,0);
  \coordinate (r2) at (0,0.1);
  \coordinate (r3) at (0.2,0.1);
  \coordinate (r4) at (-0.3,0);
  \coordinate (r5) at (0.4,0.1);

  \draw[thick] (P3) +(c1) -- +(c2) -- +(c3) -- +(c4) -- +(c1) +(c1) -- +(c3) +(c2) -- +(c4);
  \draw[thick] (P3) ++(r2) +(c1) -- +(c2) -- +(c3) -- +(c4) -- cycle;
  \draw[thick] (P3) ++(r4) +(c1) -- +(c2) -- +(c3) -- +(c4) -- cycle;
  \draw[thick] (P3) ++(r5) +(c1) -- +(c2) -- +(c3) -- +(c4) -- cycle;
\end{tikzpicture} 
    \\
Probability & 0.1882 & 0.1829 & 0.1794 & 0.1531 & 0.1260 \\ \hline
\end{tabular}
\end{table}
We see that many interactions have high posterior probabilities. 

\begin{figure}
        \begin{subfigure}[b]{0.5\textwidth}
         \vspace*{-0.1cm}
                \includegraphics[width=0.9\linewidth]{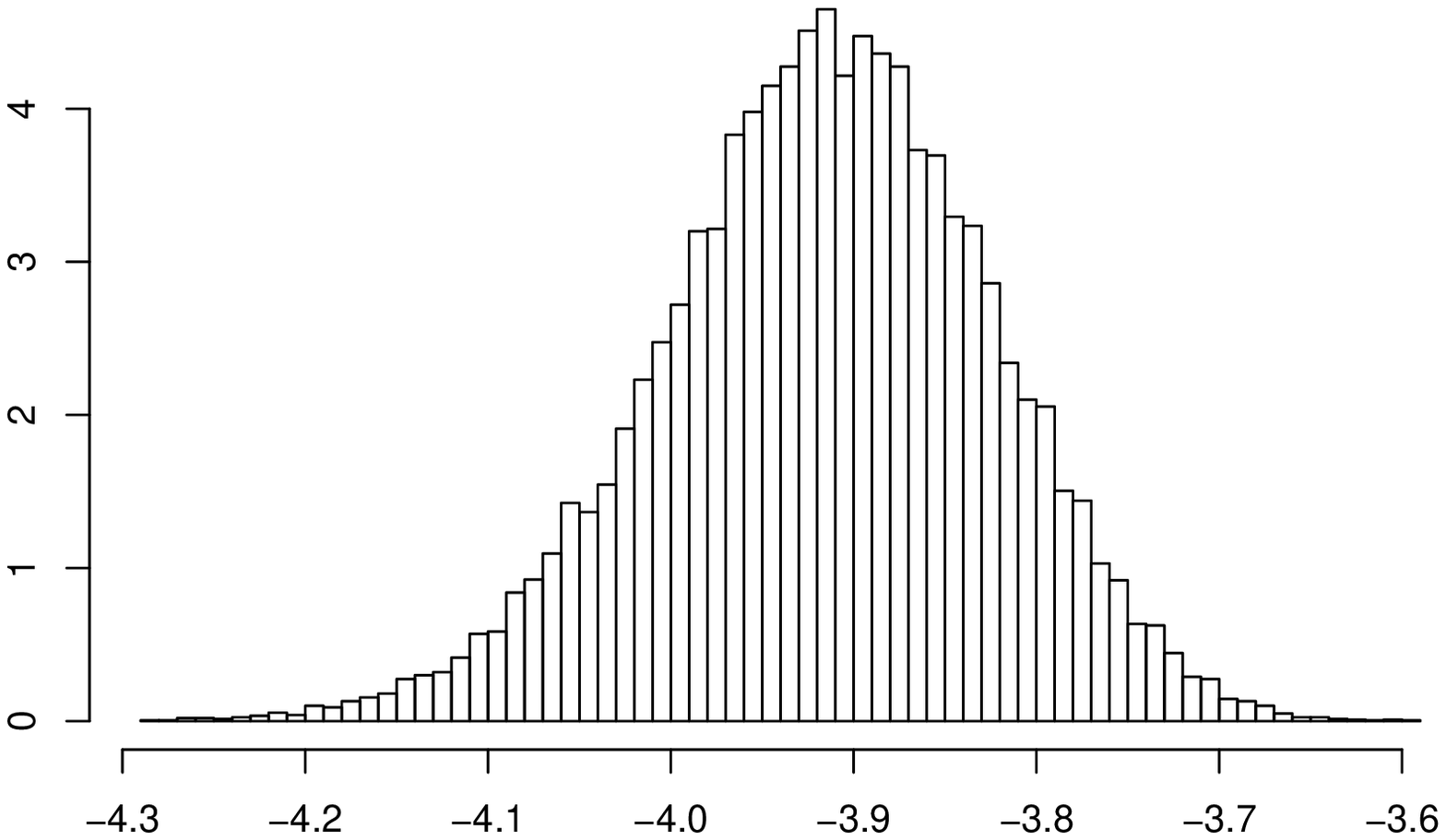}
         \vspace*{-0.5cm}
                \caption[]{$\theta\hspace*{-0.08cm}\left(\hspace*{0.02cm}
                \begin{tikzpicture}[scale=2]
	  \def\radius{38.9};	
  		\coordinate (Pempty) at (0,0);

  \coordinate (c1) at (-0.05,-0.05);
  \coordinate (c2) at (0.05,-0.05);
  \coordinate (c3) at (0.05,0.05);
  \coordinate (c4) at (-0.05,0.05);

  \draw[thick] (Pempty) +(c1) -- +(c2) -- +(c3) -- +(c4) -- +(c1) +(c1) -- +(c3) +(c2) -- +(c4);

\end{tikzpicture}
\hspace*{0.02cm}\right)$}
                \label{fig:histpar1_sisim}
        \end{subfigure}%
        \hfill
        \begin{subfigure}[b]{0.5\textwidth}
        \vspace*{-0.1cm}
                \includegraphics[width=0.9\linewidth]{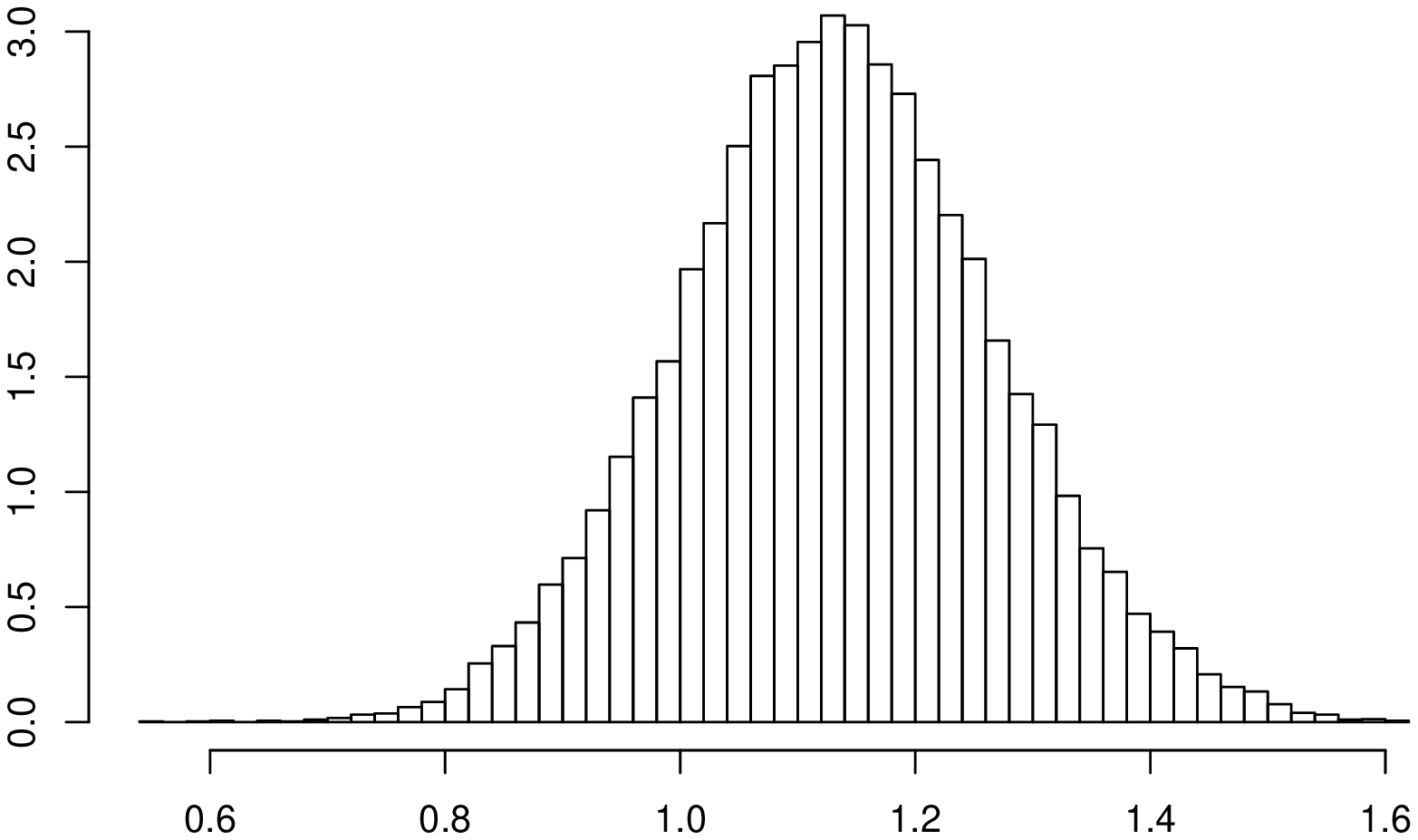}
         \vspace*{-0.5cm}
                \caption[]{$\theta\hspace*{-0.05cm}\left(\hspace*{0.02cm}
                \begin{tikzpicture}[scale=2]
  \def\radius{38.9};
  \coordinate (Pempty) at (0,0);
  \coordinate (P1) at (-1.5,1);
  \coordinate (P2) at (-0.5,1);
  \coordinate (P3) at (0.5,1);
  \coordinate (P4) at (1.5,1);
  \coordinate (P12) at (-1,2);
  \coordinate (P14) at (0,2);

  \coordinate (c1) at (-0.05,-0.05);
  \coordinate (c2) at (0.05,-0.05);
  \coordinate (c3) at (0.05,0.05);
  \coordinate (c4) at (-0.05,0.05);
  
  \coordinate (r1) at (-0.1,0);
  \coordinate (r2) at (0,0.1);
  \coordinate (r3) at (-0.1,0.1);
  \coordinate (r4) at (0.1,0.1);

  \draw[thick] (P12) +(c1) -- +(c2) -- +(c3) -- +(c4) -- +(c1) +(c1) -- +(c3) +(c2) -- +(c4);
  \draw[thick] (P12) ++(r1) +(c1) -- +(c2) -- +(c3) -- +(c4) -- cycle;

\end{tikzpicture}
\hspace*{0.02cm}\right)$}
                \label{fig:histpar2_sisim}
        \end{subfigure}
        
        \begin{subfigure}[b]{0.5\textwidth}
        \vspace*{-0.1cm}
                \includegraphics[width=0.9\linewidth]{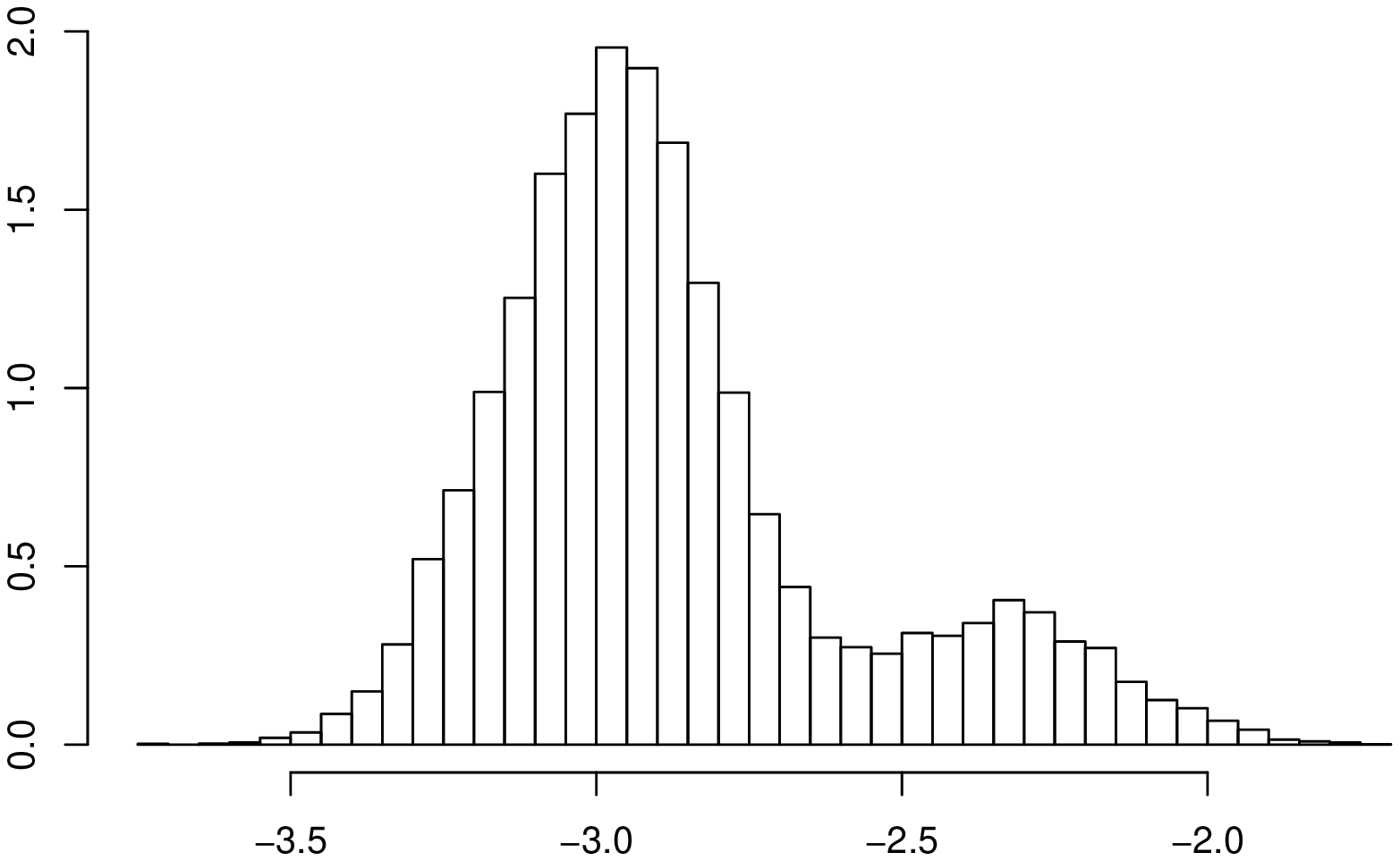}
         \vspace*{-0.5cm}
            \caption[]{$\theta\hspace*{-0.02cm}\left(\hspace*{-0.15cm}\raisebox{-0.1cm}{
               \begin{tikzpicture}[scale=2]
  \def\radius{38.9};
  \coordinate (Pempty) at (0,0);
  \coordinate (P1) at (-1.5,1);
  \coordinate (P2) at (-0.5,1);
  \coordinate (P3) at (0.5,1);
  \coordinate (P4) at (1.5,1);
  \coordinate (P12) at (-1,2);
  \coordinate (P14) at (0,2);

  \coordinate (c1) at (-0.05,-0.05);
  \coordinate (c2) at (0.05,-0.05);
  \coordinate (c3) at (0.05,0.05);
  \coordinate (c4) at (-0.05,0.05);

  \coordinate (r1) at (-0.1,0);
  \coordinate (r2) at (0,0.1);
  \coordinate (r3) at (-0.1,0.1);
  \coordinate (r4) at (0.1,0.1);  

  \draw[thick] (P2) +(c1) -- +(c2) -- +(c3) -- +(c4) -- +(c1) +(c1) -- +(c3) +(c2) -- +(c4);
  \draw[thick] (P2) ++(r2) +(c1) -- +(c2) -- +(c3) -- +(c4) -- cycle;

\end{tikzpicture}}
\right)$}
                \label{fig:histpar3_sisim}
        \end{subfigure}%
        \hfill
        \begin{subfigure}[b]{0.5\textwidth}
        \vspace*{-0.1cm}
                \includegraphics[width=0.9\linewidth]{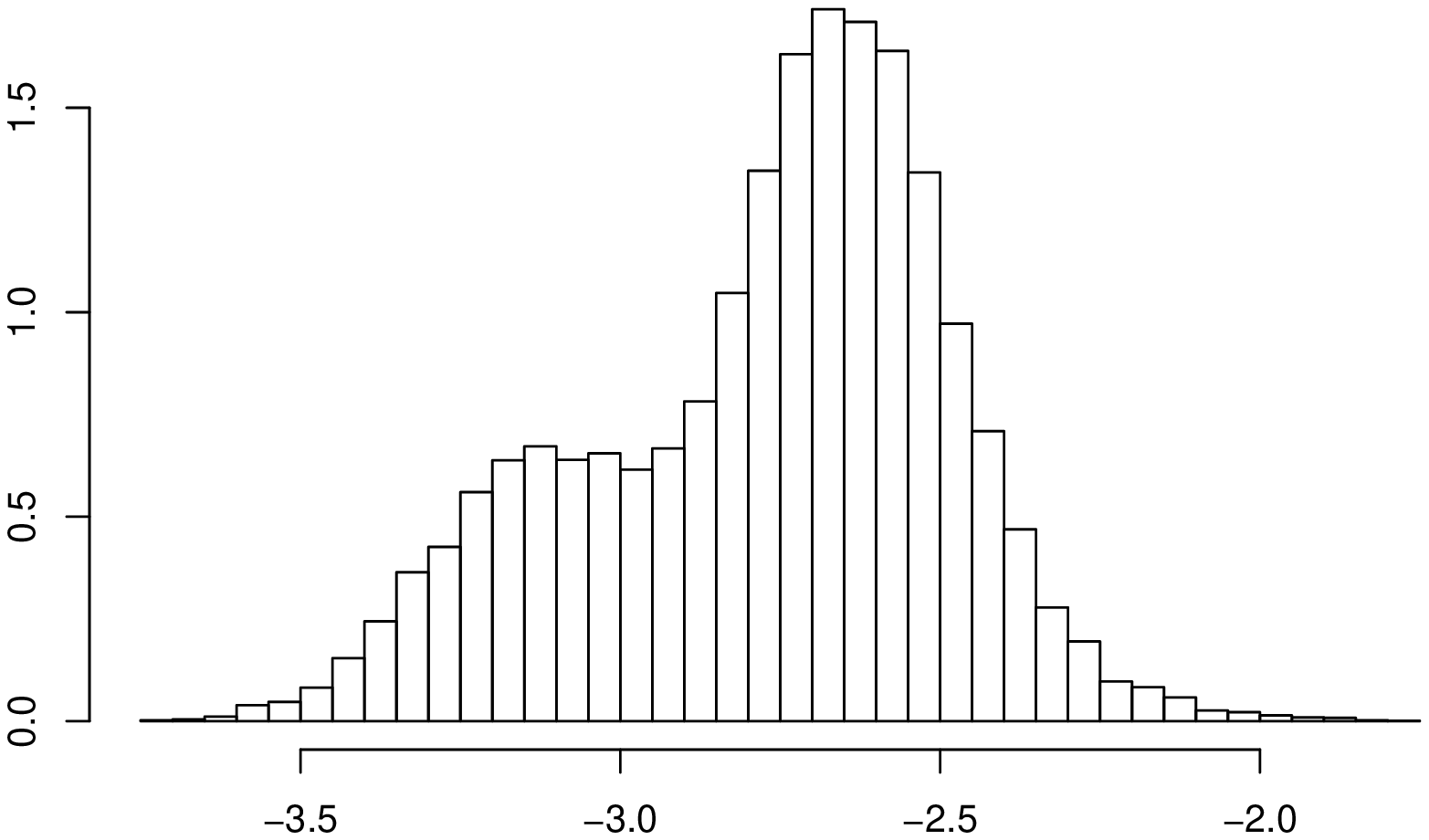}
        \vspace*{-0.5cm}
                \caption[]{$\theta\hspace*{-0.06cm}\left(\hspace*{-0.15cm}\raisebox{-0.1cm}{
              \begin{tikzpicture}[scale=2]
   \def\radius{38.9};
  \coordinate (Pempty) at (0,0);
  \coordinate (P1) at (-1.5,1);
  \coordinate (P2) at (-0.5,1);
  \coordinate (P3) at (0.5,1);
  \coordinate (P4) at (1.5,1);
  \coordinate (P12) at (-1,2);
  \coordinate (P14) at (0,2);

  \coordinate (c1) at (-0.05,-0.05);
  \coordinate (c2) at (0.05,-0.05);
  \coordinate (c3) at (0.05,0.05);
  \coordinate (c4) at (-0.05,0.05);

  \coordinate (r1) at (-0.1,0);
  \coordinate (r2) at (0,0.1);
  \coordinate (r3) at (0.2,0.1);
 
  \draw[thick] (P12) +(c1) -- +(c2) -- +(c3) -- +(c4) -- +(c1) +(c1) -- +(c3) +(c2) -- +(c4);
  \draw[thick] (P12) ++(r3) +(c1) -- +(c2) -- +(c3) -- +(c4) -- cycle;

\end{tikzpicture}}
\right)$}
                \label{fig:histpar4_sisim}
        \end{subfigure}
\hfill 
 \begin{subfigure}[b]{0.5\textwidth}
         \vspace*{-0.1cm}
                \includegraphics[width=0.9\linewidth]{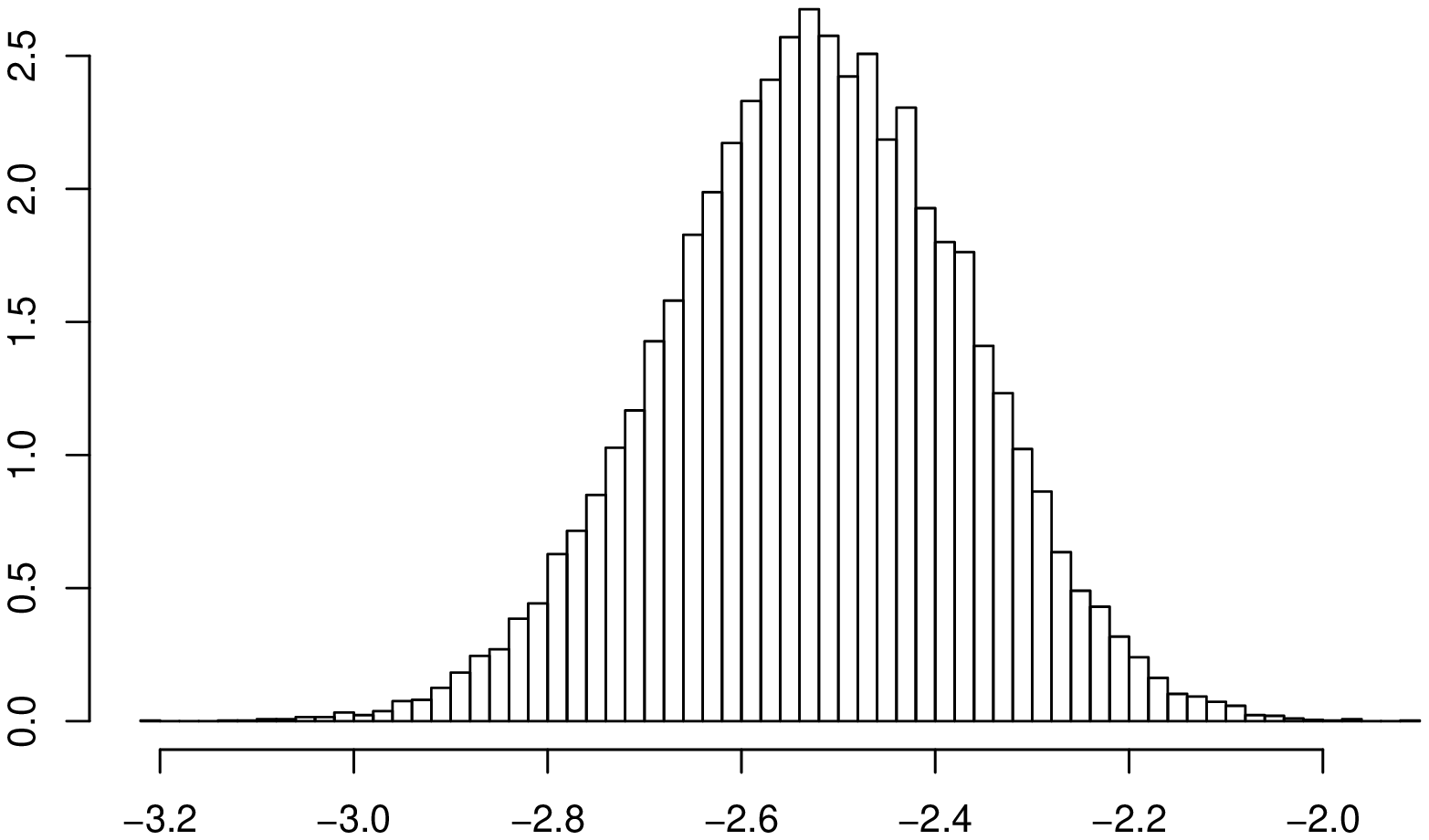}
         \vspace*{-0.5cm}
                \caption[]{$\theta\hspace*{-0.05cm}\left(\hspace*{0.02cm}
                \begin{tikzpicture}[scale=2]
  \def\radius{38.9};
  \coordinate (Pempty) at (0,0);
  \coordinate (P1) at (-1.5,1);
  \coordinate (P2) at (-0.5,1);
  \coordinate (P3) at (0.5,1);
  \coordinate (P4) at (1.5,1);
  \coordinate (P12) at (-1,2);
  \coordinate (P14) at (0,2);

  \coordinate (c1) at (-0.05,-0.05);
  \coordinate (c2) at (0.05,-0.05);
  \coordinate (c3) at (0.05,0.05);
  \coordinate (c4) at (-0.05,0.05);

  \coordinate (r1) at (-0.1,0);
  \coordinate (r2) at (0,0.1);
  \coordinate (r3) at (0.2,0.1);
  \coordinate (r4) at (-0.3,0);

  \draw[thick] (P3) +(c1) -- +(c2) -- +(c3) -- +(c4) -- +(c1) +(c1) -- +(c3) +(c2) -- +(c4);
  \draw[thick] (P3) ++(r4) +(c1) -- +(c2) -- +(c3) -- +(c4) -- cycle;

\end{tikzpicture}
\hspace*{-0.02cm}\right)$}
                \label{fig:histpar5_sisim}
        \end{subfigure}%
        \hfill
        \begin{subfigure}[b]{0.5\textwidth}
        \vspace*{-0.1cm}
                \includegraphics[width=0.9\linewidth]{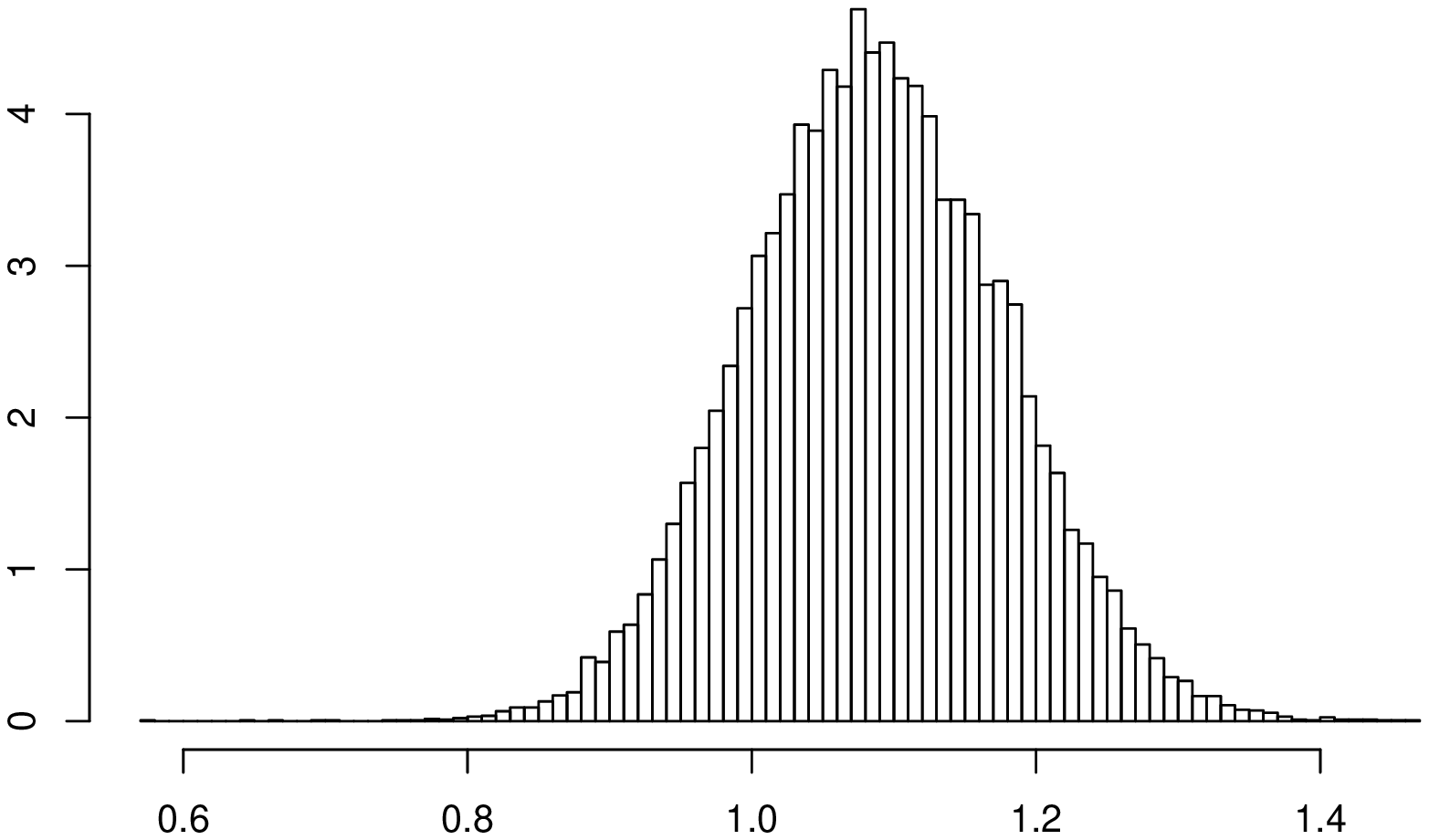}
         \vspace*{-0.5cm}
                \caption[]{$\theta\hspace*{-0.05cm}\left(\hspace*{0.02cm}
\begin{tikzpicture}[scale=2]
  \def\radius{38.9};
  \coordinate (Pempty) at (0,0);
  \coordinate (P1) at (-1.5,1);
  \coordinate (P2) at (-0.5,1);
  \coordinate (P3) at (0.5,1);
  \coordinate (P4) at (1.5,1);
  \coordinate (P12) at (-1,2);
  \coordinate (P14) at (0,2);

  \coordinate (c1) at (-0.05,-0.05);
  \coordinate (c2) at (0.05,-0.05);
  \coordinate (c3) at (0.05,0.05);
  \coordinate (c4) at (-0.05,0.05);

  \coordinate (r1) at (-0.1,0);
  \coordinate (r2) at (0,0.1);
  \coordinate (r3) at (0.2,0.1);
  \coordinate (r4) at (-0.3,0);

  \draw[thick] (P3) +(c1) -- +(c2) -- +(c3) -- +(c4) -- +(c1) +(c1) -- +(c3) +(c2) -- +(c4);
  \draw[thick] (P3) ++(r1) +(c1) -- +(c2) -- +(c3) -- +(c4) -- cycle;
  \draw[thick] (P3) ++(r4) +(c1) -- +(c2) -- +(c3) -- +(c4) -- cycle;
\end{tikzpicture}
\hspace*{0.02cm}\right)$}
                \label{fig:histpar6_sisim}
        \end{subfigure}
        
        \begin{subfigure}[b]{0.5\textwidth}
        \vspace*{-0.1cm}
                \includegraphics[width=0.9\linewidth]{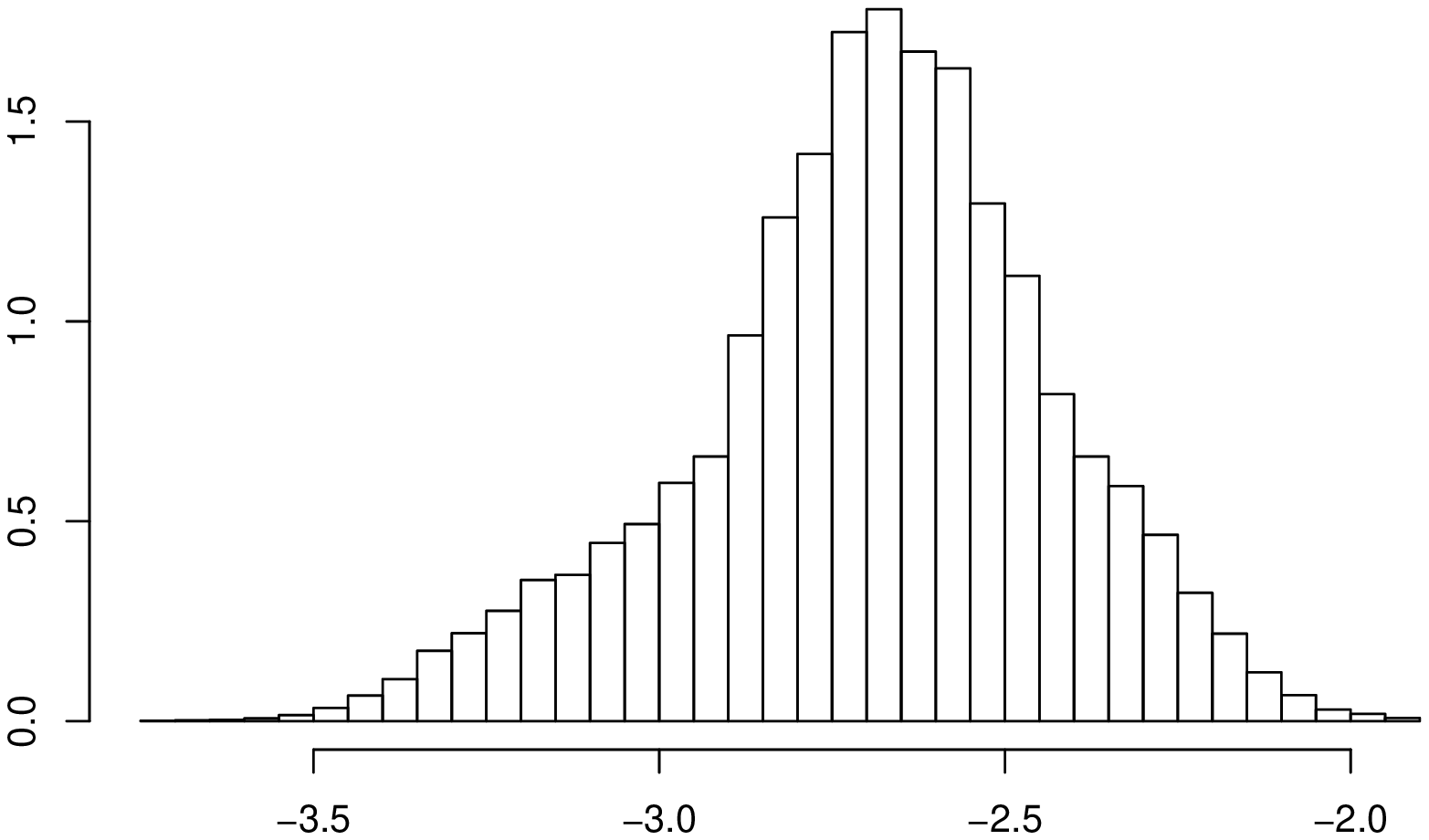}
         \vspace*{-0.5cm}
            \caption[]{$\theta\hspace*{-0.06cm}\left(\hspace*{-0.15cm}\raisebox{-0.1cm}{
               \begin{tikzpicture}[scale=2]
  \def\radius{38.9};
  \coordinate (Pempty) at (0,0);
  \coordinate (P1) at (-1.5,1);
  \coordinate (P2) at (-0.5,1);
  \coordinate (P3) at (0.5,1);
  \coordinate (P4) at (1.5,1);
  \coordinate (P12) at (-1,2);
  \coordinate (P14) at (0,2);

  \coordinate (c1) at (-0.05,-0.05);
  \coordinate (c2) at (0.05,-0.05);
  \coordinate (c3) at (0.05,0.05);
  \coordinate (c4) at (-0.05,0.05);

  \coordinate (r1) at (-0.1,0);
  \coordinate (r2) at (0,0.1);
  \coordinate (r3) at (0.2,0.1);
  \coordinate (r4) at (-0.3,0);
  \coordinate (r5) at (0.4,0.1);

  \draw[thick] (P3) +(c1) -- +(c2) -- +(c3) -- +(c4) -- +(c1) +(c1) -- +(c3) +(c2) -- +(c4);
  \draw[thick] (P3) ++(r5) +(c1) -- +(c2) -- +(c3) -- +(c4) -- cycle;

\end{tikzpicture}}
\right)$}
                \label{fig:histpar7_sisim}
        \end{subfigure}%
        \hfill
        \begin{subfigure}[b]{0.5\textwidth}
        \vspace*{-0.1cm}
                \includegraphics[width=0.9\linewidth]{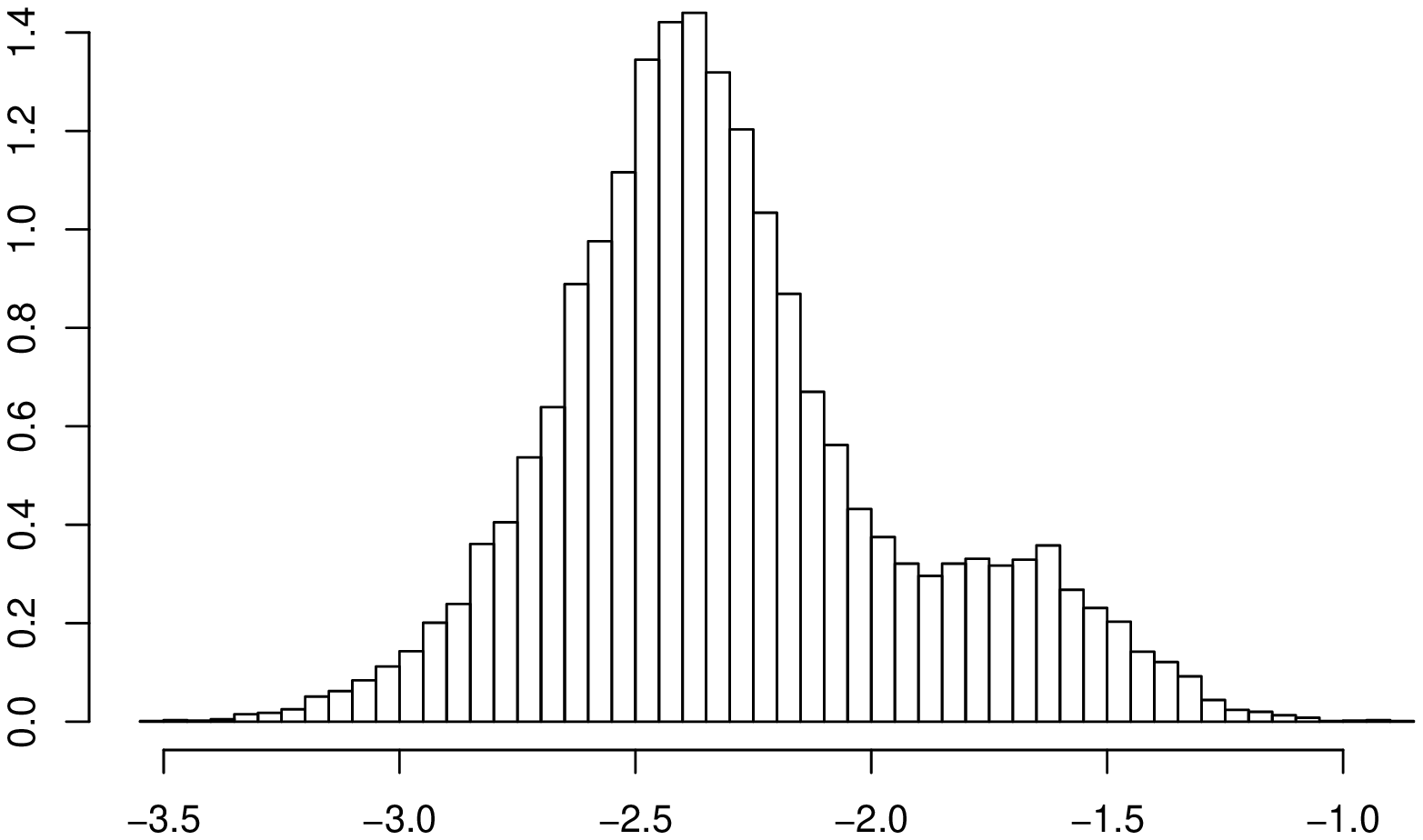}
        \vspace*{-0.5cm}
                \caption[]{$\theta\hspace*{-0.06cm}\left(\hspace*{-0.15cm}\raisebox{-0.1cm}{
              \begin{tikzpicture}[scale=2]
  \def\radius{38.9};
  \coordinate (Pempty) at (0,0);
  \coordinate (P1) at (-1.5,1);
  \coordinate (P2) at (-0.5,1);
  \coordinate (P3) at (0.5,1);
  \coordinate (P4) at (1.5,1);
  \coordinate (P12) at (-1,2);
  \coordinate (P14) at (0,2);

  \coordinate (c1) at (-0.05,-0.05);
  \coordinate (c2) at (0.05,-0.05);
  \coordinate (c3) at (0.05,0.05);
  \coordinate (c4) at (-0.05,0.05);

  \coordinate (r1) at (-0.1,0);
  \coordinate (r2) at (0,0.1);
  \coordinate (r3) at (0.2,0.1);
  \coordinate (r4) at (-0.3,0);

  \draw[thick] (P3) +(c1) -- +(c2) -- +(c3) -- +(c4) -- +(c1) +(c1) -- +(c3) +(c2) -- +(c4);
  \draw[thick] (P3) ++(r3) +(c1) -- +(c2) -- +(c3) -- +(c4) -- cycle;
  \draw[thick] (P3) ++(r4) +(c1) -- +(c2) -- +(c3) -- +(c4) -- cycle;
   \end{tikzpicture}}
\right)$}
                \label{fig:histpar8_sisim}
        \end{subfigure} 
        \caption{Sisim data set example: Histograms of the simulated parameter values $\theta(\cdot)$ for the top eight 
a posteriori most likely interactions.}\label{fig:histpar_sisim}
\end{figure}

\begin{figure}
        \begin{center}
                \begin{tikzpicture}[scale=1.75]
  \def\radius{38.9};
  \coordinate (Pempty) at (0,-1);
  \coordinate (P1) at (-3.75,1);
  \coordinate (P2) at (-2.25,1);
  \coordinate (P3) at (-0.75,1);
  \coordinate (P4) at (0.75,1);
  \coordinate (P5) at (2.25,1);
  \coordinate (P6) at (3.75,1);
  \coordinate (P13) at (-2.25,3);
  \coordinate (P15) at (-0.75,3);
  \coordinate (P35) at (0.75,3);
  \coordinate (P26) at (2.25,3);
  \coordinate (P135) at (0,5);

  \coordinate (c1) at (-0.05,-0.05);
  \coordinate (c2) at (0.05,-0.05);
  \coordinate (c3) at (0.05,0.05);
  \coordinate (c4) at (-0.05,0.05);

  \coordinate (r1) at (-0.1,0);
  \coordinate (r2) at (0,0.1);
  \coordinate (r3) at (0.2,0.1);
  \coordinate (r4) at (-0.1,0.3);
  \coordinate (r5) at (-0.3,0);
  \coordinate (r6) at (0.4,0.1);

  \node[draw,circle,inner sep=0pt,minimum size=1.4*\radius,name=Nempty] at (Pempty) {};
  \draw[thick] (Pempty) +(c1) -- +(c2) -- +(c3) -- +(c4) -- +(c1) +(c1) -- +(c3) +(c2) -- +(c4);
  
  \node[draw,circle,inner sep=0pt,minimum size=1.4*\radius,name=N1] at (P1) {};
  \draw[thick] (P1) +(c1) -- +(c2) -- +(c3) -- +(c4) -- +(c1) +(c1) -- +(c3) +(c2) -- +(c4);
  \draw[thick] (P1) ++(r1) +(c1) -- +(c2) -- +(c3) -- +(c4) -- cycle;
  \draw[thick,->] (Nempty) -- (N1);
   
  \node[draw,circle,inner sep=0pt,minimum size=1.4*\radius,name=N2] at (P2) {};
  \draw[thick] (P2) +(c1) -- +(c2) -- +(c3) -- +(c4) -- +(c1) +(c1) -- +(c3) +(c2) -- +(c4);
  \draw[thick] (P2) ++(r2) +(c1) -- +(c2) -- +(c3) -- +(c4) -- cycle;
  \draw[thick,->] (Nempty) -- (N2);
  
  \node[draw,circle,inner sep=0pt,minimum size=1.4*\radius,name=N3] at (P3) {};
  \draw[thick] (P3) +(c1) -- +(c2) -- +(c3) -- +(c4) -- +(c1) +(c1) -- +(c3) +(c2) -- +(c4);
  \draw[thick] (P3) ++(r3) +(c1) -- +(c2) -- +(c3) -- +(c4) -- cycle;
  \draw[thick,->] (Nempty) -- (N3);
  
   \node[draw,circle,inner sep=0pt,minimum size=1.4*\radius,name=N4] at (P4) {};
  \draw[thick] (P4) +(c1) -- +(c2) -- +(c3) -- +(c4) -- +(c1) +(c1) -- +(c3) +(c2) -- +(c4);
  \draw[thick] (P4) ++(r4) +(c1) -- +(c2) -- +(c3) -- +(c4) -- cycle;
  \draw[thick,->] (Nempty) -- (N4);
  
   \node[draw,circle,inner sep=0pt,minimum size=1.4*\radius,name=N5] at (P5) {};
  \draw[thick] (P5) +(c1) -- +(c2) -- +(c3) -- +(c4) -- +(c1) +(c1) -- +(c3) +(c2) -- +(c4);
  \draw[thick] (P5) ++(r5) +(c1) -- +(c2) -- +(c3) -- +(c4) -- cycle;
  \draw[thick,->] (Nempty) -- (N5);
  
   \node[draw,circle,inner sep=0pt,minimum size=1.4*\radius,name=N6] at (P6) {};
  \draw[thick] (P6) +(c1) -- +(c2) -- +(c3) -- +(c4) -- +(c1) +(c1) -- +(c3) +(c2) -- +(c4);
  \draw[thick] (P6) ++(r6) +(c1) -- +(c2) -- +(c3) -- +(c4) -- cycle;
  \draw[thick,->] (Nempty) -- (N6);
  
  \node[draw,circle,inner sep=0pt,minimum size=1.4*\radius,name=N13] at (P13) {};
  \draw[thick] (P13) +(c1) -- +(c2) -- +(c3) -- +(c4) -- +(c1) +(c1) -- +(c3) +(c2) -- +(c4);
  \draw[thick] (P13) ++(r1) +(c1) -- +(c2) -- +(c3) -- +(c4) -- cycle;
  \draw[thick] (P13) ++(r3) +(c1) -- +(c2) -- +(c3) -- +(c4) -- cycle;
  \draw[thick,->] (N1) -- (N13);
  \draw[thick,->] (N3) -- (N13);
  
   \node[draw,circle,inner sep=0pt,minimum size=1.4*\radius,name=N15] at (P15) {};
  \draw[thick] (P15) +(c1) -- +(c2) -- +(c3) -- +(c4) -- +(c1) +(c1) -- +(c3) +(c2) -- +(c4);
  \draw[thick] (P15) ++(r1) +(c1) -- +(c2) -- +(c3) -- +(c4) -- cycle;
  \draw[thick] (P15) ++(r5) +(c1) -- +(c2) -- +(c3) -- +(c4) -- cycle;
  \draw[thick,->] (N1) -- (N15);
  \draw[thick,->] (N5) -- (N15);
  
   \node[draw,circle,inner sep=0pt,minimum size=1.4*\radius,name=N35] at (P35) {};
  \draw[thick] (P35) +(c1) -- +(c2) -- +(c3) -- +(c4) -- +(c1) +(c1) -- +(c3) +(c2) -- +(c4);
  \draw[thick] (P35) ++(r3) +(c1) -- +(c2) -- +(c3) -- +(c4) -- cycle;
  \draw[thick] (P35) ++(r5) +(c1) -- +(c2) -- +(c3) -- +(c4) -- cycle;
  \draw[thick,->] (N3) -- (N35);
  \draw[thick,->] (N5) -- (N35);
  
   \node[draw,circle,inner sep=0pt,minimum size=1.4*\radius,name=N26] at (P26) {};
  \draw[thick] (P26) +(c1) -- +(c2) -- +(c3) -- +(c4) -- +(c1) +(c1) -- +(c3) +(c2) -- +(c4);
  \draw[thick] (P26) ++(r2) +(c1) -- +(c2) -- +(c3) -- +(c4) -- cycle;
  \draw[thick] (P26) ++(r6) +(c1) -- +(c2) -- +(c3) -- +(c4) -- cycle;
  \draw[thick,->] (N2) -- (N26);
  \draw[thick,->] (N6) -- (N26);
  
  \node[draw,circle,inner sep=0pt,minimum size=1.4*\radius,name=N135] at (P135) {};
  \draw[thick] (P135) +(c1) -- +(c2) -- +(c3) -- +(c4) -- +(c1) +(c1) -- +(c3) +(c2) -- +(c4);
  \draw[thick] (P135) ++(r1) +(c1) -- +(c2) -- +(c3) -- +(c4) -- cycle;
  \draw[thick] (P135) ++(r3) +(c1) -- +(c2) -- +(c3) -- +(c4) -- cycle;
   \draw[thick] (P135) ++(r5) +(c1) -- +(c2) -- +(c3) -- +(c4) -- cycle;
  \draw[thick,->] (N13) -- (N135);
  \draw[thick,->] (N15) -- (N135);
  \draw[thick,->] (N35) -- (N135);
  
\end{tikzpicture}
\end{center}

        \caption{Sisim data set example: The a posteriori most likely model. The 
estimated posterior probability for this model is $0.13802$.}\label{fig:dag_sisim}
\end{figure}
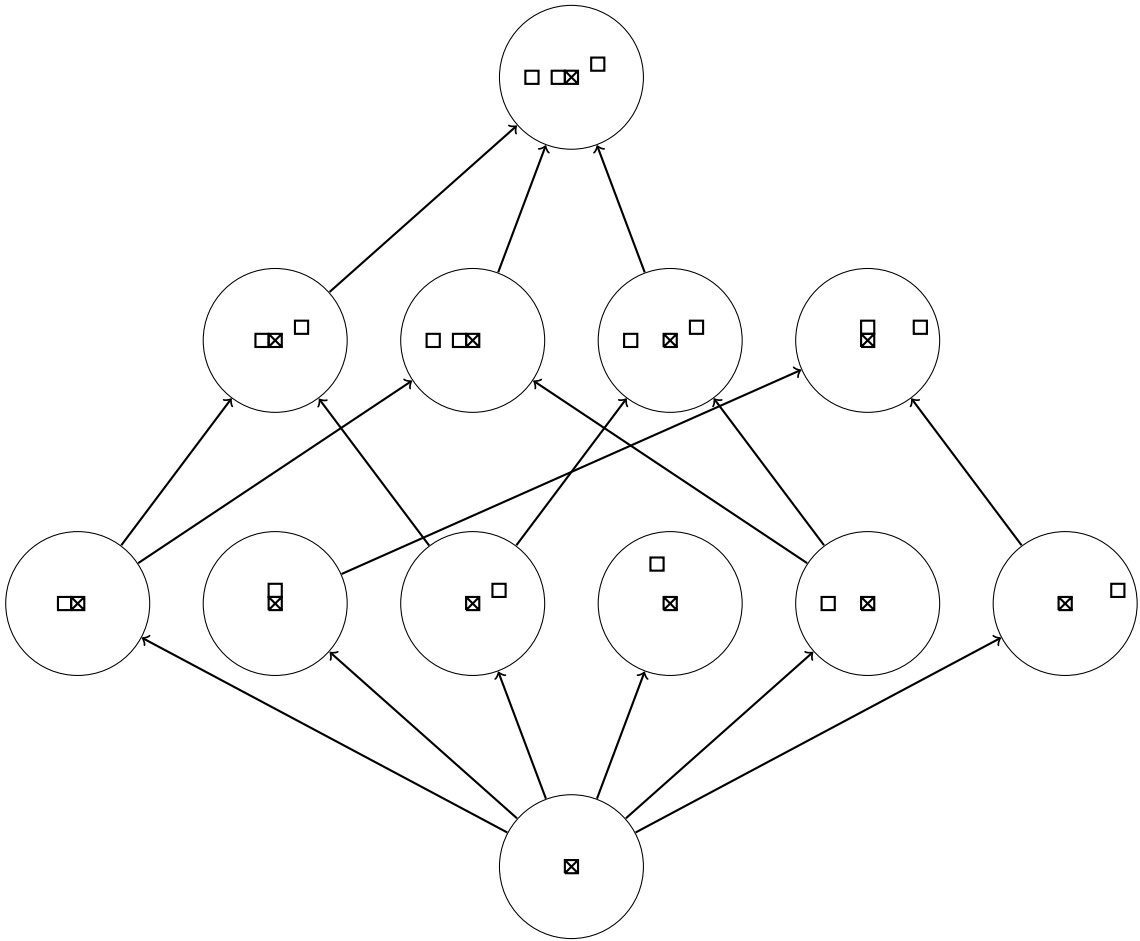

We also estimate the a posteriori marginal distributions for the parameter values $\theta(\cdot)$ corresponding to the top 
eight most likely interactions. Figure \ref{fig:histpar_sisim} depicts the histograms of the 
simulated parameter values $\theta(\cdot)$.
From the simulation we also estimate the posterior probability for each of the possible models. The most probable model is 
shown in Figure \ref{fig:dag_sisim}. This model has posterior probability equal to $0.13802$. 
The remaining probability mass is spread out on a very large number of models.

\begin{figure}
        \begin{subfigure}[b]{0.5\textwidth}
        	\vspace*{-0.5cm}
                \includegraphics[width=\linewidth]{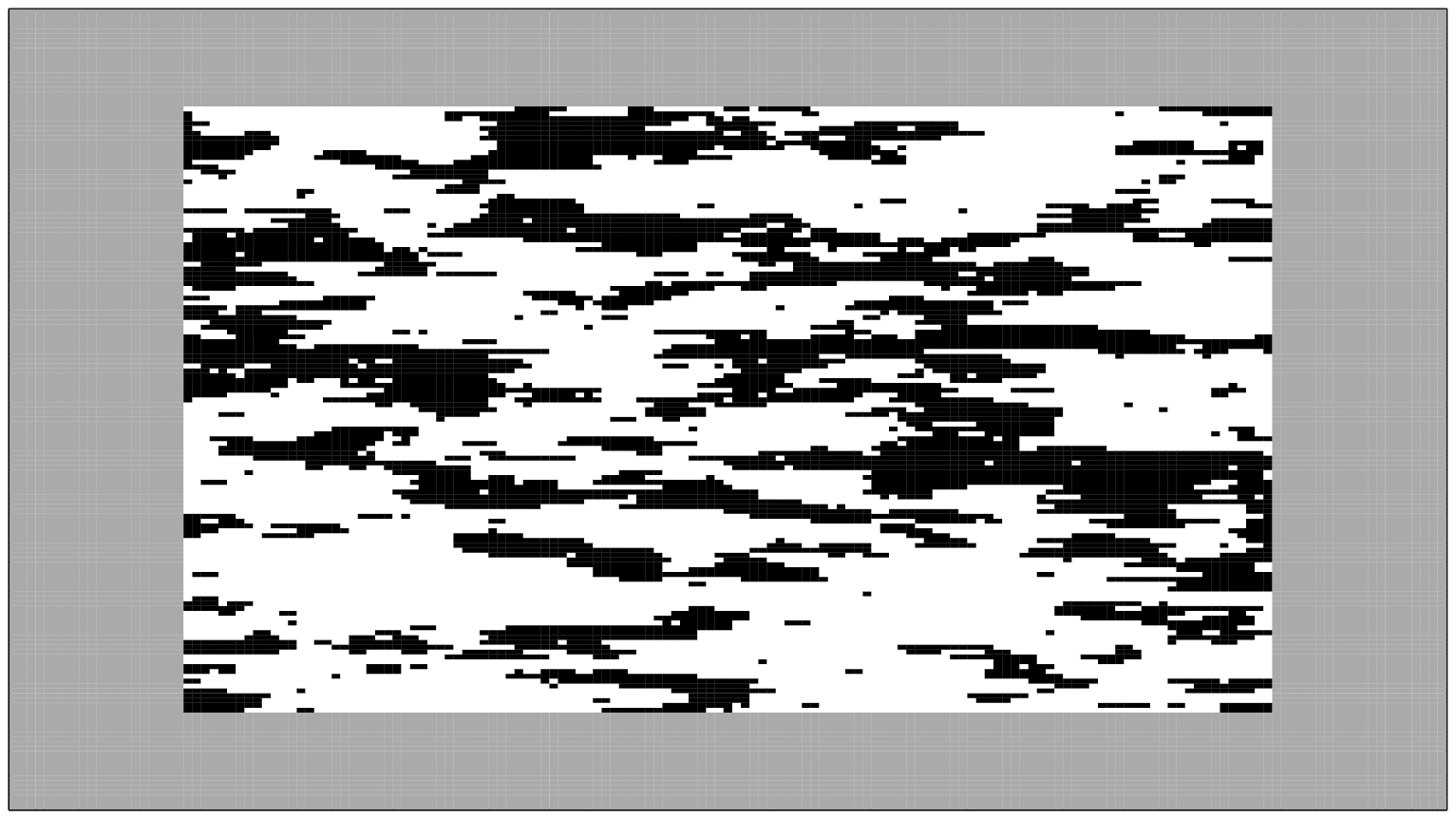}
         \vspace*{-1.5cm}
        \end{subfigure}%
        \hfill
        \begin{subfigure}[b]{0.5\textwidth}
        \vspace*{-0.5cm}
                \includegraphics[width=\linewidth]{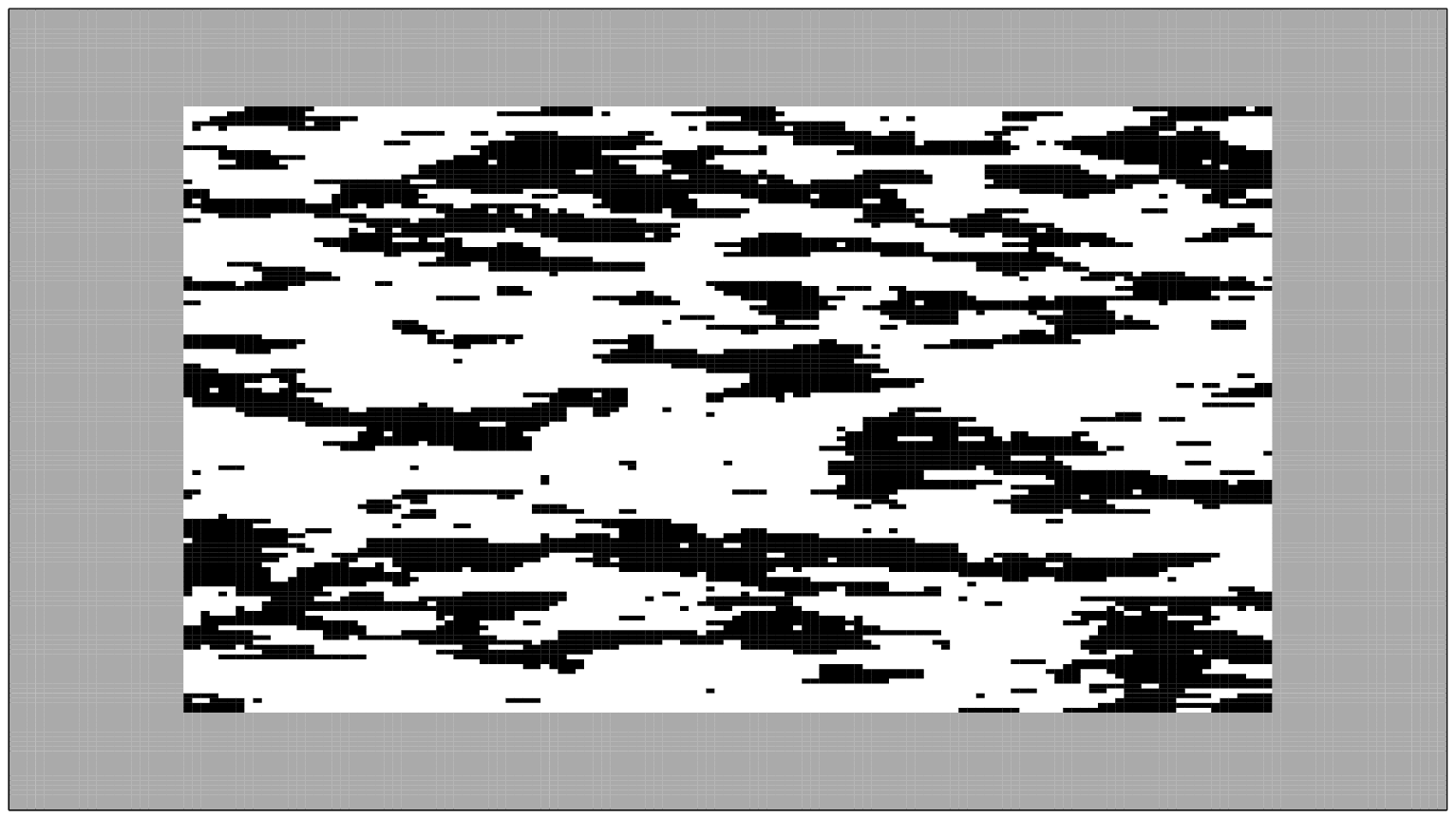}
        \vspace*{-1.5cm}
        \end{subfigure}
     
        \begin{subfigure}[b]{0.5\textwidth}
        \vspace*{-0.1cm}
                \includegraphics[width=\linewidth]{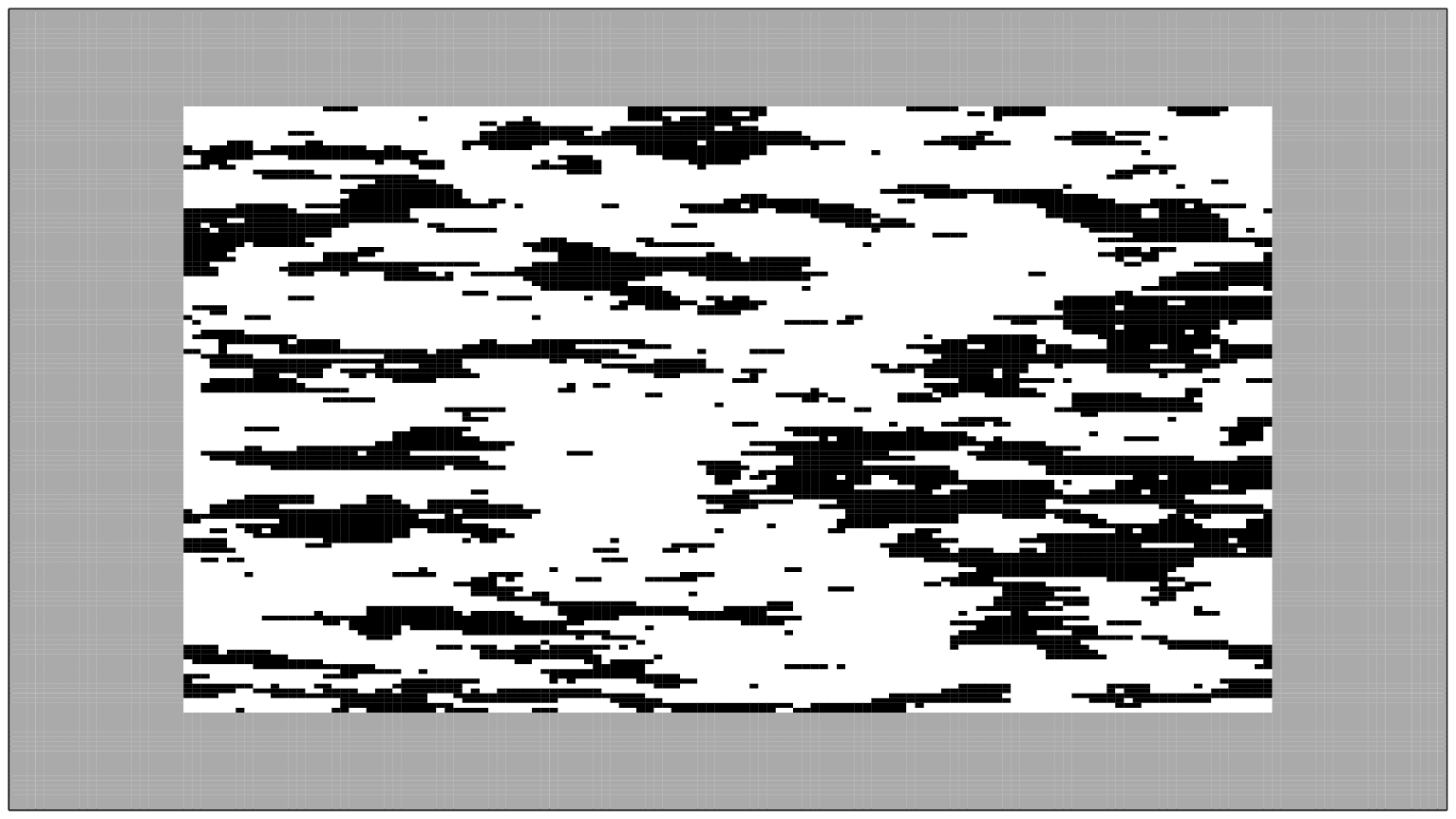}
        \vspace*{-1.5cm}
        \end{subfigure}%
        \hfill
        \begin{subfigure}[b]{0.5\textwidth}
         \vspace*{-0.1cm}
                \includegraphics[width=\linewidth]{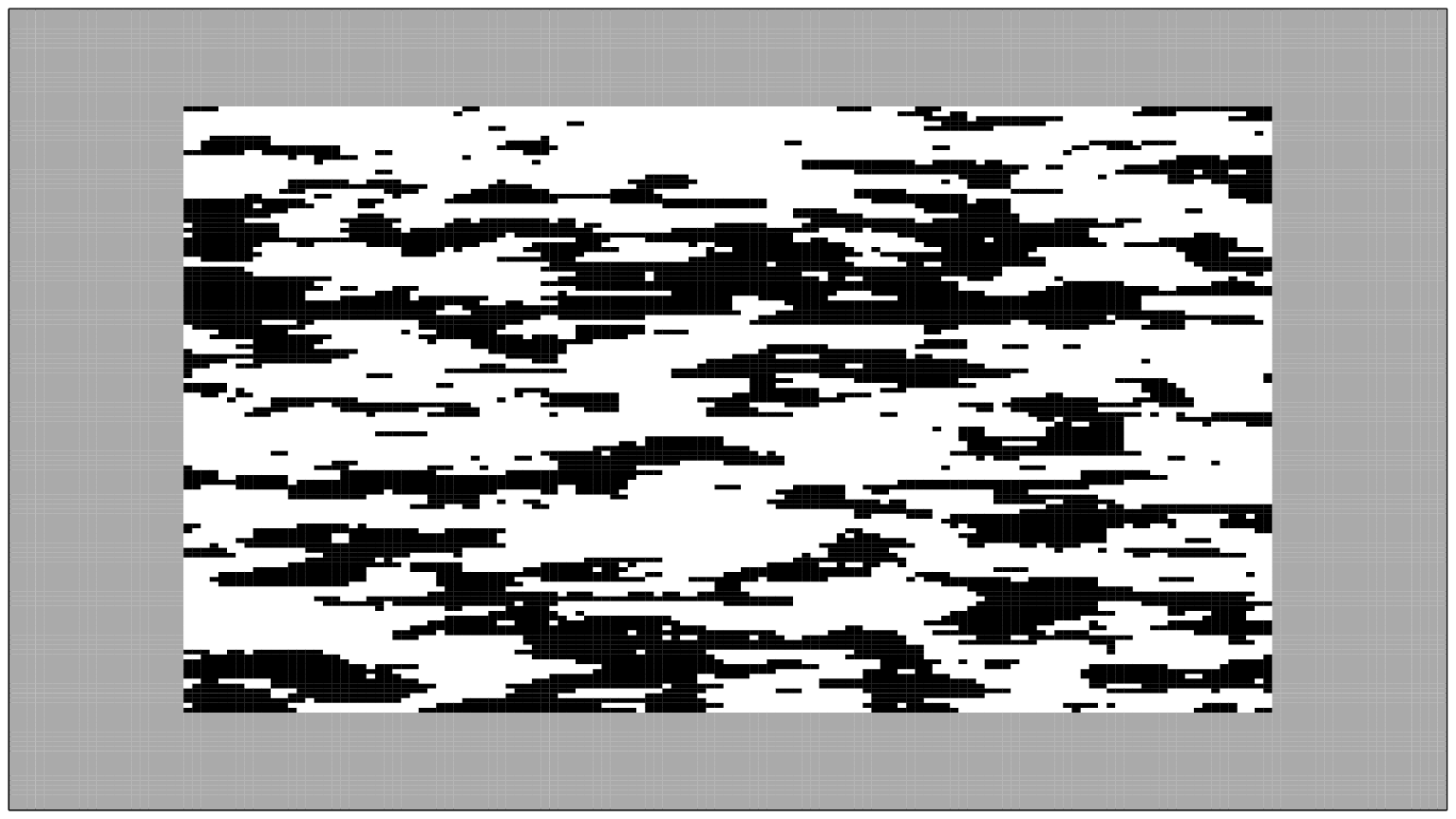}
        \vspace*{-1.5cm}
        \end{subfigure}      
        \caption{Sisim data set example: Four Markov mesh model realizations where the models used are randomly sampled 
from all models simulated in the RJMCMC (after the specified burn-in). The color coding is the same as in Figure 
\ref{fig:sisim}.}\label{fig:real_sisim}
\end{figure}

\begin{figure}
\begin{tabular}{@{}c@{}c@{}c@{}}
\\[-0.9cm]
  \includegraphics[width=0.33\linewidth,height=0.20\linewidth]{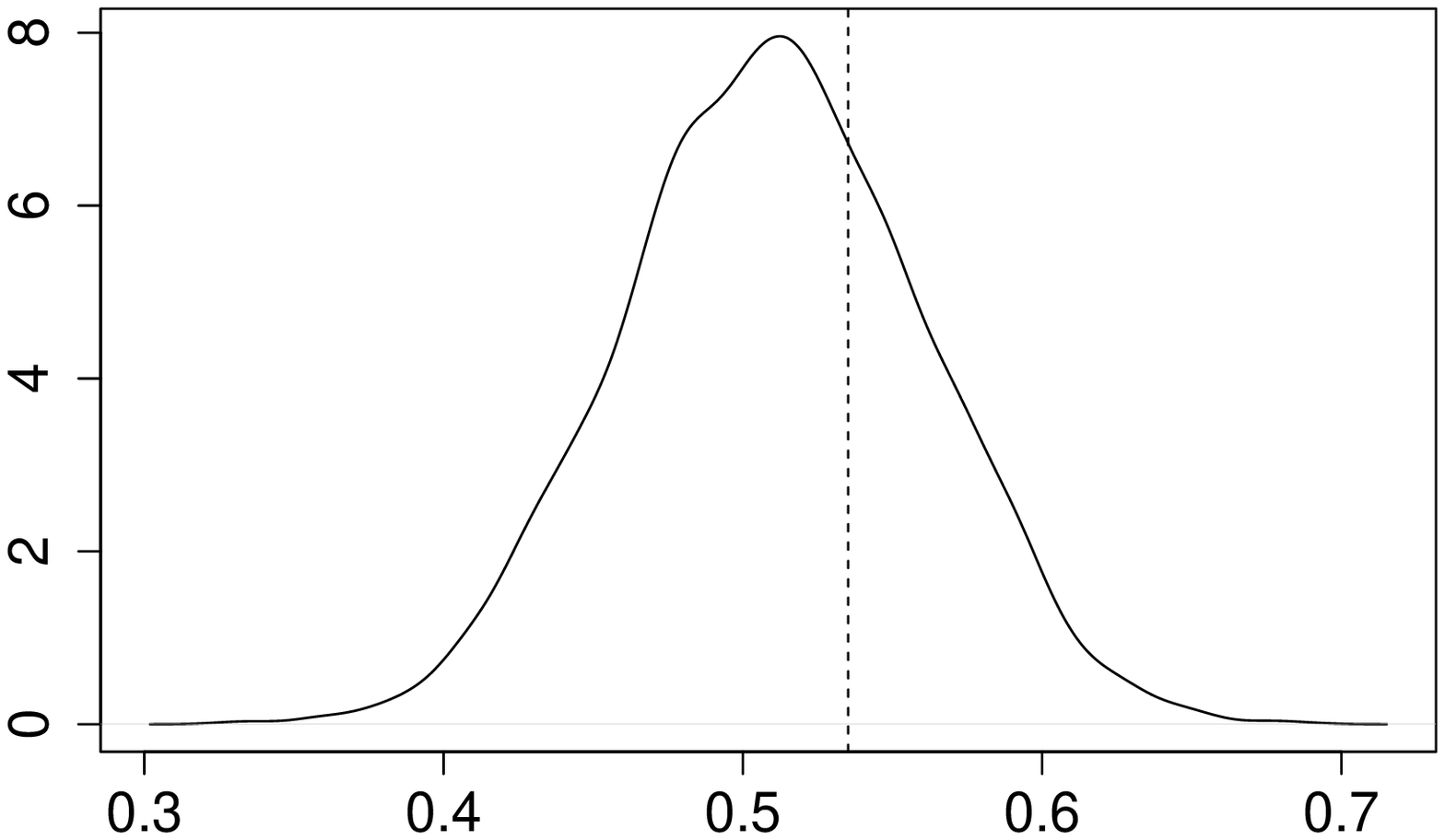} &
  \includegraphics[width=0.33\linewidth,height=0.20\linewidth]{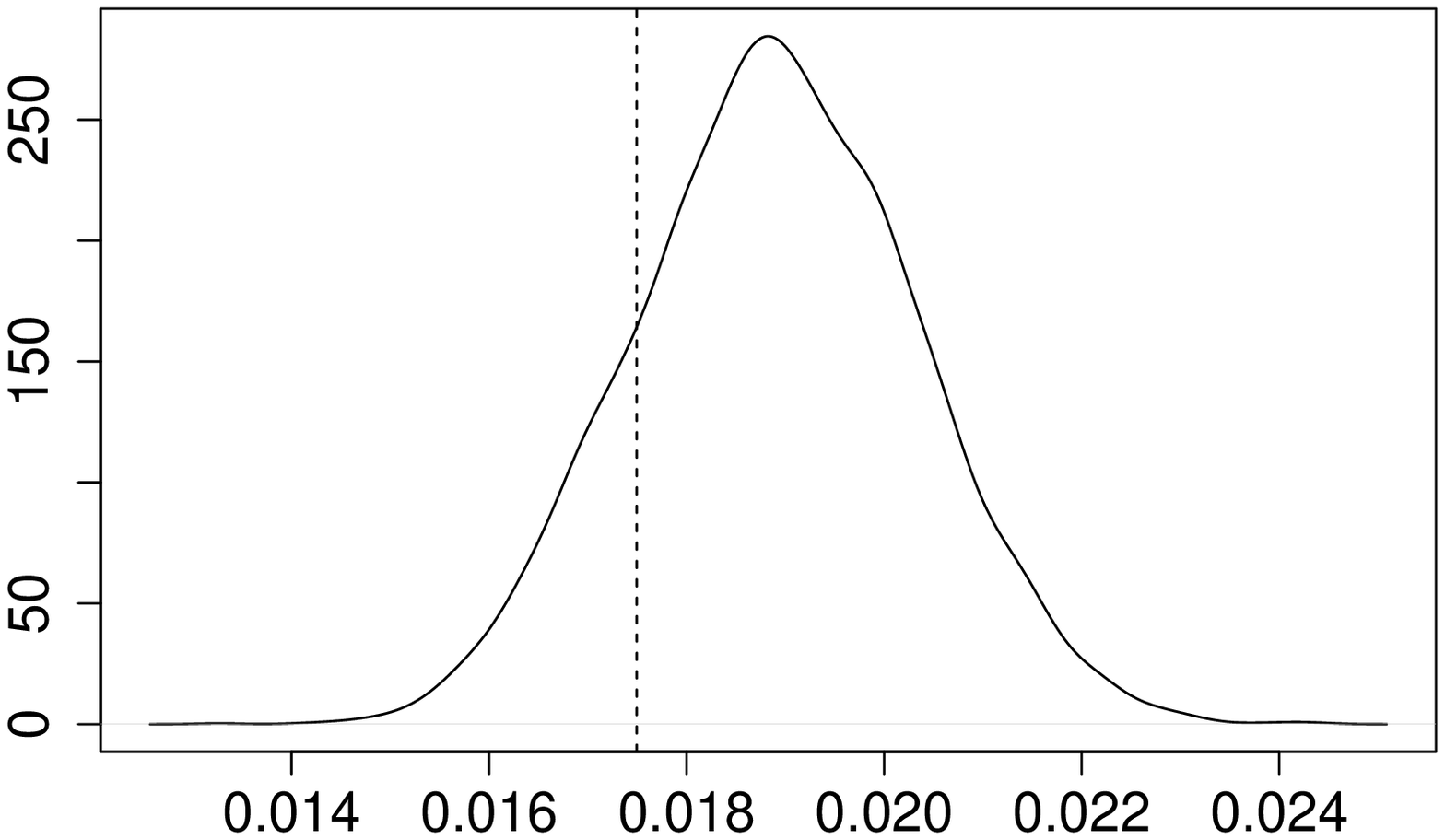} &
  \includegraphics[width=0.33\linewidth,height=0.20\linewidth]{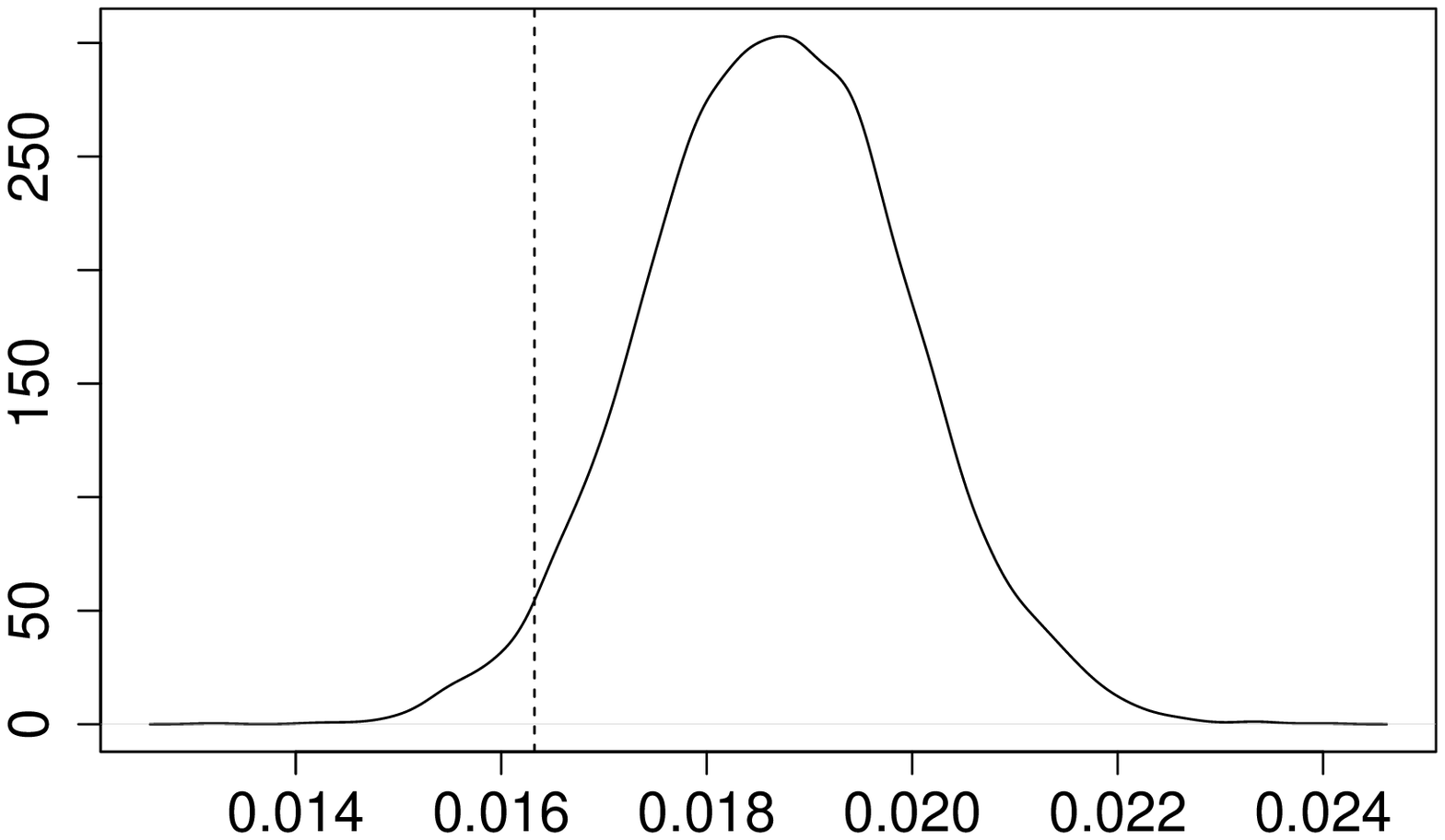} \\[-0.4cm]
(a)  \begin{tikzpicture}[scale=1.5]
  \def\radius{38.9};
  \coordinate (Pempty) at (0,0);
  \coordinate (P1) at (-1.5,1);
  \coordinate (P2) at (-0.5,1);
  \coordinate (P3) at (0.5,1);
  \coordinate (P4) at (1.5,1);
  \coordinate (P12) at (-1,2);
  \coordinate (P14) at (0,2);

  \coordinate (c1) at (-0.05,-0.05);
  \coordinate (c2) at (0.05,-0.05);
  \coordinate (c3) at (0.05,0.05);
  \coordinate (c4) at (-0.05,0.05);

  \coordinate (r1) at (-0.1,0);
  \coordinate (r2) at (0,0.1);
  \coordinate (r3) at (-0.1,0.1);
  \coordinate (r4) at (0.1,0.1);

   \draw[thick] (P12) ++(r3) +(c1) -- +(c2) -- +(c3) -- +(c4) -- cycle;
   \draw[thick] (P12) ++(r2) +(c1) -- +(c2) -- +(c3) -- +(c4) -- cycle;
   \draw[thick] (P12) ++(r1) +(c1) -- +(c2) -- +(c3) -- +(c4) -- cycle;
   \draw[thick] (P12) ++(Pempty) +(c1) -- +(c2) -- +(c3) -- +(c4) -- cycle;
   
\end{tikzpicture}   &          
(b) \begin{tikzpicture}[scale=1.5]
  \def\radius{38.9};
  \coordinate (Pempty) at (0,0);
  \coordinate (P1) at (-1.5,1);
  \coordinate (P2) at (-0.5,1);
  \coordinate (P3) at (0.5,1);
  \coordinate (P4) at (1.5,1);
  \coordinate (P12) at (-1,2);
  \coordinate (P14) at (0,2);

  \coordinate (c1) at (-0.05,-0.05);
  \coordinate (c2) at (0.05,-0.05);
  \coordinate (c3) at (0.05,0.05);
  \coordinate (c4) at (-0.05,0.05);

  \coordinate (r1) at (-0.1,0);
  \coordinate (r2) at (0,0.1);
  \coordinate (r3) at (-0.1,0.1);
  \coordinate (r4) at (0.1,0.1);

   \draw[thick] (P12) ++(r3) +(c1) -- +(c2) -- +(c3) -- +(c4) -- cycle;
   \draw[thick] (P12) ++(r2) +(c1) -- +(c2) -- +(c3) -- +(c4) -- cycle;
   \draw[thick] (P12) ++(r1) +(c1) -- +(c2) -- +(c3) -- +(c4) -- cycle;
   \draw[thick, fill=black] (P12) ++(Pempty) +(c1) -- +(c2) -- +(c3) -- +(c4) -- cycle;
   
\end{tikzpicture}    &         
(c) \begin{tikzpicture}[scale=1.5]
  \def\radius{38.9};
  \coordinate (Pempty) at (0,0);
  \coordinate (P1) at (-1.5,1);
  \coordinate (P2) at (-0.5,1);
  \coordinate (P3) at (0.5,1);
  \coordinate (P4) at (1.5,1);
  \coordinate (P12) at (-1,2);
  \coordinate (P14) at (0,2);

  \coordinate (c1) at (-0.05,-0.05);
  \coordinate (c2) at (0.05,-0.05);
  \coordinate (c3) at (0.05,0.05);
  \coordinate (c4) at (-0.05,0.05);

  \coordinate (r1) at (-0.1,0);
  \coordinate (r2) at (0,0.1);
  \coordinate (r3) at (-0.1,0.1);
  \coordinate (r4) at (0.1,0.1);

   \draw[thick] (P12) ++(r3) +(c1) -- +(c2) -- +(c3) -- +(c4) -- cycle;
   \draw[thick] (P12) ++(r2) +(c1) -- +(c2) -- +(c3) -- +(c4) -- cycle;
   \draw[thick, fill=black] (P12) ++(r1) +(c1) -- +(c2) -- +(c3) -- +(c4) -- cycle;
   \draw[thick] (P12) ++(Pempty) +(c1) -- +(c2) -- +(c3) -- +(c4) -- cycle;
   
\end{tikzpicture}     \\[-0.05cm]        
  \includegraphics[width=0.33\linewidth,height=0.20\linewidth]{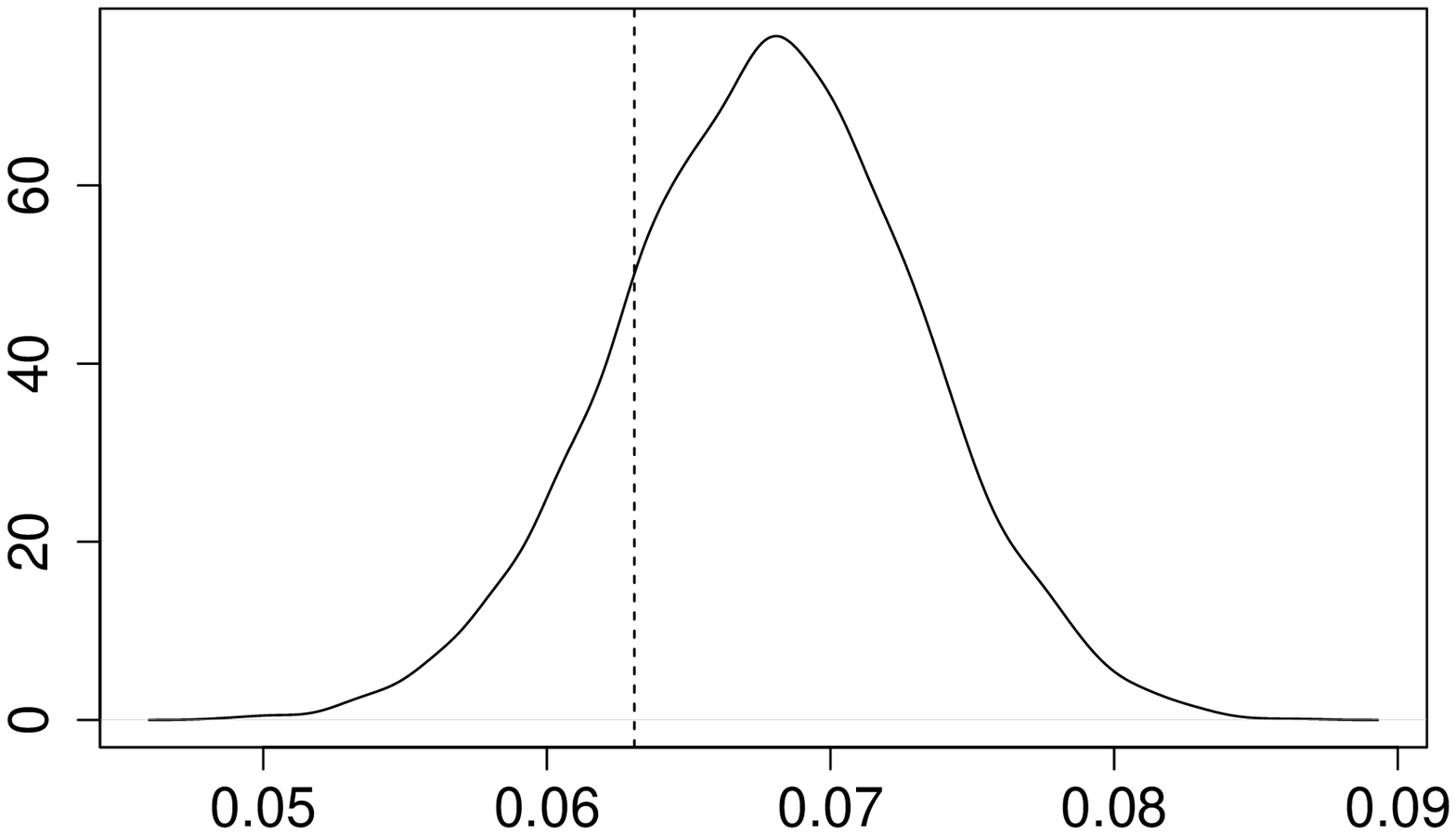} &
  \includegraphics[width=0.33\linewidth,height=0.20\linewidth]{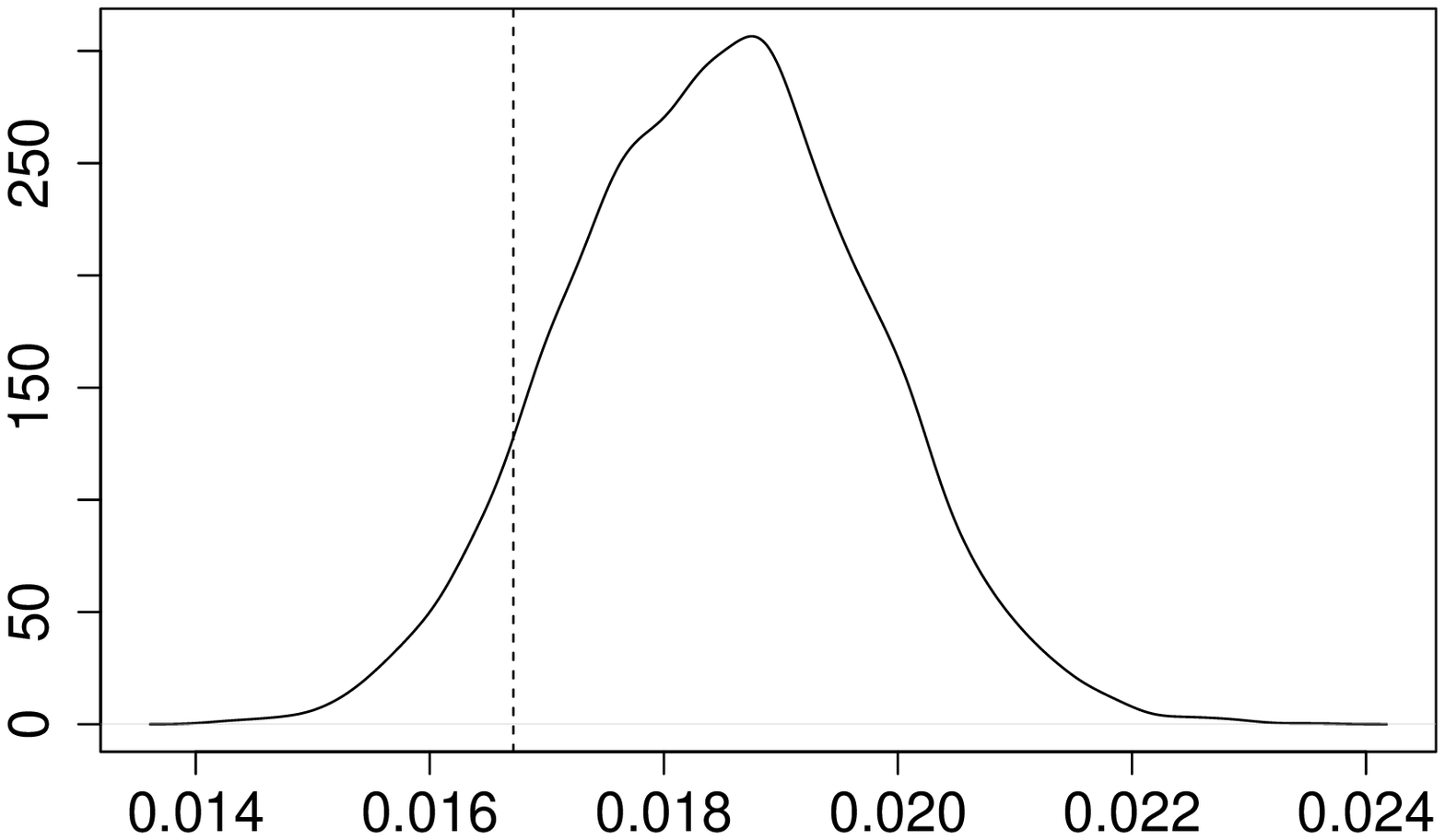} &
  \includegraphics[width=0.33\linewidth,height=0.20\linewidth]{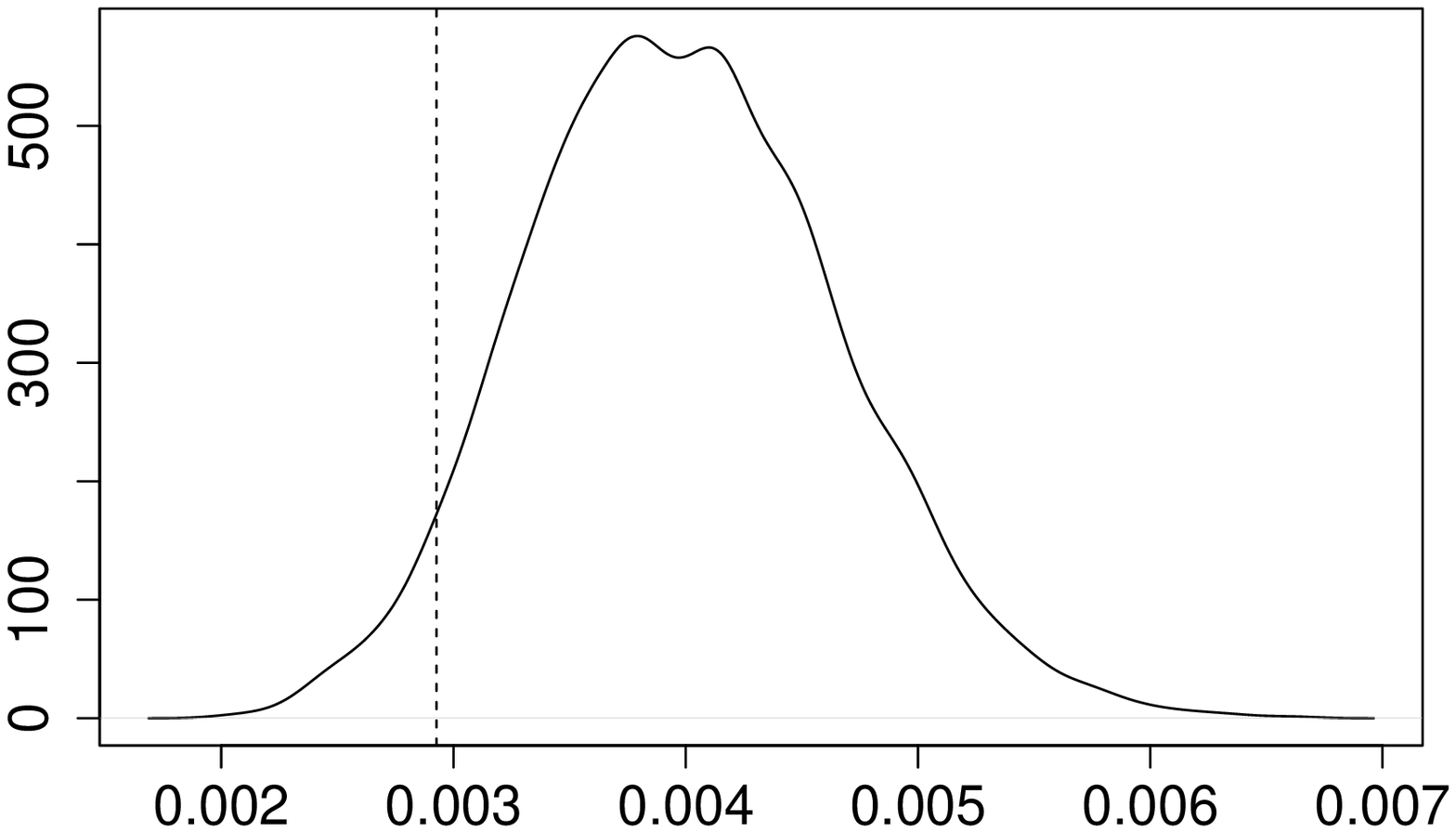} \\[-0.4cm]
(d)  \begin{tikzpicture}[scale=1.5]
  \def\radius{38.9};
  \coordinate (Pempty) at (0,0);
  \coordinate (P1) at (-1.5,1);
  \coordinate (P2) at (-0.5,1);
  \coordinate (P3) at (0.5,1);
  \coordinate (P4) at (1.5,1);
  \coordinate (P12) at (-1,2);
  \coordinate (P14) at (0,2);

  \coordinate (c1) at (-0.05,-0.05);
  \coordinate (c2) at (0.05,-0.05);
  \coordinate (c3) at (0.05,0.05);
  \coordinate (c4) at (-0.05,0.05);

  \coordinate (r1) at (-0.1,0);
  \coordinate (r2) at (0,0.1);
  \coordinate (r3) at (-0.1,0.1);
  \coordinate (r4) at (0.1,0.1);

   \draw[thick] (P12) ++(r3) +(c1) -- +(c2) -- +(c3) -- +(c4) -- cycle;
   \draw[thick] (P12) ++(r2) +(c1) -- +(c2) -- +(c3) -- +(c4) -- cycle;
   \draw[thick, fill=black] (P12) ++(r1) +(c1) -- +(c2) -- +(c3) -- +(c4) -- cycle;
   \draw[thick, fill=black] (P12) ++(Pempty) +(c1) -- +(c2) -- +(c3) -- +(c4) -- cycle;
   
\end{tikzpicture}             &
(e)  \begin{tikzpicture}[scale=1.5]
  \def\radius{38.9};
  \coordinate (Pempty) at (0,0);
  \coordinate (P1) at (-1.5,1);
  \coordinate (P2) at (-0.5,1);
  \coordinate (P3) at (0.5,1);
  \coordinate (P4) at (1.5,1);
  \coordinate (P12) at (-1,2);
  \coordinate (P14) at (0,2);

  \coordinate (c1) at (-0.05,-0.05);
  \coordinate (c2) at (0.05,-0.05);
  \coordinate (c3) at (0.05,0.05);
  \coordinate (c4) at (-0.05,0.05);

  \coordinate (r1) at (-0.1,0);
  \coordinate (r2) at (0,0.1);
  \coordinate (r3) at (-0.1,0.1);
  \coordinate (r4) at (0.1,0.1);

   \draw[thick] (P12) ++(r3) +(c1) -- +(c2) -- +(c3) -- +(c4) -- cycle;
   \draw[thick, fill=black] (P12) ++(r2) +(c1) -- +(c2) -- +(c3) -- +(c4) -- cycle;
   \draw[thick] (P12) ++(r1) +(c1) -- +(c2) -- +(c3) -- +(c4) -- cycle;
   \draw[thick] (P12) ++(Pempty) +(c1) -- +(c2) -- +(c3) -- +(c4) -- cycle;
   
\end{tikzpicture}             &
(f)  \begin{tikzpicture}[scale=1.5]
  \def\radius{38.9};
  \coordinate (Pempty) at (0,0);
  \coordinate (P1) at (-1.5,1);
  \coordinate (P2) at (-0.5,1);
  \coordinate (P3) at (0.5,1);
  \coordinate (P4) at (1.5,1);
  \coordinate (P12) at (-1,2);
  \coordinate (P14) at (0,2);

  \coordinate (c1) at (-0.05,-0.05);
  \coordinate (c2) at (0.05,-0.05);
  \coordinate (c3) at (0.05,0.05);
  \coordinate (c4) at (-0.05,0.05);

  \coordinate (r1) at (-0.1,0);
  \coordinate (r2) at (0,0.1);
  \coordinate (r3) at (-0.1,0.1);
  \coordinate (r4) at (0.1,0.1);

   \draw[thick] (P12) ++(r3) +(c1) -- +(c2) -- +(c3) -- +(c4) -- cycle;
   \draw[thick, fill=black] (P12) ++(r2) +(c1) -- +(c2) -- +(c3) -- +(c4) -- cycle;
   \draw[thick] (P12) ++(r1) +(c1) -- +(c2) -- +(c3) -- +(c4) -- cycle;
   \draw[thick, fill=black] (P12) ++(Pempty) +(c1) -- +(c2) -- +(c3) -- +(c4) -- cycle;
   
\end{tikzpicture}          \\[-0.05cm]
  \includegraphics[width=0.33\linewidth,height=0.20\linewidth]{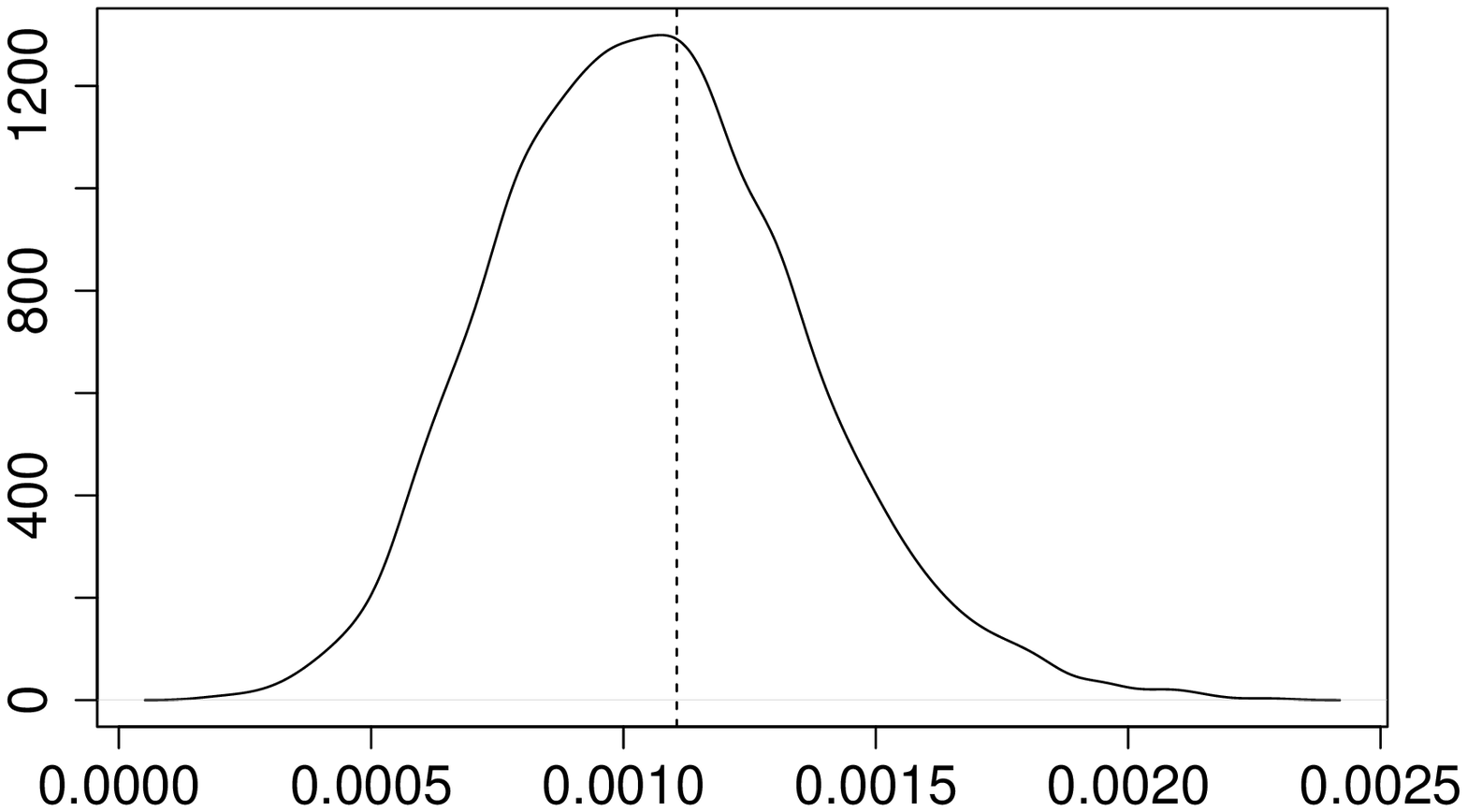} &
  \includegraphics[width=0.33\linewidth,height=0.20\linewidth]{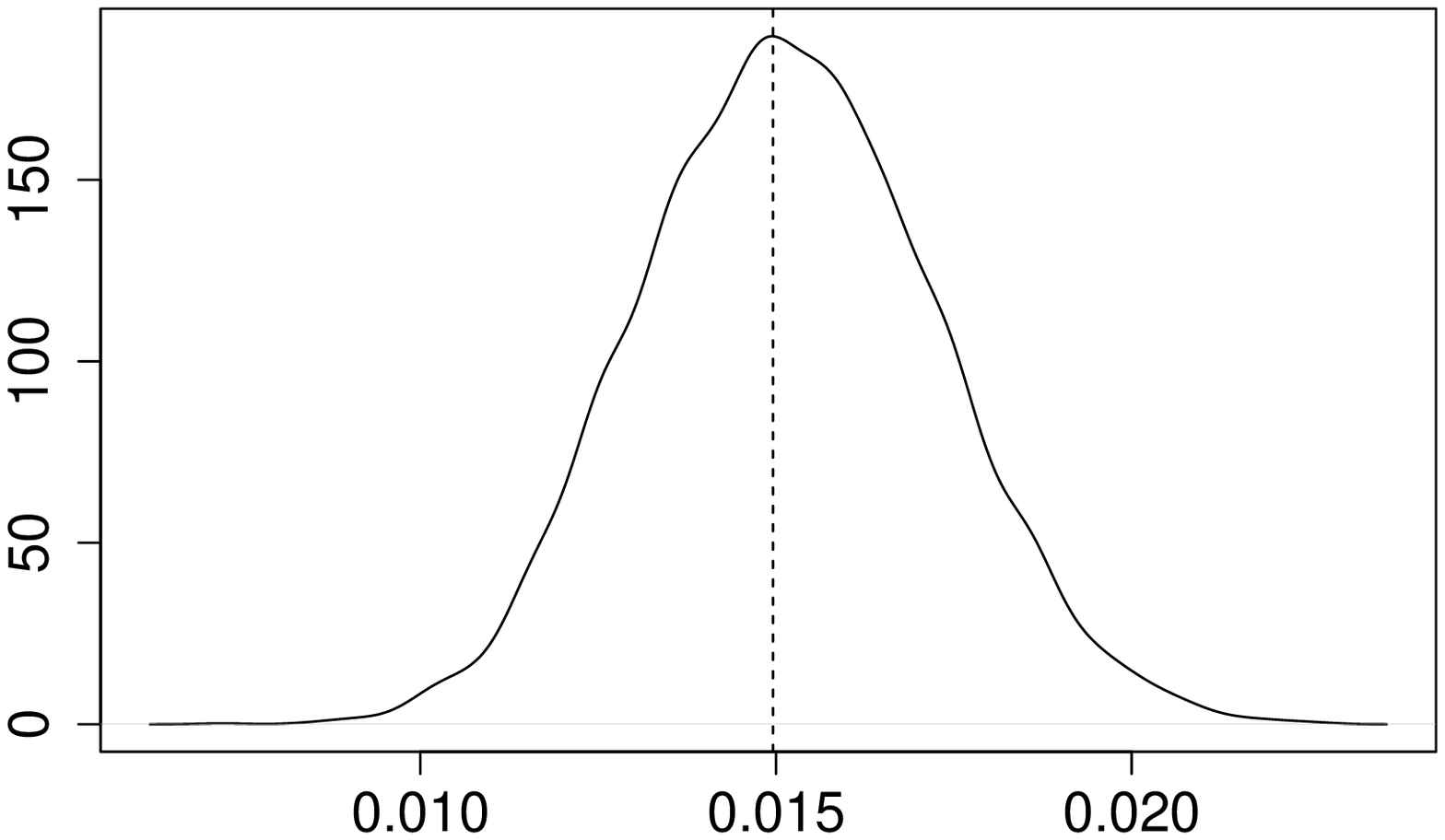} &
  \includegraphics[width=0.33\linewidth,height=0.20\linewidth]{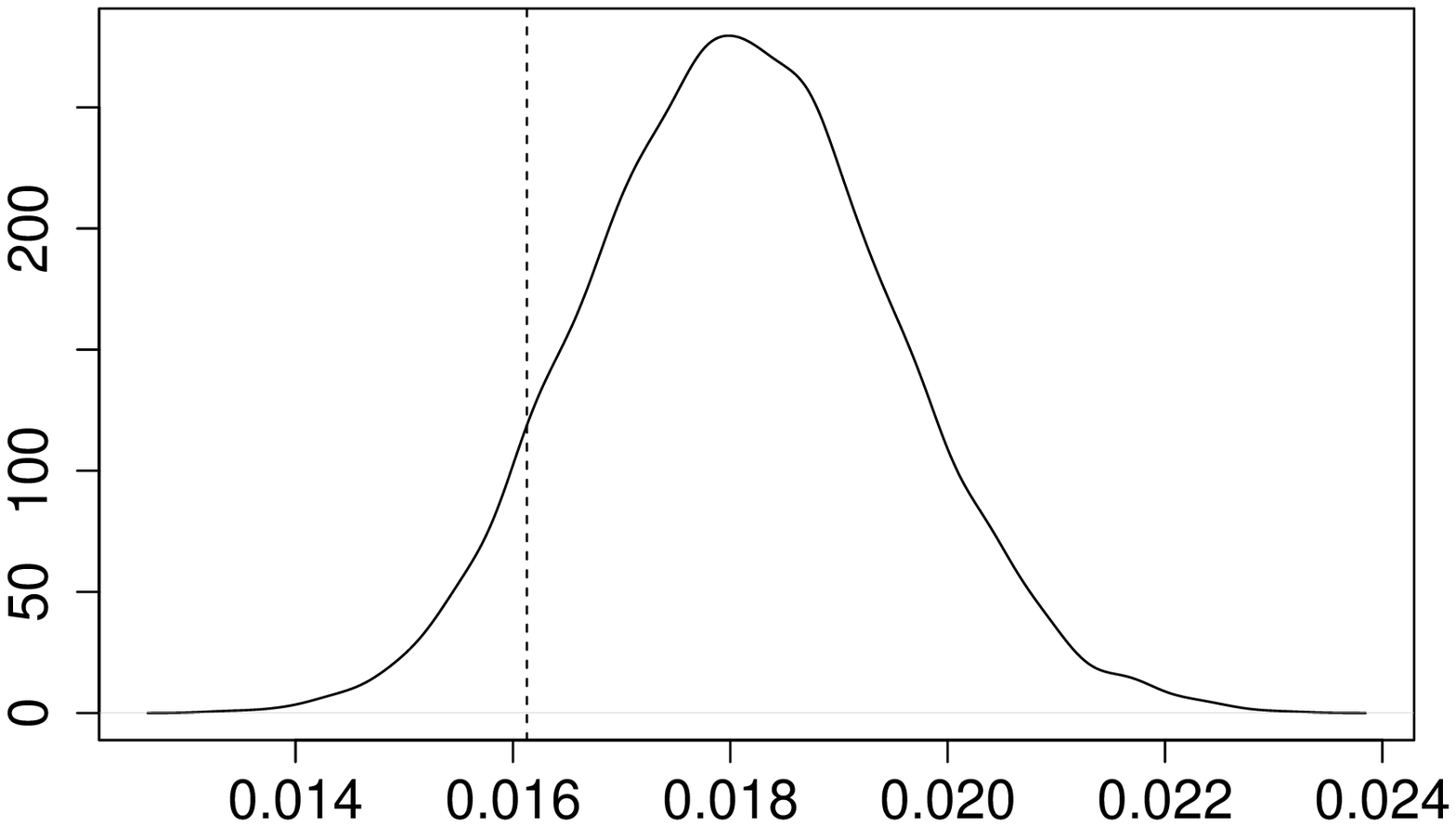} \\[-0.4cm]
(g)  \begin{tikzpicture}[scale=1.5]
  \def\radius{38.9};
  \coordinate (Pempty) at (0,0);
  \coordinate (P1) at (-1.5,1);
  \coordinate (P2) at (-0.5,1);
  \coordinate (P3) at (0.5,1);
  \coordinate (P4) at (1.5,1);
  \coordinate (P12) at (-1,2);
  \coordinate (P14) at (0,2);

  \coordinate (c1) at (-0.05,-0.05);
  \coordinate (c2) at (0.05,-0.05);
  \coordinate (c3) at (0.05,0.05);
  \coordinate (c4) at (-0.05,0.05);

  \coordinate (r1) at (-0.1,0);
  \coordinate (r2) at (0,0.1);
  \coordinate (r3) at (-0.1,0.1);
  \coordinate (r4) at (0.1,0.1);

   \draw[thick] (P12) ++(r3) +(c1) -- +(c2) -- +(c3) -- +(c4) -- cycle;
   \draw[thick, fill=black] (P12) ++(r2) +(c1) -- +(c2) -- +(c3) -- +(c4) -- cycle;
   \draw[thick, fill=black] (P12) ++(r1) +(c1) -- +(c2) -- +(c3) -- +(c4) -- cycle;
   \draw[thick] (P12) ++(Pempty) +(c1) -- +(c2) -- +(c3) -- +(c4) -- cycle;
   
\end{tikzpicture}             &
(h)  \begin{tikzpicture}[scale=1.5]
  \def\radius{38.9};
  \coordinate (Pempty) at (0,0);
  \coordinate (P1) at (-1.5,1);
  \coordinate (P2) at (-0.5,1);
  \coordinate (P3) at (0.5,1);
  \coordinate (P4) at (1.5,1);
  \coordinate (P12) at (-1,2);
  \coordinate (P14) at (0,2);

  \coordinate (c1) at (-0.05,-0.05);
  \coordinate (c2) at (0.05,-0.05);
  \coordinate (c3) at (0.05,0.05);
  \coordinate (c4) at (-0.05,0.05);

  \coordinate (r1) at (-0.1,0);
  \coordinate (r2) at (0,0.1);
  \coordinate (r3) at (-0.1,0.1);
  \coordinate (r4) at (0.1,0.1);

   \draw[thick] (P12) ++(r3) +(c1) -- +(c2) -- +(c3) -- +(c4) -- cycle;
   \draw[thick, fill=black] (P12) ++(r2) +(c1) -- +(c2) -- +(c3) -- +(c4) -- cycle;
   \draw[thick, fill=black] (P12) ++(r1) +(c1) -- +(c2) -- +(c3) -- +(c4) -- cycle;
   \draw[thick, fill=black] (P12) ++(Pempty) +(c1) -- +(c2) -- +(c3) -- +(c4) -- cycle;
   
\end{tikzpicture}             &
(i)  \begin{tikzpicture}[scale=1.5]
  \def\radius{38.9};
  \coordinate (Pempty) at (0,0);
  \coordinate (P1) at (-1.5,1);
  \coordinate (P2) at (-0.5,1);
  \coordinate (P3) at (0.5,1);
  \coordinate (P4) at (1.5,1);
  \coordinate (P12) at (-1,2);
  \coordinate (P14) at (0,2);

  \coordinate (c1) at (-0.05,-0.05);
  \coordinate (c2) at (0.05,-0.05);
  \coordinate (c3) at (0.05,0.05);
  \coordinate (c4) at (-0.05,0.05);

  \coordinate (r1) at (-0.1,0);
  \coordinate (r2) at (0,0.1);
  \coordinate (r3) at (-0.1,0.1);
  \coordinate (r4) at (0.1,0.1);

   \draw[thick, fill=black] (P12) ++(r3) +(c1) -- +(c2) -- +(c3) -- +(c4) -- cycle;
   \draw[thick] (P12) ++(r2) +(c1) -- +(c2) -- +(c3) -- +(c4) -- cycle;
   \draw[thick] (P12) ++(r1) +(c1) -- +(c2) -- +(c3) -- +(c4) -- cycle;
   \draw[thick] (P12) ++(Pempty) +(c1) -- +(c2) -- +(c3) -- +(c4) -- cycle;
   
\end{tikzpicture}             \\[-0.05cm]
  \includegraphics[width=0.33\linewidth,height=0.20\linewidth]{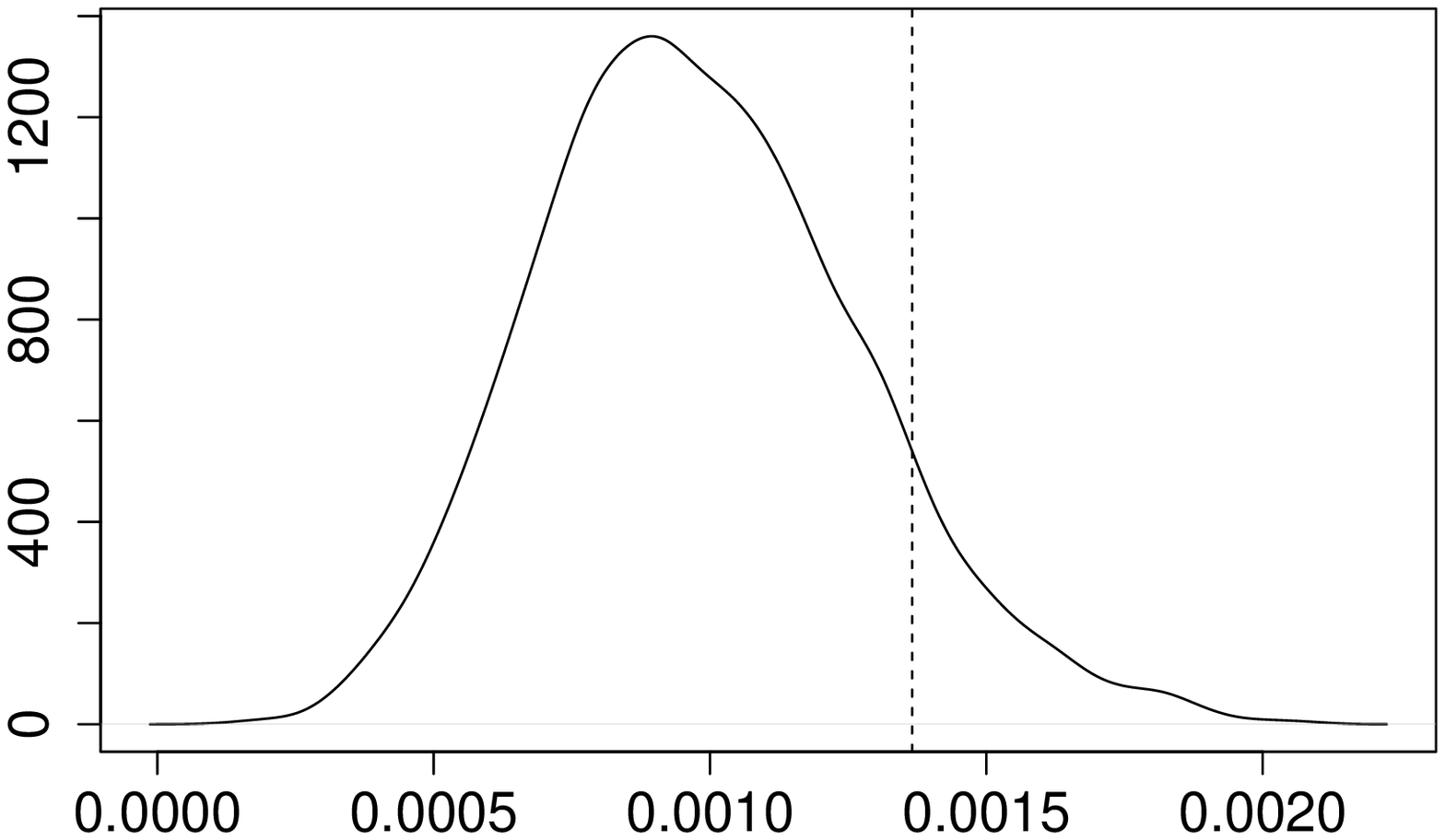} &
  \includegraphics[width=0.33\linewidth,height=0.20\linewidth]{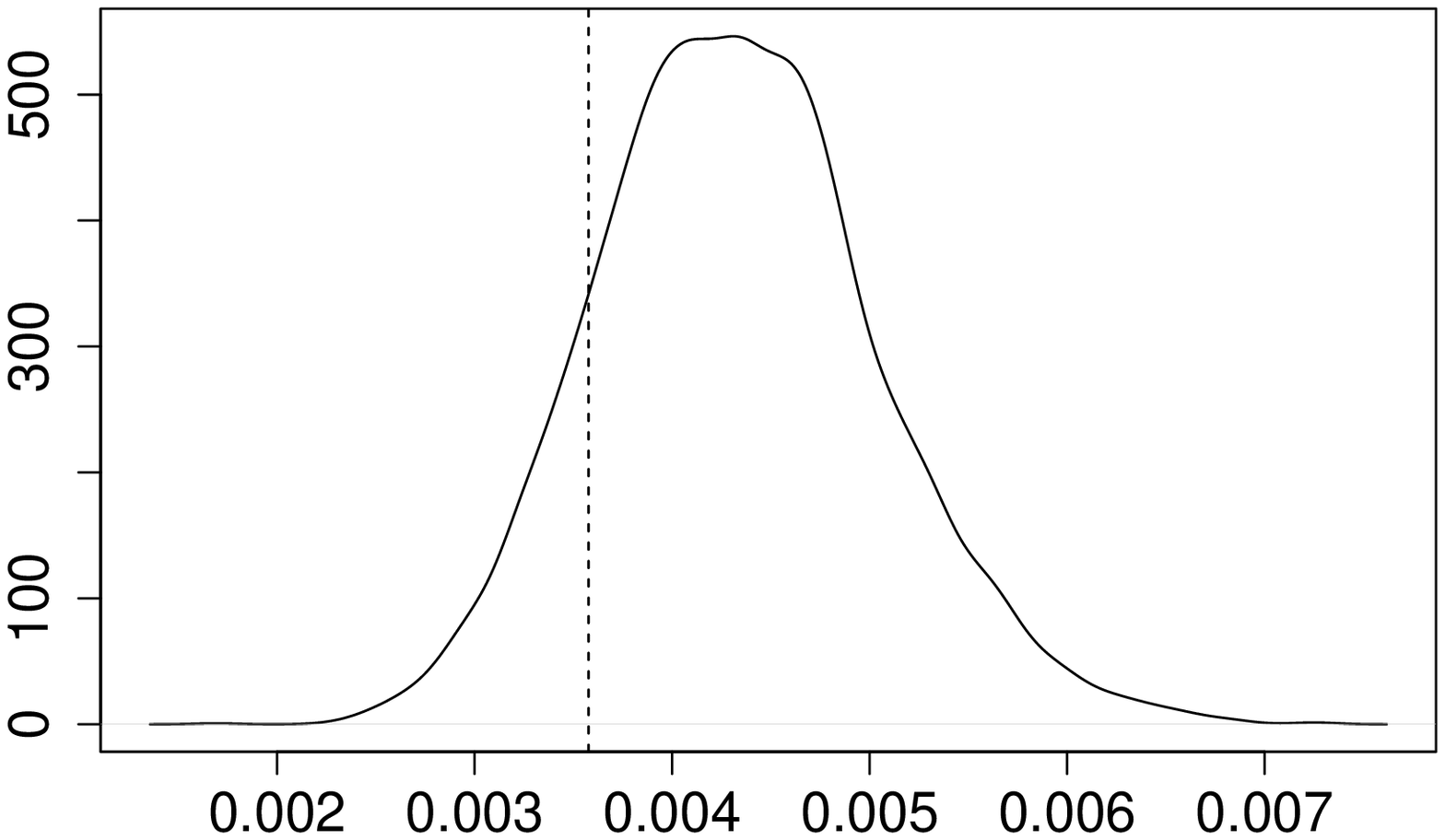} &
  \includegraphics[width=0.33\linewidth,height=0.20\linewidth]{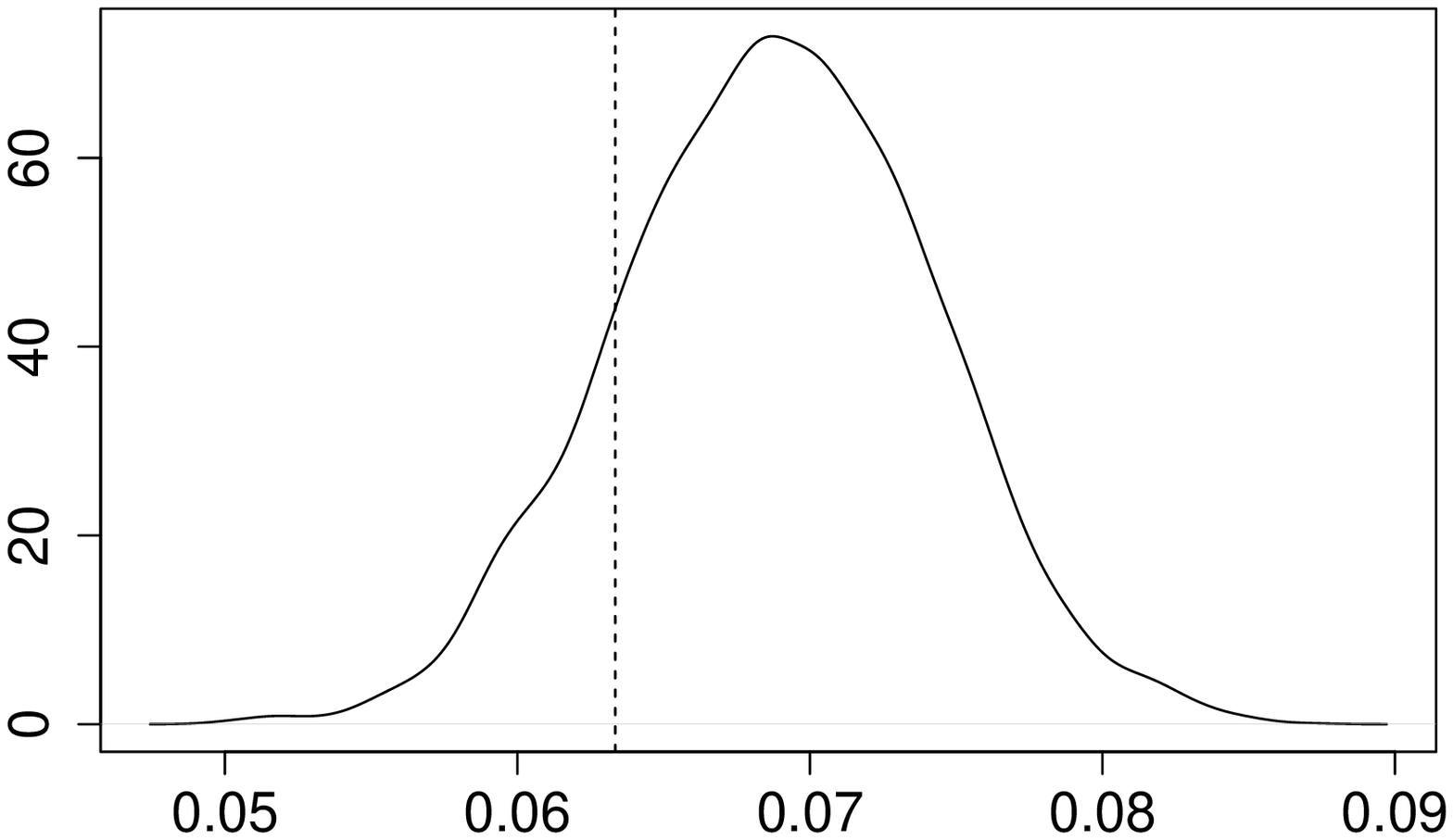} \\[-0.4cm]
(j)  \begin{tikzpicture}[scale=1.5]
  \def\radius{38.9};
  \coordinate (Pempty) at (0,0);
  \coordinate (P1) at (-1.5,1);
  \coordinate (P2) at (-0.5,1);
  \coordinate (P3) at (0.5,1);
  \coordinate (P4) at (1.5,1);
  \coordinate (P12) at (-1,2);
  \coordinate (P14) at (0,2);

  \coordinate (c1) at (-0.05,-0.05);
  \coordinate (c2) at (0.05,-0.05);
  \coordinate (c3) at (0.05,0.05);
  \coordinate (c4) at (-0.05,0.05);

  \coordinate (r1) at (-0.1,0);
  \coordinate (r2) at (0,0.1);
  \coordinate (r3) at (-0.1,0.1);
  \coordinate (r4) at (0.1,0.1);

   \draw[thick, fill=black] (P12) ++(r3) +(c1) -- +(c2) -- +(c3) -- +(c4) -- cycle;
   \draw[thick] (P12) ++(r2) +(c1) -- +(c2) -- +(c3) -- +(c4) -- cycle;
   \draw[thick] (P12) ++(r1) +(c1) -- +(c2) -- +(c3) -- +(c4) -- cycle;
   \draw[thick, fill=black] (P12) ++(Pempty) +(c1) -- +(c2) -- +(c3) -- +(c4) -- cycle;
   
\end{tikzpicture}             &
(k)  \begin{tikzpicture}[scale=1.5]
  \def\radius{38.9};
  \coordinate (Pempty) at (0,0);
  \coordinate (P1) at (-1.5,1);
  \coordinate (P2) at (-0.5,1);
  \coordinate (P3) at (0.5,1);
  \coordinate (P4) at (1.5,1);
  \coordinate (P12) at (-1,2);
  \coordinate (P14) at (0,2);

  \coordinate (c1) at (-0.05,-0.05);
  \coordinate (c2) at (0.05,-0.05);
  \coordinate (c3) at (0.05,0.05);
  \coordinate (c4) at (-0.05,0.05);

  \coordinate (r1) at (-0.1,0);
  \coordinate (r2) at (0,0.1);
  \coordinate (r3) at (-0.1,0.1);
  \coordinate (r4) at (0.1,0.1);

   \draw[thick, fill=black] (P12) ++(r3) +(c1) -- +(c2) -- +(c3) -- +(c4) -- cycle;
   \draw[thick] (P12) ++(r2) +(c1) -- +(c2) -- +(c3) -- +(c4) -- cycle;
   \draw[thick, fill=black] (P12) ++(r1) +(c1) -- +(c2) -- +(c3) -- +(c4) -- cycle;
   \draw[thick] (P12) ++(Pempty) +(c1) -- +(c2) -- +(c3) -- +(c4) -- cycle;
   
\end{tikzpicture}             &
(l)  \begin{tikzpicture}[scale=1.5]
  \def\radius{38.9};
  \coordinate (Pempty) at (0,0);
  \coordinate (P1) at (-1.5,1);
  \coordinate (P2) at (-0.5,1);
  \coordinate (P3) at (0.5,1);
  \coordinate (P4) at (1.5,1);
  \coordinate (P12) at (-1,2);
  \coordinate (P14) at (0,2);

  \coordinate (c1) at (-0.05,-0.05);
  \coordinate (c2) at (0.05,-0.05);
  \coordinate (c3) at (0.05,0.05);
  \coordinate (c4) at (-0.05,0.05);

  \coordinate (r1) at (-0.1,0);
  \coordinate (r2) at (0,0.1);
  \coordinate (r3) at (-0.1,0.1);
  \coordinate (r4) at (0.1,0.1);

   \draw[thick, fill=black] (P12) ++(r3) +(c1) -- +(c2) -- +(c3) -- +(c4) -- cycle;
   \draw[thick, fill=black] (P12) ++(r2) +(c1) -- +(c2) -- +(c3) -- +(c4) -- cycle;
   \draw[thick] (P12) ++(r1) +(c1) -- +(c2) -- +(c3) -- +(c4) -- cycle;
   \draw[thick] (P12) ++(Pempty) +(c1) -- +(c2) -- +(c3) -- +(c4) -- cycle;
   
\end{tikzpicture}             \\[-0.05cm]
  \includegraphics[width=0.33\linewidth,height=0.20\linewidth]{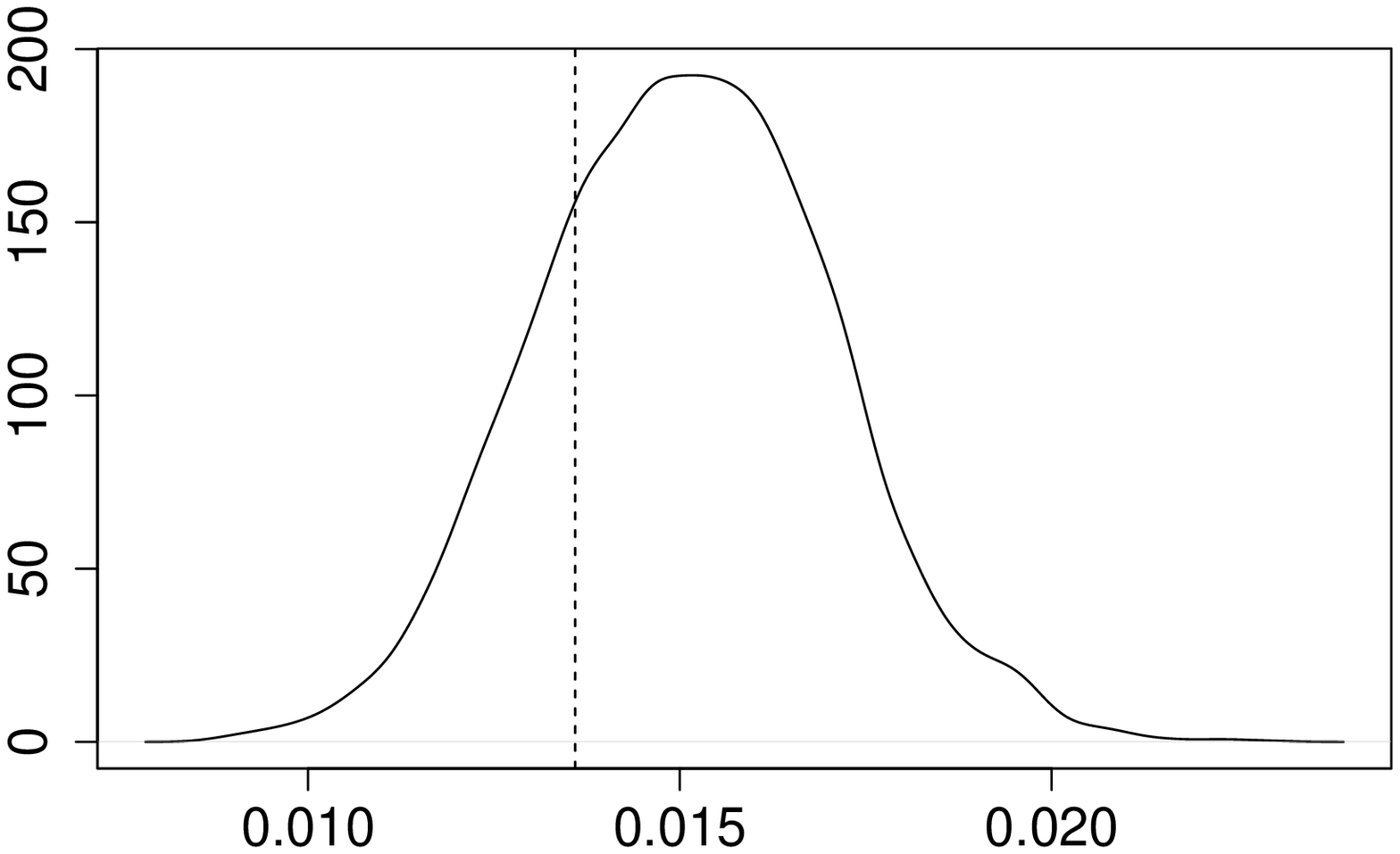} &
  \includegraphics[width=0.33\linewidth,height=0.20\linewidth]{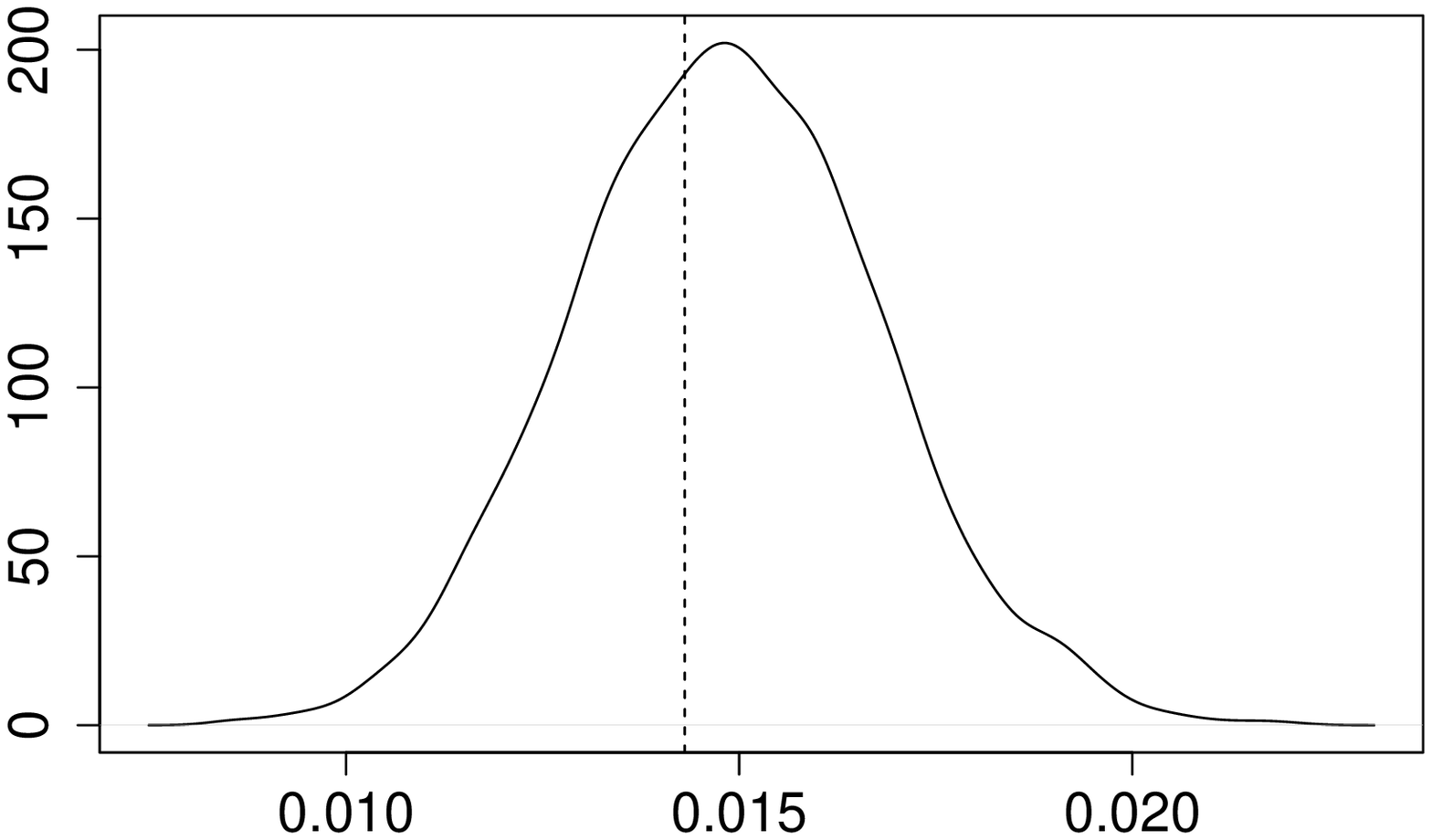} &
  \includegraphics[width=0.33\linewidth,height=0.20\linewidth]{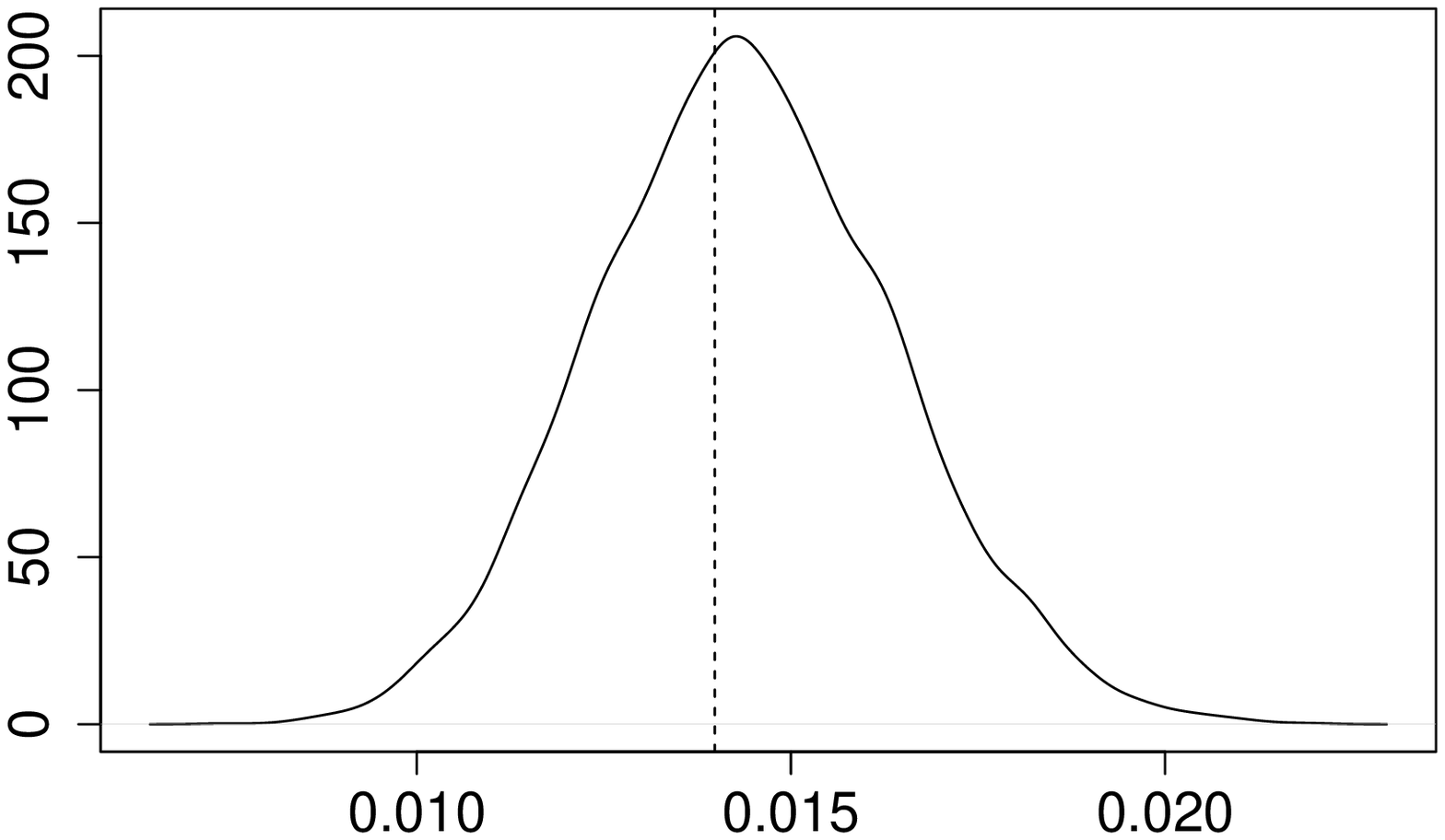} \\[-0.4cm]
(m)  \begin{tikzpicture}[scale=1.5]
  \def\radius{38.9};
  \coordinate (Pempty) at (0,0);
  \coordinate (P1) at (-1.5,1);
  \coordinate (P2) at (-0.5,1);
  \coordinate (P3) at (0.5,1);
  \coordinate (P4) at (1.5,1);
  \coordinate (P12) at (-1,2);
  \coordinate (P14) at (0,2);

  \coordinate (c1) at (-0.05,-0.05);
  \coordinate (c2) at (0.05,-0.05);
  \coordinate (c3) at (0.05,0.05);
  \coordinate (c4) at (-0.05,0.05);

  \coordinate (r1) at (-0.1,0);
  \coordinate (r2) at (0,0.1);
  \coordinate (r3) at (-0.1,0.1);
  \coordinate (r4) at (0.1,0.1);

   \draw[thick, fill=black] (P12) ++(r3) +(c1) -- +(c2) -- +(c3) -- +(c4) -- cycle;
   \draw[thick] (P12) ++(r2) +(c1) -- +(c2) -- +(c3) -- +(c4) -- cycle;
   \draw[thick, fill=black] (P12) ++(r1) +(c1) -- +(c2) -- +(c3) -- +(c4) -- cycle;
   \draw[thick, fill=black] (P12) ++(Pempty) +(c1) -- +(c2) -- +(c3) -- +(c4) -- cycle;
   
\end{tikzpicture}             &
(n)  \begin{tikzpicture}[scale=1.5]
  \def\radius{38.9};
  \coordinate (Pempty) at (0,0);
  \coordinate (P1) at (-1.5,1);
  \coordinate (P2) at (-0.5,1);
  \coordinate (P3) at (0.5,1);
  \coordinate (P4) at (1.5,1);
  \coordinate (P12) at (-1,2);
  \coordinate (P14) at (0,2);

  \coordinate (c1) at (-0.05,-0.05);
  \coordinate (c2) at (0.05,-0.05);
  \coordinate (c3) at (0.05,0.05);
  \coordinate (c4) at (-0.05,0.05);

  \coordinate (r1) at (-0.1,0);
  \coordinate (r2) at (0,0.1);
  \coordinate (r3) at (-0.1,0.1);
  \coordinate (r4) at (0.1,0.1);

   \draw[thick, fill=black] (P12) ++(r3) +(c1) -- +(c2) -- +(c3) -- +(c4) -- cycle;
   \draw[thick, fill=black] (P12) ++(r2) +(c1) -- +(c2) -- +(c3) -- +(c4) -- cycle;
   \draw[thick] (P12) ++(r1) +(c1) -- +(c2) -- +(c3) -- +(c4) -- cycle;
   \draw[thick, fill=black] (P12) ++(Pempty) +(c1) -- +(c2) -- +(c3) -- +(c4) -- cycle;
   
\end{tikzpicture}             &
(o)  \begin{tikzpicture}[scale=1.5]
  \def\radius{38.9};
  \coordinate (Pempty) at (0,0);
  \coordinate (P1) at (-1.5,1);
  \coordinate (P2) at (-0.5,1);
  \coordinate (P3) at (0.5,1);
  \coordinate (P4) at (1.5,1);
  \coordinate (P12) at (-1,2);
  \coordinate (P14) at (0,2);

  \coordinate (c1) at (-0.05,-0.05);
  \coordinate (c2) at (0.05,-0.05);
  \coordinate (c3) at (0.05,0.05);
  \coordinate (c4) at (-0.05,0.05);

  \coordinate (r1) at (-0.1,0);
  \coordinate (r2) at (0,0.1);
  \coordinate (r3) at (-0.1,0.1);
  \coordinate (r4) at (0.1,0.1);

   \draw[thick, fill=black] (P12) ++(r3) +(c1) -- +(c2) -- +(c3) -- +(c4) -- cycle;
   \draw[thick, fill=black] (P12) ++(r2) +(c1) -- +(c2) -- +(c3) -- +(c4) -- cycle;
   \draw[thick, fill=black] (P12) ++(r1) +(c1) -- +(c2) -- +(c3) -- +(c4) -- cycle;
   \draw[thick] (P12) ++(Pempty) +(c1) -- +(c2) -- +(c3) -- +(c4) -- cycle;
   
\end{tikzpicture}             \\[-0.05cm]
 &
  \includegraphics[width=0.33\linewidth,height=0.20\linewidth]{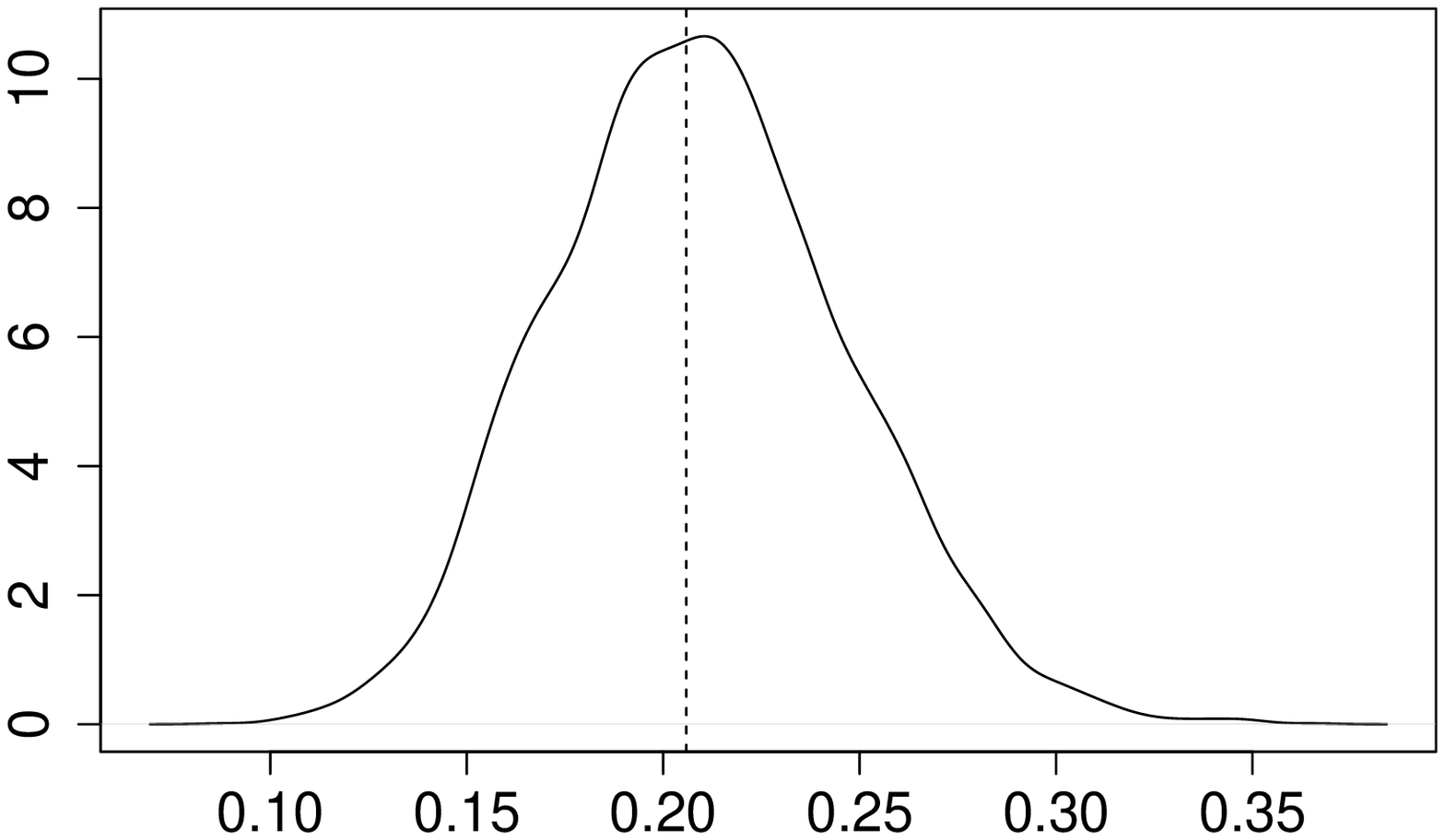} &
 \\[-0.4cm]
 & (p)  \begin{tikzpicture}[scale=1.5]
  \def\radius{38.9};
  \coordinate (Pempty) at (0,0);
  \coordinate (P1) at (-1.5,1);
  \coordinate (P2) at (-0.5,1);
  \coordinate (P3) at (0.5,1);
  \coordinate (P4) at (1.5,1);
  \coordinate (P12) at (-1,2);
  \coordinate (P14) at (0,2);

  \coordinate (c1) at (-0.05,-0.05);
  \coordinate (c2) at (0.05,-0.05);
  \coordinate (c3) at (0.05,0.05);
  \coordinate (c4) at (-0.05,0.05);

  \coordinate (r1) at (-0.1,0);
  \coordinate (r2) at (0,0.1);
  \coordinate (r3) at (-0.1,0.1);
  \coordinate (r4) at (0.1,0.1);

   \draw[thick, fill=black] (P12) ++(r3) +(c1) -- +(c2) -- +(c3) -- +(c4) -- cycle;
   \draw[thick, fill=black] (P12) ++(r2) +(c1) -- +(c2) -- +(c3) -- +(c4) -- cycle;
   \draw[thick, fill=black] (P12) ++(r1) +(c1) -- +(c2) -- +(c3) -- +(c4) -- cycle;
   \draw[thick, fill=black] (P12) ++(Pempty) +(c1) -- +(c2) -- +(c3) -- +(c4) -- cycle;
   
\end{tikzpicture}             
& 
\end{tabular}
        \caption{Sisim data set example: Estimated posterior marginal densities for each of the possible $16$ configurations 
in a $2\times 2$ block of nodes. Corresponding values computed from the sisim data set is shown as a vertical dotted line. 
The configuration corresponding to an estimated density is shown below each figure, where black and white nodes
represent one and zero, respectively.}\label{fig:frac_sisim}
\end{figure}

As in the cancer mortality data set example, we also now generate realizations from 
the simulated Markov mesh models. Figure \ref{fig:real_sisim} contains 
realizations simulated from four randomly chosen models simulated in the Markov chain (after the specified burn-in).
As in Figure \ref{fig:sisim}, showing the observed data set, black and white nodes $v$ represent $x_v=1$ and $0$, respectively. 
Also now we estimate the distribution of values in a $2\times2$ block of nodes.
Figure \ref{fig:frac_sisim} 
shows the estimated density for each of the $16$ configurations. The corresponding fractions 
for the observed data set are marked by vertical dotted lines. Note that for most of these distributions, the 
corresponding fractions for the observed data set are centrally located in the distribution. 
The exceptions are (c), and partly (e), (f), (i) and (j), where the observed quantities are more in the tail of the distribution.

\begin{figure}
        \begin{center}
                \begin{tikzpicture}[scale=1.75]
  \def\radius{38.9};
  \coordinate (Pempty) at (0,-1);
  \coordinate (P1) at (-3.75,1);
  \coordinate (P2) at (-2.25,1);
  \coordinate (P3) at (-0.75,1);
  \coordinate (P4) at (0.75,1);
  \coordinate (P5) at (2.25,1);
  \coordinate (P6) at (3.75,1);
  \coordinate (P12) at (-3,3);
  \coordinate (P15) at (-1.5,3);
  \coordinate (P25) at (0,3);
  \coordinate (P26) at (1.5,3);
   \coordinate (P56) at (3,3);
  \coordinate (P125) at (-0.75,5);
  \coordinate (P256) at (0.75,5);

  \coordinate (c1) at (-0.05,-0.05);
  \coordinate (c2) at (0.05,-0.05);
  \coordinate (c3) at (0.05,0.05);
  \coordinate (c4) at (-0.05,0.05);

  \coordinate (r1) at (-0.1,0);
  \coordinate (r2) at (0,0.1);
  \coordinate (r3) at (0.2,0.1);
  \coordinate (r4) at (-0.1,0.3);
  \coordinate (r5) at (-0.3,0);
  \coordinate (r6) at (0.4,0.1);

  \node[draw,circle,inner sep=0pt,minimum size=1.4*\radius,name=Nempty] at (Pempty) {};
  \draw[thick] (Pempty) +(c1) -- +(c2) -- +(c3) -- +(c4) -- +(c1) +(c1) -- +(c3) +(c2) -- +(c4);
  
  \node[draw,circle,inner sep=0pt,minimum size=1.4*\radius,name=N1] at (P1) {};
  \draw[thick] (P1) +(c1) -- +(c2) -- +(c3) -- +(c4) -- +(c1) +(c1) -- +(c3) +(c2) -- +(c4);
  \draw[thick] (P1) ++(r1) +(c1) -- +(c2) -- +(c3) -- +(c4) -- cycle;
  \draw[thick,->] (Nempty) -- (N1);
   
  \node[draw,circle,inner sep=0pt,minimum size=1.4*\radius,name=N2] at (P2) {};
  \draw[thick] (P2) +(c1) -- +(c2) -- +(c3) -- +(c4) -- +(c1) +(c1) -- +(c3) +(c2) -- +(c4);
  \draw[thick] (P2) ++(r2) +(c1) -- +(c2) -- +(c3) -- +(c4) -- cycle;
  \draw[thick,->] (Nempty) -- (N2);
  
  \node[draw,circle,inner sep=0pt,minimum size=1.4*\radius,name=N3] at (P3) {};
  \draw[thick] (P3) +(c1) -- +(c2) -- +(c3) -- +(c4) -- +(c1) +(c1) -- +(c3) +(c2) -- +(c4);
  \draw[thick] (P3) ++(r3) +(c1) -- +(c2) -- +(c3) -- +(c4) -- cycle;
  \draw[thick,->] (Nempty) -- (N3);
  
   \node[draw,circle,inner sep=0pt,minimum size=1.4*\radius,name=N4] at (P4) {};
  \draw[thick] (P4) +(c1) -- +(c2) -- +(c3) -- +(c4) -- +(c1) +(c1) -- +(c3) +(c2) -- +(c4);
  \draw[thick] (P4) ++(r4) +(c1) -- +(c2) -- +(c3) -- +(c4) -- cycle;
  \draw[thick,->] (Nempty) -- (N4);
  
   \node[draw,circle,inner sep=0pt,minimum size=1.4*\radius,name=N5] at (P5) {};
  \draw[thick] (P5) +(c1) -- +(c2) -- +(c3) -- +(c4) -- +(c1) +(c1) -- +(c3) +(c2) -- +(c4);
  \draw[thick] (P5) ++(r5) +(c1) -- +(c2) -- +(c3) -- +(c4) -- cycle;
  \draw[thick,->] (Nempty) -- (N5);
  
   \node[draw,circle,inner sep=0pt,minimum size=1.4*\radius,name=N6] at (P6) {};
  \draw[thick] (P6) +(c1) -- +(c2) -- +(c3) -- +(c4) -- +(c1) +(c1) -- +(c3) +(c2) -- +(c4);
  \draw[thick] (P6) ++(r6) +(c1) -- +(c2) -- +(c3) -- +(c4) -- cycle;
  \draw[thick,->] (Nempty) -- (N6);
  
  \node[draw,circle,inner sep=0pt,minimum size=1.4*\radius,name=N12] at (P12) {};
  \draw[thick] (P12) +(c1) -- +(c2) -- +(c3) -- +(c4) -- +(c1) +(c1) -- +(c3) +(c2) -- +(c4);
  \draw[thick] (P12) ++(r1) +(c1) -- +(c2) -- +(c3) -- +(c4) -- cycle;
  \draw[thick] (P12) ++(r2) +(c1) -- +(c2) -- +(c3) -- +(c4) -- cycle;
  \draw[thick,->] (N1) -- (N12);
  \draw[thick,->] (N2) -- (N12);
  
   \node[draw,circle,inner sep=0pt,minimum size=1.4*\radius,name=N15] at (P15) {};
  \draw[thick] (P15) +(c1) -- +(c2) -- +(c3) -- +(c4) -- +(c1) +(c1) -- +(c3) +(c2) -- +(c4);
  \draw[thick] (P15) ++(r1) +(c1) -- +(c2) -- +(c3) -- +(c4) -- cycle;
  \draw[thick] (P15) ++(r5) +(c1) -- +(c2) -- +(c3) -- +(c4) -- cycle;
  \draw[thick,->] (N1) -- (N15);
  \draw[thick,->] (N5) -- (N15);
  
   \node[draw,circle,inner sep=0pt,minimum size=1.4*\radius,name=N25] at (P25) {};
  \draw[thick] (P25) +(c1) -- +(c2) -- +(c3) -- +(c4) -- +(c1) +(c1) -- +(c3) +(c2) -- +(c4);
  \draw[thick] (P25) ++(r2) +(c1) -- +(c2) -- +(c3) -- +(c4) -- cycle;
  \draw[thick] (P25) ++(r5) +(c1) -- +(c2) -- +(c3) -- +(c4) -- cycle;
  \draw[thick,->] (N2) -- (N25);
  \draw[thick,->] (N5) -- (N25);
  
   \node[draw,circle,inner sep=0pt,minimum size=1.4*\radius,name=N26] at (P26) {};
  \draw[thick] (P26) +(c1) -- +(c2) -- +(c3) -- +(c4) -- +(c1) +(c1) -- +(c3) +(c2) -- +(c4);
  \draw[thick] (P26) ++(r2) +(c1) -- +(c2) -- +(c3) -- +(c4) -- cycle;
  \draw[thick] (P26) ++(r6) +(c1) -- +(c2) -- +(c3) -- +(c4) -- cycle;
  \draw[thick,->] (N2) -- (N26);
  \draw[thick,->] (N6) -- (N26);
  
  \node[draw,circle,inner sep=0pt,minimum size=1.4*\radius,name=N56] at (P56) {};
  \draw[thick] (P56) +(c1) -- +(c2) -- +(c3) -- +(c4) -- +(c1) +(c1) -- +(c3) +(c2) -- +(c4);
  \draw[thick] (P56) ++(r5) +(c1) -- +(c2) -- +(c3) -- +(c4) -- cycle;
  \draw[thick] (P56) ++(r6) +(c1) -- +(c2) -- +(c3) -- +(c4) -- cycle;
  \draw[thick,->] (N5) -- (N56);
  \draw[thick,->] (N6) -- (N56);
  
  \node[draw,circle,inner sep=0pt,minimum size=1.4*\radius,name=N125] at (P125) {};
  \draw[thick] (P125) +(c1) -- +(c2) -- +(c3) -- +(c4) -- +(c1) +(c1) -- +(c3) +(c2) -- +(c4);
  \draw[thick] (P125) ++(r1) +(c1) -- +(c2) -- +(c3) -- +(c4) -- cycle;
  \draw[thick] (P125) ++(r2) +(c1) -- +(c2) -- +(c3) -- +(c4) -- cycle;
   \draw[thick] (P125) ++(r5) +(c1) -- +(c2) -- +(c3) -- +(c4) -- cycle;
  \draw[thick,->] (N12) -- (N125);
  \draw[thick,->] (N15) -- (N125);
  \draw[thick,->] (N25) -- (N125);
  
   \node[draw,circle,inner sep=0pt,minimum size=1.4*\radius,name=N256] at (P256) {};
  \draw[thick] (P256) +(c1) -- +(c2) -- +(c3) -- +(c4) -- +(c1) +(c1) -- +(c3) +(c2) -- +(c4);
  \draw[thick] (P256) ++(r2) +(c1) -- +(c2) -- +(c3) -- +(c4) -- cycle;
  \draw[thick] (P256) ++(r5) +(c1) -- +(c2) -- +(c3) -- +(c4) -- cycle;
   \draw[thick] (P256) ++(r6) +(c1) -- +(c2) -- +(c3) -- +(c4) -- cycle;
  \draw[thick,->] (N25) -- (N256);
  \draw[thick,->] (N26) -- (N256);
  \draw[thick,->] (N56) -- (N256);
  
\end{tikzpicture}
\end{center}

        \caption{Sisim data set example: The estimated a posteriori most probable model in the second 
          cluster of models.}\label{fig:dag_sisim1}
\end{figure}
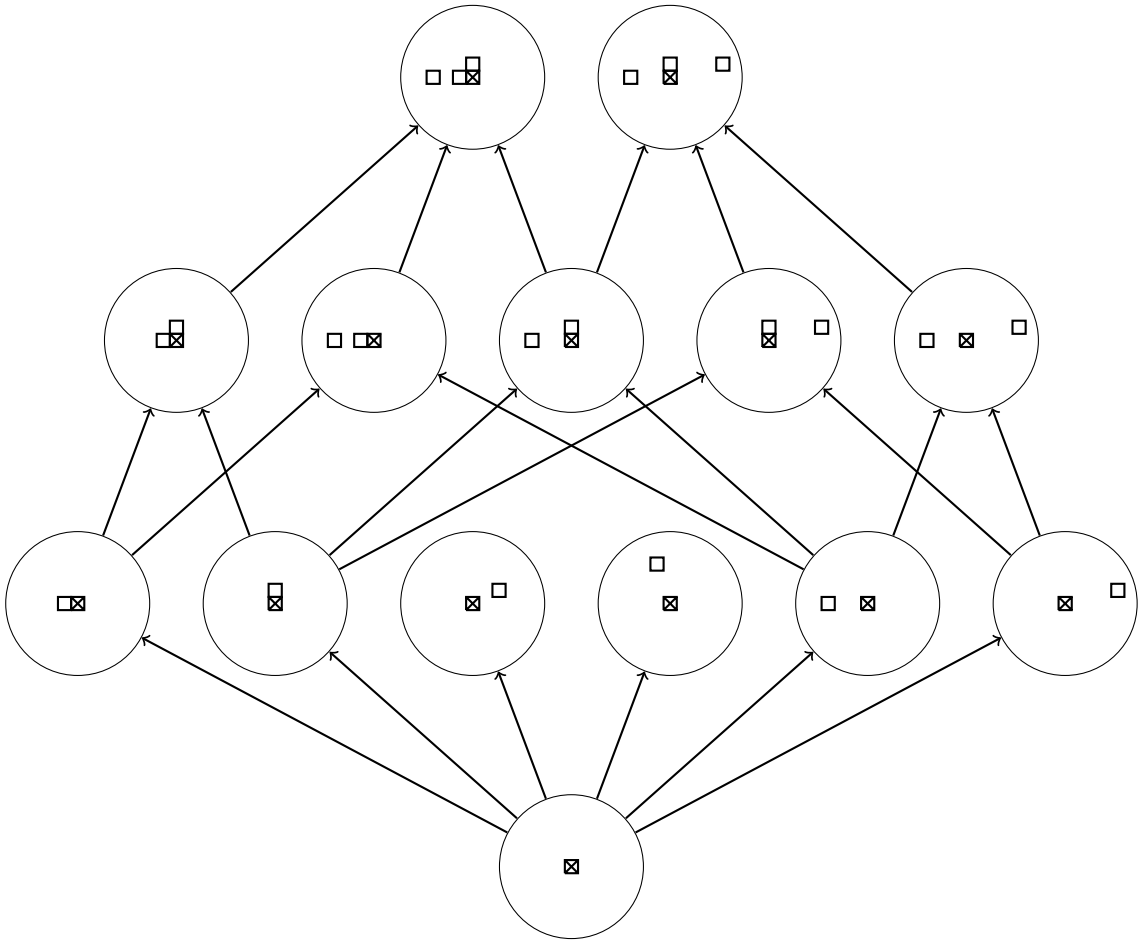

\begin{figure}
                	\vspace*{-0.6cm}
                \begin{center}
                \includegraphics[width=0.7\linewidth,height=0.35\linewidth]{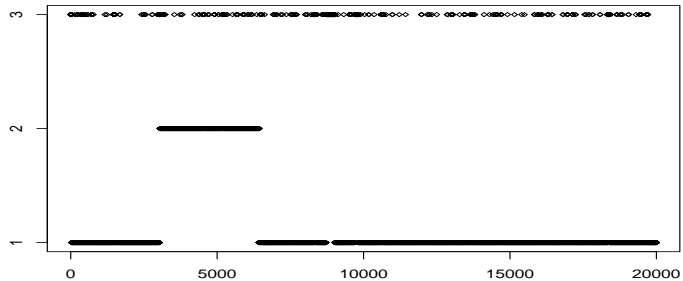}
                \end{center}
                	\vspace*{-1.2cm}
                \caption{Sisim data set example: Trace plot visited clusters for the subsampled models.
                  On the $y$-axis $1$ and $2$ represent the first and second clusters of models found,
                  respectively, and $3$ represents all remaining models.}
                \label{fig:modelIndex}
\end{figure}

In the cancer mortality data set example, essentially all of the posterior probability mass was concentrated in a few
models. In the sisim data set example, the probability mass is spread out on a very large number of models. In 
particular, as also discussed above, the most probable model has a posterior probability estimated to be as 
low as $0.13802$. Using the 
simulated models to understand the posterior model distribution is then more difficult. As a first step in describing
the posterior model distribution, our focus here is on whether it has one or several modes. To do this we 
first need to define what we should mean by a mode in this complicated model space. We start by defining two models to be 
neighbors if one of them can be obtained from the other by including one extra interaction. Thus, our proposal
distribution in Section \ref{sec:propose interactions}, proposing to change the set of active interactions, is 
always generating a potential new model that is a neighbor of the current model. To explore whether we have several modes
in our posterior distribution, we first subsample the simulated Markov chain, keeping a realization every
$50$ iterations after the burn-in period. This leave us with 
$20\,000$ realizations.  From these we first find the most frequent model, 
visualized in Figure \ref{fig:dag_sisim}, and then all neighbor models to this most probable model, all 
neighbor models to the neighbors, and so on until the process stops. The sum of the estimated 
posterior probabilities of the models in the resulting cluster of models is $0.80755$,
giving a clear indication that the posterior model distribution have more than one mode. To find a second 
mode we limit the attention to the simulated models that was not included in the first cluster of models and repeat
the process. Thus, we first find the a posteriori most probable model not included in the first model cluster. This
model is shown in Figure \ref{fig:dag_sisim1}. 
Then we find all neighbors of this model,
all neighbors of the neighbors and so on. The estimated posterior probability in this second cluster of models is
$0.146563$. Thus, these two first clusters contain more than $95\%$ of the simulated models,
and we therefore choose not to search for a third cluster. Knowing that we have two important clusters or 
modes it is natural to reconsider 
the convergence and mixing properties of our Markov chain. Figure \ref{fig:modelIndex}  
shows a trace plot of the visited clusters for the subsampled models, where $1$  and $2$ on the $y$-axis represent the 
first and second clusters found, respectively, and $3$ represent all remaining models. We then see that the second cluster is in fact 
visited only once, giving a clear indication of poor mixing. We should thereby not trust the estimated probabilities
for the two clusters given above, but that the chain is first moving from the first cluster to the second and thereafter
back again clearly shows that both of them have a significant posterior probability mass.

\section{\label{sec:closing remarks}Closing remarks}
In this article we propose a prior distribution for a binary Markov mesh model. The specification of a Markov mesh model has
three parts. First a sequential neighborhood is specified, next the parametric form of the conditional distributions
is defined, and finally we assign values to the parameters. We formulate prior distributions for all these three 
parts. To favor parsimonious models, our prior in particular assigns positive prior probabilities for some interaction
parameters to be exactly zero. A corresponding prior formulation has previously been proposed for Markov random fields
\citep{art157}. The advantage of using it for a Markov mesh model is that an explicit and easy to compute expression is
available for the resulting posterior distribution, whereas the posterior based on a Markov random field will
include the computationally intractable normalizing constant of the Markov random field.

To sample from the resulting posterior distribution when conditioning on an observed scene, 
we adopt the RJMCMC setup. We propose an algorithm based on the 
combination of two proposal distributions, a Gibbs proposal for the parameter values and a reversible jump proposal 
changing the sequential neighborhood and parametric form of the conditional distributions.

To explore the performance of the specified prior distribution and the corresponding RJMCMC posterior simulation 
algorithm, we consider two scenes. The first is an observed cancer mortality map data with small spatial coupling
between neighboring nodes. For this scene the RJMCMC algorithm converges quickly and has good mixing properties. 
Most of the posterior mass ends up in models with only two nodes in the sequential neighborhood. 
The second scene we tried is a frequently used
scene in the geostatistical community. It has more spatial continuity than the first scene. The 
convergence of the RJMCMC algorithm becomes much slower when conditioning on this scene. In particular the posterior
seems to have at least two modes and the mixing between the modes is slow. Our simulation results
indicate that the a posteriori most likely model has six nodes in the sequential neighborhood and the conditional
distributions has a parametric form with as much as twelve parameters. This shows that the specified prior is 
flexible in that the model complexity favored by the corresponding posterior adapts to the the complexity of 
the scene conditioned on.

In this article we have focused on binary Markov mesh models and thereby binary scenes. Our strategy for prior
specification and posterior simulation, however, can easily be extended to a situation with more than two colors. 
The main challenge in this generalization does not lie in the specification of the prior, but is computational
in that one should expect the convergence and mixing of a corresponding RJMCMC algorithm to be slower for a 
multi-color model. A direction for future research is therefore to improve the proposal 
distributions to obtain better convergence and mixing for the RJMCMC algorithm, both in the binary and 
multi-color cases. In particular we think a promising direction here is to define an MCMC algorithm where 
several Metropolis--Hastings proposals can be generated in parallel and where the proposals may have 
added and removed more than just one interaction relative to the current model.

\bibliographystyle{jasa}
\bibliography{mybib}

\appendix

\section{\label{app:logConcave}Log-concavity of the full conditional for $\alpha$}
In this appendix we prove that the full conditional $f(\alpha|\tau,\Lambda,\{\theta(\lambda)+\alpha\Delta(\lambda):\lambda\in\Lambda\},x)$
defined in \eqref{eq:fullAlpha} is log-concave, so that we can use adaptive rejection sampling to generate samples from it. Defining
$g(\alpha) = \ln\left[ f(\alpha|\tau,\Lambda,\{\theta(\lambda)+\alpha\Delta(\lambda):\lambda\in\Lambda\},x)\right]$
and using \eqref{eq:fullAlpha} we have
\begin{equation}
\begin{split}
g(\alpha) &= \ln \left[  f(\{\theta(\lambda) + \alpha\Delta(\lambda):\lambda\in\Lambda\}|\tau,\Lambda)\right]\\
&+ \ln \left[ f(x|\tau,\Lambda,\{ \theta(\lambda)+\alpha\Delta(\lambda):\lambda\in\Lambda\})\right]  + C,
\end{split}
\end{equation}
where $C$ is the logarithm of the normalizing constant in \eqref{eq:fullAlpha}. Inserting expressions for 
the prior and likelihood in \eqref{eq:prior_Theta} and \eqref{eq:mmm}, respectively, we get
\begin{equation}
\begin{split}
g(\alpha) &=  \sum_{\lambda\in\Lambda} \left[ c(\lambda) +\theta(\lambda)+\alpha\Delta(\lambda) 
-2\ln\left(1+e^{\theta(\lambda)+\alpha\Delta(\lambda)}\right) - 
\frac{(\theta(\lambda)+\alpha\Delta(\lambda))^2}{2\sigma^2}\right] \\
&+ \sum_{v\in\chi} \left[ x_v (\theta(\xi(x)\cap(\tau\oplus v)) + \alpha \Delta(\xi(x)\cap(\tau\oplus v))) \right. \\
&\hspace*{1.1cm}- \left.\ln\left(1+e^{ \theta(\xi(x)\cap(\tau\oplus v)) + \alpha \Delta(\xi(x)\cap(\tau\oplus v))}\right)\right]
+ C.
\end{split}
\end{equation}
Grouping terms of the same functional form, we get
\begin{equation}\label{eq:g}
\begin{split}
g(\alpha) &= C_0 + C_1 \alpha - \frac{1}{2\sigma^2}\sum_{\lambda\in\Lambda} (\theta(\lambda)+\alpha\Delta(\lambda))^2 
- 2\sum_{\lambda\in\Lambda} \ln\left(1+e^{\theta(\lambda)+\alpha\Delta(\lambda)}\right)\\
&- \sum_{v\in\chi} \ln\left(1+e^{ \theta(\xi(x)\cap(\tau\oplus v)) + \alpha \Delta(\xi(x)\cap(\tau\oplus v))}\right),
\end{split}
\end{equation}
where 
\begin{equation}
C_0 = C + \sum_{\lambda\in\Lambda}\theta(\lambda) +\sum_{v\in\chi}x_v\theta(\xi(x)\cap (\tau\oplus v)) \mbox{ and }
C_1 = \sum_{\lambda\in\Lambda}\Delta(\lambda) + \sum_{v\in\chi} \Delta(\xi(x)\cap (\tau\oplus v))
\end{equation}
are constants as a function of $\alpha$. The second derivative of the constant and linear terms in 
\eqref{eq:g} are of course zero. Since the coefficients of the quadratic terms are all negative, the second 
derivative of all of these are less or equal to zero, and unless $\Delta(\lambda)$ equals zero for 
all $\lambda\in\Lambda$ the second derivative of the sum of these terms is even strictly less than zero. 
The remaining terms in \eqref{eq:g}
all have the same functional form 
as a function of $\alpha$, namely
\begin{equation}
h(\alpha) = - a \ln \left(1+e^{b+c\alpha}\right),
\end{equation}
a term in the sum over $\lambda\in\Lambda$ has $a=2$, $b=\theta(\lambda)$ and $c=\Delta(\lambda)$, whereas a term in the 
sum over $v\in\chi$ has $a=1$, $b=\theta(\xi(x)\cap (\tau\oplus v))$ and $c=\Delta(\xi(x)\cap (\tau\oplus v))$. To prove 
that the second derivative of all of these terms are negative, it is thereby sufficient to show that $h^{\prime\prime}(\alpha) < 0$
for all $a>0$ and $\alpha,b,c\in\mathbb{R}$. Simple differentiation gives
\begin{equation}
h^{\prime\prime}(\alpha) = - \frac{ac^2 e^{b+c\alpha}}{\left(1+e^{b+c\alpha}\right)^2}.
\end{equation}
Thus, $h^{\prime\prime}(\alpha)<0$ for all $a>0$ and $\alpha,b,c\in\mathbb{R}$, and thereby $g(\alpha)$ is concave and the full conditional 
$f(\alpha|\tau,\Lambda,\{\theta(\lambda)+\alpha\Delta(\lambda):\lambda\in\Lambda\},x)$ is log-concave.

\section{\label{app:determinant}Jacobian determinant for the proposal in 
Section \ref{sec:proposeRemove}}

The Jacobi determinant for our removing an active interaction from $\Lambda$ proposal is $\det(A)$, where $A$ is defined by
\eqref{eq:1-1}. The exact form of the matrix $A$ depends on how we define the vectors $\theta$ and $\theta^\star$ used in \eqref{eq:1-1}.
The vector $\theta$ should contain the set of current parameters $\{\theta(\lambda):\lambda\in\Lambda\}$, but so far we 
have not specified what order to use when arranging this set of parameters into the vector $\theta$. Correspondingly,
we have not specified what order to use when arranging the set of potential new parameters 
$\{\theta^\star(\lambda):\lambda\in\Lambda^\star\}$ into the vector $\theta^\star$. However, even if the elements of 
$A$ depends on how we construct $\theta$ and $\theta^\star$, the absolute value of the determinant of $A$ is the same 
for all arrangements of $\theta$ and $\theta^\star$.
To find $\det(A)$ we arrange the vector $\theta$ so that parameters corresponding to lower order interactions comes first. 
The first element of the vector $\theta$ is thereby $\theta(\emptyset)$, thereafter follows parameters corresponding to 
the first order interactions $\{\theta(\lambda):\lambda\in\Lambda, |\lambda|=1\}$ (in an arbitrary order), then all parameters
corresponding to second order interactions $\{\theta(\lambda):\lambda\in\Lambda, |\lambda|=2\}$ (again in an arbitrary order), 
and so on. We arrange $\theta^\star$ correspondingly, parameters corresponding to lower order interactions comes first.

As also touched on in Section \ref{sec:proposeRemove}, the transformation in \eqref{eq:1-1} can be done in three steps.
First $\theta$ is transformed into a vector $\beta$ of the corresponding current interaction parameters 
$\{\beta(\lambda):\lambda\in\Lambda\}$. This relation is given in \eqref{eq:m2} and is in particular linear so we can 
write
\begin{equation}
\beta = A_1\theta.
\end{equation}
Arranging also the vector $\beta$ so that lower order interactions comes first, it is easy to see from 
\eqref{eq:m2} that $A_1$ is a lower triangular matrix with all diagonal elements equal to one. Thus $\det(A_1)=1$.
The second step in the transformation is to use \eqref{eq:proposeRemove} to define a vector $\beta^\star$ containing
the set of potential new interaction parameters $\{\beta^\star(\lambda):\lambda\in\Lambda^\star\}$. As the proposal 
is to remove an interaction, the number of elements in $\beta^\star$ is one less than the number of elements in 
$\beta$. To obtain a one-to-one relation as required in the reversible jump setup, we
include the current value $\beta(\lambda^\star)$ in a vector together with $\beta^\star$. We let $\beta(\lambda^\star)$ be
the last element in the vector and we arrange also the vector $\beta^\star$ so that lower order interaction 
parameters come first. As the relation in \eqref{eq:proposeRemove} is linear we can then write 
\begin{equation}\label{eq:A2}
\left[\begin{array}{c}\beta^\star \\ \beta(\lambda^\star)\end{array}\right] = A_2\beta,
\end{equation}
where the elements of the square matrix $A_2$ is defined by \eqref{eq:proposeRemove}. To find the determinant of $A_2$, let 
$r$ denote the number of elements in $\beta$ before $\beta(\lambda^\star)$, so that element number $r+1$ in $\beta$ 
is $\beta(\lambda^\star)$. From \eqref{eq:proposeRemove} it then follows that $A_2$ has the block structure,
\begin{equation}
A_2 = \left[ \begin{array}{cc}
I_{r\times r} & A_2^{12} \\
0_{(|\Lambda|-r)\times r} & A_2^{22} \end{array} \right],
\end{equation}
where $I_{r\times r}$ is the $r\times r$ identity matrix, $A_2^{12}$ is an $r\times (|\Lambda|-r)$ matrix, $0_{(|\Lambda|-r)\times r}$
is a $(|\Lambda|-r)\times r$ matrix of only zeros, and $A_2^{22}$ is the $(|\Lambda|-r)\times (|\Lambda|-r)$ permutation matrix
where the elements $(i,i+1)$ for $i=1,\ldots,|\Lambda|-r-1$ and $(|\Lambda|-r,1)$ equals one and all other elements are 
zero. Thereby we have $\det(A_2) = \det(I_{r\times r}) \cdot \det(A_2^{22}) = \det(A_2^{22})$, and as $A_2^{22}$ is a
permutation matrix its determinant is plus or minus one. Thus, $|\det(A_2)|=1$. The third step in 
the transformation from $\theta$ to $\theta^\star$ is to use \eqref{eq:m1} to transform the vector of potential new 
interaction parameters, $\beta^\star$, to a corresponding vector $\theta^\star$ of potential new parameter values. As the relation in 
\eqref{eq:m1} is also linear, we can write
\begin{equation}
\left[\begin{array}{c} \theta^\star \\ \beta(\lambda^\star) \end{array} \right] = A_3
\left[\begin{array}{c} \beta^\star \\ \beta(\lambda^\star) \end{array}\right],
\end{equation}
where the elements of the matrix $A_3$ is defined by \eqref{eq:m1}. Recalling
that we have arranged the elements in both $\theta^\star$ and $\beta^\star$ so 
that parameters corresponding to lower order interactions come first, it 
is easy to see from \eqref{eq:m1} that $A_3$ is an upper triangular
matrix with all diagonal elements equal to one. Thus $\det(A_3)=1$. Setting the 
three steps in the transformation together we have $A=A_1 A_2 A_3$ and thereby $|\det(A)| = |\det(A_1)| 
\cdot |\det(A_2)|\cdot |\det(A_3)| = 1$.

\end{document}